\documentclass[aps,prd,longbibliography,nofootinbib]{revtex4-2}

\usepackage{graphicx}
\usepackage{dcolumn}
\usepackage{amsmath,amssymb,mathtools,epsfig}
\usepackage{paralist}
\usepackage{siunitx}
\usepackage{multirow}
\usepackage{suffix}
\usepackage{tabularx}
\usepackage{tabulary} 
\usepackage{hyperref}
\usepackage{comment}
\hypersetup{colorlinks=true,linkcolor=magenta,citecolor=blue,urlcolor=blue}

\usepackage{amsthm}
\usepackage{enumitem}
\setlist[itemize]{leftmargin=1.4em,topsep=2pt,itemsep=2pt,parsep=0pt}
\setlist[enumerate]{leftmargin=1.6em,topsep=2pt,itemsep=2pt,parsep=0pt}

\newtheorem{theorem}{Theorem}
\newtheorem*{remark}{Remark}

\newcommand{\EigenNumber}{E \equiv \mu_{\mathrm{eff}}\,L/\ln s}
\newcommand{\EigenGate}{\ensuremath{\EigenNumber < 1}}
\newcommand{\EigenFail}{\ensuremath{\EigenNumber \ge 1}}

\usepackage[final]{microtype}
\usepackage{siunitx}      
\usepackage{mathtools}
\usepackage{amsthm}
\usepackage[nameinlink,capitalise]{cleveref} 
\usepackage{booktabs}
\usepackage{algorithm}
\usepackage{algpseudocode} 
\usepackage{tikz}          

\newcommand{\mPhi}{m_{\Phi}}                        
\newcommand{\mE}{m_{E}}                             
\newcommand{\mChi}{m_{\chi}}                        
\newcommand{\mG}{m_{G}}                             
\newcommand{\mC}{m_{C}}                             
\newcommand{\mcyc}[1]{m^{\{\mathrm{#1}\}}_{\mathrm{cyc}}} 

\usepackage{tikz}
\usepackage{pgfplots}
\pgfplotsset{compat=newest} 

\usepackage{tikz}
\usetikzlibrary{positioning,arrows.meta}

\usepackage{tikz}
\usetikzlibrary{arrows.meta,positioning,fit,calc}
\usepackage{pgfplots}
\pgfplotsset{compat=1.18}
\usepackage{siunitx}
\usepackage{booktabs}

\definecolor{Ink}{RGB}{30,30,30}
\definecolor{Accent}{RGB}{0,102,204} 

\tikzset{
  box/.style={draw=Ink, rounded corners, thick, align=left,
              inner sep=6pt, fill=black!3},
  arrow/.style={-{Stealth[length=3.5mm,width=2.4mm]}, very thick, draw=Ink},
}

\pgfplotsset{
  myplot/.style={
    width=\linewidth, height=0.5\linewidth,
    line width=1pt, tick style={line width=0.6pt, draw=black!70},
    axis line style={draw=black!70},
    grid=both, minor grid style={line width=.1pt, draw=black!10},
    major grid style={line width=.2pt, draw=black!15},
    label style={font=\small}, tick label style={font=\small},
    legend style={font=\small, draw=none, fill=white, fill opacity=0.85,
                  text opacity=1, at={(0.98,0.02)}, anchor=south east},
    every axis plot/.append style={Ink},
    cycle list={{Ink, solid}, {Accent, thick}, {black!50, dashed}, {black!70, dotted}}
  }
}

\begin{document}

\title{Quantitative Nonequilibrium Pathway from Fundamental Physics \\ to the Emergence and Persistence of Exoplanetary Biospheres}

\author{Slava G. Turyshev}
\affiliation{Jet Propulsion Laboratory, California Institute of Technology,\\
4800 Oak Grove Drive, Pasadena, CA 91109-0899, USA}

\date{\today}

\begin{abstract}
We present a physics-based framework that runs from fundamental interactions and constants to biospheres, using a sequence of quantitative nonequilibrium thresholds (``gates''). Each gate is an inequality in measurable variables---free-energy flux, reaction--transport rates, replication fidelity, coding capacity, ecological closure, and climate feedback gains. Crucially, the gate vector is anchored in fundamental physics: dimensionless constants (e.g., the fine-structure constant $\alpha$ and mass ratio $m_e/m_p$), nuclear resonance placements (e.g., the $^{12}$C Hoyle state), and statistical mechanics (Landauer’s bound $k_BT\ln 2$) fix the energetic, kinetic, and information-theoretic margins that propagate through the gates. This anchoring lets us propagate sensitivities of the constants into biosphere-level metrics (net primary productivity (NPP), cycle-closure ratios, and climate feedback gain), yielding an end-to-end map from constants to biospheres. The framework is predictive: it yields testable inequalities, margin rankings, and population-level correlations between stellar/planetary boundary conditions/biosphere feasibility. It does not claim point-predictions of life prevalence; rather, it specifies which gate margins are observable-bounded versus prior-dominated under explicitly stated chemistry/solvent families and forward models. Darwinian dynamics (heritable variation under selection) appears mid-pipeline; the end of the pipeline is a planet-scale biosphere capable of sustaining positive NPP and closing elemental cycles over geologic time. Questions of prevalence are secondary; our primary objective was to establish a constructive physics$\to$chemistry$\to$biology$\to$genetics$\to$ecosystems pipeline with testable margins and observables. As a result, we recast abiogenesis and biosphere persistence as a gate vector of falsifiable inequalities and map their margins to exoplanet observables, turning the problem into a phase diagram with explicit, testable slack.

\end{abstract}

\maketitle

\tableofcontents


\section{Introduction}\label{sec:intro}

The Universe is governed by the same fundamental physics everywhere \cite{TuryshevFunPhys2025}; consequently, chemistry and information processing obey universal constraints. This observation does more than motivate a search for life beyond Earth---it restricts how life can be realized. Electronic structure fixes covalent and hydrogen-bond energies; statistical mechanics fixes the minimal work for information erasure; astrophysics fixes irradiation, hazards, and volatile budgets. At $T=300\,\mathrm{K}$ one has $k_{\mathrm B}T\simeq 0.026\,\mathrm{eV}$ and $k_{\mathrm B}T\ln 2\simeq 1.73\,\mathrm{kJ\,mol^{-1}}$ per erased bit, while typical C--C/C--H/O--H/P--O bonds have $E_b\!\sim\!3.5$--$5\,\mathrm{eV}$ and hydrogen bonds $E_{\mathrm{HB}}\!\gtrsim\!0.1\,\mathrm{eV}$. These numerical separations already permit template pairing, amphiphile assembly, and energy-budgeted proofreading across broad solvent/temperature ranges and anchor the quantitative thresholds used here.

Our objective is constructive and predictive: to derive, from fundamental physics through planetary context, a compact sequence of nonequilibrium thresholds (\emph{gates}) that a driven chemical system must satisfy to enter first the Darwinian regime (heritable variation under selection) and ultimately the biosphere regime (sustained net primary productivity, closed elemental cycles, stabilized climate/chemistry). Each gate is an inequality in measurable variables; taken together they define a phase diagram for life in the space of environmental controls. Where the inequalities hold with margin on relevant timescales, chemistry becomes biological and, with additional constraints, biospheric; where any one fails, abiogenesis stalls or biospheres do not persist.

The gates are coupled by conservation laws and information--thermodynamic trade-offs. Surplus work reduces copying error via kinetic proofreading; spatial localization raises effective rate constants and network gain while suppressing leakage; coding slack trades off against dissipation and empirically measured translation confusion matrices. These couplings generate sharp transitions---eigenvalue sign changes in linearized dynamics or percolation of autocatalytic sets---rendering the framework falsifiable in laboratory reactors and interpretable via exoplanet observables.

Our emphasis is biospheres: Darwinian dynamics is necessary but not sufficient; a biosphere additionally requires ecological closure and long-lived feedback stability. The form of the gate inequalities is solvent-agnostic: changing solvent or temperature shifts numerical values (through $E_b/k_{\mathrm B}T$, diffusivities, permeabilities, and error rates), not the structure of the constraints. Consequently, while pigments, alphabets, or compartment chemistries may vary across worlds, the gating physics---and thus the classes of biospheres one should expect---are strongly constrained.

Because the aim is predictive, we map gate \emph{margins}---how far each inequality is from failure---to observables: stellar spectral energy distributions and activity histories constrain usable work and volatile retention; atmospheric redox disequilibria and seasonal variability bear on productivity and closed cycles \cite{Olson2018}; surface reflectance and polarimetry inform pigments and compartment-forming chemistry; mineralogical context informs catalytic density and network gain. Formal definitions and symbols are introduced later; here we refer only to their physical roles.

The gates trade completeness for testability: each inequality is conservative, local in its inputs, and designed to be checked in reactors or from remote observables. We do not attempt full ecosystem simulation; instead, we require margin sufficiency gate by gate and report how uncertainties map to failure modes and falsifiers.

In making explicit how fundamental physics constrains and defines viable biospheres and their observables, our objectives are fourfold: (i) to construct a constants$\to$gates framework that casts abiogenesis and biosphere persistence as inequalities with explicit, dimensionless margins; (ii) to derive an information--energetics upper bound on net primary productivity (NPP) that subtracts the minimal work per bit and an explicit proofreading cost; (iii) to formulate a translation feasibility condition as a rate--capacity test with finite--blocklength corrections; and (iv) to develop a retrieval procedure---log-sensitivity accounting, uncertainty propagation, and a defined set of falsification tests---that maps observations to gate margins. (All quantities and notation are introduced in the sections that follow.)

Our overarching objective is to develop a gate-based framework built on the following components:
{}
\begin{itemize}
\item \textit{Gate-vector formalism:} a minimal set of quantitative, testable necessary conditions
for biosphere persistence, expressed as margins that can be compared across environments.
\item \textit{End-to-end coupling:} explicit propagation of stellar/planetary/chemical parameters and
fundamental constants into gate margins and bottleneck ranking.
\item \textit{Information as a physical constraint:} incorporation of Landauer/Shannon limits,
proofreading dissipation, and finite-blocklength coding into ecological productivity and
translation feasibility.
\item \textit{Exoplanet-facing diagnostics:} mapping of each gate to observables and falsifiers
to support empirical prioritization.
\end{itemize}

This paper is a \emph{framework/perspective} that organizes necessary conditions for sustained biospheres
into a small set of quantitative ``gates'' and dimensionless margins. We do \emph{not} attempt:
(i) a comprehensive review of abiogenesis mechanisms, (ii) a complete survey of all proposed life-detection
biosignatures, or (iii) end-to-end exoplanet retrievals of all gate parameters. Instead, we provide (a) a unified gate vector with portable margins, (b) quantitative worked gates (including the coding feasibility Gate~10 and an information-thermodynamic productivity ceiling Gate~11),
and (c) sensitivity propagation from physical anchors to biosphere observables.

This paper is a framework/perspective that organizes necessary conditions for sustained biospheres
into a compact set of quantitative ``gates'' and dimensionless margins. Section~\ref{sec:antecedents}
situates the gate construction relative to prior work; here we proceed directly to the constructive program:
we define the gate vector, make the couplings explicit, and map gate margins to laboratory falsifiers and
exoplanet-facing observables.

The remainder of the paper formalizes this program: Sections~\ref{sec:gate1}--\ref{sec:gate11-12} derive a sequence of falsifiable gate inequalities that connect fundamental physics to biosphere longevity via the constants$\to$gates composition in (\ref{eq:constants_to_biosphere}) and the anchors summarized in Table~\ref{tab:anchors}. Gate~1 (Sec.~\ref{sec:gate1}) establishes chemical possibility from bonding physics and information thermodynamics; Gates~2--4 (Secs.~\ref{sec:gate2}--\ref{sec:gate4}) set cosmological and planetary boundary conditions and compute usable work; Gates~5--8 (Secs.~\ref{sec:gate5}--\ref{sec:gate8}) quantify driven network chemistry, fidelity, autocatalytic closure, and compartmental transport; Gate~9 (Sec.~\ref{sec:gate9}) gives the eigenvalue criterion for Darwinian dynamics; Gate~10 (Sec.~\ref{sec:gate10}) treats coding feasibility as a noisy channel; Gates~11--12 (Sec.~\ref{sec:gate11-12}) formulate ecological persistence and climate stabilization, including the NPP bound in (\ref{eq:PhiNPP}). Section~\ref{sec:obsmap} connects gate margins to exoplanet observables and develops biosphere archetypes by stellar type and solvent. Section~\ref{sec:methods} describes inference of margins from observations and laboratory constraints, and reports log-sensitivity accounting via Eq.~(\ref{eq:sensitivity}). Section~\ref{sec:validation} sets out predictions and decisive falsifiers; Section~\ref{sec:discussion} concludes with implications for target selection and for the universals versus contingencies of life in the Universe.

\subsection{Foundations and Relation to Prior Work}\label{sec:antecedents}

Our gate-based program---which casts abiogenesis and biosphere persistence as a vector of quantitative, falsifiable inequalities with margins that map to observables---sits at the intersection of nonequilibrium thermodynamics, information theory, origins-of-life chemistry, protocell physics, planetary escape and geophysics, and exoplanet biosignatures \cite{Landauer1961,Bennett1982,Shannon1948,CoverThomas2006}. In contrast to purely qualitative narratives, the present framework formalizes pass/fail thresholds (power, kinetic gain, confinement, fidelity, channel capacity, ecological closure, climate gain) and ties them to measurable quantities and astronomical priors. (Definitions of the gate vector, margins, and the gate$\leftrightarrow$observable mapping are given in the main text. Acronyms are expanded at first use and collected for convenience in Table~\ref{tab:abbrev}.)

\paragraph{Nonequilibrium information and the cost of accuracy:}
Foundational results link information processing to dissipation: Landauer’s bound for bit erasure and Bennett’s thermodynamics of computation \cite{Landauer1961,Bennett1982}. In biochemical copying and decoding, kinetic proofreading (Hopfield, Ninio) trades energy for reduced error \cite{Hopfield1974,Ninio1975}. We treat these as explicit budget terms, coupling the power margin to achievable replication and translation fidelities (Gates~4, 6, 10).

\paragraph{Error thresholds and quasispecies:}
Eigen’s quasispecies theory and the associated error threshold relate maintainable genomic length to per-site error and selective advantage \cite{Eigen1971,EigenSchuster1977}. Our Gate~6 adopts this bound in Eigen-number form ${\cal E}=\mu L/\ln s<1$ and adds a kinetic viability constraint tying copying speed to loss, which is crucial in open, leaky reactors and flow environments.

\paragraph{Autocatalytic closure and network amplification:}

From Kauffman’s arguments for reflexively autocatalytic and food-generated (RAF) sets and defined catalytic reaction system (CRS) models, through modern prebiotic syntheses, autocatalysis has been treated as a topological property with percolation-like onset \cite{Kauffman1986,HordijkSteel2004,HordijkSteel2012}. We build on this by requiring not just RAF topology but also kinetic supercriticality over realistic loss ($\lambda_{\mathrm{kin}}>0$) and net exergonic flux---a joint stoichiometric/kinetic/thermodynamic test (Gate~7).
 
\paragraph{Compartments, transport control, and protocells:}
Vesicles and coacervates provide molecular selection, retention, and throughput consistent with early life \cite{Szostak2001,Hanczyc2003,MonnardDeamer2002,AumillerKeating2017,ChenSzostak2004}. Prior work measured permeability, partitioning, and growth--division instabilities; we fold these into a compact Gate~8 triple: confinement ($\tau_{\mathrm{ret}}/\tau_r$), feed throughput, and leak-retention inequalities that explicitly budget maintenance power for gradient leaks.

\paragraph{Code, capacity, and degeneracy under noise:}
Shannon capacity places quantitative limits on reliable symbol transmission; Tlusty and others framed the genetic code as an error-tolerant mapping on a noisy channel with degeneracy absorbing likely confusions \cite{Shannon1948,CoverThomas2006,Tlusty2008}. Gate~10 formalizes a code slack $\chi=C_{\rm cap}/R_{\mathrm{code}}>1$ and couples it to energetic costs of proofreading and to finite-blocklength penalties \cite{Polyanskiy2010}.

\paragraph{Prebiotic feedstocks and geochemical/photochemical drivers:}
Photoredox and geochemical routes to ribonucleotides, lipids, and key intermediates (e.g., cyanide/nitriles) have been demonstrated under plausible ultraviolet (UV)/aqueous conditions \cite{Sutherland2009,Ritson2012,Patel2015,RanjanSasselov2017}. Alkaline hydrothermal vents and serpentinization offer redox work and natural pH gradients \cite{MartinRussell2007,LaneMartin2010}. We recast these as source/sink budgets and activation-power terms (Gate~5) that must exceed hydrolysis, photolysis, dilution, and adsorption at functional thresholds.

\paragraph{Volatile retention, escape, and the radius valley:}
Classical and modern escape formalisms (energy-limited, recombination-limited; Roche effects) quantify integrated loss under evolving stellar extreme ultraviolet (XUV) history irradiation \cite{Watson1981,Erkaev2007,Ribas2005}. Observationally, the small-planet radius valley and survival of thin atmospheres encode mass-loss histories \cite{OwenWu2017,LopezFortney2014,Fulton2017}; exospheric H and metastable He~$1083$\,nm detections provide direct escape diagnostics \cite{VidalMadjar2003,Spake2018,OklopcicHirata2018}. Our Gate~3 bundles these into volatile-retention and surface-access inequalities that precede all downstream chemistry.

\paragraph{Galactic ecology and hazards:}
The Galactic Habitable Zone literature connects metallicity (targets, solids) and transient hazards (SNe, GRBs) to survival windows \cite{Lineweaver2004,Gonzalez2001,PiranJimenez2014}. We adopt these as priors on quiet-time intervals and solids budgets (Gate~2), which propagate to disks, escape, and power availability.

\paragraph{Biosignatures, disequilibria, and seasonal coherence:}
Atmospheric disequilibrium as a life diagnostic has roots in Hitchcock \& Lovelock and has been sharpened in modern retrievals \cite{HitchcockLovelock1967,Sagan1993,KrissansenTotton2018}. Oxygen false positives and context-aware strategies are well developed \cite{Meadows2017}. Surface reflectance edges (``red edge'') and their host-star dependence have been modeled as pigment diagnostics \cite{Seager2005,Kiang2007}. Gate~11 turns these into quantitative NPP and cycle-closure metrics; Gate~12 requires small climate feedback gains consistent with buffered secondary atmospheres and reservoir co-existence \cite{Walker1981,WatsonLovelock1983,CatlingKasting2017}.

\paragraph{Synthesis and departures:}
Across these threads, prior work established individual constraints: fidelity limits, autocatalytic closure, compartment transport, planetary escape, and biosignature observables. Our contribution is to (i) cast each into a falsifiable inequality with a scaleless margin, (ii) make the couplings explicit (e.g., proofreading\,$\leftrightarrow$\,power, localization$\,\leftrightarrow$\,gain, coding\,$\leftrightarrow$\,channel cost), and (iii) map margins to exoplanet observables for inference and target selection. This yields a compact physics$\to$chemistry$\to$biology$\to$biosphere pipeline whose bottlenecks and predictions can be tested in reactors and across exoplanet populations.

\paragraph{Assembly theory and agnostic biosignatures:} Complementary to our gate vector, Assembly Theory formalizes object history and complexity to quantify selection and has been proposed—and demonstrated with mass spectrometry—as an agnostic life‑detection approach \cite{Sharma2023_AT_Nature,Marshall2021_AT_NatComm}.
Whereas Assembly Theory quantifies selection via object histories and molecular assembly indices, the present work derives an environment‑to‑biosphere pipeline of falsifiable inequalities with margins that map directly to exoplanet observables and climate‑scale persistence tests.

\subsection{Scope and modeling assumptions}\label{sec:assumptions}

Several individual gate inequalities correspond to well-established results (e.g.\ Landauer bounds, Eigen error thresholds, Shannon rate--capacity limits, RAF percolation thresholds, and next-generation-matrix criteria). The novelty here is (i) organizing them into a single \emph{gate vector} with dimensionless margins, (ii) expressing each gate in terms of observable-facing order parameters suitable for exoplanet inference, and (iii) providing a consistent sensitivity chain from fundamental constants to biosphere-scale observables.

\begin{table}[t]
\caption{Abbreviations used throughout the paper.}
\label{tab:abbrev}
\centering
\begin{tabular}{ll}
\hline
Acronym & Meaning \\
\hline\hline
AGN & active galactic nucleus \\
ALMA & Atacama Large Millimeter/submillimeter Array \\
BBN & Big Bang nucleosynthesis \\
BRDF & bidirectional reflectance distribution function \\
CIA & collision-induced absorption \\
CME & coronal mass ejection \\
CHNOPS & biogenic elements C, H, N, O, P, S \\
CRN & chemical reaction network \\
CRS & catalytic reaction system \\
CSTR & continuous stirred-tank reactor \\
EUV / FUV & extreme/far ultraviolet \\
GRB & gamma-ray burst \\
GCM & general circulation model \\
HZ & habitable zone \\
ISM & interstellar medium \\
MCMC & Markov chain Monte Carlo \\
NGM & next-generation matrix \\
NIR & near-infrared \\
NPP & net primary productivity \\
NTP & nucleoside triphosphate \\
OLR & outgoing longwave radiation \\
PAR & photosynthetically available radiation \\
RAF & reflexively autocatalytic and food-generated \\
RBE & relative biological effectiveness \\
SED & spectral energy distribution \\
SN/SNe & supernova/supernovae \\
UV  / XUV & ultraviolet  /  X-ray + extreme ultraviolet \\
\hline
\end{tabular}
\end{table}

We use conservative, testable inequalities rather than full ecosystem simulations. Throughout, unless stated otherwise, we assume:
(i) catastrophic-reset hazards are approximated as Poisson processes with an effective kill criterion;
(ii) early-time network growth is assessed by linearization about a dilute state ($x\simeq 0$);
(iii) copying errors are treated as independent and approximately stationary over a replication event;
(iv) translation is approximated as a memoryless channel with an empirically determined confusion matrix (or a $k$-ary symmetric proxy);
(v) climate stability is assessed by linear feedback about a reference state, valid for small perturbations and away from bifurcations. These assumptions define the regime of validity for the gate margins and can be relaxed in system-specific models.

\subsection{Definitions, Order Parameters and Gate Vector}\label{sec:notation}

To avoid notational repetition later, we collect once the symbols used throughout. Let $\theta$ collect environmental and chemical controls: stellar spectrum/flux $F_\star$ and evolving X-ray and extreme-ultraviolet (XUV) irradiation, metallicity $Z$ and mineralogy, volatile inventory, solvent properties $(T,\epsilon)$, transport coefficients $(D,P)$, error kernels for copying/translation, geometry, and turnover times. We concentrate the physics into a set of order parameters (see Table ~\ref{tab:parameters}) that capture power surplus, kinetic amplification, transport/localization, heritable fidelity, and coding capacity. These order parameters are defined as:
{}
\begin{equation}
\Phi \equiv \frac{P_{\rm in}}{P_{\rm maint}+P_{\rm synth}},\qquad \quad
\mathcal{G} \equiv \lambda_{\rm kin}\,\tau_{\rm env},\qquad \quad
\mathcal{E} \equiv \frac{\mu L}{\ln s},\qquad \quad
\mathcal{C} \equiv \frac{\tau_{\rm ret}}{\tau_r},
\qquad \quad
\chi \equiv \frac{C_{\rm cap}}{R_{\rm code}},
\label{eq:orderparams}
\end{equation}
where we introduced the following quantities  
\begin{itemize}
\item 
$\Phi$  is the power margin  with  $P_{\mathrm{in}}$ being usable free-energy input; $P_{\mathrm{maint}}+P_{\mathrm{synth}}$ is the power to maintain gradients and synthesize key species; 

\item $J$ is the kinetic Jacobian of the driven chemical reaction network (CRN) linearized near a dilute state and
$\boldsymbol{\Lambda}\equiv{\rm diag}(\ell_1,\ldots,\ell_{|X|})$ collects effective first-order losses. We use $\lambda_{\rm kin}$ to denote the \emph{dominant linear growth exponent}, which for Metzler/nonnegative matrices this coincides with the Perron root and is real \cite{FarinaRinaldi2000,BermanPlemmons1994}
\begin{equation}
\lambda_{\rm kin}\equiv \lambda_{\max}(J-\boldsymbol{\Lambda}),
\qquad
\lambda_{\max}(A)\equiv \max_i \Re\,\lambda_i(A),
\label{eq:lambda_kin_def}
\end{equation}
(the spectral abscissa\footnote{\emph{Eigenvalue conventions:} For a square matrix $A$ we define the \emph{spectral abscissa} $\lambda_{\max}(A)\equiv \max_i \Re\lambda_i(A)$ and the \emph{spectral radius} $\rho(A)\equiv \max_i |\lambda_i(A)|$. Continuous-time linearized dynamics $\dot{x}=Ax$ grow iff $\lambda_{\max}(A)>0$, whereas discrete-time maps $x_{t+1}=Ax_t$ grow iff $\rho(A)>1$. Accordingly, Gate~7 uses $\lambda_{\rm kin}=\lambda_{\max}(J-\Lambda)$ for kinetic onset, and uses $\rho(\mathbf{B})$ only for (nonnegative) next-generation/branching matrices.}). We also use ${\cal G}\!=\!\lambda_{\rm kin}\tau_{\mathrm{env}}$ as a dimensionless amplification when needed, with $\tau_{\mathrm{env}}$ the environmental forcing/refresh time; We use $\rho(\cdot)$ only for the \emph{spectral radius} of nonnegative (next-generation / branching) matrices (e.g.\ $B$ in Gate~7). For continuous-time linearized kinetics $\dot x = A x$, growth/decay is controlled by $\lambda_{\max}(A)\equiv \max_i \Re\,\lambda_i(A)$; we denote $\lambda_{\mathrm{kin}}\equiv \lambda_{\max}(J-\Lambda)$. The Jeans escape parameter is written $\lambda_J$ (Gate~3), while climate feedback parameters are $\lambda_0,\lambda_{\mathrm{fb}}$ (Gate~12);

\item ${\cal E}$  is the quasispecies Eigen number where $\mu$ is per-site copy error for an informational polymer of length $L$ under selective advantage $s$; 

\item ${\cal C}$  is the confinement/retention number linking boundary exchange to reaction:
$\tau_{\rm ret}\equiv \min_i\tau_{{\rm exch},i}$ is the characteristic retention time against loss across the compartment boundary (e.g., $\tau_{{\rm exch},i}=V/(A P_i)$ for permeability-limited leakage of key species $i$), and $\tau_r$ is a representative timescale for bottleneck reaction(s) that must complete before those species are lost. Internal homogenization $\tau_{\rm mix}\sim a^2/D_{\rm eff}$ is used only as an auxiliary diagnostic in Gate~8 when sustained internal gradients are invoked; and 

\item $\chi$ is the channel slack comparing translation-channel capacity $C_{\mathrm{cap}}$ to required coding rate $R_{\mathrm{code}}$.  
\end{itemize}

\begin{table}[t]
\caption{Symbols and order parameters used throughout. Scales are indicative and context-dependent.}
\label{tab:parameters}
\centering
\begin{tabular}{l l l}
\hline
Symbol & Meaning & Typical scale \\
\hline\hline
$\Phi$ & power margin  & $0.1$--$10$ \\
$\lambda_{\rm kin}$ & dominant linear growth exponent (spectral abscissa) of the linearized CRN
& $-O(10)$ to $+O(10)\ \mathrm{s}^{-1}$ \\
${\cal G}$& kinetic gain (dimensionless), ${\cal G} \equiv \lambda_{\rm kin}\tau_{\rm env}$ & $0$ to $O(10)$ \\
$\mathcal{C}$ & confinement/compositional--transport ratio  & $10^{-2}$--$10^{2}$ \\
$\mathcal{E}$ & Eigen number (error threshold)  & $10^{-2}$--$1$ \\
$C_{\mathrm{cap}}$ & channel capacity (per symbol)  & $0.5$--$2$ $\text{bits symbol}^{-1}$\\
$\chi$ & channel slack  & $0.3$--$5$ \\
$G_{\mathrm{climate}}$ & climate feedback gain & $<1$ (stable) \\
$D$ & molecular diffusivity & $10^{-12}$--$10^{-9}$ m$^2$\,s$^{-1}$ \\
$P$ & membrane permeability & $10^{-10}$--$10^{-7}$ m\,s$^{-1}$ \\
\hline
\end{tabular}
\end{table}

These are not independent. Surplus work ($\Phi>1$) funds proofreading that lowers $\mu$ and hence $\cal E$; increased localization (large $\cal C$) raises effective rate constants and the kinetic growth exponent $\lambda_{\mathrm{kin}}$ (Gate~7) while suppressing diffusive losses; coding slack $\chi$ trades off against dissipation and empirically measured confusion matrices. The thresholds associated with these couplings are sharp---eigenvalue crossings in linearized dynamics or percolation of autocatalytic sets---and therefore falsifiable in controlled reactors and, in favorable cases, \emph{partially constrainable} by exoplanet observables
\emph{under explicit forward models and degeneracy assumptions} (Sec.~\ref{sec:methods}--\ref{sec:validation}).

The biosphere-relevant pass/fail conditions are gathered in the gate vector
\begin{equation}
\mathcal{V} \equiv 
\Bigg\{
\begin{aligned}
&\Phi>1,\ \lambda_{\rm kin}>0,\ p_{\mathrm{cat}}>p_c,\ \mathcal{C}\gtrsim 1,\ J_{\mathrm{in}}\ge J_{\mathrm{cons}},\ J_{\mathrm{leak}}\le J_{\mathrm{make}},\\
&\mathcal{E}<1,\ \chi>1,\ \mathrm{NPP}>0,\ \text{cycles close},\ G_{\mathrm{climate}}<1
\end{aligned}\Bigg\}.
\label{eq:gatevector}
\end{equation}
Astrophysical prerequisites (adequate element production, hazard survival, volatile retention) are implicit antecedents to~\eqref{eq:gatevector}. A world lies in the biosphere phase if all entries in $\mathcal{V}$ hold with margin (Table~\ref{tab:gate-atlas}).
  
We use \eqref{eq:gatevector}, to define the dimensionless gate margins $\mPhi\!=\!\Phi$, $m_{\cal C}\!=\!{\cal C}\!=\!\tau_{\rm ret}/\tau_r$, $m_{\cal E}\!=\!1/{\cal E}$, $\mChi\!=\!C_{\rm cap}/R_{\rm code}$, and  $\mG\!=\!1-G_{\rm climate}$, together with cycle-closure margins $\mcyc{X}\!=\!\tau_{\rm loss,X}/\tau_{\rm recycle,X}$. Passing the full gate vector requires each gate inequality to be satisfied with a safety factor. For ratio-type margins we target $\mPhi,\,m_{\cal C},\,m_{\cal E},\,\mChi\!, ...>1$ (often $\gg 1$), whereas for stability-slack margins (e.g.\ $m_G\equiv 1-G_{\rm climate}$) the requirement is $m_G>0$ and, for robustness, not too close to zero.

Let $M_X$ be the standing inventory of element $X$ in the active biosphere reservoir.
Define $\tau_{\rm loss,X}\equiv M_X/J_{\rm loss,X}$ and $\tau_{\rm recycle,X}\equiv M_X/J_{\rm proc,X}$,
where $J_{\rm proc,X}$ is the biologically mediated processing/recycling flux.
Then
\[
m_{\rm cyc,X}\equiv \frac{\tau_{\rm loss,X}}{\tau_{\rm recycle,X}}
= \frac{M_X/J_{\rm loss,X}}{M_X/J_{\rm proc,X}}
= \frac{J_{\rm proc,X}}{J_{\rm loss,X}}.
\]
We apply $J_{\rm recycle,X}\equiv J_{\rm proc,X}$ and therefore use the equivalent flux-ratio form in Table~\ref{tab:gate-atlas}. Passing the full gate vector requires all margins $>1$ with a safety factor; unless otherwise noted we target $\mPhi,\,\mC,\,\mE,\,\mChi\!\gtrsim\!1.5$ and $\mG\!\gtrsim\!0.2$ for robustness.

 As our intent is to go from fundamental physics interactions and constants to biospheres, let $\mathbf{c}$ denote the vector of fundamental-physics constants or anchors,
\[
\mathbf{c} \equiv \big(\alpha,\; m_e/m_p,\; G_{\mathrm N}m_p^2/(\hbar c),\; \ldots\big),
\]
which determine atomic/molecular scales and stellar/planetary boundary conditions. The end-to-end pipeline may be written as a composition
\begin{equation}
\mathbf{c}
\xrightarrow{\text{Gate 1}}
\big(\Xi_b,\Xi_{\mathrm{pair}},\Xi_{\mathrm{agg}}\big)
\;\xrightarrow{\text{Gates 2--4}}
\Phi
\xrightarrow{\text{Gates 5--9}}\;
\big({\cal E}, \lambda_{\rm kin},{\cal C}\big)
\xrightarrow{\text{Gate 10}}
\chi
\xrightarrow{\text{Gates 11--12}}
\Big(\mathrm{NPP},\,\{m_{\mathrm{cyc},X}\},\,G_{\mathrm{climate}}\Big),
\label{eq:constants_to_biosphere}
\end{equation}
which we use later to carry out sensitivity accounting from fundamental physical interactions and constants to biosphere observables within the Milky Way galaxy via the chain rule. \emph{Guide to the pipeline:} Gate~1 (Sec.~\ref{sec:gate1}), Gates~2--4 (Secs.~\ref{sec:gate2}--\ref{sec:gate4}), Gates~5--9 (Secs.~\ref{sec:gate5}--\ref{sec:gate9}), Gate~10 (Sec.~\ref{sec:gate10}), and Gates~11--12 (Secs.~\ref{sec:gate11-12}) establish each arrow in (\ref{eq:constants_to_biosphere}).

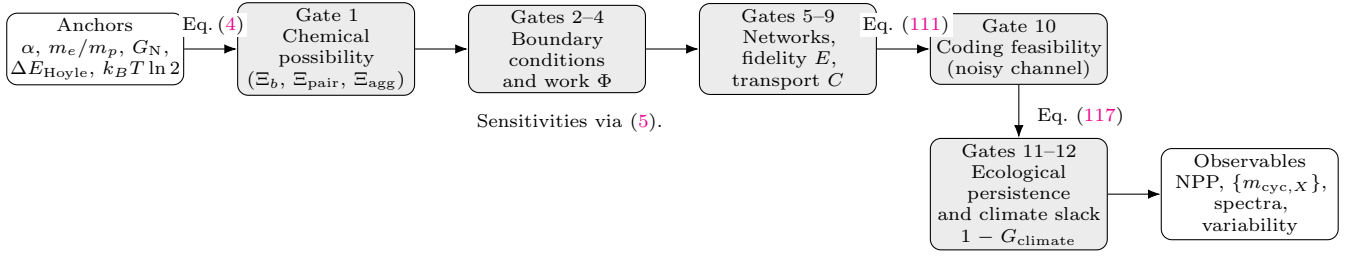
\begin{figure}[t]
\centering
\begin{tikzpicture}[
  >=Latex,
  font=\scriptsize,
  node distance=7mm and 7mm,
  box/.style={draw, rounded corners, align=center, inner sep=2pt, text width=22mm, minimum height=9mm},
  group/.style={box, fill=gray!15},
  linklabel/.style={fill=white, inner sep=1pt, font=\scriptsize}
]
\node[box]   (anchors) {Anchors\\ $\alpha,\, m_e/m_p,\, G_{\mathrm N},$\, $\Delta E_{\rm Hoyle},\, k_BT\ln 2$};

\node[group, right=of anchors] (g1)  {Gate 1\\ Chemical possibility\\ $(\Xi_b,\,\Xi_{\mathrm{pair}},\,\Xi_{\mathrm{agg}})$};
\node[group, right=of g1]      (g24) {Gates 2--4\\ Boundary conditions\\ and work $\Phi$};
\node[group, right=of g24]     (g59) {Gates 5--9\\ Networks, fidelity $E$,\\ transport $C$};
\node[group, right=of g59]     (g10) {Gate 10\\ Coding feasibility\\ (noisy channel)};

\node[group, below=of g10]     (g1112) {Gates 11--12\\ Ecological persistence\\ and climate slack $1-G_{\mathrm{climate}}$};
\node[box, right=of g1112]     (obs)   {Observables\\ NPP, $\{m_{\mathrm{cyc},X}\}$,\\ spectra, variability};

\draw[->] (anchors.east) -- (g1.west)
  node[linklabel, above, pos=0.55, yshift=1.5mm] {Eq.\,(\ref{eq:constants_to_biosphere})};
\draw[->] (g1.east)  -- (g24.west);
\draw[->] (g24.east) -- (g59.west);
\draw[->] (g59.east) -- (g10.west)
  node[linklabel, above, pos=0.55, yshift=1.5mm] {Eq.~(\ref{eq:gate10-box})};
\draw[->] (g10.south) -- (g1112.north)
  node[linklabel, right, pos=0.6, xshift=2mm] {Eq.~(\ref{eq:PhiNPP})};
\draw[->] (g1112.east) -- (obs.west);

\node[anchor=north west] at ([yshift=-1.5mm]g24.south west)
  {Sensitivities via (\ref{eq:sensitivity}).};

\end{tikzpicture}
\caption{Overview of the program: constants\,$\to$\,gates\,$\to$\,observables pipeline Eq.~(\ref{eq:constants_to_biosphere}); sensitivities use Eq.~(\ref{eq:sensitivity}). Order parameters (see \Cref{tab:parameters}) propagate through the gates to biosphere metrics. Coding feasibility is set by the rate--capacity inequality Eq.~\eqref{eq:gate10-box}; the NPP ceiling uses Eq.~\eqref{eq:PhiNPP};  sensitivity accounting follows the chain rule Eq.~\eqref{eq:sensitivity}.
  }
\label{fig:pipeline}
\end{figure}

\begin{table*}[t]
\vskip -10pt
\caption{Gate Atlas: Inequalities (with margin definitions in parentheses), key lab falsifiers, and astronomical observables that can help constrain/bound each gate. See Sec.~\ref{sec:validation} for consolidated tests.}
\label{tab:gate-atlas}
\centering
\begingroup
\setlength{\tabcolsep}{1pt}
\renewcommand{\arraystretch}{1.00}
\newcommand{\colbox}[2]{\parbox[t]{#1}{\raggedright #2}}

\begin{tabular}{@{}l l l l l@{}}
\hline
\# &
\colbox{2.5cm}{Name} &
\colbox{5.0cm}{Inequality / Margin} &
\colbox{4.8cm}{Lab falsifiers} &
\colbox{4.8cm}{Astronomical observables} \\
\hline\hline\addlinespace[3pt]
1 & \colbox{2.5cm}{Fundamental physics $\to$ chemical possibility}
  & \colbox{5.0cm}{Feasible alphabet/backbone and pairing windows:  $E_b/k_{\mathrm{B}}T$ and $E_{\mathrm{HB}}/k_{\mathrm{B}}T$ within viability bands; solvent permits templating}
  & \colbox{4.8cm}{Backbone instability in target solvent; no base‐pair selectivity; photochemistry yields $\ll$ required}
  & \colbox{4.8cm}{Stellar SED/UV; host metallicity $Z$; ISM/cosmic‐ray field; planet bulk composition (density)} \\\addlinespace[3pt]
2 & \colbox{2.5cm}{Astrophysical context \& solids}
  & \colbox{5.0cm}{Elemental/solid budget adequate: $Z \ge Z_{\min}$, $\Sigma_{\mathrm{solid}}>\Sigma_{\min}$; irradiation hazards below limits}
  & \colbox{4.8cm}{Measured flare/CME/RBE rates exceed survival window; insufficient refractory inventory in condensates}
  & \colbox{4.8cm}{Host $Z$, age, activity/flare stats; disk/planet occurrence; galactocentric environment} \\\addlinespace[3pt]
3 & \colbox{2.5cm}{Volatile retention (surface access)}
  & \colbox{5.0cm}{Escape/Jeans parameter large for key volatiles:  $\lambda \equiv GM_p m/(k_{\mathrm{B}}T_{\mathrm{exo}}R_p)\gg 1$; \ margin $m_{\lambda}\!\equiv\!\lambda/20$}
  & \colbox{4.8cm}{Exobase heating drives hydrodynamic escape; measured escape $>$ source for H/He/H$_2$O}
  & \colbox{4.8cm}{Ly$\alpha$, He 1083\,nm, metastable lines; bulk density/radius; retrieved high‐altitude $T_{\mathrm{exo}}$} \\\addlinespace[3pt]
4 & \colbox{2.5cm}{Power / exergy}
  & \colbox{5.0cm}{$\Phi=P_{\mathrm{in}}/(P_{\mathrm{maint}}+P_{\mathrm{synth}}) > 1$; \ margin $m_{\Phi}\!=\!\Phi$}
  & \colbox{4.8cm}{Calorimetry shows $P_{\mathrm{in}} < P_{\mathrm{maint}}+P_{\mathrm{synth}}$; redox/photonic budgets insufficient}
  & \colbox{4.8cm}{Day--night/seasonal phase curves; energy‐balance closure; strong atmospheric disequilibria} \\\addlinespace[3pt]
5 & \colbox{2.5cm}{Feedstock continuity \& leakage}
  & \colbox{5.0cm}{Mass/flux balance: $J_{\mathrm{make}}\!\ge\!J_{\mathrm{loss}}$, $J_{\mathrm{leak}}\!\le\!J_{\mathrm{make}}$;
    margin $m_{\mathrm{stoich}}\!=\!\min(J_{\mathrm{make}}/J_{\mathrm{loss}},\,J_{\mathrm{make}}/J_{\mathrm{leak}})$}
  & \colbox{4.8cm}{Measured deposition/washout or leakage exceeds supply; no steady state in chemostats}
  & \colbox{4.8cm}{Outgassing vs.\ escape rates; aerosol deposition; boundary‐layer fits; volatile inventories} \\\addlinespace[3pt]
 6 & \colbox{2.5cm}{Fidelity (Eigen bound)}
  & \colbox{5.0cm}{Error threshold satisfied:
    $\mathcal{E}=\mu L/\ln s<1$; \ margin $m_E\!\equiv\!1/\mathcal{E}$; \quad proofreading work $W_{\mathrm{proof}}\!\gtrsim\!k_{\mathrm{B}}T\ln r$}
  & \colbox{4.8cm}{Catastrophic mutational meltdown at required $L$; energy budget can’t support needed proofreading}
  & \colbox{4.8cm}{Indirect: sustained disequilibria requiring long‐lived templates; (no direct exoplanet probe yet)} \\\addlinespace[3pt]
7 & \colbox{2.5cm}{Autocatalysis \& thermodynamics}
  & \colbox{5.0cm}{RAF exists with sufficient catalysis density:  $\lambda_{\rm kin}>0$,
    $p_{\mathrm{cat}}>p_c\!\sim\!c\,\ln|R|/|R|$, a feasible flux $v\!\ge\!0$ with $\sum_i v_i\Delta G_i<0$}
  & \colbox{4.8cm}{No RAF despite high $p_{\mathrm{cat}}$ and foods; cycles thermodynamically uphill}
  & \colbox{4.8cm}{Complex organics far from equilibrium; redox ladder occupancy; aerosol/polymer spectral features} \\\addlinespace[3pt]
8 & \colbox{2.5cm}{Confinement / compartment geometry}
  & \colbox{5.0cm}{Exists geometry with adequate confinement:
    $\mathcal{C}=\tau_{\rm ret}/\tau_r \gtrsim 1$; \ margin $m_{\mathcal{C}}\!=\!\mathcal{C}$;
    permeabilities $P_i$ in a viable range (so $\tau_{{\rm exch},i}=V/(A P_i)$ exceeds bottleneck $\tau_r$).}
  & \colbox{4.8cm}{Over‐mixing (washout) or too‐leaky membranes prevent sustained growth}
  & \colbox{4.8cm}{Cloud/haze microphysics; droplet/film scales; inferred eddy $K_{zz}$; surface porosity indicators} \\\addlinespace[3pt]
 9 & \colbox{2.5cm}{Onset of Darwinian dynamics} 
  & \colbox{5.0cm}{
$  \lambda_{\max}(BQ-D)>0; N_e s\gtrsim1, S_b = N_e U_b \gtrsim 1,\ \ s\,\tau_{\mathrm{env}}\gg 1$ \quad (margins: $m_r,\, m_{\rm drift}, \ m_{\rm ben}$)} 
  & \colbox{4.8cm}{Chemostat growth curves subcritical; linearized dynamics stable ($\lambda_{\max}(BQ-D)\le 0$) despite ample feedstock}
  & \colbox{4.8cm}{Retrieved production vs.\ loss proxies; photochemical column budgets; lifetime vs.\ mixing constraints} \\\addlinespace[3pt]
10 & \colbox{2.5cm}{Translation channel (rate--capacity)}
  & \colbox{5.0cm}{Slack $>\!1$:
    $\chi=C_{\mathrm{cap}}/R_{\mathrm{code}}>1$; \ margin $m_{\chi}\!=\!\chi$; \quad
    $C_{\mathrm{cap}}=\log_2 k - H_k(\varepsilon)$, $R_{\mathrm{code}} = (1/n)\log_2 m $}
  & \colbox{4.8cm}{Error spectra show $\chi\!\le\!1$; code cannot support required alphabet/degeneracy under noise}
  & \colbox{4.8cm}{Alphabet‐size proxies; error‐tolerant coding signatures (degeneracy patterns, if any)} \\\addlinespace[3pt]
11 & \colbox{2.5cm}{Productivity \& cycles closure}
  & \colbox{5.0cm}{$\mathrm{NPP}>0$; \ biogeochemical cycles close:   $m_{\mathrm{cyc},\{C,N,P\}}\!\equiv\!J_{\mathrm{recycle}}/J_{\mathrm{loss}}\!\gg\!1$}
  & \colbox{4.8cm}{Mesocosms yield $\mathrm{NPP}\!\le\!0$ or cycles remain open (recycle $\not\gg$ loss)}
  & \colbox{4.8cm}{Reflectance/thermal phase curves; seasonal lag/area; multi‐species disequilibria co‐sustained} \\\addlinespace[3pt]
12 & \colbox{2.5cm}{Climate feedback stability}
  & \colbox{5.0cm}{$G_{\mathrm{climate}}<1$; \ margin $m_G\!\equiv\!1-G_{\mathrm{climate}}$}
  & \colbox{4.8cm}{Perturbation--response shows runaway/positive feedback dominance ($G_{\mathrm{climate}}\!\ge\!1$)}
  & \colbox{4.8cm}{Emission bands/continuum (lapse rate, water window); albedo--temperature co‐variability} \\\addlinespace[3pt]
\hline
\end{tabular}
\endgroup

\begin{minipage}{\linewidth}\footnotesize
\emph{Units:} $\Phi$, $\mathcal{C}$, $\chi$, $G_{\mathrm{climate}}$, and all $m_{\cdot}$ are dimensionless; $D$ is [$\mathrm{m}^{2}\,\mathrm{s}^{-1}$]; membrane permeability $P$ is [$\mathrm{m}\,\mathrm{s}^{-1}$]; powers/fluxes follow the text.
\end{minipage}
\vskip -10pt
\end{table*}

As a visual summary of the constants$\to$gates$\to$observables mapping, Fig.~\ref{fig:pipeline} provides the  relevant schematic, while Table~\ref{tab:parameters} summarizes the quantities used throughout and Table~\ref{tab:gate-atlas} provides the Gate Atlas, while Table~\ref{tab:gate_identifiability-margins} discusses remote identifiability of gate margins.

Because the biosphere observables are a composition of gate margins driven by the fundamental-physics anchors (\ref{eq:constants_to_biosphere}),
we will report \emph{dimensionless log-sensitivities} of biosphere metrics to constants.
For net primary productivity (NPP), the chain rule gives
\begin{equation}
\frac{\partial \ln \mathrm{NPP}}{\partial \ln \alpha}
=\sum_{g}
\frac{\partial \ln \mathrm{NPP}}{\partial \ln m_g}
\cdot
\frac{\partial \ln m_g}{\partial \ln \alpha},
\qquad
m_g\in\big\{\Phi,\,{\cal E},\,{\cal C},\,\chi,\,m_{\mathrm{cyc},X},\,1-G_{\mathrm{climate}}\big\}.
\label{eq:sensitivity}
\end{equation}
We include $1-G_{\mathrm{climate}}$ (the climate-stability slack) so that a positive derivative denotes improved stabilizing margin. Here, $\Phi$ is the available power density, $\cal E$ the effective error/discrimination margin, $\cal C$ the compositional/transport ratio, $\chi$ the translation--channel slack (Gate~10), and $m_{\mathrm{cyc},X}$ the elemental cycle-closure margins for species $X$. This identity is used throughout to propagate uncertainties or hypothetical shifts in $\alpha$ into biosphere-scale responses.

In practice, most observables constrain \emph{combinations} of gate margins and are subject to degeneracies; inference is therefore posed as a Bayesian retrieval with explicit priors and uncertainty propagation (Sec.~\ref{sec:methods}).

\begin{table}[t]
\caption{Remote identifiability of gate margins. ``Direct'' means bounded from observables with weak priors; ``Model-dependent'' means only via forward models with strong nuisance parameters; ``Prior-dominated'' means laboratory priors dominate; ``Not identifiable'' means no meaningful remote constraint.}
\label{tab:gate_identifiability-margins}
\centering
\begin{tabular}{llll}
\hline
Gate & Margin(s) & Primary remote proxy & Identifiability \\
\hline\hline
2  & $m_Z, m_M, m_S, m_t$ & host $Z$, age/activity, Galactic context & Direct / Model-dependent \\
3  & $m_{\rm vol}, m_\lambda, m_{\rm env}, m_F$ & $M_p,R_p$, escape lines, XUV history & Model-dependent \\
4  & $m_\Phi$ (or $m_{\Phi,\mathrm{eff}}$) & SED+albedo+phase curves, $\Delta G_{\rm diseq}$ & Model-dependent \\
5  & $\Psi_{\min}$ (proxy) & precursors/haze/deposition, turnover times & Model-dependent \\
6  & $m_{\cal E}, m_{\rm copy}$ & energy-consistent bounds + lab priors on $\mu$ & Prior-dominated \\
7  & $m_{\rm kin}, m_p$ (proxy) & turnover vs photochemical timescales; mineralogy & Prior-dominated \\
8  & $m_{\cal C}, m_{\rm in}, m_{\rm leak}$ & liquids/microphysics + transport analogs & Prior-dominated \\
9  & $m_r, m_{\rm drift}, m_{\rm ben}$ & indirect via Gates 10--12 only & Not identifiable \\
10 & $m_\chi$ & energy-consistent bounds + lab confusion matrices & Prior-dominated \\
11 & $m_{\rm NPP}, m_{{\rm cyc},X}$ & disequilibrium + variability coherence & Model-dependent \\
12 & $m_G, m_{\rm tip}, b$ & multi-epoch stability + energy-balance fits & Model-dependent \\
\hline
\end{tabular}
\end{table}

Thus, assuming base-pair discrimination scales with the Rydberg ($\Delta\Delta G_{\rm pair}\propto E_{\rm Ry}\propto \alpha^2$), $\mu_{\rm eq}\simeq g\,e^{-\Delta\Delta G_{\rm pair}/k_BT}$ gives  $\partial\ln\mu_{\rm eq}/\partial\ln\alpha \simeq -2\,\Delta\Delta G_{\rm pair}/k_BT$.  At fixed target $\mu$, the proofreading work $W_{\rm proof}=k_BT\ln(\mu_{\rm eq}/\mu)$ yields $\partial W_{\rm proof}/\partial\ln\alpha \simeq k_BT\,\partial\ln\mu_{\rm eq}/\partial\ln\alpha$. Because Gate~10’s $C_{\rm cap}(\varepsilon_b)$ increases as $\varepsilon_b$ decreases with available $W_{\rm proof}$, one obtains $\partial\ln\chi/\partial\ln\alpha > 0$ at fixed $(k,n,m)$. Via (\ref{eq:PhiNPP}), a smaller $N\!\dot{}_{\rm copy}\,[k_BT\ln 2(\log_2 k_p-H_{k_p}(\mu)) + W_{\rm proof}]$ reduces the subtraction from $(P_{\rm in}-P_{\rm maint})$, thus $\partial\ln{\rm NPP}/\partial\ln\alpha>0$ at fixed fluxes. This example makes (\ref{eq:sensitivity}) concrete.

Because these relations are governed by conservation laws and information thermodynamics, the \emph{form} of \eqref{eq:orderparams}--\eqref{eq:gatevector} is solvent-agnostic: solvent identity and temperature only shift the numerical values of transport and error parameters, not the inequalities themselves. Passing Gates~1--9 establishes the Darwinian regime; Gates~10--12 extend to scalable coding and biosphere stabilization. Our emphasis is the full pipeline to biospheres and its observational consequences.

As the gate vector \(\cal  V\) translates physical preconditions into measurable \emph{margins}, the framework provides a predictive basis for triaging exoplanets. Rather than labeling orbits as merely ``habitable'', we evaluate whether a target’s \emph{retrievable} margins---power \(\Phi\), localization \(\cal C\), kinetic gain \(\lambda_{\rm kin}\), fidelity \(\cal E\), coding slack \(\chi\), productivity/cycle closure (Gate~11), and feedback stability (Gate~12)---comfortably exceed unity on relevant timescales. This converts the search into a phase diagram with explicit slack, establishes concrete falsifiers for each link in the chain, and yields a quantitative rationale for target selection before expensive follow-up and potential detection of exo-life \cite{Turyshev:2025-exoPL}.

\paragraph*{Scope of inference (identifiability).}
Several gate margins are not directly retrievable from remote sensing (notably replication errors $\mu$ and
translation confusion matrices), and are therefore treated as \emph{bounds} derived from energy-consistency
and laboratory priors rather than as fully data-driven posteriors. Accordingly, whenever we use the term
\emph{predictive} we mean \emph{predictive constraints}: inequalities, bottleneck rankings, and falsifiers
that are conditional on explicitly stated priors and forward models (Sec.~\ref{sec:methods}), not unconditional
claims of biosphere prevalence.

\emph{Limits.} With only Earth as a working example, our claims must remain conditional. The value of the framework is not ontological but predictive: it yields falsifiable margins and retrievals that can be supported or refuted as exoplanet data and laboratory tests improve.

\section{Gate 1: Fundamental Physics $\rightarrow$ Chemical Possibility}
\label{sec:gate1}

Gate~1 asserts that the observed values of the fundamental constants and the structure of the interactions jointly provide (i) a periodic table with abundant CHNOPS (i.e., C, H, N, O, P, S) elements, (ii) atomic and molecular energy scales that permit robust covalent backbones and directional noncovalent interactions at accessible temperatures, and (iii) thermodynamic headroom to encode and process information with controllable error. When these physical preconditions are met within at least one solvent/temperature window, chemistry possesses the raw \emph{degrees of freedom} required by all downstream gates.

\subsection{Standard Model, constants, and nuclear astrophysics}

The Standard Model and general relativity \cite{TuryshevFunPhys2025} set interactions and cosmic expansion, fixing elemental inventories and energy quanta. Dimensionless constants (e.g., $\alpha$, $m_e/m_p$) and nuclear resonance placements (e.g., the $^{12}$C Hoyle state) must allow abundant C/O production.  Table~\ref{tab:anchors} lists the primary anchors in fundamental physics, the immediate controls they set, and the downstream gate margins and biosphere-level consequences they influence.

The characteristic atomic energy scale is set by the Rydberg,
\begin{equation}
E_{\mathrm{Ry}}=\tfrac12\,\alpha^2 m_e c^2 \simeq 13.6~\mathrm{eV},
\end{equation}
with chemical bond energies $E_b$ of order $(0.2$--$0.4)\,E_{\mathrm{Ry}}$. The stability of nuclei and the availability of bio-relevant elements follow from the strong and electromagnetic interactions together with cosmic expansion. Two nuclear constraints are critical:
\begin{enumerate}
\item \emph{Deuteron binding.} The existence of a bound deuteron ($B_d\simeq 2.22$~MeV) enables Big Bang and stellar nucleosynthesis beyond H and where the $^{12}$C Hoyle resonance enables efficient triple-$\alpha$ synthesis \cite{Oberhummer2000,Epelbaum2013}.  If the deuteron becomes unbound, Big Bang nucleosynthesis (BBN) yields almost no He and stellar nucleosynthesis pathways change \cite{Dmitriev2004}. Substantial reductions of $B_d$ would suppress primordial and stellar pathways to He, C, and O.

\item \emph{Triple-$\alpha$ resonance.} The $^{12}$C Hoyle resonance state at $E_x\simeq 7.65$~MeV relative to the ground state enhances the $3\alpha\rightarrow{}^{12}$C rate in He-burning stars. Carbon production depends sensitively on the Hoyle state; modest shifts in nuclear levels significantly alter C/O yields \cite{Oberhummer2000}. Modest shifts in this resonance (tens to $\sim\!100$~keV) appreciably alter the C/O yield balance; the observed constants yield abundant C and O across stellar populations.
\end{enumerate}
Gate~1 does not require detailed \emph{anthropic} bounds; it requires only that the realized constants produce a periodic table with long-lived isotopes of C, H, N, O, P, and S and that stellar yields make these elements cosmically common. This condition is empirically satisfied.\footnote{{\it Context:} While Gate 1 does not require anthropic bounds, broader discussions of fine‑tuning and anthropic reasoning in
cosmology provide useful context for how changes in constants would cascade to chemistry and habitability \cite{LivioRees2005_Science, LivioRees2018_CUP, Beer2004_SolarSystem_MNRAS}.}

\emph{Observer selection.} Our use of constants does not presume that only this set permits life. Anthropic and multiverse perspectives (e.g., \cite{LivioRees2018_CUP}) remind us that selection effects can make such arguments circular. Here we treat constants as empirical inputs for this universe and ask what \emph{testable} consequences follow for gate margins.

\begin{table*}[t]
\vskip -10pt
\centering
\caption{Fundamental-physics anchors and their downstream gate and biosphere consequences. Entries list the primary order parameters or margins impacted and the qualitative biosphere-level effect.}
\label{tab:anchors}

\begingroup
\setlength{\tabcolsep}{6pt}
\renewcommand{\arraystretch}{1.00}
\newcommand{\colbox}[2]{\parbox[t]{#1}{\raggedright #2}}

\begin{tabular}{@{}l l l@{}}
\hline
\colbox{3.2cm}{Anchor (fundamental)} &
\colbox{4.8cm}{Direct controls} &
\colbox{6.8cm}{Gates $\to$ biosphere consequence} \\
\hline\hline
$\alpha$ (fine-structure) &
\colbox{4.8cm}{$E_{\rm Ry}\!\propto\!\alpha^2 m_e c^2$; bond energies $E_b$; H-bond scale $E_{\rm HB}$} &
\colbox{6.8cm}{Gate 1: $(\Xi_b,\Xi_{\rm pair})$; $\Rightarrow$ lower $\mu$ (Gate 6), larger $\chi$ (Gate 10); boosts feasible NPP via Eq.~(\ref{eq:PhiNPP}).} \\\addlinespace[3pt]
$m_e/m_p$ &
\colbox{4.8cm}{Vibrational/rotational scales; viscosities $\eta$; diffusivities $D$} &
\colbox{6.8cm}{Gate 8: $C=\tau_{\rm ret}/\tau_r$; Gate 4 via rate constants; affects Gate 11 turnover times.} \\\addlinespace[3pt]
Hoyle-state offset $\Delta E_{\rm Hoyle}$ &
\colbox{4.8cm}{Stellar C/O yields, hence solids/volatiles} &
\colbox{6.8cm}{Gate 2: $Z$, solids; Gate 3 volatile retention; prerequisites for Gates 4--12.} \\\addlinespace[3pt]
$G_\mathrm{N}$ &
\colbox{4.8cm}{$g$, escape (Jeans) parameter $\lambda_J$, scale height $H$} &
\colbox{6.8cm}{Gate 3: volatile retention; Gate 4: P--T structure; Gate 12: reservoir buffering masses.} \\\addlinespace[3pt]
$k_B T\ln 2$ (Landauer, \cite{Landauer1961}) &
\colbox{4.8cm}{Minimal erasure work; proofreading work $W_{\rm proof}$} &
\colbox{6.8cm}{Gates 6,10: sets $\mu$ and $\epsilon_b$ at given $\Phi$; enters biosphere power budget Eq.~(\ref{eq:PhiNPP}).} \\\addlinespace[3pt]
\hline
\end{tabular}
\endgroup
\vskip -10pt
\end{table*}

\subsection{Atomic structure, bonding scales, and solvent windows}

A minimal chemical feasibility condition at temperature $T$ is
\begin{equation}
E_b \gg k_{\mathrm B} T \ \text{for covalent bonds (C--C, C--H, O--H, P--O)},\qquad E_{\mathrm{HB}}\gtrsim 0.1~\mathrm{eV}.
\label{eq:chemfeasible}
\end{equation}
At $T=\SI{300}{K}$, $k_{\mathrm B} T\simeq \SI{0.026}{eV}$; typical covalent bonds are $3.5$--$5~\mathrm{eV}$. These scales enable template pairing and amphiphile assembly. 

For chemistry to furnish templating, catalysis, and compartment formation, several energy-scale inequalities must hold in at least one solvent window $\mathcal{W}_T\subset\mathbb{R}_+$:
\begin{align}
\text{(covalent backbone)}\qquad~~~
\Xi_b &\equiv \min_{b\in\mathcal{B}}\frac{E_b}{k_{\mathrm B} T} \gg 1 \quad (\text{e.g., }\Xi_b\gtrsim 50), 
\label{eq:gate1-backbone}\\[4pt]
\text{(directional pairing)}\qquad 
\Xi_{\mathrm{pair}} & \equiv \frac{\Delta\Delta G_{\mathrm{pair}}}{k_{\mathrm B} T} \gtrsim 4\text{--}6,
\label{eq:gate1-pair}\\[4pt]
\text{(amphiphile assembly)}\qquad~ 
\Xi_{\mathrm{agg}} &\equiv \frac{\lvert\Delta G_{\mathrm{agg}}\rvert}{k_{\mathrm B} T} \gtrsim 5.
\label{eq:gate1-assembly}
\end{align}
Here $\mathcal{B}$ is the set of backbone bonds (C--C, C--N, C--O, P--O for aqueous organophosphates); $\Delta\Delta G_{\mathrm{pair}}$ is the free-energy advantage of correct over incorrect pairing for a single site (relevant to templating); and $\Delta G_{\mathrm{agg}}$ is the free energy of transferring an amphiphile from bulk solvent into an aggregate (micelle/vesicle/coacervate). At $T=300$~K, $k_{\mathrm B} T\simeq 0.026$~eV. Typical covalent bonds in organics are $E_b\sim 3.5$--$5$~eV, giving $\Xi_b\sim 130$--$190$; hydrogen bonds span $E_{\mathrm{HB}}\sim 0.05$--$0.3$~eV, supporting $\Xi_{\mathrm{pair}}\!\gtrsim\!4$ when sequence context contributes cooperative stabilization. For hydrophobic assembly in water, the transfer free energy per methylene is $\sim\!1.9$~kJ\,mol$^{-1}$; chains of length $n\!\gtrsim\!8$ yield $\lvert\Delta G_{\mathrm{agg}}\rvert \gtrsim 15$~kJ\,mol$^{-1}\!\approx\!6\,k_{\mathrm B} T$ at 300~K, enabling submicron vesicles.

\textit{Sequence context.} The discrimination margin $\Xi_{\rm pair}$ is sequence- and salt-dependent 
(stacking and ionic screening); we therefore report ranges for $\Xi_{\rm pair}$ under realistic ionic-strength priors.

The solvent identity enters through dielectric response, hydrogen-bonding propensity, viscosity, and transport ($\epsilon$, $E_{\mathrm{HB}}$, $\eta$, $D$). The \emph{form} of \eqref{eq:gate1-backbone}--\eqref{eq:gate1-assembly} is solvent-agnostic; only the numerical margins shift. In hydrocarbons at $T\sim 90$--$150$~K, $k_{\mathrm B} T$ decreases, increasing $\Xi_b$ and $\Xi_{\mathrm{pair}}$ at fixed $E_b$ and $E_{\mathrm{HB}}$; lower dielectric screening, however, disfavors charged oligomer retention and shifts feasible chemistries toward neutral backbones and different pairing interactions.

\subsection{Information thermodynamics: minimal costs for fidelity}

As erasure and accuracy are physical, the minimal dissipation scales with the mutual information \cite{Landauer1961,Bennett1982}, and kinetic proofreading trades energy for reduced error \cite{Hopfield1974,Ninio1975}, quantified by mutual information and bounded thermodynamically in cyclic information-processing devices \cite{Parrondo2015,Still2012}. 
In particular, irreversible erasure dissipates $k_{\mathrm B} T\ln 2$ per bit \cite{Landauer1961}.  Copying and coding are physical processes. For a $k$-ary symmetric decision with symbol error $\varepsilon$, the mutual information per symbol is $I=\log_2 k - H_k(\varepsilon)$ with $H_k$ the $k$-ary entropy. The minimal work to \emph{irreversibly} erase or reliably establish this information is bounded \cite{England2013} by\footnote{We emphasize that the $k_BT\ln2\,I$ term is used as a conservative \emph{floor} on dissipation for
biochemical copying/proofreading architectures with reset steps; we do not claim it is universally
tight for arbitrarily reversible implementations.
}
\begin{equation}
W_{\min} \ge k_{\mathrm B} T \ln 2 \, I.
\label{eq:landauer-info}
\end{equation}
For templated polymerization with per-site error $\mu$ over an alphabet of size $k_p$, the information per site is $I_{\text{site}}=\log_2 k_p - H_{k_p}(\mu)$; the minimal dissipation for copying length $L$ with accuracy $\mu$ is therefore
\begin{equation}
W_{\min}^{\text{copy}} \gtrsim L\,k_{\mathrm B} T \ln 2\,\big[\log_2 k_p - H_{k_p}(\mu)\big].
\label{eq:copy-cost}
\end{equation}
Kinetic proofreading lowers $\mu$ below the equilibrium misincorporation $\mu_{\mathrm{eq}}$ at a cost of additional chemical driving $\Delta\mu_{\mathrm{chem}}$ (e.g., NTP hydrolysis). Stochastic thermodynamics yields a generic tradeoff: to reduce error by a factor $r\!=\!\mu_{\mathrm{eq}}/\mu$ one must dissipate at least
\begin{equation}
W_{\mathrm{proof}} \gtrsim k_{\mathrm B} T \ln r\quad \text{per decision},
\label{eq:proof-cost}
\end{equation}
with tighter bounds when time constraints and finite discrimination are included. Eqs.~\eqref{eq:copy-cost}--\eqref{eq:proof-cost} couple Gate~1 to later gates: the existence of feasible copying and coding at a given temperature requires both (i) a chemical alphabet with sufficient $\Delta\Delta G_{\mathrm{pair}}$ Eq.~(\ref{eq:gate1-pair}) and (ii) a power budget (Gate~4) that can fund the dissipation implied by the desired $\mu$ (Gate~6) and channel slack (Gate~10). In other words, proofreading trades energy for accuracy, explicitly coupling Gate~4 (power) to Gate~6 (fidelity) and Gate~10 (translation).

\subsection{Quantitative Gate~1 statement and margins}
Gate~1 is passed if there exists a solvent/temperature interval $\mathcal{W}_T$ and a candidate chemical alphabet/backbone such that
\begin{equation}
\boxed
{
\Xi_b \ge \zeta_b,\qquad 
\Xi_{\mathrm{pair}}\ge \zeta_{\mathrm{pair}},\qquad 
\Xi_{\mathrm{agg}}\ge \zeta_{\mathrm{agg}}
}
\label{eq:gate1-box}
\end{equation}
with conservative thresholds $\zeta_b\!\sim\!50$, $\zeta_{\mathrm{pair}}\!\sim\!4$--$6$ (for equilibrium site discrimination $\mu_{\mathrm{eq}}\!\lesssim\!10^{-2}$--$10^{-3}$), and $\zeta_{\mathrm{agg}}\!\sim\!5$. Margins are the left--right differences in \eqref{eq:gate1-box}. For Earth-like aqueous conditions at 300~K, representative margins are $\Xi_b/\zeta_b\!\sim\!2$--$4$, $\Xi_{\mathrm{pair}}/\zeta_{\mathrm{pair}}\!\sim\!1$--$2$ (sequence/context dependent), and $\Xi_{\mathrm{agg}}/\zeta_{\mathrm{agg}}\!\sim\!1$--$2$ for common fatty-acid amphiphiles.

Thus, for \(k_p=4\) (\(g=3\)) and \(\Delta\Delta G_{\rm pair}=6\,k_B T\), 
\(\mu_{\rm eq}\simeq 3\,e^{-6}\approx 7.5\times 10^{-3}\), consistent with the Gate~6 baseline (no proofreading).
Reaching \(\mu=10^{-4}\) then requires \(W_{\rm proof}\gtrsim k_B T\ln(\mu_{\rm eq}/\mu)\approx 4.3\,k_B T\) per site; 
see Eq.~(\ref{eq:mu-eq}) and Gate~6.

\subsection{Diagnostics, falsifiers, and observational corollaries}
\paragraph*{Laboratory diagnostics.}
(i) \emph{Alphabet discrimination:} directly measure $\Delta G_{\mathrm{pair}}$ and $\Delta G_{\mathrm{mismatch}}$ for candidate backbones/solvents to determine $\Xi_{\mathrm{pair}}$; (ii) \emph{Assembly thermodynamics:} determine $\Delta G_{\mathrm{agg}}$ from critical aggregation concentrations and its temperature dependence to bound $\Xi_{\mathrm{agg}}$; (iii) \emph{Backbone stability:} measure hydrolysis rates and activation energies to infer an effective $\Xi_b$ under the intended solvent, $T$, and ionic strength.

\paragraph*{Falsifiers.}
Gate~1 fails if, in all plausible solvent windows, either (a) $E_b/k_{\mathrm B} T$ cannot exceed $\zeta_b$ for any robust backbone, or (b) $\Delta\Delta G_{\mathrm{pair}}/k_{\mathrm B} T<\zeta_{\mathrm{pair}}$ even with cooperative effects, or (c) no amphiphile/coacervate chemistry achieves $\Xi_{\mathrm{agg}}\ge \zeta_{\mathrm{agg}}$. Any such failure blocks templating, localization, or both, and stalls the pipeline independent of downstream power or network structure.

\paragraph*{Observational corollaries.}
Where Gate~1 holds in the interstellar medium and protoplanetary disks, one expects widespread organics (aliphatic/aromatic C--H features at $3.3\,\mu$m and $6$--$8\,\mu$m) and amphiphile precursors. On planets within $\mathcal{W}_T$, surface--atmosphere spectra should reveal solvents and functional-group inventories consistent with large $\Xi_b$ and viable $\Xi_{\mathrm{pair}}$ (e.g., water bands with organic overtone structure for aqueous worlds, hydrocarbon signatures for cryogenic worlds). These are necessary but not sufficient conditions; Gates~2--4 determine whether usable work and boundary conditions are available to \emph{drive} the chemistry.

Gate~1 formalizes the non-negotiable physical preconditions for life-like chemistry as inequalities on energy scales and information costs. It is satisfied whenever the constants of nature, together with at least one solvent window, provide large bond-energy/thermal ratios, sufficient site discrimination for templating, and favorable assembly thermodynamics. The resulting margins propagate into the fidelity and coding gates via Eqs.~\eqref{eq:copy-cost}--\eqref{eq:proof-cost} and set the quantitative targets for power delivery in Gate~4.

\textit{Bridge to biospheres.}
Because the same anchors that set $(\Xi_b,\Xi_{\rm pair},\Xi_{\rm agg})$ also set the dissipation scale $k_BT\ln 2$ and proofreading costs, Gate~1 margins propagate forward to the biosphere inequalities via Eqs.~(\ref{eq:constants_to_biosphere}) and (\ref{eq:PhiNPP}). In practice, larger $\Xi_{\rm pair}$ at fixed $T$ reduces $\mu_{\rm eq}$, lowering the proofreading work needed to achieve a target $E$ (Gate~6), which increases the energetic slack available for NPP and for climate buffering (Gates~11--12).

Consolidated criteria appear in Sec.~\ref{sec:validation}. Gate~1 pass/fail hinges on Eqs.~(\ref{eq:gate1-backbone})--(\ref{eq:gate1-assembly}) and the summary inequality~(\ref{eq:gate1-box}); observational proxies are the abundance‐controlled margins $\Xi_b$, $\Xi_{\mathrm{pair}}$, and $\Xi_{\mathrm{agg}}$.
{Fundamental-physics priors and sensitivities} are discussed in Sec.~\ref{sec:FP-priors}--\ref{sec:methods-sensitivity} (log-sensitivity propagation via (\ref{eq:constants_to_biosphere})--(\ref{eq:sensitivity})).

\section{Gate 2: Cosmological Boundary Conditions and Galactic Ecology}
\label{sec:gate2}

Gate~2 asks whether a location in a galaxy, over a specified time window, simultaneously (i) supplies sufficient heavy elements and solids to enable rocky-planet assembly and chemically rich surfaces, and (ii) avoids sterilizing or strongly erosive transients at a rate that would reset or abort subsequent gates. Unlike Gate~1, which concerns local energy scales, Gate~2 is statistical and environmental: it constrains \emph{where and when} target worlds are likely to exist and remain quiescent long enough for driven chemistry to proceed.

\subsection{Metal enrichment and target availability}

Cosmic star-formation and chemical evolution yield metallicity $Z(R,t)$ across galactic radii $R$ and times $t$. Planet formation requires $Z$ above a rocky floor. Photochemical pathways to activated precursors are illustrated in \cite{RitsonSutherland2012,Patel2015,Sutherland2016}.

Let $Z(R,t)$ denote the metallicity at galactocentric radius $R$ and cosmic time $t$ (we use $Z_\odot$ for the solar value). A minimal requirement for rocky-planet assembly is a solids reservoir above the streaming-instability threshold. Writing the dust-to-gas mass ratio as $D(Z)=\xi_d\,Z$ with dust-to-metal ratio $\xi_d\in(0,1)$ (empirically of order unity in metal-rich disks), the midplane solids-to-gas ratio satisfies
\begin{equation}
Z_{\mathrm{mid}}(R,t)\equiv \frac{\rho_{\mathrm{solid}}}{\rho_{\mathrm{gas}}}\approx \frac{D(Z)}{\Pi}\,\mathcal{S}(R,t),
\label{eq:midplaneZ}
\end{equation}
where $\Pi\!\sim\!H_{\mathrm{gas}}/H_{\mathrm{solid}}$ is the ratio of gas to solid scale heights and $\mathcal{S}$ encodes local concentration by drift and trapping. Streaming instability requires $Z_{\mathrm{mid}}\ge Z'_{\mathrm{crit}}$ (typical values $10^{-2}$--$10^{-1}$ at the midplane depending on turbulence and particle Stokes number). A conservative \emph{metallicity floor} for rocky targets is therefore
\begin{equation}
Z(R,t)\ \ge\ Z_{\mathrm{rocky\;floor}}\equiv \frac{Z'_{\mathrm{crit}}\Pi}{\xi_d\,\mathcal{S}(R,t)}.
\label{eq:Zfloor}
\end{equation}
Inequality~\eqref{eq:Zfloor} bundles grain growth and vertical settling into $\mathcal{S}$ and $\Pi$; it is deliberately agnostic about specific disk physics (Gate~3 will refine it). It states that cosmological chemical evolution must have progressed far enough at $(R,t)$ to furnish the raw solid mass needed for rapid planetesimal formation.

A complementary (and often more directly usable) form is a mass budget for solids in a protoplanetary disk:
\begin{equation}
M_{\mathrm{solid}}(R,t)\simeq D(Z)\,M_{\mathrm{disk}}(R,t)\ \ge\ M_{\mathrm{solid}}^{\min},
\label{eq:solidmass}
\end{equation}
where $M_{\mathrm{solid}}^{\min}$ is the minimum solid's mass required to assemble at least one terrestrial planet before disk dispersal. Combining \eqref{eq:solidmass} with empirical $M_{\mathrm{disk}}$--$M_\star$ relations and the age--metallicity relation provides a practical check of ``target availability'' at $(R,t)$.

\subsection{Hazard kernels and survival}
\label{sec:Haz-ker}

Sterilizing or chemistry-resetting transients define a \emph{hazard kernel} $h(R,t)$, the rate (per unit time) at which events capable of catastrophic atmospheric or surface-dose perturbations occur within the relevant kill radius (ozone depletion and biospheric disruption from nearby supernovae and GRBs \cite{Gehrels2003,Thomas2005} and etc.) To leading order the kernel is the sum of contributions:
\begin{equation}
h(R,t)=h_{\tt SN}(R,t)+h_{\tt GRB}(R,t)+h_{\tt AGN}(R,t)+h_{\mathrm{cluster}}(R,t)+\cdots,
\label{eq:hkern}
\end{equation}
with
\begin{align}
h_{\tt SN}(R,t)&\simeq \Sigma_{\tt SN}(R,t)\,\pi r_{\mathrm{kill,\tt SN}}^2,\qquad
h_{\tt GRB}(R,t)\simeq \Sigma_{\tt GRB}(R,t)\,f_b\,\pi r_{\mathrm{kill,\tt GRB}}^2,\qquad
h_{\tt AGN}(R,t)\simeq \nu_{\tt AGN}(t)\,\Theta \big(r_{\mathrm{kill,\tt AGN}}-R\big),
\end{align}
where $\Sigma$ denotes the surface rate density of events, $r_{\mathrm{kill}}$ the distance within which atmospheric chemistry or surface dose exceeds critical thresholds (ozone depletion, nitration, UV flash, cosmic-ray fluence), $f_b$ the GRB beaming fraction, and $\Theta$ a step function representing a central engine hazard zone \cite{Gehrels2003,Thomas2005}. The ``birth-cluster'' term $h_{\mathrm{cluster}}$ accounts for early irradiation (FUV/EUV) and dynamical encounters that truncate disks or strip atmospheres; it is transient (cluster dissolution time $\tau_{\mathrm{cl}}\sim 10$--$100$~Myr) but relevant to Gate~3.

Sterilizing transients (i.e., supernovae (SNe), gamma-ray bursts (GRBs), and active galactic nuclei (AGN)) contribute a hazard rate $h(R,t)$. Survival over a quiescent window $[t_0,t_1]$ is the Poisson factor
\begin{equation}
S_{\mathrm{haz}}(R; t_0,t_1)=\exp\!\Big[-\int_{t_0}^{t_1} h(R,t)\,dt\Big].
\label{eq:hazard}
\end{equation}
Two refinements are often useful. First, not all events are equally damaging; a dose--response function $p_{\mathrm{kill}}(d)$ can be folded into an effective $r_{\mathrm{kill}}(E)$ and then convolved with the event energy distribution. Second, $h_{\mathrm{GRB}}$ depends on metallicity because long GRBs preferentially occur in low-$Z$ hosts; this couples \eqref{eq:Zfloor} and \eqref{eq:hazard} in a favorable way for mid-disk, metal-rich regions. Because long-GRB rates are suppressed in metal-rich hosts, $h_{\rm GRB}(R,t)$ inherits a negative metallicity dependence; this couples favorably to the solids prior in (\ref{eq:Zfloor}) and should be included when mapping $m_S$.

{\it Working priors and sensitivity:}
At \(R\approx 8\,\mathrm{kpc}\), adopt \(r_{\rm kill,SN}=8\)--\(10\,\mathrm{pc}\) and long‑GRB beaming \(f_b\sim 10^{-2}\)--\(10^{-1}\) at low \(Z\), giving local \(h_{\rm SN}\sim 10^{-9}\)--\(10^{-8}\,\mathrm{yr}^{-1}\).
Report
\[
m_S \equiv \frac{S_{\rm haz}}{S_{\min}},\qquad 
\frac{\partial \ln S_{\rm haz}}{\partial r_{\rm kill}} = -2\pi \!\int \Sigma(R,t)\, r_{\rm kill}\, dt,
\]
to show survival sensitivity to \(r_{\rm kill}\) and event‑rate priors.

\subsection{Quiet-time requirement and the galactic ecology window}

Let $\tau_{\mathrm{chem}}$ be the time needed to establish driven prebiotic cycles above loss (Gates~4--7) and $\tau_{\mathrm{hered}}$ the time to reach Darwinian dynamics with selectable length $L$ (Gates~6--9). Gate~2 demands a \emph{quiet interval}
\begin{equation}
\Delta t_{\mathrm{quiet}}\equiv t_1-t_0 \ \ge\ \tau_{\mathrm{chem}}+\tau_{\mathrm{hered}}
\quad\text{with}\quad
S_{\mathrm{haz}}(R;t_0,t_1)\ \ge\ S_{\min},
\label{eq:quiet}
\end{equation}
for some survival floor $S_{\min}$ (e.g., 0.5--0.9, depending on how much resetting the downstream gates can tolerate). Eq.~\eqref{eq:quiet} formalizes the intuitive notion of a ``galactic habitable zone'' as a contour in $(R,t)$ where Eqs.~\eqref{eq:Zfloor}, \eqref{eq:solidmass}, and \eqref{eq:hazard} jointly hold.

\subsection{Quantitative Gate~2 statement and margins}
Gate~2 is passed at $(R,t_0\!\to\!t_1)$ if there exist times $t_0<t_1$ with
\begin{equation}
\boxed
{
Z(R,t_0)\ge Z_{\mathrm{rocky\;floor}},\qquad 
M_{\mathrm{solid}}(R,t_0)\ge M_{\mathrm{solid}}^{\min},\qquad
S_{\mathrm{haz}}(R;t_0,t_1)\ge S_{\min},\qquad
\Delta t_{\mathrm{quiet}}\ge \tau_{\mathrm{chem}}+\tau_{\mathrm{hered}}
}
\label{eq:gate2-box}
\end{equation}
The \emph{margins} are $m_Z\!\equiv\!Z/Z_{\mathrm{rocky\;floor}}$, $m_M\!\equiv\!M_{\mathrm{solid}}/M_{\mathrm{solid}}^{\min}$, $m_S\!\equiv\!S_{\mathrm{haz}}/S_{\min}$, and $m_t\!\equiv\!\Delta t_{\mathrm{quiet}}/(\tau_{\mathrm{chem}}+\tau_{\mathrm{hered}})$. These margins propagate to Gate~3 by setting priors on disk mass and lifetime, to Gate~4 by shaping the stellar spectral energy distribution and activity history, and to all later gates via the probability of catastrophic resets (see Table~\ref{tab:gate2-hazard-priors}).

\begin{table*}[t]
\vskip -10pt
\caption{Gate~2 hazard priors by Galactic context (illustrative; order-of-magnitude).
Rates combine Eqs.~(\ref{eq:hkern})--\eqref{eq:hazard} with $r_{\rm kill,SN}$ priors in Sec.~\ref{sec:Haz-ker}.
Long-GRBs are metallicity-suppressed in the thin disk.}
\label{tab:gate2-hazard-priors}
\centering
\begingroup
\setlength{\tabcolsep}{2pt}
\renewcommand{\arraystretch}{1.00}
\newcommand{\colbox}[2]{\parbox[t]{#1}{\raggedright #2}}
\begin{tabular}{@{}l l l l l l l@{}}
\hline
\colbox{3.3cm}{Context} &
\colbox{1.0cm}{$r_{\rm kill,SN}$ [pc]} &
\colbox{1.8cm}{$\Sigma_{\rm SN}$ [$\mathrm{yr}^{-1}\,\mathrm{kpc}^{-2}$]} &
\colbox{1.6cm}{$h_{\rm SN}$ [$\mathrm{yr}^{-1}$]} &
\colbox{1.6cm}{$f_b$ (GRB)} &
\colbox{1.2cm}{$h_{\rm GRB}$ [$\mathrm{yr}^{-1}$]} &
\colbox{6.0cm}{Notes} \\
\hline\hline
\colbox{3.3cm}{Thin disk ($R\!\sim\!8$\,kpc)} &
\colbox{1.0cm}{$8$--$10$} &
\colbox{1.8cm}{$(2\text{--}4)\!\times\!10^{-5}$} &
\colbox{1.6cm}{$(6\text{--}9)\!\times\!10^{-9}$} &
\colbox{1.6cm}{$10^{-2}$--$10^{-1}$} &
\colbox{1.2cm}{$\lesssim\!10^{-10}$} &
\colbox{6.0cm}{Local priors; metal-rich GRB suppression} \\
\colbox{3.3cm}{Thick disk} &
\colbox{1.0cm}{$8$--$10$} &
\colbox{1.8cm}{$(1\text{--}2)\!\times\!10^{-5}$} &
\colbox{1.6cm}{$(3\text{--}5)\!\times\!10^{-9}$} &
\colbox{1.6cm}{$10^{-2}$--$10^{-1}$} &
\colbox{1.2cm}{$\lesssim\!10^{-10}$} &
\colbox{6.0cm}{Older stars; lower core-collapse rate} \\
\colbox{3.3cm}{Inner bulge ($R\!\lesssim\!3$\,kpc)} &
\colbox{1.0cm}{$8$--$10$} &
\colbox{1.8cm}{$\gtrsim\!10^{-4}$} &
\colbox{1.6cm}{$\gtrsim\!3\!\times\!10^{-8}$} &
\colbox{1.6cm}{$10^{-2}$--$10^{-1}$} &
\colbox{1.2cm}{$\gtrsim\!10^{-10}$} &
\colbox{6.0cm}{High transient density} \\
\hline
\end{tabular}
\endgroup
\vskip -10pt
\end{table*}

\subsection{Worked order-of-magnitude estimates (Milky Way archetype)}

For an illustrative solar-circle location ($R\sim 8$~kpc) in a quiescent epoch: (i) metallicity $Z\!\sim\!Z_\odot$ and dust-to-metal $\xi_d\!\sim\!0.5$--1 imply, for $\Pi/\mathcal{S}\!\sim\!3$--10 and $Z'_{\mathrm{crit}}\!\sim\!0.02$, that $Z_{\mathrm{rocky\;floor}}\!\lesssim\!0.3$--$0.6\,Z_\odot$; hence $m_Z\!\gtrsim\!1$ is plausible. (ii) With $M_{\mathrm{disk}}\!\sim\!(0.5$--$5)\%\,M_\star$ and $D\!\sim\!10^{-2}$--$10^{-1}$, one finds $M_{\mathrm{solid}}$ comfortably above a few Earth masses, giving $m_M\!\gtrsim\!1$. (iii) For the hazard kernel, a thin-disk supernova surface rate $\Sigma_{\mathrm{SN}}$ combined with $r_{\mathrm{kill,SN}}\!\sim\!8$--$10$~pc yields $h_{\mathrm{SN}}$ corresponding to one potentially disruptive event per $\gtrsim\!10^8$--$10^9$~yr locally; long-GRB terms are suppressed at near-solar metallicity and further reduced by beaming ($f_b\!\ll\!1$). Over a Gyr quiet interval, one expects $m_S\!\sim\!\mathcal{O}(1)$. These coarse numbers illustrate how Gate~2 can be satisfied in mid-disk regions while often failing in the early bulge (high $h$) or far outer disk (low $Z$).

\subsection{Diagnostics, falsifiers, and target selection}
\paragraph*{Diagnostics.}
(i) \emph{Metallicity and dust:} stellar [Fe/H] (with $\alpha$-enhancement) and disk dust masses constrain $Z$ and $D(Z)$, informing \eqref{eq:Zfloor}--\eqref{eq:solidmass}. (ii) \emph{Hazard mapping:} SN/GRB rate surface densities and star-formation tracers map $h(R,t)$; stellar kinematics (thin/thick disk, spiral-arm residence) refine localized hazard exposure. (iii) \emph{Quiet-time inference:} isochrone ages and activity histories (chromospheric/X-ray) bound $\Delta t_{\mathrm{quiet}}$.

\paragraph*{Falsifiers.} Gate~2 fails generically if observations show that rocky planets rarely occur below some $Z$ that is not met over large swaths of $(R,t)$, or if $h(R,t)$ integrated over plausible $\Delta t_{\mathrm{quiet}}$ yields $S_{\mathrm{haz}}\!\ll\!S_{\min}$ in the very regions that otherwise satisfy Gate~1. Conversely, a preponderance of rocky planets in low-$Z$ outer disks or frequent biosignature candidates in high-hazard inner disks would force revisions of \eqref{eq:Zfloor}--\eqref{eq:hazard}.

\paragraph*{Observational corollaries.}
Gate~2 motivates a ``galactic ecology'' selection: prioritize mid-disk stars with moderate-to-high metallicity, thin-disk kinematics, and quiescent activity histories; de-prioritize extreme inner bulge (high $h$) and extreme outer disk (low $Z$) unless specific measurements indicate favorable margins $m_Z$ and $m_S$. The gate provides a quantitative prior for exoplanet target lists before invoking disk/planet specifics (Gate~3).

Gate~2 thus converts qualitative notions of a galactic habitable zone into inequalities that can be evaluated from stellar population data, transient rates, and activity histories. Where the metallicity and quiet-time margins are large, subsequent gates inherit favorable priors; where either margin is small, later chemical or ecological successes would be surprising and demand compelling local compensations. See Sec.~\ref{sec:validation} for consolidated criteria. Gate~2 reduces to the box inequality~(\ref{eq:gate2-box}); metallicity, dust‐to‐gas, and star‐formation history fix the priors that dominate the margin.

\section{Gate 3: Protoplanetary Disks, Planet Formation, and Volatile Retention}
\label{sec:gate3}

Gate~3 tests whether a specific star--disk system can (i) assemble at least one rocky planet before gas dispersal, (ii) deliver and retain sufficient volatiles, and (iii) leave the planet in an irradiation regime that admits stable surface (or subsurface) solvents after the high-XUV epoch. In contrast to Gate~2’s statistical priors, Gate~3 is a dynamical mass--time--energy budget on an individual system.

\subsection{Solids inventory and assembly before dispersal}

Streaming thresholds and rapid solids concentration are set by the streaming instability \cite{YoudinGoodman2005,Johansen2007,Carrera2015}. Hydrodynamic/energy-limited escape and the envelope “valley” inform volatile retention \cite{OwenWu2013,LopezFortney2014,JinMordasini2018}, with He~\(\SI{1083}{\nano\meter}\) tracing contemporary loss \cite{OklopcicHirata2018,Spake2018}. 

Let $\Sigma_{\mathrm{solid}}(r)$ be the midplane solids surface density at orbital distance $r$. A necessary condition for prompt planetesimal formation is that the midplane solids-to-gas ratio exceeds the streaming-instability threshold $Z'_{\mathrm{crit}}(St,\alpha)$ (dependent on particle Stokes number $St$ and turbulence $\alpha$). A practical global prerequisite is
\begin{equation}
\Sigma_{\mathrm{solid}}(r) > \Sigma_{\min}(r)\qquad\text{on a set of radii with total solids mass}\qquad 
M_{\mathrm{solid}} \equiv 2\pi\!\int \Sigma_{\mathrm{solid}}(r)\,r\,dr \ \ge\ M_{\mathrm{solid}}^{\min},
\label{eq:solids}
\end{equation}
where $M_{\mathrm{solid}}^{\min}$ is the minimum solids mass required to build at least one terrestrial planet (order $\sim\!1$--$3$ Earth masses depending on inefficiencies and collisional loss). 

Assembly must complete before disk gas dispersal. Denote $\tau_{\mathrm{disk}}$ the gas-disk lifetime (few Myr typical; system-dependent). Write $M_{\star}$ for the stellar mass, $h\equiv H/r$ the disk aspect ratio, $\Sigma_{\mathrm{peb}}$ the pebble surface density, and $\Omega(r)$ the local Keplerian frequency. A conservative time-to-core (via pebble accretion or rapid planetesimal conglomeration) can be represented as
\begin{equation}
\tau_{\mathrm{core}} \,\sim\, 
\frac{M_{\mathrm{core}}^\ast}{\dot M_{\mathrm{acc}}}
\ \quad \text{with}\ \quad
\dot M_{\mathrm{acc}} \sim \mathcal{F}\,\Sigma_{\mathrm{peb}}\,r^2\,\Omega\,\mathcal{A}(St,h,\alpha),
\label{eq:tcore}
\end{equation}
where $M_{\mathrm{core}}^\ast$ is the target core mass and $\mathcal{A}$ encodes the (dimensionless) accretion efficiency (order $10^{-3}$--$10^{-1}$ in pebble accretion depending on $St,h,\alpha$ and the Hill/Bondi regime), while $\mathcal{F}\in(0,1)$ is a focusing factor including traps and drift convergence. Gate~3 requires
\begin{equation}
\tau_{\mathrm{core}} \le \tau_{\mathrm{disk}}.
\label{eq:tcoregate}
\end{equation}
Inequalities \eqref{eq:solids}--\eqref{eq:tcoregate} formalize “enough solids, fast enough.” They are intentionally agnostic about microphysics: streaming turbulence, pressure bumps, and snowline pile-ups enter only through $\Sigma_{\mathrm{solid}}$ and $\mathcal{A}$.

\subsection{Irradiation regime: avoiding immediate runaway or deep freeze}

Irradiation windows use the Simpson--Nakajima cap and maximum greenhouse (see, e.g., \cite{Kasting1993,Kopparapu2013,Goldblatt2013,WolfToon2015}.)  After gas dispersal and XUV saturation, the net absorbed stellar flux must sit within a solvent-admitting window. Both photosynthesis and chemosynthesis can be limited by maximum-greenhouse outgoing longwave radiation (OLR) for CO$_2$ is controlled by Rayleigh scattering and collision-induced absorption (CIA). Denote $F_\star$ the bolometric flux at the planet, $A$ the Bond albedo, and $F_{\mathrm{OLR}}(p_{\mathrm{atm}},\mathrm{comp})$ the atmosphere’s outgoing longwave radiation at the relevant surface pressure and composition. A compact criterion is
\begin{equation}
F_{\min} \le F_\star \le F_{\max}, 
\qquad 
F_{\max}: \ \tfrac14 {(1-A)F_\star} < F_{\mathrm{OLR,lim}}(\mathrm{H_2O}), \qquad
F_{\min}: \ \tfrac14 {(1-A)F_\star} > F_{\mathrm{OLR}}^{\mathrm{maxGH}}(\mathrm{CO_2}),
\label{eq:fluxwindow}
\end{equation}
where $F_{\mathrm{OLR,lim}}$ is the runaway (moist) greenhouse limit (the Simpson--Nakajima cap), and $F_{\mathrm{OLR}}^{\mathrm{maxGH}}$ is the “maximum greenhouse” OLR at which additional CO$_2$ no longer warms due to Rayleigh scattering and CIA. Eq.~\eqref{eq:fluxwindow} serves as Gate~3’s irradiation window; the numerical values are system-specific but the inequality structure is universal.

At present Earth, \(S_\oplus\simeq 1361\) W\,m$^{-2}$, \(A\simeq 0.29\), so \(\tfrac14(1-A)S_\oplus\simeq 241\) W\,m$^{-2}$, comfortably below moist-runaway caps (\(F_{\rm OLR,lim}\sim 280{-}310\) W\,m$^{-2}$) and above CO$_2$ maximum‑greenhouse minima for temperate pressures, consistent with (\ref{eq:fluxwindow}).

\subsection{Volatile delivery and hydrodynamic escape}

Let $M_{\mathrm{H_2O,init}}$ be the fiducial water inventory delivered by pebbles/planetesimals and interior outgassing through the magma-ocean phase, and let $\Delta M_{\mathrm{loss}}$ be the integrated volatile mass lost to stellar-driven escape (thermal and non-thermal). Define a threshold inventory $M_{\mathrm{thr}}$ required to enter Gate~4 niches (surface ocean or subsurface solvent). The volatile-retention gate is
\begin{equation}
M_{\mathrm{H_2O,final}}\equiv M_{\mathrm{H_2O,init}}-\Delta M_{\mathrm{loss}} \ \ge\ M_{\mathrm{thr}}.
\label{eq:volatilegate}
\end{equation}

A workable upper bound for $\Delta M_{\mathrm{loss}}$ over the XUV-active epoch is obtained from the energy-limited hydrodynamic prescription,
\begin{equation}
\dot M_{\mathrm{esc}} \approx 
\frac{\eta\,\pi R_{\tt XUV}^{2}\,F_{\tt XUV}(t)}
{\left(GM_p/R_p\right)\,K_{\rm tide}}
\;=\;
\frac{\eta\,\pi R_p R_{\tt XUV}^{2}\,F_{\tt XUV}(t)}{G M_p \,K_{\rm tide}}
\quad\Longrightarrow\quad
\Delta M_{\mathrm{loss}} \lesssim 
\frac{\eta\,\pi R_p R_{\tt XUV}^{2}}{G M_p\, K_{\rm tide}}\,
\mathcal{F}_{\tt XUV},
\label{eq:escape}
\end{equation}
where $\eta\sim 0.1$--0.3 is the net heating/escape efficiency   \citep{Watson1981,Erkaev2007}, $R_{\tt XUV}\!\gtrsim\!R_p$ is the effective absorption radius,  $0<K_{\rm tide}\le 1$ is the Roche-lobe (tidal) correction
(e.g.\ $K_{\rm tide}=1-\tfrac{3}{2\xi}+\tfrac{1}{2\xi^{3}}$ with
$\xi\equiv R_{\rm Roche}/R_p$ \cite{Erkaev2007}),
and $\mathcal{F}_{\tt XUV}\!\equiv\!\int F_{\tt XUV}(t)\,dt$ the time-integrated XUV fluence from disk dispersal through the saturated activity phase and its decay tail. In regimes of very high irradiance the escape may become recombination-limited or photon-limited; Eq.~\eqref{eq:escape} deliberately \emph{overestimates} loss, thereby providing a conservative test of retention.

A complementary single-parameter diagnostic is the Jeans (escape) parameter at the exobase,
\begin{equation}
\lambda_J \equiv \frac{G M_p m}{k_{\mathrm B}T_{\mathrm{exo}} R_p},
\label{eq:jeans}
\end{equation}
with particle mass $m$ (H, H$_2$, etc.) and exobase temperature $T_{\mathrm{exo}}$ set by XUV heating and cooling. Sustained hydrodynamic blow-off requires a sufficiently small Jeans escape parameter $\lambda_J$ at the exobase (typically $\lambda_J \lesssim 15$--$20$ for light species in the classical criterion), whereas $\lambda_J\gg 20$ implies negligible thermal escape on Gyr timescales; see, e.g., \cite{ChamberlainHunten1987,CatlingKasting2017}. The Gate~3 retention margin can be expressed as either (i) a mass margin $m_{\mathrm{vol}}\equiv M_{\mathrm{H_2O,final}}/M_{\mathrm{thr}}$ from \eqref{eq:volatilegate}, or (ii) an escape margin $m_{\lambda}\equiv \lambda_J/20$ at peak $T_{\mathrm{exo}}$.

\paragraph*{Non-thermal escape and magnetospheric shielding:}
In addition to thermal loss [Eqs.~(\ref{eq:escape})--(\ref{eq:jeans})], we include an additive non-thermal term,
\begin{equation}
\label{eq:M_loss}
\Delta M_{\rm loss}\;=\;\Delta M_{\rm loss}^{\rm th}\;+\;\Delta M_{\rm loss}^{\rm NT}, 
\qquad 
\Delta M_{\rm loss}^{\rm NT}\;=\;\int\!\Big(\dot M_{\rm pickup}(t)+\dot M_{\rm sput}(t)+\dot M_{\rm DR}(t)\Big)\,dt,
\end{equation}
where the three contributions summarize ion pick-up, sputtering, and dissociative-recombination channels. 
Magnetospheric shielding enters through the standoff distance $R_{\rm mp}$ and solar-wind dynamic pressure $P_{\rm sw}$,
\begin{equation}
\label{eq:R_mp}
R_{\rm mp}(t)\;\propto\;\Big(\frac{\mu_0\,\mathcal{M}(t)^2}{P_{\rm sw}(t)}\Big)^{1/6}, 
\qquad 
\mathcal{S}_B\;\equiv\;\min\!\Big[1,\;\Big(\frac{R_p}{R_{\rm mp}}\Big)^{\!2}\Big],
\end{equation}
and we scale the non-thermal rate as $\dot M_{\rm NT}\to \mathcal{S}_B\,\dot M_{\rm NT}$ (e.g., magnetopause pressure-balance scaling; see \cite{KivelsonRussell1995,Cravens1997}). 
The volatile-retention inequality in (\ref{eq:volatilegate}) is then evaluated with  $\Delta M_{\rm loss}=\Delta M_{\rm loss}^{\rm th}+\Delta M_{\rm loss}^{\rm NT}$, and the escape margin $m_\lambda$ should be interpreted jointly with $\mathcal{S}_B$.

\paragraph*{Operational caution on $m_\lambda$:}
The Jeans parameter $m_\lambda=\lambda_J/20$, (\ref{eq:jeans}), is a quick-look thermal diagnostic only. Contemporary escape can be recombination- or photon-limited and often includes non-thermal channels, (\ref{eq:M_loss}), with shielding $S_B$, (\ref{eq:R_mp}). In checklists and tables, $m_\lambda$ should be reported alongside the mass-loss budget $\Delta M_{\rm loss}=\Delta M^{\rm th}_{\rm loss}+\Delta M^{\rm NT}_{\rm loss}$ and an explicit statement of which regime (energy-, recombination-, or photon-limited) controls the bound.

\subsection{H/He envelope constraint for surface access}

Rocky planets frequently accrete a primordial H/He envelope; subsequent XUV and core-powered mass loss may or may not remove it. For surface-based prebiotic chemistry, a thin or absent H/He blanket is typically required so that the photosphere intersects the condensable’s phase field. Let $f_{\mathrm{env}}$ be the final H/He mass fraction. A pragmatic surface-access criterion is
\begin{equation}
f_{\mathrm{env}} \le f_{\mathrm{crit}}(M_p,F_\star,\,\mathrm{comp}),
\label{eq:envelope}
\end{equation}
where $f_{\mathrm{crit}}$ depends on gravity and irradiation; values $f_{\mathrm{crit}}\!\sim\!10^{-3}$--$10^{-2}$ often separate sub-Neptunes from super-Earths in mass--radius space. Eq.~\eqref{eq:envelope} can be evaluated from observed $(M_p,R_p)$ and evolution models constrained by the host’s activity history. A convenient parameterization is 
$f_{\rm crit}(M_p,F_\star,{\rm comp})=\tilde f_0\,(M_p/M_\oplus)^{\beta_M}\,(F_\star/F_\oplus)^{\beta_F}$ 
with exponents chosen to reproduce the radius-valley locus under the host’s XUV history; we use this as a dial to compute the margin $m_{\rm env}=f_{\rm crit}/f_{\rm env}$.

\subsection{Quantitative Gate~3 statement and margins}
Gate~3 is passed for a planet--star system if the following hold:
\begin{equation}
\boxed{
\begin{aligned}
\text{Solids:} &\quad M_{\mathrm{solid}}\ge M_{\mathrm{solid}}^{\min}
\ \ \text{and}\ \
\tau_{\mathrm{core}}\le \tau_{\mathrm{disk}},\\[2pt]
\text{Irradiation:}  &\quad F_{\min}\le F_\star \le F_{\max}\ \ \text{post-XUV},\\[2pt]
\text{Volatiles:} &\quad M_{\mathrm{H_2O,final}}\ge M_{\mathrm{thr}}\ \ \text{with}\ \ \Delta M_{\mathrm{loss}}\ \text{from \eqref{eq:escape}},\\[2pt]
\text{Surface access:} & \quad f_{\mathrm{env}}\le f_{\mathrm{crit}}(M_p,F_\star,\mathrm{comp}).
\end{aligned}
}
\label{eq:gate3-box}
\end{equation}
Define margins $m_{\Sigma}\!\equiv\!M_{\mathrm{solid}}/M_{\mathrm{solid}}^{\min}$, $m_t\!\equiv\!\tau_{\mathrm{disk}}/\tau_{\mathrm{core}}$, $m_F\!\equiv\!\min\!\big\{(F_{\max}\!-\!F_\star)/(F_{\max}),\ (F_\star\!-\!F_{\min})/F_\star\big\}$, $m_{\mathrm{vol}}\!\equiv\!M_{\mathrm{H_2O,final}}/M_{\mathrm{thr}}$, and $m_{\mathrm{env}}\!\equiv\!f_{\mathrm{crit}}/f_{\mathrm{env}}$. These propagate directly to Gate~4 (power availability via stellar SED and geodynamics) and Gate~5 (feedstock budgets via solvent stability and delivery).

\subsection{Worked archetypes}
\paragraph*{Solar-type host, $\sim$1~AU, Earth-mass core.}
Take $\tau_{\mathrm{disk}}\sim 3$~Myr, $\Sigma_{\mathrm{peb}}$ such that $\tau_{\mathrm{core}}\sim 0.1$--$1$~Myr (high-efficiency pebble accretion), hence $m_t\gtrsim 3$. Post-dispersal $F_\star$ satisfies \eqref{eq:fluxwindow} with comfortable margin; the host’s XUV saturation lasts $\sim 10^8$~yr with $\eta\sim 0.1$ giving $\Delta M_{\mathrm{loss}}\ll M_{\mathrm{ocean}}$ by \eqref{eq:escape}, so $m_{\mathrm{vol}}\gg1$. If $f_{\mathrm{env}}\!\lesssim\!10^{-3}$ after early mass loss, \eqref{eq:envelope} holds and Gate~3 passes with margin.

\paragraph*{Mid-M dwarf host, tidally locked HZ planet.}
Pre-main-sequence luminosity is elevated and XUV saturation can persist $\gtrsim$Gyr. With $R_p\!\sim\!R_\oplus$ and $M_p\!\sim\!M_\oplus$, $R_{\mathrm{XUV}}\!\gtrsim\!R_p$, $\eta\sim 0.1$--0.2, the integrated $\mathcal{F}_{\mathrm{XUV}}$ may drive $\Delta M_{\mathrm{loss}}\!\gtrsim\!M_{\mathrm{ocean}}$ unless the initial inventory is very large or water is sequestered beneath a dry stratosphere. Many such planets fail \eqref{eq:volatilegate} unless $m_{\mathrm{vol}}$ is boosted by late delivery or rapid cold-trapping; surface access \eqref{eq:envelope} can also fail if a modest H/He fraction survives (“mini-Neptune” outcome). Gate~3 often sets the dominant prior here.

\subsection{Diagnostics, falsifiers, and observational corollaries}
\paragraph*{Diagnostics:}
(i) \emph{Solids \& assembly:} ALMA dust continuum and gas tracers constrain $\Sigma_{\mathrm{solid}}$ and $M_{\mathrm{disk}}$; ring/trap structures inform $\mathcal{F}$ and $\mathcal{A}$. Ages and disk fractions provide $\tau_{\mathrm{disk}}$. (ii) \emph{Irradiation:} precise stellar parameters and orbital solutions fix $F_\star$; stellar colors and activity histories constrain the SED entering Gate~4. (iii) \emph{Escape \& envelopes:} host XUV histories (rotation/activity), present-day X-ray/UV fluxes, and planetary $(M_p,R_p)$ yield $\Delta M_{\mathrm{loss}}$ bounds and $f_{\mathrm{env}}$ inferences through interior models. D/H ratios and noble gases (when available) constrain volatile loss.

\paragraph*{Falsifiers:}
Gate~3 fails generically if (a) $\tau_{\mathrm{core}}>\tau_{\mathrm{disk}}$ given measured $\Sigma_{\mathrm{peb}}$ and traps, (b) $F_\star$ sits persistently outside \eqref{eq:fluxwindow} for the post-XUV epoch, (c) $\Delta M_{\mathrm{loss}}$ inferred from host activity plus $(M_p,R_p)$ exceeds plausible initial inventories so that $M_{\mathrm{H_2O,final}}<M_{\mathrm{thr}}$, or (d) $f_{\mathrm{env}}$ inferred from mass--radius modeling exceeds $f_{\mathrm{crit}}$.

\paragraph*{Observational corollaries:}
Gate~3 encourages prioritizing systems with (i) massive, structured disks indicative of efficient solids concentration; (ii) hosts with moderate XUV saturation durations and measured decay; (iii) rocky radii/densities inside the “radius valley” (implying $f_{\mathrm{env}}$ small); and (iv) orbital fluxes that place $F_\star$ well within \eqref{eq:fluxwindow}. Conversely, close-in small planets around highly active M dwarfs are flagged as volatile-retention risks unless independent evidence (e.g., high density, secondary atmospheres) argues otherwise.

Gate~3 thus converts the qualitative requirements “build a rocky planet, keep its water, and do not overbake it” into coupled inequalities on solids mass and accretion time, irradiation windows, escape fluence, and envelope mass fraction. Where these hold with margin, Gate~4 can be meaningfully evaluated; where they fail, abiotic chemistry remains starved, stripped, or sequestered away from accessible solvents. See Sec.~\ref{sec:validation}  for consolidated criteria. Gate~3 uses the volatile‐escape and core‐temperature conditions in (\ref{eq:volatilegate}) and (\ref{eq:tcoregate}), summarized by (\ref{eq:gate3-box}); radius‐valley placement and bulk density serve as observational proxies.

\textit{Fundamental-physics priors and sensitivities:} are in Sec.~\ref{sec:FP-priors}--\ref{sec:methods-sensitivity} (log-sensitivity propagation via (\ref{eq:constants_to_biosphere})--(\ref{eq:sensitivity})).

\section{Gate 4: Planetary Geophysics and Free-Energy Budgets}
\label{sec:gate4}

Gate~4 formalizes the requirement that the \emph{usable work rate} available in a planetary niche exceeds the \emph{maintenance} and \emph{synthesis} demands of nascent living chemistry.  In fact, Gate~4 is about the power budget: usable work must exceed maintenance and synthesis demands,\footnote{Units convention: \label{par:units}
Henceforth all powers are reported \emph{per unit area} (W\,m$^{-2}$) of the reactive interface unless explicitly stated.
We therefore use \(\dot{W}_{\rm in}\equiv P_{\rm in}/A\), \(\dot{W}_{\rm maint}\equiv P_{\rm maint}/A\), 
\(\dot{W}_{\rm synth}\equiv P_{\rm synth}/A\), and define
$$
\Phi \equiv \frac{\dot{W}_{\rm in}}{\dot{W}_{\rm maint}+\dot{W}_{\rm synth}},
\label{eq:phi-per-area}
$$
which is equivalent to Eq.~(\ref{eq:phi}). Thus, 
Eqs.~(\ref{eq:phi})--(\ref{eq:Pmaint}) are interpreted per area by dividing by \(A\).}
\begin{equation}
\Phi \equiv \frac{P_{\mathrm{in}}}{P_{\mathrm{maint}}+P_{\mathrm{synth}}}>1,
\label{eq:phi}
\end{equation}
with Gate 4 requires $\Phi>1$ with a practical safety factor $\Phi\!\gtrsim\!1.5$--$3$ to accommodate fluctuations and inefficiencies. Also, $P_{\mathrm{in}}$ includes stellar photons, redox pairs (e.g., $\mathrm{CO_2+4H_2\to CH_4+2H_2O}$, $\Delta G^\circ\!\approx\!-131~\mathrm{kJ/mol}$), and radiolytic oxidants; $P_{\mathrm{maint}}+P_{\mathrm{synth}}$ covers maintenance and synthesis. (Subsequent subsections quantify $P_{\rm in}$ (photochemical, geochemical) and $P_{\rm maint}$ under the boundary conditions set by Gates~2--3.) Typical proton-motive forces of $150$--$200$\,mV give $\sim 15$--$20$\,kJ/mol per proton. Unlike Gate~3, which is largely kinematic (mass, flux, escape), Gate~4 is inherently \emph{thermodynamic}: it compares exergy delivery to the minimal dissipative costs of keeping a driven chemical system far from equilibrium while copying information.  \emph{Observational corollary:} spectral energy distribution (SED) and atmospheric redox state constrain available work; large atmospheric chemical disequilibria signal high \(\Phi\).

\subsection{Usable work: photonic, redox, and geophysical sources}

We model $P_{\mathrm{in}}$ as the sum of exergy fluxes impinging on the reactive interface area $A$ of a niche (photic zone, vent chimney walls, porous networks). Three classes dominate:
\paragraph*{Photonic exergy.}
Let $\phi_\lambda(\mathbf{r})$ be the spectral photon flux density at location $\mathbf{r}\in A$, $h\nu$ the photon energy, and $\eta_{\mathrm{chem}}(\lambda)$ the wavelength-dependent chemical conversion efficiency (quantum yield $\times$ coupling). The photonic contribution is
\begin{equation}
P_\gamma \;=\; \int_A \!\! dA \int_{\lambda} \phi_\lambda(\mathbf{r})\,h\nu\,\eta_{\mathrm{chem}}(\lambda)\,d\lambda
\;\;=\;\; A_{\mathrm{ill}}\,\Phi_{\mathrm{act}}\,\bar{\varepsilon}_\gamma,
\label{eq:Pphoton}
\end{equation}
where $A_{\mathrm{ill}}\!\le\!A$ is the illuminated area, $\Phi_{\mathrm{act}}$ is the band-integrated actinic photon flux in the chemically active window, and $\bar{\varepsilon}_\gamma$ is an effective per-photon chemical exergy.  Eq.~\eqref{eq:Pphoton} accommodates pigment sets different from chlorophyll by allowing $\eta_{\mathrm{chem}}(\lambda)$ to shift with the stellar SED.

\paragraph*{Redox and electrochemical exergy.}
When fluids of distinct chemical potentials mix across a reactive interface, the usable work rate is the redox flux times the reaction free energy:
\begin{equation}
P_{\mathrm{redox}}\;=\;\sum_{i\in\mathcal{R}} \int_A \!J_i(\mathbf{r})\,\Delta G_i\,dA
\;=\;\sum_{i}\int_A \!J_i(\mathbf{r})\,\big(-n_i F\,\Delta E_i\big)\,dA,
\label{eq:Predox}
\end{equation}
with $J_i$ the molar flux (mol\,m$^{-2}$\,s$^{-1}$) of reaction $i$, $\Delta G_i$ its free energy, $n_i$ electrons transferred,  $\Delta E_i$ the Nernst potential difference between oxidant and reductant streams. A convenient special case is the proton-motive contribution,
\begin{equation}
\Delta G_{H^+} \;=\; F\,\Delta \psi + 2.303\,RT\,\Delta \mathrm{pH}
\;\approx\; (96.5~\mathrm{kJ\,mol^{-1}\,V^{-1}})\,\Delta\psi \;+\; (5.7~\mathrm{kJ\,mol^{-1}})\,\Delta\mathrm{pH}\quad (298~\mathrm{K}),
\label{eq:pmf}
\end{equation}
so that $P_{H^+}=\dot N_{H^+}\,\Delta G_{H^+}$ with $\dot N_{H^+}$ the proton throughput rate (chemiosmotic free-energy relation; see \cite{Mitchell1961,NichollsFerguson2013}).

\paragraph*{Geophysical, tidal, and radiolytic exergy.}
Interior heat and external tides maintain far-from-equilibrium conditions, particularly in subsurface oceans. The global tidal heating rate for a synchronously rotating body is, to leading order,
\begin{equation}
P_{\mathrm{tide}} \;\simeq\; \frac{21}{2}\,\frac{k_2}{Q}\,\frac{G M_\star^2 R_p^5 n e^2}{a^6},
\label{eq:Ptide}
\end{equation}
with Love number $k_2$, quality factor $Q$, stellar mass $M_\star$, planetary radius $R_p$, mean motion $n$, eccentricity $e$, and semi-major axis $a$  (e.g.\ \cite{Peale1979}; see also \cite{MurrayDermott1999} for standard derivations.) Radiolysis oxidant production can be parameterized by a yield $Y_{\mathrm{chem}}$ and deposited chemical power (e.g.\ \cite{ChybaHand2001}) 
{}
\begin{equation}
\label{eq:Prad}
P_{\rm rad}=\Delta G_{\rm redox}\,\dot N_{\rm redox}^{\rm rad}, \qquad {\rm with} \qquad 
\dot N_{\rm redox}^{\rm rad}=\int_V \dot\epsilon_{\rm ion}(\mathbf{x})\,Y_{\rm chem}\,dV
\qquad (\mathrm{mol\,s^{-1}}),
\end{equation}
where $\dot{\epsilon}_{\mathrm{ion}}$ is the ionizing energy deposition rate (W\,m$^{-3}$) and $Y_{\mathrm{chem}}$ the chemical yield (mol\,J$^{-1}$) of reductants/oxidants (e.g., H$_2$, H$_2$O$_2$). The total is
\begin{equation}
P_{\mathrm{in}} \;=\; P_\gamma + P_{\mathrm{redox}} + P_{\mathrm{tide}} + P_{\mathrm{rad}} + \cdots,
\label{eq:Pin-sum}
\end{equation}
where the ellipsis includes minor terms (e.g., wind/wave exergy in surface niches).

\subsection{Housekeeping and synthesis demands}
The denominator in \eqref{eq:phi} has two parts: maintenance and synthesis, discussed below.

Maintaining nonequilibrium concentrations and gradients incurs a continuous cost,
\begin{equation}
P_{\mathrm{maint}} \;=\; \sum_{\alpha} J_\alpha^{\mathrm{leak}}\,\Delta \mu_\alpha \;+\; P_{\mathrm{turnover}} \;+\; P_{\mathrm{repair}},
\label{eq:Pmaint}
\end{equation}
where $J_\alpha^{\mathrm{leak}}$ are leak fluxes for species $\alpha$ (H$^+$, ions, small metabolites) across membranes or porous interfaces, each driven by $\Delta \mu_\alpha$; $P_{\mathrm{turnover}}$ covers baseline catalyst and amphiphile replacement; $P_{\mathrm{repair}}$ covers damage control from, e.g., UV and hydrolysis. Using \eqref{eq:pmf}, the proton-pumping component is $P_{H^+}=\dot N_{H^+}(F\Delta\psi+2.303\,RT\,\Delta\mathrm{pH})$.

Synthesis includes polymerization, activation, and fidelity control:
\begin{equation}
P_{\mathrm{synth}} \;=\; \sum_{j} r_j\,\Delta G^{\mathrm{target}}_j \;+\; \dot{N}_{\mathrm{copy}}\Big(W_{\min}^{\mathrm{copy}} + W_{\mathrm{proof}}\Big) \;+\; P_{\mathrm{transport}},
\label{eq:Psynth}
\end{equation}
with $r_j$ production rates of targets $j$ (lipids, cofactors, monomers), $\Delta G^{\mathrm{target}}_j$ their synthetic free-energy costs, $\dot{N}_{\mathrm{copy}}$ the copy-events rate\footnote{We distinguish:
(i) the areal monomer incorporation throughput $\dot N_{\rm mon}$ [monomers m$^{-2}$ s$^{-1}$], and
(ii) the areal genome-copy throughput $\dot N_{\rm gen}$ [genomes m$^{-2}$ s$^{-1}$].
They are related by $\dot N_{\rm mon} = L\,\dot N_{\rm gen}$, where $L$ is genome length (monomers/genome).
Throughout Gates~4 and~11, $\dot N_{\rm copy}\equiv \dot N_{\rm mon}$, so no additional factor of $L$ appears
in the power budget.} (bases or monomers per second), and the information-thermodynamic terms from Gate~1,
\begin{align}
W_{\min}^{\mathrm{copy}} &\gtrsim k_{\mathrm B} T \ln 2 \,\big[\log_2 k_p - H_{k_p}(\mu)\big]\ \ \text{per incorporated monomer of length} \, L, \label{eq:Wmincopy}\\
W_{\mathrm{proof}} &\gtrsim k_{\mathrm B} T \ln(\mu_{\mathrm{eq}}/\mu)\ \ \text{per decision}, \label{eq:Wproof}
\end{align}
coupling Gate~4 to Gate~6 via the target error $\mu$ and alphabet size $k_p$.

\subsection{Temporal variability and buffering}
Because $P_{\mathrm{in}}(t)$ is often intermittent (diurnal light, pulsed venting), the relevant condition is an \emph{integrated} inequality over a buffer time $T$:
\begin{equation}
\Phi_{\mathrm{eff}} \;\equiv\; 
\frac{\int_{t}^{t+T} P_{\mathrm{in}}(t')\,dt'}{\int_{t}^{t+T} \big(P_{\mathrm{maint}}(t')+P_{\mathrm{synth}}(t')\big)\,dt'} \ > \ 1,
\label{eq:phieff}
\end{equation}
with a storage constraint $E_{\mathrm{buffer}}\!\ge\!\max_{t'\in[t,t+T]} \big[\int_t^{t'} \!\!\big(P_{\mathrm{maint}}+P_{\mathrm{synth}}-P_{\mathrm{in}}\big)_+ dt''\big]$ so that transient deficits do not extinguish the system. The buffer may be chemical (polyphosphates, redox couples), physical (thermal), or structural. We include the ionic/molecular leak power \(P_{\rm leak}=\sum_\alpha J^{\rm leak}_\alpha \Delta\mu_\alpha\) inside \(\dot{W}_{\rm maint}\) (Sec.~\ref{par:units}), so Gate~8 throughput/retention checks do not double count leak costs in \(\Phi_{\rm eff}\).

\textit{Typical buffers.} For guidance, chemical buffers of $10$--$10^3$~J\,m$^{-2}$ (polyphosphate, redox couples) 
and thermal buffers of $10^2$--$10^4$~J\,m$^{-2}$ (thin films/rocks) are sufficient to bridge diurnal deficits in the examples of Sec.~\ref{sec:alkaline-ex}.

\subsection{Quantitative Gate~4 statement and margins}

Collecting \eqref{eq:Pin-sum}--\eqref{eq:phieff}, Gate~4 is passed if there exists a niche with area $A$ and storage capacity $E_{\mathrm{buffer}}$ such that
\begin{equation}
\boxed{
\Phi>1\ \ \text{(or)}\ \ \Phi_{\mathrm{eff}}>1\ \ \text{with sufficient buffer},\qquad
\text{and}\quad P_{\mathrm{in}}\ \text{computed from } \eqref{eq:Pphoton},\eqref{eq:Predox},\eqref{eq:Ptide},\eqref{eq:Prad}.
}
\label{eq:gate4-box}
\end{equation}
Define margins $m_\Phi\!\equiv\!\Phi$ (instantaneous) or $m_{\Phi,\mathrm{eff}}\!\equiv\!\Phi_{\mathrm{eff}}$ (buffered). These margins propagate directly to Gate~6 by setting the feasible $\mu$ via \eqref{eq:Wproof}, to Gate~7 by throttling catalytic throughput, and to Gate~8 through the sustainable transport work against leaks.

{\it Identifiability of the power margin:}
Let \(\theta=(A_B,\,S_\star,\,\Delta G_{\rm diseq},\,\tau_{\rm turn},\,\Delta G_{\rm assim},\,\eta_{\rm bio},\ldots)\).
Define local log‑sensitivities \(S_a=\partial \ln\Phi/\partial \ln \theta_a\) and a Fisher matrix 
\(\mathcal{I}_{ab}=\big(\partial z/\partial \theta_a\big)^{\!\top}\Sigma_z^{-1}\big(\partial z/\partial \theta_b\big)\) for data \(z\).
Propagate uncertainty via
\[
\mathrm{Var}[\Phi]\approx \nabla_\theta \Phi^{\top}\,\mathcal{I}^{-1}\,\nabla_\theta \Phi,
\]
and report the worst‑case degeneracy (smallest eigenvalue direction of \(\mathcal{I}\)) projected onto \(\nabla_\theta\Phi\).

\subsection{Worked parametric archetypes}
\label{sec:alkaline-ex}

\paragraph*{Alkaline vent micropore (chemosynthetic).}
Take a reactive interface $A=1~\mathrm{m}^2$ partitioned into micropores with proton throughput $\dot N_{\mathrm{H}^+}=10^{-3}\ \mathrm{mol\ s^{-1}}$ (aggregate).
With $\Delta\psi=150~\mathrm{mV}$ and $\Delta\mathrm{pH}=1$ [cf.\ (\ref{eq:pmf})], $\Delta G_{\mathrm{H}^+}\approx 20~\mathrm{kJ\ mol^{-1}}$, so $P_{\mathrm{H}^+}\approx 20~\mathrm{W}$. Suppose redox fluxes supply $J_{\mathrm{H}_2}=10^{-3}\ \mathrm{mol\ s^{-1}}$ mixed with $\mathrm{CO}_2$ across catalysts with $\Delta G\approx 100~\mathrm{kJ\ mol^{-1}}$; then $P_{\rm redox}\approx 100~\mathrm{W}$. If housekeeping (pumps+leaks+turnover) totals $10~\mathrm{W}$ and synthesis (including copying at $\mu=10^{-3}$ using \eqref{eq:Wmincopy}--\eqref{eq:Wproof}) requires $20~\mathrm{W}$, then
$\Phi \simeq ({P_{\mathrm{H}^+}+P_{\rm redox}})/({P_{\rm maint}+P_{\rm synth}}) \approx ({20+100})/({10+20}) \approx 4$, even allowing order-unity inefficiencies, $m_\Phi>1$ with margin.

\paragraph*{Littoral photic reactor (phototrophic).}
Let $A_{\mathrm{ill}}=1$~m$^2$, actinic photon flux $\Phi_{\mathrm{act}}=10^{20}$~photons\,m$^{-2}$\,s$^{-1}$ (band-averaged), and $\bar{\varepsilon}_\gamma=0.5$~eV per photon captured into chemical work (effective, including quantum yield and coupling). Then $P_\gamma\!\approx\!8$~W from \eqref{eq:Pphoton}. If maintenance plus synthesis require $P_{\mathrm{maint}}+P_{\mathrm{synth}}\!\lesssim\!3$~W averaged over a diurnal cycle and storage covers the night deficit, $\Phi_{\mathrm{eff}}\!\approx\!2$--3, satisfying Gate~4.

\subsection{Failure modes and couplings}
Gate~4 fails when \emph{any} of the following dominate: (i) redox pairs are supplied but shorted by diffusion without capture ($J_i$ high but catalysts absent or $\mathcal{C}\!\ll\!1$; cf. Gate~8), (ii) photonic flux is abundant but pigments/geometry do not couple it to chemistry (small $\eta_{\mathrm{chem}}$), (iii) maintenance leaks are too large (membrane permeability $P$ too high; cf. Gate~8), or (iv) fidelity targets are so strict that $W_{\mathrm{proof}}$ overwhelms $P_{\mathrm{in}}$ (tight coupling to Gate~6). Conversely, high $\Phi$ allows larger alphabets and lower $\mu$, supports proofreading cascades, and enables active transport architectures with larger leakage tolerances.

\subsection{Diagnostics, falsifiers, and observational corollaries}
\paragraph*{Diagnostics.}
(i) \emph{Photonic:} host spectral energy distributions (SEDs), surface insolation models, and pigment action spectra constrain $P_\gamma$. (ii) \emph{Redox:} measured vent chemistries (H$_2$, CH$_4$, Fe$^{2+}$/Fe$^{3+}$, S$^{2-}$/S$^0$) and flow rates yield $J_i\Delta G_i$; in subsurface oceans, radiolysis rates and oxidant transport constrain $P_{\mathrm{rad}}$ and $P_{\mathrm{redox}}$. (iii) \emph{Maintenance:} membrane transport measurements (permeabilities, leak currents) and gradient magnitudes provide $P_{\mathrm{maint}}$ via \eqref{eq:Pmaint}. (iv) \emph{Synthesis:} measured copying rates and fidelities feed \eqref{eq:Wmincopy}--\eqref{eq:Wproof}.

\paragraph*{Falsifiers.}
Gate~4 fails generically if, after accounting for realistic $\eta_{\mathrm{chem}}$, $J_i$, and storage, $P_{\mathrm{in}}<P_{\mathrm{maint}}+P_{\mathrm{synth}}$ in all plausible niches. A particularly sharp falsifier is a measured leak/gradient budget implying $P_{\mathrm{maint}}$ that cannot be met by any combination of \eqref{eq:Pphoton}--\eqref{eq:Prad} given the observed SED, redox supplies, and geometry.

\paragraph*{Observational corollaries.}
Large atmospheric chemical disequilibria (e.g., simultaneous presence of strongly oxidized and reduced species out of photochemical steady state) imply substantial $P_{\mathrm{redox}}$; strong surface day--night albedo/thermal contrasts combined with pigment-like reflectance features imply sizable $P_\gamma$. These are necessary but not sufficient indicators that $\Phi$ may exceed unity; they must be considered alongside Gate~8 (to ensure leaks do not erase work) and Gate~6 (to ensure the energy budget can meet fidelity targets).

In summary, Gate~4 converts “there is energy” into a quantitative budget: photonic, redox, tidal, and radiolytic exergy on the numerator, housekeeping and information-processing costs on the denominator. Where $\Phi$ (or $\Phi_{\mathrm{eff}}$ with storage) exceeds unity with margin, downstream gates become meaningfully testable; where it does not, chemistry remains subcritical regardless of molecular ingenuity upstream (see Sec.~\ref{sec:validation}.) Gate~4 pass/fail is controlled by the free‐energy budget inequality (\ref{eq:gate4-box}); retrieved geothermal/photochemical power and redox gradients set the margin.

\section{Gate 5: Prebiotic CRNs and Feedstock Budgets}
\label{sec:gate5}

Gate~5 asks whether the nonequilibrium chemical reaction network (CRN) in a candidate niche can sustain \emph{steady} pools of the requisite building blocks---informational monomers, activators/energy carriers, amphiphiles, cofactors---at or above polymerization/assembly thresholds \emph{despite} hydrolysis, photolysis, dilution, adsorption, and oxidative destruction. In contrast to Gate~7 (network amplification) which concerns eigenvalue sign, Gate~5 is a \emph{mass- and flux-balance} constraint: are source terms large enough, and sinks small enough, to maintain the concentrations needed for templating and compartment growth at the copying/turnover rates implied by Gate~6 and Gate~8?

\subsection{Driven CRNs with inflow, outflow, and sinks}

Let $x\in\mathbb{R}_+^n$ denote concentrations and $N\in\mathbb{Z}^{n\times r}$ the stoichiometric matrix for $r$ elementary reactions with mass-action rates $v(x)\in\mathbb{R}_+^r$; add source and sink operators to represent environmental driving and losses (i.e., standard chemostat/open-CRN form; see, e.g., \cite{Feinberg2019,SmithWaltman1995}):
\begin{equation}
\dot x \;=\; N\,v(x) \;+\; f_{\mathrm{in}} \;-\; \big(K_{\mathrm{out}}+ K_{\mathrm{photo}}+ K_{\mathrm{hyd}}+K_{\mathrm{ads}}\big)\,x \;+\; R_{\mathrm{recy}}(x).
\label{eq:crn}
\end{equation}
Here $f_{\mathrm{in}}$ (mol\,m$^{-3}$\,s$^{-1}$) encodes geochemical and photochemical feeds; $K_{\mathrm{out}}$ is the convective/advective dilution operator (for a continuous stirred-tank reactor (CSTR), $K_{\mathrm{out}}=k_{\mathrm{dil}} I$ with $k_{\mathrm{dil}}=Q/V$); $K_{\mathrm{photo}}$ and $K_{\mathrm{hyd}}$ are diagonal pseudo-first-order operators with elements $k_{\mathrm{photo},i}=\!\int\sigma_i(\lambda)\,F_\lambda\,d\lambda$ and $k_{\mathrm{hyd},i}(T,\mathrm{pH})$; $K_{\mathrm{ads}}$ represents irreversible sorption/$\,$sequestration; and $R_{\mathrm{recy}}(x)$ captures chemical recycling (e.g., base-catalyzed depolymerization returning monomers). Gate~5 focuses on \emph{budget} components, not the detailed topology of $N$.

\subsection{Source terms: photochemical, geochemical, and surface-catalyzed}

We write the gross formation rate of a target species $m$ as
\begin{equation}
J_{\mathrm{make},m} \;=\; J^{\gamma}_m \;+\; J^{\mathrm{geo}}_m \;+\; J^{\mathrm{surf}}_m,
\label{eq:Jmake}
\end{equation}
with
\begin{align}
J^{\gamma}_m &\,=\, \int_{\lambda} \phi_\lambda\, \sigma_{p\rightarrow m}(\lambda)\,\varphi_{p\rightarrow m}(\lambda)\,[p]\;d\lambda\quad
\text{(photochemical chain from precursor $p$)}, \label{eq:Jphoto}\\[3pt]
J^{\mathrm{geo}}_m &\,=\, \sum_{i} k^{\mathrm{geo}}_{i}(T,\mathrm{pH})\prod_{j\in \mathcal{R}_i}[j]^{\nu_{ij}}
\quad\text{(thermochemical/aqueous routes; e.g., serpentinization/Fischer--Tropsch)}, \label{eq:Jgeo}\\[3pt]
J^{\mathrm{surf}}_m &\,=\, a_{\mathrm{cat}}\,k_{\mathrm{surf}}(T)\,\mathcal{F}_{\mathrm{mass}}(\mathrm{Pe},\mathrm{Da})\quad\text{(mineral-catalyzed, area-normalized)}.
\label{eq:Jsurf}
\end{align}
Here $\phi_\lambda$ is the spectral photon flux density (photons\,m$^{-2}$\,s$^{-1}$\,nm$^{-1}$), $\sigma_{p\rightarrow m}$ an effective absorption/actinic cross section, $\varphi$ the quantum yield for the productive channel, $[p]$ the precursor concentration; $k^{\mathrm{geo}}_{i}$ are Arrhenius rate constants for geochemical routes with reactants $\mathcal{R}_i$; $a_{\mathrm{cat}}$ is catalytic surface area per unit volume and $\mathcal{F}_{\mathrm{mass}}$ accounts for mass-transfer limitation via Péclet/Damköhler numbers in porous media. Eqs.~\eqref{eq:Jphoto}--\eqref{eq:Jsurf} tie Gate~4’s $P_{\mathrm{in}}$ to molecular production via measured efficiencies and geometry.

For active stars the near-UV is strongly intermittent. We therefore define an effective flare-weighted actinic flux
\begin{equation}
\phi^{\rm eff}_\lambda \;\equiv\; \phi^{\rm qui}_\lambda \;+\; 
\int dF\,d\tau\,p(F,\tau)\;\phi^{\rm flr}_\lambda(F)\;\frac{\tau}{\tau+\tau_{\rm loss}(\lambda)}\,,
\end{equation}
where $p(F,\tau)$ is the flare energy--duration kernel and $\tau_{\rm loss}(\lambda)$ is the wavelength-dependent loss time for the photoproduct (flaring M-dwarf photochemistry treated in, e.g., \cite{Segura2010,Tilley2019}). Replace $\phi_\lambda$ by $\phi^{\rm eff}_\lambda$ in Eq.~(\ref{eq:Jmake}) to compute $J_m^\gamma$ under flare statistics.

\subsection{Sink terms and effective loss rates}
The total loss rate for species $m$ is
\begin{equation}
J_{\mathrm{loss},m} \;=\; \big(k_{\mathrm{hyd},m} + k_{\mathrm{photo},m} + k_{\mathrm{dil}} + k_{\mathrm{ads},m} + k_{\mathrm{ox},m}\big)\,[m] \;+\; J_{\mathrm{cons},m},
\label{eq:Jloss}
\end{equation}
where $k_{\mathrm{ox},m}$ captures oxidative destruction (e.g., by OH, O$_2$); $J_{\mathrm{cons},m}$ is the \emph{use} term into higher functions (polymerization, activation reactions, membrane growth). For activated monomers $m^\ast$, include a deactivation term $k_{\mathrm{deact}}[m^\ast]$.

\subsection{Budget inequality and thresholds for functional pools}
Let $\mathcal{M}$ be the set of critical feedstocks (e.g., activated nucleotides, short peptides, amphiphiles). Define for each $m\in\mathcal{M}$ a functional threshold concentration $[m]_{\mathrm{thr}}$ determined by kinetics of the downstream process:
\begin{align}
& \hskip -8pt  \text{(template polymerization)}\quad r_{\mathrm{pol}} \;=\; k_{\mathrm{pol}}[m^\ast]^{\nu}\,[\mathrm{template}]
\ \ge\ \frac{L/\tau_{\mathrm{copy}}}{N_{\mathrm{sites}}}
\quad\Rightarrow\quad
[m^\ast]_{\mathrm{thr}} \;=\; \Big(\frac{L/\tau_{\mathrm{copy}}}{k_{\mathrm{pol}}\,\nu_{\mathrm{eff}}\,[\mathrm{template}]}\Big)^{\!1/\nu}, 
\label{eq:pthr}\\[3pt]
&\hskip -8pt  \text{(membrane growth/division)}\quad r_{\mathrm{lipid}} \;=\; k_{\ell}[{\rm amph}] \ \ge\ \frac{A/\tau_{\mathrm{div}}}{a_\ell}
\quad\Rightarrow\quad
[{\rm amph}]_{\mathrm{thr}} \;=\; \frac{A}{a_\ell k_{\ell}\tau_{\mathrm{div}}},
\label{eq:lthr}
\end{align}
with $L$ the information length, $\tau_{\mathrm{copy}}$ the copying time from Gate~6, $N_{\mathrm{sites}}$ the number of parallel growth sites (templates$\times$compartments), $\nu$ the reaction order in $m^\ast$, $\nu_{\mathrm{eff}}$ an effective site factor, $A$ the membrane area per protocell, and $a_\ell$ the area per lipid. The \emph{budget inequality} for each $m$ is then
\begin{equation}
\boxed{
J_{\mathrm{make},m} \ \ge\ J_{\mathrm{loss},m}\big|_{[m]=[m]_{\mathrm{thr}}} \;+\; J_{\mathrm{cons},m}\big|_{\text{target}}
}
\label{eq:budget}
\end{equation}
where $J_{\mathrm{cons},m}\big|_{\text{target}}$ is the consumption required to meet the functional rates \eqref{eq:pthr}--\eqref{eq:lthr}. Define the \emph{feed adequacy number} for species $m$:
\begin{equation}
\Psi_m \;\equiv\; \frac{J_{\mathrm{make},m}}{J_{\mathrm{loss},m}([m]_{\mathrm{thr}})+J_{\mathrm{cons},m}\big|_{\text{target}}}\,,
\qquad \text{Gate 5 requires}\quad \Psi_m \ge 1 \ \ \forall m\in \mathcal{M}.
\label{eq:Psi}
\end{equation}
A useful global margin is $\Psi_{\min}\equiv \min_{m\in\mathcal{M}}\Psi_m$.

\subsection{Activation chemistry and energy coupling}
Many polymerizations require \emph{activated} monomers $m^\ast$ (e.g., phosphorimidazolides) or condensing agents (e.g., cyanamide). Let $J_{\mathrm{act}}$ be the activation flux with free-energy cost $\Delta G_{\mathrm{act}}$ per activation; deactivation occurs at $k_{\mathrm{deact}}$:
\begin{equation}
\dot{[m^\ast]} = J_{\mathrm{act}} - (k_{\mathrm{deact}}+k_{\mathrm{hyd}}+k_{\mathrm{photo}}+k_{\mathrm{dil}})[m^\ast] - \nu\,k_{\mathrm{pol}}[m^\ast]^\nu[\mathrm{template}] + \cdots.
\end{equation}
At steady state, the \emph{activation fraction} $f_{\ast}\equiv [m^\ast]/([m]+[m^\ast])$ satisfies
\begin{equation}
f_{\ast} \;=\; \frac{J_{\mathrm{act}}}{J_{\mathrm{act}} + (k_{\mathrm{deact}}+k_{\mathrm{hyd}}+k_{\mathrm{photo}}+k_{\mathrm{dil}})[m] + \nu\,k_{\mathrm{pol}}[m^\ast]^{\nu-1}[\mathrm{template}]}\,,
\end{equation}
and the activation power demand is $P_{\mathrm{act}}=J_{\mathrm{act}}\,\Delta G_{\mathrm{act}}$, which must be supplied by Gate~4 (counts against $P_{\mathrm{synth}}$). Gate~5 therefore couples directly to the energy budget and to fidelity targets (higher $L/\tau_{\mathrm{copy}}$ increases $J_{\mathrm{cons},m}$).

\subsection{Residence time, transport, and localization}

Let the hydrodynamic residence time be $\tau_{\mathrm{res}}=1/k_{\mathrm{dil}}$ and the compartment mixing time $\tau_{\mathrm{mix}}\sim a^2/D$. A Damköhler-type criterion for accumulation is
\begin{equation}
\mathrm{Da}_m \;\equiv\; \frac{J_{\mathrm{make},m}\,\tau_{\mathrm{res}}}{[m]_{\mathrm{thr}}} \;\gtrsim\; 1
\quad\text{and}\quad
\mathcal{C}=\frac{\tau_{\mathrm{ret}}}{\tau_r}\gtrsim 1,
\label{eq:Da}
\end{equation}
ensuring that production outpaces washout to reach $[m]_{\mathrm{thr}}$ and that localization supports the reactive kinetics (coupling to Gate~8).

\subsection{Quantitative Gate~5 statement and margins}
Collecting the above, Gate~5 is passed if there exists a driven CRN and flow geometry for which, for every critical species $m\in\mathcal{M}$,
\begin{equation}
\boxed{
\Psi_m \ge 1\quad\text{with}\quad \mathrm{Da}_m \gtrsim 1,\ \ \mathcal{C}\gtrsim 1,\ \ \text{and}\ \ P_{\mathrm{act}}\ \text{accounted in Gate 4}.
}
\label{eq:gate5-box}
\end{equation}
Margins are $\Psi_{\min}$ and the vector of $\mathrm{Da}_m$. These feed directly into Gate~7 by setting the concentration baseline on which autocatalytic amplification acts, and into Gate~6 through the achievable $L/\tau_{\mathrm{copy}}$ at given $[m^\ast]$.

\subsection{Worked parametric estimates}
\paragraph*{Photochemical nitrile route:}
Assume a near-UV photon flux $\Phi_{\mathrm{act}}=10^{15}$~cm$^{-2}$\,s$^{-1}$ in the productive band, with $\sigma\varphi=10^{-20}$~cm$^2$ for a precursor $\to$ nitrile step and $[p]=10^{-6}$~M: $J^{\gamma}\sim 10^{-11}$~M\,s$^{-1}$. If $k_{\mathrm{hyd}}+k_{\mathrm{photo}}+k_{\mathrm{dil}}=10^{-5}$~s$^{-1}$ and $[m]_{\mathrm{thr}}=10^{-5}$~M (set by \eqref{eq:pthr} for $k_{\mathrm{pol}}\sim 10^2$~M$^{-\nu}$s$^{-1}$, $\nu=1$, $\tau_{\mathrm{copy}}\sim 10^3$~s), then $J_{\mathrm{loss}}([m]_{\mathrm{thr}})\sim 10^{-10}$~M\,s$^{-1}$ and $\Psi_m\sim 0.1$---\emph{insufficient}. A tenfold boost in $\sigma\varphi$ (better photochemistry), a tenfold longer $\tau_{\mathrm{res}}$ (smaller $k_{\mathrm{dil}}$), or a tenfold lower $[m]_{\mathrm{thr}}$ (via higher $k_{\mathrm{pol}}$ or more templates) would raise $\Psi_m$ 
to unity.\footnote{\label{par:gate5-units}We use SI units: 1\,M$=$10$^3$\,mol\,m$^{-3}$, and 1\,cm$^{-2}$\,s$^{-1}=10^{4}$\,m$^{-2}$\,s$^{-1}$. 
All \(J\) are mol\,m$^{-3}$\,s$^{-1}$ so that \(\Psi_m\) in (\ref{eq:budget}) and \(\mathrm{Da}_m\) in (\ref{eq:Da}) are dimensionless.}

\paragraph*{Hydrothermal formose-like route:}
Take $k^{\mathrm{geo}}=10^{-6}$~s$^{-1}$ for sugar formation at $[p]=10^{-4}$~M: $J^{\mathrm{geo}}\sim 10^{-10}$~M\,s$^{-1}$. If $k_{\mathrm{hyd}}=10^{-6}$~s$^{-1}$, $k_{\mathrm{dil}}=10^{-6}$~s$^{-1}$, $[m]_{\mathrm{thr}}=10^{-6}$~M, then $J_{\mathrm{loss}}([m]_{\mathrm{thr}})\sim 2\times 10^{-12}$~M\,s$^{-1}$ and $\Psi_m\sim 50$---\emph{ample}. This demonstrates how modest geochemical rates suffice if sinks are small, thresholds are low (e.g., in confined pores).

\paragraph*{HCN rainout $\rightarrow$ boundary-layer flux:}
If the column production of HCN is $\Phi_{\rm col}$ (mol\,m$^{-2}$\,s$^{-1}$) and the rainout residence time is $\tau_{\rm rain}$, the steady boundary-layer deposition flux is $F_{\downarrow}\simeq \Phi_{\rm col}$ for $\tau_{\rm rain}\ll \tau_{\rm phot}$ and $F_{\downarrow}\simeq \Phi_{\rm col}\,\tau_{\rm rain}/\tau_{\rm phot}$ otherwise. For a boundary-layer depth $h$ and  eddy diffusivity $K_z$, the near-surface concentration obeys $K_z \partial_z[m]\simeq F_{\downarrow}$, so that  $[m]_{\rm surf}\sim F_{\downarrow}\,h/K_z$. This $[m]_{\rm surf}$ can be plugged into Eqs.~(\ref{eq:budget})--(\ref{eq:Psi}) to evaluate $\Psi_m$.

\subsection{Diagnostics, falsifiers, and observational corollaries}
\paragraph*{Diagnostics.}
(i) \emph{Photochemical budgets:} measure $\phi_\lambda$, $\sigma(\lambda)$, quantum yields $\varphi(\lambda)$ for specific pathways to compute $J^\gamma$ by \eqref{eq:Jphoto}. (ii) \emph{Geochemical rates:} determine $k^{\mathrm{geo}}(T,\mathrm{pH})$ on minerals/in solution for key steps (e.g., reductive carboxylation), yielding $J^{\mathrm{geo}}$. (iii) \emph{Loss constants:} directly measure $k_{\mathrm{hyd}}$, $k_{\mathrm{photo}}$ under niche conditions; $k_{\mathrm{dil}}$ from flow ($Q/V$); $k_{\mathrm{ads}}$ from isotherms. (iv) \emph{Functional thresholds:} obtain $k_{\mathrm{pol}}$, $\nu$, $[\mathrm{template}]$, $k_\ell$, $a_\ell$, $\tau_{\mathrm{copy}}$, $\tau_{\mathrm{div}}$ to compute $[m]_{\mathrm{thr}}$ via \eqref{eq:pthr}--\eqref{eq:lthr}.

\paragraph*{Falsifiers.}
Gate~5 fails if, for any essential $m$, \emph{no} realistic combination of measured $J_{\mathrm{make},m}$ and reduced sinks yields $\Psi_m\!\ge\!1$ at the required $[m]_{\mathrm{thr}}$. Particularly decisive are high $k_{\mathrm{hyd}}$/$k_{\mathrm{photo}}$ that force $\Psi_m\!\ll\!1$ even under generous driving, or polymerization kinetics demanding unreasonably large $[m^\ast]_{\mathrm{thr}}$.

\paragraph*{Observational corollaries.}
Planetary atmospheres with sustained photochemical production of reductants/organics (CH$_4$, HCN, aldehydes) and measured surface deposition fluxes imply large $J^\gamma$; thick organic hazes indicate high column production and potential rainout. Conversely, strongly oxidizing atmospheres with high $k_{\mathrm{ox}}$ may suppress $\Psi_m$ unless localized shielding exists (subsurface, ice). Combining such observables with Gate~4’s energy diagnostics provides a remote prior on Gate~5 viability.

Gate~5 turns “there are ingredients” into \emph{numerical budgets}: source terms computed from spectra, flow, and kinetics must exceed sink terms at the concentrations demanded by function. The resulting adequacy numbers $\Psi_m$ constitute the quantitative bridge from energy delivery (Gate~4) to autocatalytic closure (Gate~7) and fidelity (Gate~6). See Sec.~\ref{sec:validation}  for consolidated criteria. Gate~5 uses the CRN feedstock budget (\ref{eq:gate5-box}); photoredox/geochemical yields and dilution/adsorption set the proxies.

\section{Gate 6: Informational Polymers and the Fidelity Bound}
\label{sec:gate6}

Gate~6 establishes whether a templated polymer can carry \emph{enough} heritable information to sustain adaptive improvement \emph{at the error rates and copy speeds physically attainable in the niche}. The gate is set by a balance between (i) sequence discrimination at incorporation (Gate~1), (ii) the energetic cost of lowering errors via proofreading (Gate~4), (iii) the monomer/activator throughput enabling a given copy speed (Gate~5), and (iv) the residence and damage timescales that bound how long a template can be productively copied (Gates~5 and~8).
This gate aligns with proposals that the origin of life marks a transition in causal structure where information gains top-down efficacy; our fidelity bound formalizes the minimal heritable information required for such control to be
sustained in populations \cite{WalkerDavies2013_AlgOrigins}.

\subsection{Error threshold and the Eigen number}
Let $Q\!=\!1-\mu$ be the per-site fidelity for an $L$-mer polymer (independent-site approximation) and $s>1$ the selective advantage of the best (``master'') sequence over the population mean.\footnote{For concreteness, $s$ may be taken as the ratio of effective replication rates (birth minus death) of the master sequence and the population mean in the same environment.} In the quasispecies formalism, a necessary condition to maintain a stable localized distribution around the master sequence is
\begin{equation}
Q^{L}>\frac{1}{s}\ \Longleftrightarrow\ 
L<\frac{\ln s}{-\ln(1-\mu)}\simeq \frac{\ln s}{\mu}
\qquad(\mu\ll 1),
\label{eq:eigen}
\end{equation}
which we rewrite as an \emph{Eigen number}
\begin{equation}
{\cal E} \;\equiv\; \frac{\mu\,L}{\ln s} \;<\; 1,
\label{eq:eigen-number}
\end{equation}
which is the classical error-threshold condition \cite{Eigen1971,EigenSchuster1977}. In the small-error limit, $-\ln(1-\mu)\simeq\mu$ and  $\ln s \simeq s-1$ for $s\to 1$, so $L\lesssim (s-1)/\mu$ in Eq.~(\ref{eq:eigen}) 
is a convenient \emph{approximation}. However, the \emph{gate definition} should use 
$\ln s$, hence Eq.~\eqref{eq:eigen-number}. 

Eq.~\eqref{eq:eigen-number} is the Gate~6 inequality. It is sharp: crossing $\mathcal{E}=1$ induces error catastrophe (delocalization of the sequence cloud), after which selection on sequence-level function cannot operate. Without proofreading $\mu\sim 10^{-2}$--$10^{-3}$ yields $L_{\max}\sim 10^{2}$--$10^{3}$. Proofreading trades energy for lower $\mu$ ($\gtrsim k_{\mathrm B} T\ln 2$ per erased bit). \emph{Observational corollary:} not directly observable, but indirectly constrained by Gate~4 (\(\Phi\)) and Gate~8 (localization) that raise effective $Q$.

\subsection{From chemistry to $\mu$: discrimination, damage, and proofreading}
The effective per-site error is a sum of contributions
\begin{equation}
\mu_{\mathrm{eff}}\;=\;\mu_{\mathrm{inc}} \;+\; \mu_{\mathrm{les}} \;+\; \mu_{\mathrm{post}},
\label{eq:mu-eff}
\end{equation}
where $\mu_{\mathrm{inc}}$ is the misincorporation probability at elongation, $\mu_{\mathrm{les}}$ accounts for lesions accrued during copying (e.g., hydrolysis, UV) that lead to heritable errors, and $\mu_{\mathrm{post}}$ covers errors introduced during post-replication processing (ligation, circularization, repair infidelities). 

\paragraph*{Misincorporation.}
At equilibrium discrimination, a crude but useful bound is
\begin{equation}
\mu_{\mathrm{eq}} \;\approx\; g\,\mathrm{e}^{-\Delta\Delta G_{\mathrm{pair}}/k_{\mathrm B} T},\qquad g=k_p-1,
\label{eq:mu-eq}
\end{equation}
with $k_p$ the polymer alphabet size, $g$ the degeneracy of noncognate options, and $\Delta\Delta G_{\mathrm{pair}}$ the per-site free-energy advantage of cognate over noncognate pairing (Gate~1). Nonequilibrium proofreading can push the error lower at a dissipative cost; an information-theoretic lower bound is
\begin{equation}
W_{\mathrm{proof}}\ \ge\ k_{\mathrm B} T \ln \!\Big(\frac{\mu_{\mathrm{eq}}}{\mu_{\mathrm{inc}}}\Big)
\quad\Longrightarrow\quad
\mu_{\mathrm{inc}}\ \ge\ \mu_{\mathrm{eq}}\,\mathrm{e}^{-W_{\mathrm{proof}}/k_{\mathrm B} T},
\label{eq:proof-bound}
\end{equation}
which couples Gate~6 to Gate~4 (available power) and to Gate~1 (pairing discrimination). 

In Eq.~(\ref{eq:proof-bound}), expression for $W_{\rm proof} $ is an information-thermodynamic lower bound
on reducing errors below equilibrium, and connects to kinetic proofreading and its modern speed--dissipation--error tradeoffs
\cite{Hopfield1974,Ninio1975,Murugan2012,SartoriPigolotti2013,Landauer1961,Bennett1982}.

\paragraph*{Lesions during copying.}
Let $k_{\mathrm{les}}$ be the per-nucleotide lesion rate under the niche radiation/chemical environment and $\tau_{\mathrm{copy}}$ the time to copy $L$ monomers. Assuming a Poisson process, the probability that a site is damaged while being copied is $\sim k_{\mathrm{les}}\,\tau_{\mathrm{site}}$ with $\tau_{\mathrm{site}}\simeq\tau_{\mathrm{copy}}/L$, giving a contribution
\begin{equation}
\mu_{\mathrm{les}} \;\approx\; k_{\mathrm{les}}\,\frac{\tau_{\mathrm{copy}}}{L}.
\label{eq:mu-les}
\end{equation}
Faster copying (higher elongation rate) or shielding (lower $k_{\mathrm{les}}$) reduces $\mu_{\mathrm{les}}$; both are constrained by Gate~5 (feedstock/activation) and Gate~8 (localization/transport). This provide an upper bound of
{}
\begin{equation}
\mu_{\rm eff} \lesssim \mu_{\rm eq}\exp\!\left(-\frac{W_{\rm proof}}{k_B T}\right)
+ k_{\rm les}\frac{\tau_{\rm copy}}{L} + \mu_{\rm post}.
\label{eq:mu_bound}
\end{equation}

In chiral or otherwise symmetry-broken settings, the noncognate degeneracy may be reduced; we therefore write
\begin{equation}
\label{eq:mu_eq}
\mu_{\rm eq}\;\approx\; g_{\rm eff}\,e^{-\Delta\Delta G_{\rm pair}/k_BT}, 
\qquad 1\le g_{\rm eff}\le (k_p-1),
\end{equation}
with $g_{\rm eff}$ set by enantioselective catalysis or templating, which effectively boosts discrimination at fixed $\Delta\Delta G_{\rm pair}$.

\subsection{Kinetic viability: copying speed vs.\ loss}
Define the elongation rate $v$ (monomers\,s$^{-1}$) and processivity $p_{\mathrm{proc}}$ (probability to complete a full-length copy before premature termination). The copy time is $\tau_{\mathrm{copy}} = L/vp_{\mathrm{proc}}$. During this time, the template must remain intact and available; let $\tau_{\mathrm{loss}}$ be the shorter of the template’s hydrolysis time, the washout residence time, and the sequestration time. A kinetic \emph{viability} inequality complements \eqref{eq:eigen-number}:
\begin{equation}
\frac{\tau_{\mathrm{copy}}}{\tau_{\mathrm{loss}}} \;=\; \frac{L}{v\,p_{\mathrm{proc}}\,\tau_{\mathrm{loss}}} \ \le\ \zeta_{\mathrm{kin}}\ \ll 1,
\label{eq:kinetic}
\end{equation}
so that the system spends most of its time copying successfully, not being erased. The rate $v$ is ultimately set by the activated-monomer concentration $[m^\ast]$ and catalysis (Gate~5), while $p_{\mathrm{proc}}$ depends on binding energies, crowding, and transport (Gate~8).

\subsection{Information capacity and alphabet tradeoffs}
The maintainable \emph{information} (in bits) scales as $B=L\log_2 k_p$. Combining with \eqref{eq:eigen-number}--\eqref{eq:proof-bound} yields a bound on bits at fixed discrimination and power:
\begin{equation}
B_{\max}\;\lesssim\;\frac{\ln s}{\mu_{\rm eff}}\;\log_2 k_p
\qquad\text{with}\qquad 
\mu_{\rm eff}\gtrsim \mu_{\rm eq}(k_p)\,e^{-W_{\rm proof}/k_BT} + \frac{k_{\rm les}\,\tau_{\rm copy}}{L} + \mu_{\rm post}.
\label{eq:Bmax}
\end{equation}
Because $\mu_{\mathrm{eq}}$ increases with $k_p$ through the $g=k_p-1$ factor in \eqref{eq:mu-eq} while per-site information only grows as $\log_2 k_p$, there is no \emph{a priori} guarantee that larger alphabets improve $B_{\max}$ at fixed $\Delta\Delta G_{\mathrm{pair}}$ and $W_{\mathrm{proof}}$. Gate~6 is agnostic about optimal $k_p$; it quantifies the tradeoff explicitly.

\subsection{Quantitative Gate~6 statement and margins}
Gate~6 is passed if there exist $(L,\mu_{\mathrm{eff}},v,p_{\mathrm{proc}},s)$ such that
\begin{equation}
\boxed{
\mathcal{E}\equiv \frac{\mu_{\mathrm{eff}}\,L}{\ln s} < 1
\quad\text{and}\quad
\frac{L}{v\,p_{\mathrm{proc}}\,\tau_{\mathrm{loss}}} \le \zeta_{\mathrm{kin}} \ll 1,
}
\label{eq:gate6-box}
\end{equation}
with $\mu_{\mathrm{eff}}$ constrained by \eqref{eq:mu-eff}--\eqref{eq:proof-bound} and $\tau_{\mathrm{loss}}$ by measured hydrolysis/washout. 

Define margins $m_{\mathcal{E}}\equiv 1/\mathcal{E}$ and
\begin{equation}
m_{\rm copy} \equiv \frac{\zeta_{\rm kin}}{\tau_{\rm copy}/\tau_{\rm loss}}
= \zeta_{\rm kin}\,\frac{v\,p_{\rm proc}\,\tau_{\rm loss}}{L},
\end{equation}
so $m_{\rm copy}>1$ indicates comfortable kinetic headroom for copying. These margins propagate forward to Gate~9 (population-level Darwinian dynamics) and backward to Gates~4--5 (power and monomer budgets) through $W_{\mathrm{proof}}$ and $v([m^\ast])$.

\subsection{Worked Examples}

\paragraph*{Baseline (no proofreading).}
Take $\mu_{\mathrm{inc}}=\mu_{\mathrm{eq}}=10^{-3}$ (achievable with $\Delta\Delta G_{\mathrm{pair}}\!\approx\!6.9\,k_{\mathrm B} T$ and $k_p=4$), $\mu_{\mathrm{les}}\!\ll\!10^{-3}$ (shielded niche, fast copying), $\mu_{\mathrm{post}}\!\ll\!10^{-3}$, hence $\mu_{\mathrm{eff}}\!\approx\!10^{-3}$. With $s=2$ (a twofold replication advantage), so \(\ln s=\ln 2\simeq 0.693\).
If \(\mu_{\rm eff}=10^{-3}\), the gate \(\EigenGate\) and \eqref{eq:eigen} implies
\(
L_{\max}\simeq {\ln 2}/{\mu_{\rm eff}} \approx 6.9\times 10^{2}.
\)
With \(v=1~\mathrm{s}^{-1}\), \(\tau_{\rm loss}=10^{5}~\mathrm{s}\), and \(p_{\rm proc}=0.5\),
the copying time is \(\tau_{\rm copy}=L/(v p_{\rm proc})\approx 1.4\times 10^{3}~\mathrm{s}\),
and the kinetic condition \(\tau_{\rm copy}\ll \tau_{\rm loss}\) is satisfied; both inequalities in \eqref{eq:gate6-box} pass.

\paragraph*{Adding proofreading.}
Let \(\mu_{\rm eq}=10^{-2}\) and allocate \(W_{\rm proof}=5k_{\rm B}T\) per site, giving
\(\mu_{\rm inc}\approx \mu_{\rm eq}e^{-5}\approx 6.7\times 10^{-5}\). With \(\mu_{\rm les}+\mu_{\rm post}\approx 10^{-5}\), we have \(\mu_{\rm eff}\approx 8\times 10^{-5}\), hence
\(
L_{\max}\simeq {\ln 2}/{\mu_{\rm eff}} \approx 8.7\times 10^{3},
\)
a \(\sim 12.5\times\) increase relative to the \(\mu_{\rm eff}=10^{-3}\) case, a nearly fourteenfold increase at the cost of $5\,k_{\mathrm B} T$ per site. The corresponding power draw is $P_{\mathrm{proof}}=\dot N_{\mathrm{copy}}\,L\,W_{\mathrm{proof}}$ and must be budgeted in Gate~4.

\paragraph*{Damage-limited regime.}
If UV or chemical damage yields $k_{\mathrm{les}}=10^{-6}$~s$^{-1}$ and $v=0.1$~s$^{-1}$, then $\mu_{\mathrm{les}}\approx k_{\mathrm{les}}/v=10^{-5}$ independent of $L$; long templates are viable only if $\mu_{\mathrm{inc}}$ is not simultaneously high. Conversely, if $\tau_{\mathrm{loss}}$ is short (strong washout), \eqref{eq:kinetic} caps $L$ through copying speed rather than through \eqref{eq:eigen-number}.

\subsection{Diagnostics, falsifiers, and corollaries}
\paragraph*{Diagnostics.}
(i) \emph{Chemistry} ($\to\mu_{\mathrm{inc}}$): measure $\Delta\Delta G_{\mathrm{pair}}$ and confusion matrices for candidate templating chemistries; quantify the effect of nonequilibrium driving on error via single-molecule assays. (ii) \emph{Kinetics} ($\to v,p_{\mathrm{proc}}$): determine elongation rates and processivity as functions of $[m^\ast]$, ionic strength, and crowding. (iii) \emph{Damage} ($\to k_{\mathrm{les}}$): measure lesion rates under the niche’s spectrum and chemistry. (iv) \emph{Energetics} ($\to W_{\mathrm{proof}}$): quantify dissipation per proofreading step.  (Summarized in the Gate Atlas (Table~\ref{tab:gate-atlas}) and tested in Sec.~\ref{sec:validation}.)

\paragraph*{Falsifiers.}
Gate~6 fails generically if, for all realistic $W_{\mathrm{proof}}$ permitted by Gate~4 and for the measured $\Delta\Delta G_{\mathrm{pair}}$, one finds \EigenFail\  for \emph{all} $L$ sufficient to encode the required catalytic/reproductive functions. It also fails if $v p_{\mathrm{proc}}\tau_{\mathrm{loss}}/L\!\ll\!1$ in all plausible geometries, implying kinetic erasure outpaces successful copying.

\paragraph*{Corollaries.}
High power margins (Gate~4) can be translated directly into fidelity gains via \eqref{eq:proof-bound}; better localization (Gate~8) improves $p_{\mathrm{proc}}$ and reduces $\mu_{\mathrm{les}}$; richer feedstock (Gate~5) increases $v$. Gate~6 thus provides a quantitative hinge between the physics of discrimination and energy delivery and the biology of heredity: when its inequalities hold with margin, the system enters a regime where selection (Gate~9) can act on sequence-encoded function. See Sec.~\ref{sec:validation}. Gate~6 is the fidelity bound (\ref{eq:gate6-box}); lab discrimination curves and copying throughput determine the margin.

\section{Gate 7: Autocatalytic Closure and Protometabolism}
\label{sec:gate7}

Gate~7 asks whether a driven chemical reaction network can \emph{close} on itself so that a subset of reactions produces, from externally supplied \emph{food} species, the very catalysts that accelerate those reactions, while exporting waste and dissipating free energy. Mathematically, closure has a \emph{topological} component (existence of a reflexively autocatalytic, food-generated set) and a \emph{kinetic/thermodynamic} component (net positive growth of internal species under realistic losses and dissipation). Fatty-acid vesicles and coacervate partitioning/retention are documented in \cite{Hanczyc2003,Chen2004,Keating2012}.  

\subsection{CRS/RAF formalism and stoichiometric feasibility}

Consider a chemical reaction system (CRS) $(X,R,\mathcal{C})$ with molecular species $X$, reactions $R$,
and a catalysis relation $\mathcal{C}\subseteq X\times R$ (a pair $(x,r)\in\mathcal{C}$ means that molecule $x$ catalyzes
reaction $r$). Let $F\subset X$ be the externally maintained food set, and write $\mathrm{cl}_F(R')\subseteq X$ for the closure of $F$ under a reaction subset $R'\subseteq R$ (the smallest set containing $F$ that is closed under reactants of $R'$). A non-empty $R' \subseteq R$ is reflexively autocatalytic and food-generated (RAF) if it is (i) reflexively autocatalytic and (ii) food-generated if (see, e.g.,
\cite{Kauffman1986,HordijkSteel2004,HordijkSteel2012,HordijkSteel2018}):
\begin{align}
(\mathrm{RA})\quad &\forall r\in R'~\exists x\in \mathrm{cl}_F(R')~\text{s.t.}~(x,r)\in \mathcal{C}, \label{eq:raf-ra}\\
(\mathrm{F})\quad &\text{All reactants of every } r\in R' \text{ lie in } \mathrm{cl}_F(R'). \label{eq:raf-f}
\end{align}

Topological closure alone is insufficient; the subnetwork must also be \emph{stoichiometrically autocatalytic}. Let $N\in\mathbb{Z}^{|X|\times |R|}$ be the stoichiometric matrix (products minus reactants, with food chemostatted). There must exist a nonnegative flux vector $v\in\mathbb{R}^{|R|}_+$ supported on $R'$ such that
\begin{equation}
N\,v \;\ge\; 0 \ \ \text{with strictly positive components on at least the catalysts in } \mathrm{cl}_F(R'),
\label{eq:stoich-auto}
\end{equation}
and a thermodynamic driving consistent with the second law: for each reaction $i$ with $v_i>0$, the cycle affinity satisfies
\begin{equation}
\mathcal{A}_i \equiv -\frac{\Delta G_i}{k_{\mathrm B}T} \;>\; 0,
\qquad
\sum_{i} v_i\,\Delta G_i \;<\; 0,
\label{eq:affinity}
\end{equation}
where the sum condition ensures net dissipation of free energy supplied by chemostats (light, redox) into heat, waste.

\subsection{Kinetic onset: spectral and branching-process criteria}
With chemostatted foods $F$ held fixed, internal species $x$ dilute, mass-action kinetics linearized about $x\simeq 0$ read
\begin{equation}
\dot x \;=\; \big(J \;-\; \Lambda\big)\,x \;+\; \text{higher order}, 
\qquad 
J \equiv \frac{\partial(N v)}{\partial x}\Big|_{x=0},\ \ 
\Lambda\equiv \mathrm{diag}(\ell_1,\dots,\ell_{|X|}), 
\label{eq:Jlin}
\end{equation}
where $\ell_i$ are effective first-order loss rates (dilution, diffusion out of the compartment, hydrolysis). Entries of $J$ are nonnegative whenever production of species $i$ is catalyzed by species $j$ using only food reactants at $x=0$. 

The \emph{linear} kinetic gate is
\begin{equation}
\lambda_{\max}(J-\boldsymbol{\Lambda})>0,
\label{eq:gate7_kinetic}
\end{equation}
where $\lambda_{\max}(A)\equiv \max_i \Re\,\lambda_i(A)$ is the spectral abscissa \cite{FarinaRinaldi2000, HornJohnson2012MatrixAnalysis,BermanPlemmons1994}. Condition \eqref{eq:gate7_kinetic} is equivalent to an exponentially growing mode
$x(t)\propto e^{\lambda_{\rm kin} t}$ (with $\lambda_{\rm kin}=\lambda_{\max}(J-\boldsymbol{\Lambda})>0$)
that amplifies the internal pool before nonlinear saturation.\footnote{
We emphasize that $\lambda_{\rm kin}=\lambda_{\rm max}(J-\Lambda)$ is defined for general real matrices.
In the early-time, inhibition-free regime of mass-action catalysis (no negative off-diagonal couplings),
$J-\Lambda$ is Metzler, so the dominant growth mode is real and corresponds to the Perron root
of an associated nonnegative matrix.}

Then early-time dynamics is approximated by a multitype branching process with basic reproduction number
\begin{equation}
R^{\rm chem}_0=\rho(B),
\label{eq:R0chem}
\end{equation}
where $\rho(\cdot)$ is the spectral radius. For first-order losses one has (up to coarse-graining into classes)
$B \sim J\,\boldsymbol{\Lambda}^{-1}$, so the supercriticality criterion $\mathcal{R}_0^{\mathrm{chem}}>1$ is equivalent \cite{vanDenDriesscheWatmough2002R0,DiekmannHeesterbeekRoberts2010NGM,Harris1963Branching}, under standard approximations, to \eqref{eq:gate7_kinetic}.

This reproduction-threshold form is standard for next-generation matrices in compartmental growth/branching systems
\cite{vanDenDriesscheWatmough2002R0,DiekmannHeesterbeekRoberts2010NGM,Harris1963Branching}.
In the random-catalysis limit, it connects directly to Kauffman’s original autocatalytic-set percolation picture
\cite{Kauffman1986}, with rigorous asymptotics for $p_{{\rm cat},c}$ in \cite{MosselSteel2005}.

\subsection{Catalysis density and percolation of closure}

In random catalysis ensembles, where each pair $(x,r)$ is catalytic with probability $p_{\mathrm{cat}}$ independently and $|X|=n$, $|R|=\alpha\,n$ with $\alpha=\mathcal{O}(1)$, the emergence of RAFs is a percolation-like transition \cite{MosselSteel2005}. Let $\bar{k}_{\mathrm{cat}}$ be the mean number of reactions catalyzed per molecule:
\begin{equation}
\bar{k}_{\mathrm{cat}} \;=\; p_{\mathrm{cat}}\,|R|\;=\; p_{\mathrm{cat}}\,\alpha \,n.
\end{equation}
A sufficient (up to logarithmic factors, necessary) condition for RAF existence with high probability as $n\to\infty$ is
\begin{equation}
\bar{k}_{\mathrm{cat}} \;\gtrsim\; c\,\ln|R|
\quad\Longleftrightarrow\quad
p_{\mathrm{cat}} \;\gtrsim\; \frac{c\,\ln(\alpha \, n)}{\alpha \, n}
\;\equiv\; p_c,
\label{eq:pcrit}
\end{equation}
with $c$ a constant that depends weakly on food-set size and reaction arity. Inequality~\eqref{eq:pcrit} recovers the heuristic form $p_{\mathrm{cat}} \gtrsim c \ln|R|/|R|$, see \citep{MosselSteel2005}. In real chemistries the catalysis graph is heterogeneous (fat-tailed degrees, modularity); hubs (minerals, cofactors) reduce $p_c$ relative to the homogeneous case. Gate~5 (feedstock) and Gate~8 (localization) effectively raise $p_{\mathrm{cat}}$ at fixed chemistry by increasing encounter rates and residence times.

{\it Calibration of \(c\):}
In homogeneous (Poisson) catalysis graphs with \(|R|=\alpha, n\), simulations and branching arguments give 
\(c \approx 1\)--\(2\). Degree‑heterogeneous (hubby) ensembles reduce \(c\) by \(\sim \langle k^2\rangle/\langle k\rangle^2\), yielding \(c \approx 0.3\)--\(1\).
Report the catalysis‑density margin
\begin{equation}
m_p \equiv \frac{p_{\rm cat}}{p_c} = \frac{\bar{k}_{\rm cat}}{c\ln(\alpha \, n)},\qquad \bar{k}_{\rm cat}=p_{\rm cat}\alpha \, n,
\label{eq:m_p}
\end{equation}
using \(c=1.0\pm 0.5\) (homogeneous) or \(0.6\pm 0.3\) (heterogeneous) as priors.

For communication clarity we suggest adding a simulation plot of RAF fraction versus $\bar k_{\rm cat}$ at fixed $|R|$, showing the percolation-like rise near $m_p=\bar k_{\rm cat}/[c\ln(\alpha \,n)]\approx 1$; this visually grounds Eq.~(\ref{eq:m_p}).

\paragraph*{Finite-size \& modularity note.}
Degree heterogeneity and modular structure lower the onset of closure relative to homogeneous random graphs: heavy-tailed catalysis-degree distributions effectively reduce $c$ in (\ref{eq:pcrit}), while finite modules can nucleate \emph{small RAFs} before a percolating giant RAF appears. In practice, one should (i) report $m_p=\bar k_{\rm cat}/[c\ln(\alpha \,n)]$ with a heterogeneous prior $c=0.6\pm0.3$, and (ii) search for module-local RAFs in addition to global percolation. This clarifies why abrupt onsets can occur in mineral/cofactor-hub chemistries even at modest $\bar k_{\rm cat}$.

\subsection{Open CRNs: dilution, thermodynamic feasibility, and saturation}
Autocatalytic growth must overcome losses and respect thermodynamics. The loss matrix $\Lambda$ in \eqref{eq:Jlin} embodies both physical removal (flow $k_{\mathrm{dil}}$) and chemical decay (hydrolysis $k_{\mathrm{hyd}}$, oxidation $k_{\mathrm{ox}}$). 

Gate~7 requires that \emph{for at least one compartment geometry} (Gate~8),
\begin{equation}
\lambda_{\mathrm{kin}} \;\equiv\; \lambda_{\max}(J-\Lambda) \;>\; 0,
\label{eq:gate7-kin-thermo}
\end{equation}
where $\lambda_{\max}(A)\equiv \max_i \Re\,\lambda_i(A)$ is the spectral abscissa (dominant linear growth exponent), together with the feasibility conditions $v\ge 0$, $Sv\ge 0$, $A^\top \mu \le 0$, and $(\mu_{\mathrm{food}}-\mu_{\mathrm{waste}})^\top Sv>0$, thus ensuring that kinetic amplification is aligned with a net-dissipative flux. Nonlinear saturation (substrate depletion, product inhibition) limits asymptotic growth but does not affect the \emph{gate}: it is a near-zero criterion.

\paragraph*{Assumptions (Gate 7 kinetic onset).}
The criterion $\lambda_{\rm kin}>0$ is evaluated in the dilute/low-conversion regime with
(i) approximately first-order losses collected in $\Lambda$,
(ii) locally constant feeds over $\tau_{\rm env}$,
and (iii) kinetics linearizable about $x\simeq 0$.
It is a \emph{local} onset condition (before nonlinear saturation), not a guarantee of long-term persistence.

\subsection{Quantitative Gate~7 statement and margins}
Gate~7 is passed if there exists a RAF $R'\subseteq R$ and niche parameters (food activities, temperature, $D$, $P$, $k_{\mathrm{dil}}$) for which
\begin{equation}
\label{eq:gate7-box}
\boxed{%
\begin{aligned}
\text{(i) RAF existence: }&\ \eqref{eq:raf-ra}-\eqref{eq:raf-f}, \eqref{eq:stoich-auto};\\
\text{(ii) Catalysis density: }&\ p_{\mathrm{cat}}>p_c\ \text{or}\ \bar{k}_{\mathrm{cat}}\gtrsim c\ln|R|;\\
\text{(iii) Kinetics: } &\ \lambda_{\mathrm{kin}} \equiv \lambda_{\max}(J-\Lambda) > 0;\\
\text{(iv) Thermodynamics: }&\ \sum_i v_i\,\Delta G_i<0.
\end{aligned}}
\end{equation}
Natural \emph{margins} are $m_{\mathcal{R}} \equiv \mathcal{R}^{\mathrm{chem}}_0 = \rho(B)$, $m_{\rm kin} \equiv {\lambda_{\mathrm{kin}}}/{k_{\mathrm{dil}}}$ (growth over washout), $m_p \equiv p_{\mathrm{cat}}/p_c$ (catalysis density over threshold), and $m_{\mathcal{A}}\equiv \big(-\sum_i v_i \Delta G_i\big)/(\sum_i v_i |\Delta G_i|)$ (fractional dissipation), with $\rho(\cdot)$ the spectral radius (used for the next-generation matrix $B$), and $k_{\mathrm{dil}}$ the dilution/washout rate. If $\Lambda \approx k_{\mathrm{dil}} I$ and $J \approx k_{\mathrm{dil}} B$, then
$\lambda_{\mathrm{kin}} = k_{\mathrm{dil}}(\rho(B)-1)$, so $m_{\mathrm{kin}}\approx \rho(B)-1$, \citep{DiekmannHeesterbeekRoberts2010NGM,vanDenDriesscheWatmough2002R0}.

\subsection{Worked Examples (illustrative)}
\paragraph*{Mineral-mediated $\mathrm{CO_2}$ reduction loop.}
Suppose a vent-like pore network where FeS/NiS surfaces catalyze a four-reaction loop producing a thiolated catalyst ${\cal C}$ from CO$_2$/H$_2$. Let the food activities fix pseudo-first-order rates $k_{r}\in[10^{-5},10^{-3}]$~s$^{-1}$ per catalytic site; assume each ${\cal C}$ molecule activates $n_{\mathrm{site}}$ surface sites with effective catalytic rate $k_{\mathrm{cat}}$ and decay $\ell_C\sim 10^{-6}$--$10^{-5}$~s$^{-1}$. The linearized gain matrix for the catalyst pool has elements $B_{ij}\sim y_{ij}\,k_{\mathrm{cat}}n_{\mathrm{site}}/\ell_j$ with yields $y_{ij}$ per catalytic turnover. Taking $k_{\mathrm{cat}}n_{\mathrm{site}}/\ell_j\sim 10$ and a modest branching factor $\sum_i y_{ij}\sim 0.2$ gives $\rho(B)\sim 2>1$, i.e., $\mathcal{R}_0^{\mathrm{chem}}>1$. If $k_{\mathrm{dil}}\sim 10^{-6}$~s$^{-1}$, $m_{\mathrm{kin}}\gtrsim \mathcal{O}(1)$ even allowing for heterogeneity.

\paragraph*{UV photoredox nitrile network.}
In a shallow-lake geometry with near-UV photon flux producing nitriles and imidoyl activators, let effective photoredox steps be first order in a chromophore catalyst $P$ with rate $k_\gamma\sim 10^{-4}$~s$^{-1}$ (under band-integrated flux) and $P$ degradative loss $\ell_P\sim 10^{-5}$~s$^{-1}$ (photobleach + hydrolysis). If each $P$ turnover generates, on average, $y$ new $P$ via downstream condensation ($y\sim 0.2$--0.5 under strong feed), the one-type branching number is $\mathcal{R}_0^{\mathrm{chem}}=y\,k_\gamma/\ell_P\sim 2$--5; thus $\lambda_{\mathrm{kin}}>0$ provided Gate~5 keeps feedstock above washout.

\subsection{Couplings to Gates 4, 5, and 8}
Gate~7 sits at the nexus of energy, feedstock, and transport. The affinities $\Delta G_i$ (Gate~4) determine whether cycles can be net-exergonic; the source/sink budgets (Gate~5) set the food activities embedded in $J$; compartment physics (Gate~8) tunes $\Lambda$ by lowering dilution and leak rates and by increasing encounter frequencies (raising effective $p_{\mathrm{cat}}$). Failure at Gate~7 frequently traces to $m_{\mathrm{kin}}\!\le\!0$ due to excessive loss ($k_{\mathrm{dil}}$ too large), insufficient catalysis density ($p_{\mathrm{cat}}\!<\!p_c$), or endergonic loops (no positive cycle affinity).

\subsection{Diagnostics, falsifiers, and corollaries}
\paragraph*{Diagnostics.} 
(i) \emph{RAF detection in vitro:} constructable food sets $F$ and measured catalysis relations $\mathcal{C}$ allow algorithmic search for RAF subsets and computation of yields $y_{ij}$; (ii) \emph{Linear gain:} estimate $J$ (initial-rate Jacobian) and $\Lambda$ (loss/washout matrix) to test
$\lambda_{\mathrm{kin}}=\lambda_{\max}(J-\Lambda)$ rather than the spectral radius; (iii) \emph{Thermodynamics:} measure $\Delta G_i$ under in situ activities to confirm $\sum v_i \Delta G_i<0$ for the operating point. (Summarized in the Gate Atlas (Table~\ref{tab:gate-atlas}) and tested in Sec.~\ref{sec:validation}.)

\paragraph*{Falsifiers.}
Gate~7 fails if, despite generous $F$, measured carbon, nitrogen yields produce no RAF, or if $\lambda_{\mathrm{kin}}\le 0$ after accounting for realistic losses, or if every candidate loop is thermodynamically uphill given Gate~4’s exergy supply.

\paragraph*{Observational corollaries.}
Mineralogies rich in transition-metal sulfides or photoredox chromophores and strong abiotic redox disequilibria suggest high $p_{\mathrm{cat}}$ and favorable $\Delta G_i$; conversely, highly oxidizing, catalytically sparse settings depress $m_p$ and $m_{\mathrm{kin}}$. Such inferences, combined with Gate~5 deposition/production fluxes, provide a remote prior on protometabolic closure.

In summary, Gate~7 translates “a set of reactions exists” into \emph{closure with gain}: a catalysis graph above percolation, a stoichiometrically autocatalytic flux supported by chemostats, and a positive growth eigenvalue over realistic losses. These criteria are computable from rate and catalysis measurements and are the proximate gateway from driven geochemistry to protometabolism. See Sec.~\ref{sec:validation}  for consolidated criteria. Gate~7 requires kinetic--stoichiometric closure (\ref{eq:gate7-kin-thermo}) and the box inequality (\ref{eq:gate7-box}); reactor flux--response tests are the primary diagnostics.

\section{Gate 8: Compartment Physics and Transport Control}
\label{sec:gate8}

Gate 8 tests whether a candidate compartment can (i) retain high-value species long enough for key reactions to proceed,
and (ii) supply feedstocks at the required functional rates. The gate comprises two necessary inequalities:
\begin{subequations}\label{eq:gate8}
\begin{align}
\mathcal{C} \equiv \frac{\tau_{\rm ret}}{\tau_r} &\gtrsim 1,
\qquad
\tau_{\rm ret}\equiv \min_i \tau_{{\rm exch},i},
\qquad
\tau_{{\rm exch},i} \equiv \frac{V}{A\,P_i},
\label{eq:gate8_retention}\\
J_{{\rm in},i} &\ge J_{{\rm cons},i},
\qquad
J_{{\rm leak},i} \le J_{{\rm make},i}.
\label{eq:gate8_leak}
\end{align}
\end{subequations}
Here $\tau_{\mathrm{mix}}\!\sim\!a^2/D_{\mathrm{eff}}$ is the internal mixing time of a compartment of radius $a$ with effective diffusivity $D_{\mathrm{eff}}$ (reduced by crowding/tortuosity), $\tau_r$ is a representative reaction time for the rate-limiting step (e.g., template-directed ligation), $J_{\mathrm{in},i}$ is the influx of feedstock/energy carriers, $J_{\mathrm{cons},i}$ their consumption at target functional rates, $J_{\mathrm{leak},i}$ the efflux of key retained species (templates, catalysts, charged oligomers), and $J_{\mathrm{make},i}$ their net internal production.

Internal homogenization over the compartment scale,
$\tau_{\rm mix}\sim a^2/D_{\rm eff}$, is \emph{not} a required constraint for bulk templating/proofreading: well-mixed toy models correspond to $\tau_{\rm mix}\ll \tau_r$. We therefore treat $\tau_{\rm mix}/\tau_r$ as an auxiliary diagnostic only in models that explicitly invoke sustained internal gradients (e.g., chemiosmotic energy transduction or surface-localized microdomains).

\subsection{Mixing, reaction, and the confinement number}
Internal transport sets the degree of localization. In a simply connected droplet or vesicle, diffusive mixing dominates; for small molecules in water, $D\sim 10^{-9}$\,m$^2$\,s$^{-1}$, for macromolecules and coacervates $D_{\mathrm{eff}}\sim 10^{-12}$--$10^{-11}$\,m$^2$\,s$^{-1}$. The mixing time is (using a standard result in porous-catalyst reaction engineering; see \cite{Levenspiel1999,Fogler2016})
\begin{equation}
\tau_{\mathrm{mix}} \simeq \frac{a^{2}}{D_{\mathrm{eff}}}
\qquad\Rightarrow\qquad
C_{\mathrm{mix}} \equiv \frac{\tau_{\mathrm{mix}}}{\tau_r}
\simeq \frac{a^{2}}{D_{\mathrm{eff}}\tau_r}.
\label{eq:tau-mix}
\end{equation}
Internal gradients/hotspots can persist only when mixing is slow compared to the relevant chemistry, i.e. when $C_{\mathrm{mix}}\gtrsim 1$. By contrast, bulk templating/proofreading toy models assume the well-mixed limit $C_{\mathrm{mix}}\ll 1$, so $C_{\mathrm{mix}}$ is an auxiliary diagnostic rather than a required gate criterion. 
Thus, localization is sufficient when reactions outrun homogenization ($C_{\mathrm{mix}}\gtrsim 1$), enabling gradients and hotspots to persist long enough to drive autocatalysis (Gate~7) and copying (Gate~6). In porous minerals, $D_{\mathrm{eff}}=\varepsilon D/\tau$ with porosity $\varepsilon$ and tortuosity $\tau>1$; labyrinthine geometries can raise $C_{\mathrm{mix}}$ by suppressing mixing without sacrificing access to feeds.

\paragraph*{Reactive porous networks (effectiveness factor).}
In reactive porous media, the reaction rate is reduced by internal diffusion. Define a Thiele modulus 
$\phi \equiv a\,\sqrt{k_{\rm eff}/D_{\rm eff}}$ and effectiveness $\eta_{\rm eff}=\tanh\phi/\phi$. 
Then the net rate constant is $k_{\rm net}=\eta_{\rm eff}\,k_{\rm eff}$ and the confinement number becomes 
$C_{\mathrm{mix}}=\tau_{\rm mix}/\tau_r \simeq a^2/(D_{\rm eff}\,\tau_r)$ with $\tau_r\sim 1/k_{\rm net}$.  Mass-transfer limitation at interfaces is captured by $F_{\rm mass}(Pe,Da)$ in Eq.~(\ref{eq:Jsurf}), with 
$Pe\equiv u a/D_{\rm eff}$ and $Da\equiv k_{\rm net} a/u$.

\subsection{Permeation, leakage, and retention times}
For a spherical vesicle of radius $R$, the passive diffusive flux $J_i$ (mol\,s$^{-1}$) of species $i$ across a membrane of permeability $P_i$ is
\begin{equation}
J_i \;=\; P_i\,A\,\big(c_{i,\mathrm{out}}-c_{i,\mathrm{in}}\big),\qquad A=4\pi R^2.
\label{eq:flux}
\end{equation}
The corresponding \emph{leakage rate constant} for the internal concentration $c_{i,\mathrm{in}}$ is
\begin{equation}
k_{\mathrm{leak},i}\ =\ \frac{P_i\,A}{V}\ =\ \frac{3P_i}{R}\,,\qquad V=\frac{4\pi R^3}{3},
\label{eq:k-leak}
\end{equation}
giving a retention time $\tau_{\mathrm{ret},i}=1/k_{\mathrm{leak},i}=R/(3P_i)$. Early amphiphile bilayers (fatty-acid vesicles) exhibit $P\sim 10^{-7}$--$10^{-5}$\,m\,s$^{-1}$ for neutrals (e.g., nucleosides), but $P\sim 10^{-12}$--$10^{-10}$\,m\,s$^{-1}$ for charged oligomers and ions due to Born solvation barriers. Thus a $R=5~\mu$m vesicle has $k_{\mathrm{leak}}\!\sim\!0.06$\,s$^{-1}$ for neutrals at $P=10^{-7}$\,m\,s$^{-1}$ (half-life $\sim\!12$\,s) but $k_{\mathrm{leak}}\!\sim\!6\times10^{-7}$\,s$^{-1}$ (half-life $\sim\!15$\,days) for a charged oligomer at $P=10^{-12}$\,m\,s$^{-1}$, providing natural molecular selectivity.

For phase-separated coacervates, exchange across the interface is controlled by partitioning $K_i\equiv c_{i,\mathrm{in}}/c_{i,\mathrm{out}}$ and interfacial transfer; an effective leak constant can be written as $k_{\mathrm{ex},i}\!\approx\!(D_{\mathrm{out}}/R)\,\kappa(K_i,D_{\mathrm{in}}/D_{\mathrm{out}})$, with $\kappa$ a dimensionless function (order unity for $K_i\gtrsim 10$ and $D_{\mathrm{in}}\lesssim D_{\mathrm{out}}$). Large $K_i$ values for polyanions (e.g., RNA) yield strong retention with modest exchange of small neutral feedstocks.

\subsection{Throughput: meeting consumption at target function}
The influx of a nutrient $i$ must sustain the target consumption rate $J_{\mathrm{cons},i}$ required by Gate~5 (feedstock budget) and Gate~6 (copying). For passive supply one needs
\begin{equation}
J_{\mathrm{in},i}\ =\ P_i\,A\,\big(c_{i,\mathrm{out}}-c_{i,\mathrm{in}}\big)\ \ge\ J_{\mathrm{cons},i}\ +\ k_{\mathrm{dil}}\,V\,c_{i,\mathrm{in}},
\label{eq:influx}
\end{equation}
where $k_{\mathrm{dil}}$ is the hydrodynamic dilution rate in a flow-through niche. For template copying at rate $\dot N_{\mathrm{copy}}$ (monomers\,s$^{-1}$) with stoichiometric coefficient $\nu_i$ for monomer $i$, the minimal $J_{\mathrm{cons},i}\!\approx\!\nu_i\,\dot N_{\mathrm{copy}}/N_A$; more generally, include activation and side reactions from Gate~5. In early vesicles, neutral activated monomers can be supplied rapidly (large $P_i$) while charged informational polymers remain trapped (small $P_i$), satisfying both \eqref{eq:influx} and \eqref{eq:gate8}.

\subsection{Gradients, leak power, and energetic coupling}

If a compartment exploits electrochemical gradients (e.g., $\Delta\psi$, $\Delta$pH), leak currents impose a continuous energetic burden on Gate~4. With an ionic conductance per area $g_m$ [S\,m$^{-2}$], the leak power is $P_{\mathrm{leak}}=A\,g_m\,\Delta\psi^2$, or, more generally, $P_{\mathrm{leak}}=\sum_\alpha J_\alpha^{\mathrm{leak}}\,\Delta\mu_\alpha$ summing over leaking species $\alpha$. Gate~8 then implicitly requires $P_{\mathrm{in}}$ (Gate~4) to exceed $P_{\mathrm{maint}}$ by at least $P_{\mathrm{leak}}$, and membranes/phase boundaries with sufficiently low ionic permeability to keep $g_m$ small.

\textit{Energetic coupling.} The ionic/molecular leak power $P_{\rm leak}=\sum_\alpha J^{\rm leak}_\alpha\,\Delta\mu_\alpha$ [cf.~Eq.~(\ref{eq:Pmaint})] must be carried in $P_{\rm maint}$ (Gate~4); practically, Gate~8 requires margins $\{m_C,\min_i,\;m^{\rm leak}_j\}$ \emph{and} $P_{\rm leak}\ll P_{\rm in}$.

\subsection{Concentration by partitioning and crowding}
Partitioning into coacervates or adsorption onto mineral surfaces concentrates reactants, effectively renormalizing rate constants. For a bimolecular reaction $A+B\to$ products with partition coefficients $K_A, K_B$, the in-compartment rate 
\begin{equation}
r_{\mathrm{in}}\ =\ k\,K_A K_B [A]_{\mathrm{out}}[B]_{\mathrm{out}},
\end{equation}
so that the functional threshold concentrations in Gate~5 scale down by $1/(K_A K_B)$. This directly increases $\mathcal{C}$ by lowering $\tau_r$ and raises autocatalytic gain (Gate~7) without changing external abundances.

\subsection{Geometry, growth, and division}
For vesicles, surface area $A$ and volume $V$ evolve under lipid uptake and osmotic water flux. Writing $\dot A = r_\ell$ (lipid insertion) and $\dot V = L_p A\,\Delta\Pi$ (hydraulic permeability $L_p$, osmotic driving $\Delta\Pi$), shapes become unstable to budding/pearling when the reduced volume $v\equiv (6\sqrt{\pi}V)/A^{3/2}$ falls below a threshold controlled by bending modulus $\kappa$ and area compressibility $K_A$. A sufficient condition for growth--division cycling is
\begin{equation}
\frac{\dot A}{A} \;>\; \tfrac12\,\frac{\dot V}{V}\qquad\text{and}\qquad
\Delta p = \frac{2\gamma}{R}\ \ \text{(Laplace)}\ \lesssim\ \text{lysis threshold},
\label{eq:growth-division}
\end{equation}
so that area increases faster than volume until an area-excess instability yields division. Gate~8 does not \emph{require} division, but division provides a natural route to lineage formation (Gate~9).

\subsection{Quantitative Gate~8 statement and margins}
Gate~8 is passed if there exists a compartment geometry and composition for which
\begin{equation}
\boxed{
\mathcal{C}=\frac{\tau_{\mathrm{ret}}}{\tau_r}\gtrsim 1,\qquad
P_i A (c_{i,\mathrm{out}}-c_{i,\mathrm{in}})\ \ge\ J_{\mathrm{cons},i}\ \ \forall i\in\mathcal{F},\qquad
\frac{3P_j}{R}\,n_{j,\mathrm{in}} \ \le\ J_{\mathrm{make},j}\ \ \forall j\in\mathcal{K},
}
\label{eq:gate8-box}
\end{equation}
where $\mathcal{F}$ is the set of feed species and $\mathcal{K}$ the set of keystone retained species and $
\tau_{\mathrm{ret}} \equiv \min_i \tau_{\mathrm{exch},i},
\tau_{\mathrm{exch},i} \equiv {V}/(A\,P_i)$. Natural \emph{margins} are the confinement slack $m_{\mathcal{C}}\equiv \mathcal{C}$, the throughput slack $m_{\mathrm{in},i}\equiv J_{\mathrm{in},i}/J_{\mathrm{cons},i}$, and the retention slack $m_{\mathrm{leak},j}\equiv J_{\mathrm{make},j}/(k_{\mathrm{leak},j}n_{j,\mathrm{in}})$; Gate~8 calls for all margins $\gtrsim 1$ with safety factors against fluctuations.

\subsection{Worked Examples}
\paragraph*{Fatty-acid vesicle (5\,$\mu$m).}
$R=5\times10^{-6}$\,m, $A=3.1\times10^{-10}$\,m$^2$, $V=5.2\times10^{-16}$\,m$^3$. Neutral activated monomer: $P=10^{-7}$\,m\,s$^{-1}$, $\Delta c=10~\mu$M $\Rightarrow$ $J_{\mathrm{in}}\approx P A \Delta c \approx 3.1\times10^{-19}$\,mol\,s$^{-1}\approx 1.9\times10^{5}$ molecules\,s$^{-1}$, easily exceeding a copying demand of $\mathcal{O}(10^2$--$10^3)$ monomers\,s$^{-1}$. Charged 20-mer: $P=10^{-12}$\,m\,s$^{-1}$ gives $k_{\mathrm{leak}}=3P/R\approx 6\times 10^{-7}$\,s$^{-1}$; at $c_{\mathrm{in}}=1\,\mu$M, leakage is $\sim\!0.2$ molecules\,s$^{-1}$, so $m_{\mathrm{leak}}\!\gg\!1$ for modest $J_{\mathrm{make}}$. With $D_{\mathrm{eff}}=10^{-10}$\,m$^2$\,s$^{-1}$ (small solutes), \(\tau_{\rm mix}\approx 0.25~\mathrm{s}\), hence $C_{\mathrm{mix}} \approx 2.5\times10^{-3}$--$2.5\times10^{-2}$ for $\tau_r=10$--$10^2$ s (well mixed).
For macromolecules/crowded interiors with \(D_{\rm eff}\sim 10^{-12}\,\mathrm{m}^2\mathrm{s}^{-1}\), \(\tau_{\rm mix}\approx 25~\mathrm{s}\) and $C_{\mathrm{mix}} \approx 0.25$--$2.5$, indicating that sustained gradients/hotspots may be feasible in those regimes, consistent with Gate~8 localization. \emph{Unit reminder:} $10~\mu\mathrm{M}=10^{-5}\ \mathrm{mol\ L^{-1}}=10^{-2}\ \mathrm{mol\ m^{-3}}$.

\paragraph*{Coacervate droplet (10\,$\mu$m).}
Assume $K_{\mathrm{RNA}}=100$, $D_{\mathrm{eff}}=5\times10^{-12}$\,m$^2$\,s$^{-1}$ for macromolecules: $\tau_{\mathrm{mix}}\approx 20$\,s. If template-directed ligation has $\tau_r\sim 10$\,s at the elevated internal concentration, $\mathcal{C}\sim 2$; the partitioning lowers Gate~5 thresholds by $1/K^2$ for bimolecular steps, and the effective exchange $k_{\mathrm{ex}}\sim D_{\mathrm{out}}/R\sim 10^{-4}$\,s$^{-1}$ is slow enough to retain long polymers while admitting small neutral feeds.

\paragraph*{Mineral pore ($R=50\,\mu$m, tortuous).}
With $D_{\mathrm{eff}}\sim 10^{-10}$\,m$^2$\,s$^{-1}$, $\tau_{\mathrm{mix}}\sim 25$\,s; advective refresh sets $k_{\mathrm{dil}}$, but surface adsorption further retards loss of key species (effective $k_{\mathrm{leak}}\downarrow$). The large area-to-volume ratio of pore walls decouples $J_{\mathrm{in}}$ (set by external flow and diffusion) from $k_{\mathrm{leak}}$ of macromolecules, often yielding $m_{\mathrm{in}}\!\gtrsim\!1$ and $m_{\mathrm{leak}}\!\gtrsim\!1$ simultaneously.

\subsection{Diagnostics, falsifiers, and observational corollaries}
\paragraph*{Diagnostics.}
(i) Measure $P_i$ and $D_{\mathrm{eff}}$ for candidate membranes/condensates as functions of ionic strength, pH, and temperature; compute $k_{\mathrm{leak},i}$ via \eqref{eq:k-leak}. (ii) Quantify partition coefficients $K_i$ for critical species; propagate to effective rate enhancements and Gate~5 thresholds. (iii) Determine $\tau_r$ from kinetic assays at in-compartment concentrations; evaluate $\mathcal{C}$. (iv) Characterize hydraulic permeability $L_p$, bending modulus $\kappa$, and area modulus $K_A$ to test growth--division feasibility \eqref{eq:growth-division}. (v) For porous media, estimate $D_{\mathrm{eff}}$ from porosity/tortuosity and advective exchange ($k_{\mathrm{dil}}$).

\paragraph*{Falsifiers.}
Gate~8 fails if, for any plausible geometry/composition: (a) $\mathcal{C}_{\rm ret}\ll 1$  so key reactions cannot complete before loss across the boundary (insufficient retention);
(b) $m_{\mathrm{in},i}<1$ for essential feeds even at maximal $\Delta c$ (permeation too slow or external supply too low); (c) $m_{\mathrm{leak},j}<1$ for keystone species because $P_j$ is too high (no retention) or $J_{\mathrm{make},j}$ too small; (d) leak power $P_{\mathrm{leak}}$ exceeds Gate~4 margins, collapsing gradients. (Optional gradient-based architectures: $C_{\rm mix}\gtrsim 1$ is required only when sustained internal gradients are invoked.)

\paragraph*{Observational corollaries.}
Solvent identity and salinity strongly modulate $P_i$, $D_{\mathrm{eff}}$,  $K_i$. Thus, detections of surface liquids (e.g., ocean glint, specular polarization), constraints on aerosol microphysics, and bulk ionic strength (from radio occultations or refractivity) provide priors on Gate~8 viability. Worlds with abundant surface/near-surface liquids and moderate ionic strengths favor membranes and coacervates with $m_{\mathcal{C}},m_{\mathrm{in}},m_{\mathrm{leak}}\gtrsim 1$; extremely low-temperature hydrocarbons or highly oxidizing brines may require alternative compartment chemistries to satisfy \eqref{eq:gate8-box}.

Gate~8 converts the qualitative idea of “a container that keeps the good stuff in and lets the food in” into quantitative transport and stability criteria. When its margins are healthy, it amplifies Gate~7 gain (by increasing encounter rates) and reduces Gate~6 errors (by sheltering templates and enabling processivity), thereby enabling the transition to population dynamics in Gate~9. See Sec.~\ref{sec:validation}  for consolidated criteria. Gate~8 compresses to the transport--confinement inequality (\ref{eq:gate8-box}); leakage/maintenance power and selective transport set the proxies.
 
\section{Gate 9: Onset of Darwinian Dynamics}
\label{sec:gate9}

Gate~9 marks the transition from chemically self-amplifying systems (Gate~7) to lineages that display heritable variation and differential reproduction. Two quantitative conditions must hold together:
(i) supercritical growth of absolute counts in the face of physical loss (washout, decay), and
(ii) selection strong enough to overcome stochastic drift on a quasi-stationary environmental background.

\subsection{Birth--death--mutation dynamics and the leading exponent}
Let $n_i(t)$ be the abundance of variant $i$ (sequence or protocell genotype), with per-capita birth and death rates $b_i$ and $d_i$ in a niche that also imposes dilution/removal $\delta$. Let $Q_{ji}$ be the transition probability that an offspring of type $i$ is of type $j$ ($\sum_j Q_{ji}=1$). Linearizing the population process gives
\begin{equation}
\dot{\mathbf{n}} \;=\; \big(BQ - D\big)\,\mathbf{n}, 
\qquad
B=\mathrm{diag}(b_i),\quad D=\mathrm{diag}(d_i+\delta).
\label{eq:BDM}
\end{equation}
The leading Lyapunov exponent (Malthusian growth rate) is
\begin{equation}
r^* \;\equiv\; \lambda_{\max}\!\big(BQ - D\big).
\label{eq:lambda}
\end{equation}
{Requirement~9a (supercriticality):} There exists a niche (fixed by Gates~4--8) such that
\begin{equation}
r^* \;>\; 0.
\label{eq:supercrit}
\end{equation}
When \eqref{eq:supercrit} holds, normalized frequencies $x_i=n_i/\sum_k n_k$ obey the replicator--mutator equation
\begin{equation}
\dot{x}_i \;=\; \sum_{j} x_j f_j Q_{ij} - \bar f\,x_i,\qquad
f_i \equiv b_i - d_i - \delta,\quad \bar f=\sum_j x_j f_j,
\label{eq:rep-mut}
\end{equation}
whose deterministic attractor is the leading right eigenvector of $FQ$ with $F=\mathrm{diag}(f_i)$ (standard replicator--mutator dynamics; see \cite{Eigen1971,Nowak2006,HofbauerSigmund1998}).

\subsection{Selection versus drift; beneficial supply}
Finite populations drift. Let $N_e$ be the effective size of the reproducing entities that carry heritable information (templates within compartments or the compartments themselves, depending on Gate~8). For a beneficial mutation with selection coefficient $s$ (difference in Malthusian fitness), selection resolves against drift when
\begin{equation}
N_e\,s \ \gtrsim\ 1.
\label{eq:Ne_s}
\end{equation}
Adaptive progress also requires a sufficient supply of beneficial variants. If $U_b$ is the per-genome beneficial mutation rate (a fraction of $U=\mu L$ from Gate~6), the beneficial supply rate is
\begin{equation}
\Lambda_b \;\equiv\; N_e\,U_b.
\label{eq:beneficial-supply}
\end{equation}
A practical condition for sustained adaptation in the strong-selection regime is $\Lambda_b \gtrsim 1$ per selection time $t_{\mathrm{sel}}\sim 1/s$, with clonal interference corrections when $\Lambda_b\gg 1$.

\subsection{Timescale separation and environmental stationarity}
Let $\tau_{\mathrm{env}}$ be the correlation time of the environment experienced by the lineage (resource, temperature, redox; Gates~\ref{sec:gate4}--\ref{sec:gate8}). Selection must act before the niche drifts appreciably:
\begin{equation}
\frac{1}{s} \ \ll\ \tau_{\mathrm{env}}
\quad\Longleftrightarrow\quad
s\,\tau_{\mathrm{env}} \ \gg\ 1.
\label{eq:timescale}
\end{equation}
Frequent sterilizing pulses or rapid environmental wandering can preclude lasting improvements even if \eqref{eq:supercrit} is momentarily satisfied.

\subsection{Coupling to Gate~6 and mutation load}
Write $Q=\exp(-U)\,I + \text{off-diagonals}$ with per-genome error rate $U=\mu L$. Gate~6 enforces $U<\ln s$ to maintain localization in sequence space. At the population level, mutation load reduces $r^*$ roughly by $U\,\bar{s}_d$ (mean deleterious effect $\bar{s}_d$), thereby lowering $\lambda_{\max}(BQ-D)$ and the effective selection coefficients. A meltdown regime occurs when deleterious input exceeds selection capacity (heuristically $U_d \gtrsim s_{\min}$ across most segregating sites), violating \eqref{eq:Ne_s} for advantageous classes. Hence Gate~6 (per-site $\mu$ and $L$) and Gate~4 (energy for proofreading) jointly determine whether \eqref{eq:supercrit} and \eqref{eq:Ne_s} can be met.

\subsection{Compartment-level selection}
When heritability is expressed at the protocell level (Gate~8), reproduction proceeds via growth--division cycles with imperfect assortment of internal templates. Let $W_g$ be the expected number of surviving daughter compartments per cycle from a focal compartment and $z$ a heritable compartment trait (e.g., membrane composition, catalyst density). The Price equation (see \cite{Price1970}) decomposes trait change per cycle:
\begin{equation}
\Delta \bar z \;=\; \frac{\mathrm{Cov}(W_g,z)}{\bar W_g} \;+\; \frac{\mathbb{E} \big[ W_g\,\Delta z_{\mathrm{within}}\big]}{\bar W_g}.
\label{eq:price}
\end{equation}
Darwinian dynamics at the group scale requires $\bar W_g>1$ (the group analogue of \eqref{eq:supercrit}) and nonzero heritable variance so that the covariance term is nonzero. The effective $N_e$ in \eqref{eq:Ne_s} is then the number of reproducing compartments, not the total count of templates.  In protocell lineages, imperfect assortment of internal templates reduces effective selection coefficients  (``assortment load''), lowering the covariance term in (\ref{eq:price}); we account for this by using group-level $s_g$ derived from partitioning statistics.

In a steady CSTR with volumetric dilution \(\delta\) and census \(N\), 
\(N_e\approx N/(1+c_v)\), where \(c_v\) is the coefficient of variation of offspring number over a selection time \(1/s\). 
As a conservative rule, require \(N s \gtrsim 10\) when \(c_v\sim 1\) to achieve \(N_e s\gtrsim 1\).

\subsection{Quantitative Gate~9 statement and margins}
Gate~9 is passed if, for at least one plausible heritability level (molecule or compartment) under the niche fixed by Gates~4--8,
\begin{equation}
\boxed{
\lambda_{\max}(BQ-D)>0;\qquad
N_e s \gtrsim 1,\ \ \Lambda_b = N_e U_b \gtrsim 1,\ \ s\,\tau_{\mathrm{env}}\gg 1.
}
\label{eq:gate9-box}
\end{equation}
Useful margins are $m_{r}\equiv \lambda_{\max}/\delta$ (growth over dilution), $m_{\rm drift}\equiv N_{e}s$, and $m_{\rm ben}\equiv \Lambda_{b}$.

\subsection{Worked Examples}
\paragraph*{Template-level selection in a flow reactor.}
Take $b_{\mathrm{master}}=1.0\times 10^{-3}$\,s$^{-1}$, $b_{\mathrm{mut}}=(1-s)b_{\mathrm{master}}$ with $s=0.02$, and $d_i+\delta=2.0\times 10^{-4}$\,s$^{-1}$. With Gate~6 ensuring $U=\mu L=0.3$ ($Q^L\simeq 0.74$), $\lambda_{\max}(BQ-D)\approx 6\times 10^{-4}$\,s$^{-1}>0$. If the reactor holds $N_e\simeq 10^7$ templates, then $N_e s\simeq 2\times 10^5\gg 1$; for $U_b=10^{-8}$ per genome per second, $\Lambda_b=N_e U_b\simeq 0.1$\,s$^{-1}$ ($\sim 6$ per minute), indicating ample beneficial supply (with clonal interference expected).

\paragraph*{Compartment-level selection.}
Suppose protocells divide every $\tau_{\mathrm{div}}=10^4$\,s with survival probability $p_{\mathrm{surv}}=0.7$ and mean daughters $d=2(1+\eta)$ with $\eta=0.05$ heritable enhancement. Then $\bar W_g=p_{\mathrm{surv}}\,d\simeq 1.47>1$. For a metapopulation with $N_e\simeq 10^4$ reproducing compartments, between-compartment selection differential $s_g=\eta/\tau_{\mathrm{div}}\simeq 5\times 10^{-6}$\,s$^{-1}$, one has $N_e s_g\simeq 50\gg 1$, satisfying \eqref{eq:Ne_s} at group scale even if within-compartment selection is weak.

\subsection{Diagnostics, falsifiers, and corollary}
\textit{Diagnostics.} In microfluidic continuous-flow reactors or serial dilution assays: (i) measure $b_i,d_i,\delta$ to compute $\lambda_{\max}(BQ-D)$; (ii) infer $Q$ from error spectra (Gate~6) to parameterize $BQ$; (iii) estimate $N_e$ from census sizes and turnover; (iv) measure $U$ and decompose into $U_b$/$U_d$ to evaluate $\Lambda_b$; (v) modulate $\tau_{\mathrm{env}}$ (nutrient/redox forcing) to test $s\,\tau_{\mathrm{env}}\gg 1$.

\textit{Falsifiers.} Gate~9 fails if $\lambda_{\max}(BQ-D)\le 0$ under all realistic feeds and geometries; or if $N_e s\ll 1$ at every heritability level, implying drift dominance; or if deleterious input $U_d$ collapses $r^*$ despite Gate~6 holding locally; or if $\tau_{\mathrm{env}}$ is too short for selection to act.

\textit{Observational corollary.} Darwinian dynamics is not directly remote-sensed. However, concurrent satisfaction of Gates~10--12 (scalable coding, positive net productivity, stabilized feedbacks) is strong circumstantial evidence for an underlying Darwinian engine; conversely, transient disequilibria without persistence (Gate~11 failure) suggest sub-Darwinian chemistry despite Gate~7 closure.

In sum, Gate~9 supplies the spectral and population-genetic criteria that elevate protometabolism to evolution: a positive leading growth exponent in the presence of mutation and loss, and selection strong enough to overcome drift on a quasi-stationary environment. Its margins propagate forward to Gate~10 (coding scalability) and Gate~11 (ecosystem assembly). See Sec.~\ref{sec:validation}  for consolidated criteria. Gate~9 uses the spectral‐radius criterion (\ref{eq:gate9-box}); growth--mixing separation in flow reactors provides the diagnostic handle.

\section{Gate 10: Coding, Translation, and Energy--Information Tradeoffs}
\label{sec:gate10}

Gate~10 asks whether a noisy mapping from digital templates to functional polymers can be sustained at scale under the available free-energy and error constraints \cite{HaigHurst1991,FreelandHurst1998}. The central inequality compares the communication capacity of the translation channel to the coding rate required by the chosen alphabet and code structure. Channel capacity and finite-blocklength coding bounds follow \cite{Shannon1948,CoverThomas2006,Polyanskiy2010}. 

\subsection{Channel model and gate inequality}

Let a codon be $n$ symbols drawn from a $k$-letter template alphabet; the code maps the $k^n$ codons to $m$ product types (e.g., monomers), with degeneracy $d=k^n/m$. Denote by $C_{\rm cap}$ the \emph{per-symbol} Shannon capacity (bits per template symbol) of the symbol-level readout channel and by $R_{\mathrm{code}}$ the \emph{per-symbol} rate required to convey which of $m$ product types is intended via length-$n$ codons:
\begin{equation}
C_{\rm cap} \;\;\text{[bits / symbol]},\qquad
R_{\rm code}(n,m)  \equiv \frac{\log_2 m}{n}\quad [\text{bits/symbol}],
\qquad
\chi \equiv \frac{C_{\rm cap}}{R_{\rm code}}= \frac{n\,C_{\rm cap}}{\log_2 m}.
\label{eq:CR-defs}
\end{equation}
Gate~10 requires a slack $\chi>1$ so that reliable coding is possible \emph{with overhead} for regulation and finite-blocklength penalties:
\begin{equation}
\boxed{ \chi \equiv \frac{C_{\mathrm{cap}}}{R_{\mathrm{code}}} > 1. }
\label{eq:gate10}
\end{equation}
For the $k$-ary symmetric channel with per-symbol error $\varepsilon_b$,
\begin{equation}
C_{\rm cap} \;\equiv\; \log_2 k \;-\; H_k(\varepsilon_b),\qquad
H_k(\varepsilon_b)=-(1-\varepsilon_b)\log_2(1-\varepsilon_b)-\varepsilon_b\log_2\!\frac{\varepsilon_b}{k-1}.
\label{eq:capacity-symmetric}
\end{equation}
More generally, with a confusion matrix $T(y|x)$ over symbol input $x$ and output $y$, $C_{\mathrm{cap}} = \max_{p(x)} I(X; Y)$.
We measure both channel capacity and code rate in \emph{bits per channel symbol}. Thus $\mathcal{C}_{\rm cap}(\varepsilon)\in[0,\log_2 k]$ and $R_{\rm code}=(1/n)\log_2 m$; equivalently, the dimensionless \emph{fractional} rate is $r \equiv R_{\rm code}/\log_2 k \in [0,1]$.

\paragraph*{From symbol to codon.}
If the per-symbol error is $\varepsilon_b$, the effective \emph{codon} error that flips at least one of the $n$ symbols is $\varepsilon_{\mathrm{cod}}=1-(1-\varepsilon_b)^n \simeq n\varepsilon_b$ for $\varepsilon_b\ll 1$. Capacity per codon is then $n C_{\mathrm{cap}}(\varepsilon_b)$ while the payload per codon is $\log_2 m$, and Gate~10 is equivalently $n C_{\mathrm{cap}}(\varepsilon_b)>\log_2 m$.

\subsection{Energy--accuracy--speed tradeoff}
Lowering $\varepsilon_b$ below its equilibrium discrimination value $\varepsilon_{\mathrm{eq}}$ requires dissipation. A thermodynamic bound relates the mean dissipated work per symbol, $W_{\mathrm{proof}}$, to the error reduction factor:
\begin{equation}
W_{\mathrm{proof}} \ \ge\ k_{\mathrm B}T \ln\!\Big(\frac{\varepsilon_{\mathrm{eq}}}{\varepsilon_b}\Big),
\label{eq:proof-bound-g10}
\end{equation}
with kinetic proofreading schemes achieving near this scaling \cite{Hopfield1974,Ninio1975}, trade-offs between speed, dissipation, and accuracy in proofreading/copying \cite{Murugan2012,SartoriPigolotti2013}.
 Translation throughput $\nu$ (symbols\,s$^{-1}$) therefore draws power $P_{\mathrm{proof}} \ge \nu\,W_{\mathrm{proof}}$, which must be supplied by Gate~4’s power margin. Speed also competes with accuracy: proofreading typically slows elongation, reducing $\nu$ and raising the dwell time subject to damage; optimal operation balances $C_{\mathrm{cap}}(\varepsilon_b)$ gains against $P_{\mathrm{proof}}$ and throughput costs.

\subsection{Code design under noise: degeneracy and error cost}
When $\chi$ is modest ($\gtrsim 1$), performance depends on how the $k^n$ codons are partitioned into $m$ synonym sets. Let $T_{\mathrm{cod}}(\hat x|x)$ be the codon-level confusion matrix and $W(a,b)$ a penalty (``cost'') for substituting product type $b$ when $a$ was intended (e.g., a chemical or functional distance). The expected error cost under prior $p(x)$ and mapping $M:\{1,\dots,k^n\}\to\{1,\dots,m\}$ is
\begin{equation}
\mathcal{L}[M] \;=\; \sum_{x} p(x) \sum_{\hat x} T_{\mathrm{cod}}(\hat x|x)\; W\!\big(M(x),M(\hat x)\big).
\label{eq:expected-cost}
\end{equation}
Good codes minimize $\mathcal{L}[M]$ subject to $|M^{-1}(a)|\approx d$ for all $a$, which clusters codons that are confusable (large $T_{\mathrm{cod}}$) into the same synonym class whenever those confusions are costly. This is the chemical analogue of error-correcting code design on a weighted Hamming graph. Increasing degeneracy $d$ (larger synonym sets) reduces $R_{\mathrm{code}}$ [Eq.~\eqref{eq:CR-defs}] and allows larger safety margins at fixed $\varepsilon_b$.

\paragraph*{Toy error-cost example.}
Let $k=4$, $n=3$, $m=20$, and define a chemical distance $W(a,b)$ with two tiers (``similar'' cost $w_s$ and 
``dissimilar'' cost $w_d\gg w_s$). With a symmetric per-symbol confusion $\varepsilon_b$, the expected codon-level 
confusion concentrates on Hamming-1 neighbors; mapping confusable codons to the same product minimizes 
$L[M]$ in Eq.~(\ref{eq:expected-cost}) while increasing degeneracy $d=k^n/m$. This reduces $R_{\rm code}$ [Eq.~(\ref{eq:CR-defs})] and increases slack $\chi$ at fixed $\varepsilon_b$.

\subsection{Quantitative parametrics}

Take $k=4$ letters, $n=3$ symbols per codon, and $m=20$ product types. Then
\begin{equation}
R_{\rm code} = \frac{\log_2 20}{3} \approx 1.441\ \text{bits/symbol}.
\end{equation}

\begin{table}[t]
\centering
\caption{4-ary symmetric channel. Here $R_{\rm code}=\log_2(m)/n$ with $(k,n,m)=(4,3,20)$, so
$R_{\rm code}\simeq 1.441$ bits/symbol.}
\label{tab:k4cap}
\begin{tabular}{c c c c}
\hline
$\varepsilon_b$ & $C_{\rm cap}$ (bits/symbol) & $\chi=C_{\rm cap}/R_{\rm code}$ & Verdict\\
\hline
0.01  & 1.903 & 1.32 & Pass\\
0.05  & 1.634 & 1.13 & Pass (narrow)\\
0.08  & 1.471 & 1.02 & Boundary\\
0.10  & 1.370 & 0.95 & Fail\\
0.15  & 1.160 & 0.80 & Fail\\
0.25  & 0.792 & 0.55 & Fail\\
\hline
\end{tabular}
\end{table}

Table~\ref{tab:k4cap} gives anticipated values for the 4-ary symmetric channel. With $n=3$ and $m=20$,
$R_{\rm code}=\log_2(20)/3\simeq 1.441$ bits/symbol. The boundary $\chi\simeq 1$ occurs near $\varepsilon_b\simeq 0.08$--$0.09$; $\varepsilon_b=0.10$ fails and $\varepsilon_b=0.05$ passes Gate 10 with modest slack.
If noise is higher, one must (i) decrease $\varepsilon_b$ (via additional $W_{\rm proof}$), (ii) increase $n$ (longer codons), or (iii) accept a much smaller effective product alphabet $m$.

\subsection{Link to Gates 4, 6, and 9}
Gate~6 controls the fidelity of template replication ($\mu$) and thus the maintainable genome length $L$; Gate~10 controls the fidelity of reading out stored information into functional polymers. Both draw on the same power budget: $P_{\mathrm{proof}}$ in \eqref{eq:proof-bound-g10} reduces the surplus $\Phi$ of Gate~4. Gate~9 requires sufficient beneficial supply and selection; the functional consequence of Gate~10 is to increase the accessible phenotype space (larger $m$ and proteome diversity) at a tolerable error load.

\subsection{Finite-blocklength and redundancy overhead}
\label{sec:blocklength}

Inequality $\chi>1$ is asymptotic. At finite blocklength, feasibility requires a margin above capacity. In the normal approximation for a memoryless channel for Gate 10,
{}
\begin{equation}
C_{\rm cap}(\varepsilon_b) \;\ge\; R_{\rm code}(n,m) \;+\; \Delta_{\rm sym},
\qquad
\Delta_{\rm sym} \simeq \sqrt{\frac{V}{N}}\;Q^{-1}(p_e),
\label{eq:fin-back}
\end{equation}
or equivalently,
\begin{equation}
n\,C_{\rm cap}(\varepsilon_b)\ \ge\ \log_2 m\ +\ n\,\Delta_{\rm sym},
\end{equation}
where $N$ is the number of template symbols in a codeword\footnote{Notation: $n$ denotes the biochemical codon length (bases per codon), whereas $N$ denotes the information-theoretic blocklength (symbols per codeword) used in the finite-blocklength bound.
}, $V$ is the channel dispersion, $Q^{-1}$ the Gaussian tail-inverse, and $p_e$ the target block error probability for finite blocklength coding \citep{Polyanskiy2010}. Equivalently, $\chi=C_{\rm cap}/R_{\rm code}\ge 1+\Delta_{\rm sym}/R_{\rm code}$. This backoff can be absorbed into a practical margin on $\chi$ (e.g., require $\chi\gtrsim 1.2$).

\paragraph*{Blocklength notation.}
Let $N_{\rm sym}$ denote the number of \emph{template symbols} in a codeword (finite-blocklength parameter).
If a codeword contains $N_{\rm cod}$ codons of length $n$, then $N_{\rm sym}=n\,N_{\rm cod}$.
Eq.~(\ref{eq:fin-back}) uses $N_{\rm sym}$.

Take $(k,n,m)=(4,3,20)$ so $R_{\rm code}=\log_2(20)/3\simeq 1.441$ bits/symbol, blocklength $N=300$ symbols,
and target block error $p_e=10^{-3}$. For $\varepsilon_b=0.05$, Table~\ref{tab:k4cap} gives $C_{\rm cap}\simeq 1.634$ bits/symbol, so $\chi\simeq 1.13$ asymptotically. With $V=\mathcal{O}(1)$--few bits$^2$/symbol, the normal-approximation backoff $\Delta_{\rm sym}\simeq \sqrt{V/N}\,Q^{-1}(p_e)$ is $\mathcal{O}(0.1)$ bits/symbol, implying the practical feasibility condition $C_{\rm cap}\ge R_{\rm code}+\Delta_{\rm sym}$ can be marginal at $N=300$ unless $\varepsilon_b$ is smaller (or $N$ larger). This motivates a practical budget $\chi \gtrsim 1.2$--$1.4$ depending on $(N,p_e)$.

\paragraph*{Choosing safety factors from uncertainty.}
Let $m$ be a dimensionless gate margin with uncertainty $\sigma_{\ln m}$ induced by environmental variability and
model error. Approximating $\ln m$ as Gaussian, the gate-failure probability is
$P(m<1)\approx \Phi\!\left(-\ln m/\sigma_{\ln m}\right)$.
Thus requiring a $z$-sigma buffer implies $\ln m \ge z\,\sigma_{\ln m}$, i.e.
\[
m_{\rm target}=\exp(z\,\sigma_{\ln m}).
\]
For example, $\sigma_{\ln m}\simeq 0.2$ and $z=2$ gives $m_{\rm target}\simeq 1.49$.

\subsection{Quantitative Gate~10 statement and margins}
Gate~10 is passed if there exist $(k,n,m,d,\varepsilon_b,W_{\mathrm{proof}},\nu)$ consistent with the niche such that
\begin{equation}
\hskip -4pt 
\boxed{
\chi=\frac{C_{\mathrm{cap}}(\varepsilon_b)}{R_{\mathrm{code}}(k, n, m)} > 1,~~~
P_{\mathrm{proof}}=\nu\,W_{\mathrm{proof}} \le P_{\mathrm{in}}-P_{\mathrm{maint}} \ \ (\text{Gate 4}),~~~
\mathcal{L}[M]\ \text{ minimized subject to } d=\frac{k^n}{m}.
}
\label{eq:gate10-box}
\end{equation}
Convenient margins are $m_{\chi}\equiv \chi$, the energetic slack $m_{P}\equiv (P_{\mathrm{in}}-P_{\mathrm{maint}})/P_{\mathrm{proof}}$,  the degeneracy slack $m_d\equiv d/d_{\min}$.

\begin{theorem}[Gate 10: Rate--capacity feasibility]\label{thm:gate10}
Let $k$ be the template alphabet, $n$ the codon length, $m$ the product alphabet (degeneracy $d=k^n/m$), and let the symbol channel have capacity $C_{\mathrm{cap}}$ and confusion matrix $T(y|x)$ as in \eqref{eq:CR-defs}--\eqref{eq:capacity-symmetric}. With finite-blocklength overhead $\Delta$ in \eqref{eq:fin-back} and proofreading work per symbol $W_{\mathrm{proof}}$ drawing power $P_{\mathrm{proof}}=\nu W_{\mathrm{proof}}$ from the Gate~4 surplus \eqref{eq:gate4-box}, reliable coding at payload $\log_2 m$ per codon is achievable iff
\begin{equation}
C_{\rm cap}(\varepsilon_b)\ \ge\ R_{\rm code}(k,n,m)\ +\ \Delta_{\rm sym}(N,p_e)
\quad\text{and}\quad
P_{\rm proof}\le P_{\rm in}-P_{\rm maint},
\end{equation}
where $R_{\mathrm{code}}$ is from \eqref{eq:CR-defs}, the slack $\chi\equiv C_{\mathrm{cap}}/R_{\mathrm{code}}$ in \eqref{eq:gate10-box} must exceed unity after accounting for $\Delta_{\rm sym}$.
\end{theorem}

\begin{remark}
Energy--accuracy coupling is from \eqref{eq:Wproof}: lowering $\varepsilon_b$ increases $C_{\mathrm{cap}}$ but costs dissipation. Code geometry enters via $L[M]$ in \eqref{eq:expected-cost}: when $\chi\gtrsim 1$, clustering confusable codons minimizes functional penalties at fixed $d$.
\end{remark}

\subsection{Worked examples}

\paragraph*{Early noisy channel.} Take $(k,n,m)=(4,3,20)$ so $R_{\rm code}=\log_2(20)/3\simeq 1.441$ bits/symbol, blocklength $N=300$ symbols, and target block error $p_e=10^{-3}$. For $\varepsilon_b=0.05$, Table~\ref{tab:k4cap} gives $C_{\rm cap}\simeq 1.634$
bits/symbol, so $\chi\simeq 1.13$ asymptotically. With $V=\mathcal{O}(1)$--few bits$^2$/symbol, the normal-approximation backoff $\Delta_{\rm sym}\simeq \sqrt{V/N}\,Q^{-1}(p_e)$ is $\mathcal{O}(0.1)$ bits/symbol, implying that
$C_{\rm cap}\ge R_{\rm code}+\Delta_{\rm sym}$ can be marginal at $N=300$ unless $\varepsilon_b$ is smaller (or $N$ larger). This motivates a practical budget $\chi \gtrsim 1.2$--$1.4$ depending on $(N,p_e)$ and $V$.

\paragraph*{Energy-assisted improvement.}
If equilibrium discrimination yields $\varepsilon_{\mathrm{eq}}=0.2$, allocating $W_{\mathrm{proof}}=4\,k_{\mathrm B}T$ per symbol reduces \(\varepsilon_b \approx 0.2e^{-4}\approx 3.7\times 10^{-3}\) by \eqref{eq:proof-bound-g10}. Then $C_{\rm cap}\simeq 1.96$ bits/symbol and $\chi\simeq 2.72$ for $(n,m)=(3,20)$, at a power cost $P_{\mathrm{proof}}=\nu\,4k_{\mathrm B}T$ that must fit within Gate~4’s surplus.

\paragraph*{Longer codons under moderate noise.}
For $\varepsilon_b=0.1$, $k=4$, increasing $n$ from $3$ to $4$ leaves $C_{\rm cap}$ unchanged per symbol but raises per-codon capacity from $3C_{\rm cap}\simeq 4.12$\,bits to $4C_{\rm cap}\simeq 5.49$\,bits. If one simultaneously expands $m$ from $20$ to $32$ ($\log_2 m=5$), the inequality $n C_{\mathrm{cap}} > \log_2 m$ remains satisfied with comfortable slack, provided the biochemical machinery tolerates longer codons.

\paragraph*{Finite-blocklength note.}  Using (\ref{eq:fin-back}), for $N\sim 10^2$--$10^3$ and $p_e\lesssim 10^{-3}$, the backoff typically requires a practical budget $\chi \gtrsim 1.2$--$1.4$ depending on $(N,p_e)$ and channel dispersion $V$.

\subsection{Diagnostics, falsifiers, and corollaries}
\textit{Diagnostics.} (i) Measure symbol-level confusion matrices in reconstituted translation-like systems (ribozyme or ribosomal surrogates) to compute $C_{\rm cap}$; (ii) quantify $W_{\mathrm{proof}}$ per proofreading step and elongation throughput $\nu$ to budget $P_{\mathrm{proof}}$; (iii) evaluate code assignments $M$ by minimizing \eqref{eq:expected-cost} under empirically determined $T_{\mathrm{cod}}$ and chemically motivated $W(a,b)$. (Summarized in the Gate Atlas (Table~\ref{tab:gate-atlas}) and tested in Sec.~\ref{sec:validation}.)

\textit{Falsifiers.} Gate~10 fails generically if, for all feasible $(k,n,m)$ and for all $W_{\mathrm{proof}}$ permitted by Gate~4, one finds $\chi\le 1$ once finite-blocklength backoff is included; or if no assignment $M$ can keep $\mathcal{L}[M]$ below a threshold compatible with Gate~11 productivity (i.e., translation errors overwhelm functional diversity).

\textit{Corollary.} A noisier channel pushes toward larger degeneracy $d$ and/or smaller alphabets $m$, which narrows functional diversity. Conversely, higher power margins (Gate~4) can be converted into lower $\varepsilon_b$ and thus larger $\chi$, enabling richer biosynthetic repertoires and supporting Gate~11. See Sec.~\ref{sec:validation}. Gate~10 is the coding feasibility bound (\ref{eq:gate10-box}); confusion matrices and finite‐blocklength performance set the pass/fail.

\textit{Identifiability:} Remote constraints on $\chi$ are \emph{upper bounds} from energy consistency (Sec.~\ref{sec:rep-stat}); confusion matrices are lab-side inputs. 

\section{Gates 11--12: Biospheres---Productivity, Cycles, and Climate Stability}
\label{sec:gate11-12}

Gates~11--12 elevate Darwinian populations to biospheres: sustained primary production coupled to closed elemental cycles operating far from equilibrium. Gate~11 requires positive NPP and closed elemental cycles on ecological timescales. Gate~12 requires climate--chemistry feedbacks that keep $G_{\rm climate}<1$ with buffering on evolutionary timescales requiring that biosphere--planet feedbacks regulate the environment on timescales long compared to evolutionary turnover. Together they determine whether life remains a thin chemical film or becomes a planetary process. Long-term CO2 regulation and feedback analysis follow \cite{Walker1981,Roe2009}.  

\subsection{Ecological persistence (Gate 11)}
\label{sec:gate11}

Let $x_i$ denote the biomass (mol\,C\,m$^{-2}$) of functional group $i$ (primary producers, heterotrophs, decomposers, lithotrophs), and let $\mathbf{M}=(M_{\mathrm{C}},M_{\mathrm{N}},M_{\mathrm{P}},M_{\mathrm{S}},\ldots)$ be the column-integrated elemental stocks in bioavailable reservoirs (atmosphere+ocean+soil for a surface world, ocean+ice for a subsurface world). A minimal dynamical description is
\begin{align}
\dot x_i &= \Big(\Pi_i(\mathbf{R},\mathbf{E}) - R_i(\mathbf{R},\mathbf{E}) - \delta_i\Big) x_i \;+\; \sum_{j\ne i} T_{ij}(\mathbf{R},\mathbf{E})\,x_j, \label{eq:ecosys-x}\\
\dot{\mathbf{M}} &= \mathbf{F}_{\mathrm{in}} - \mathbf{F}_{\mathrm{loss}} + \mathbf{B}(\mathbf{x},\mathbf{M},\mathbf{E}), \label{eq:ecosys-M}
\end{align}
where $\mathbf{R}$ are resource concentrations (e.g., light, $\mathrm{CO_2}$, $\mathrm{H_2}$, $\mathrm{NO_3^-}$, $\mathrm{PO_4^{3-}}$), $\mathbf{E}$ are environmental controls (temperature, pH, salinity), $\Pi_i$ is gross production per biomass, $R_i$ is respiration/maintenance, $\delta_i$ summarises nonbiological removal, $T_{ij}$ are trophic transfers, $\mathbf{F}_{\mathrm{in}}$ and $\mathbf{F}_{\mathrm{loss}}$ are nonbiological source/sink fluxes (outgassing, hydrothermal supply, escape, burial), and $\mathbf{B}$ are biological transmutation fluxes (e.g., fixation, nitrification/denitrification, sulfur oxidation/reduction).

\paragraph*{Net primary productivity (NPP) and power consistency.}
Define areal NPP (mol\,C\,m$^{-2}$\,s$^{-1}$)
\begin{equation}
\mathrm{NPP}\;=\; \sum_{i\in \mathcal{P}} \Big(\Pi_i(\mathbf{R},\mathbf{E}) - R_i(\mathbf{R},\mathbf{E})\Big)x_i,
\label{eq:NPP-def}
\end{equation}
with $\mathcal{P}$ the producer set. Power consistency ties Gate~11 to Gate~4: if $\Delta G_{\mathrm{assim}}$ is the free energy stored per mol\,C fixed (pathway-dependent) and $\eta_{\mathrm{bio}}\in(0,1)$ accounts for biosynthetic overheads, then the areal power draw by primary production is
\begin{equation}
P_{\mathrm{NPP}} \;=\; \frac{\mathrm{NPP}\cdot\Delta G_{\mathrm{assim}}}{\eta_{\mathrm{bio}}}
\;\le\; P_{\mathrm{in}}-P_{\mathrm{maint}} \quad \text{(Gate 4)}.
\label{eq:NPP-power}
\end{equation}

\paragraph*{Fundamental-physics bound on NPP.}
Combining minimal erasure and proofreading costs with the Gate~4 surplus gives an explicit upper bound on areal productivity \cite{Turyshev2025-NPP}:
\begin{equation}
\mathrm{NPP}\ \le\ 
\frac{
P_{\mathrm{in}}-P_{\mathrm{maint}}
-\dot N_{\mathrm{copy}} \Big(
k_{B}T\ln 2\,\big(\log_{2}k_{p}-H_{k_{p}}(\mu)\big)
+ W_{\mathrm{proof}}
\Big)
}{
\Delta G_{\mathrm{assim}}/\eta_{\mathrm{bio}}
}.
\label{eq:PhiNPP}
\end{equation}
Here $\dot N_{\rm copy}$ is the \emph{areal} monomer-incorporation throughput (monomers m$^{-2}$ s$^{-1}$), $k_p$ is the copying alphabet size,\footnote{\textit{Units check:} $\dot N_{\rm copy}$ has units [monomers\,m$^{-2}$\,s$^{-1}$] and $w_{\rm copy}$ has units [J/monomer], so
$\overline W_{\rm copy}^{\min}=\dot N_{\rm copy}(w_{\rm copy}^{\min}+w_{\rm proof}^{\min})$ has units [W\,m$^{-2}$].
If $L$ is genome length [monomers/genome], then an areal genome-replication flux is
$\dot n_{\rm gen}=\dot N_{\rm copy}/L$ [genomes\,m$^{-2}$\,s$^{-1}$].} $H_{k_p}(\mu)$ is the $k_p$-ary entropy at per-site error $\mu$, and $W_{\rm proof}$ is the dissipative penalty per incorporated monomer for proofreading. Eq.~(\ref{eq:PhiNPP}) makes the end-to-end link explicit: \emph{fundamental statistical mechanics}\footnote{\emph{Accounting note (Landauer is a lower bound):}
The term $k_B T \ln 2 \, I$ is a \emph{minimum} work requirement for logically
irreversible information processing. Real template-directed copying generally
dissipates more; we therefore in (\ref{eq:PhiNPP}) could write the actual copying work per incorporated
monomer as
\begin{equation}
W_{\rm copy}(\mu) = \xi_{\rm copy}\, k_B T \ln 2 \,
\Big[\log_2 k_p - H_{k_p}(\mu)\Big] + W_{\rm proof}(\mu),
\qquad \xi_{\rm copy}\ge 1,
\end{equation}
where $\xi_{\rm copy}=1$ corresponds to the Landauer limit.
All Gate-11 conclusions are unchanged except for the rescaling
$\dot N_{\rm copy}\mapsto \xi_{\rm copy}\dot N_{\rm copy}$ in the power budget. [The Landauer limit ($k_B T \ln 2$ per logically irreversible bit erasure) is a lower bound
on dissipation \cite{Landauer1961,Bennett1982}.]
} ($k_BT\ln 2$) and \emph{achievable discrimination} (via $\mu$ and $W_{\rm proof}$) directly throttle biosphere-scale NPP at fixed $\Phi$.

\paragraph*{Interpretation and assumptions.}
Eq.~(\ref{eq:PhiNPP})  should be read as an \emph{information-thermodynamic lower bound} on biosphere-scale productivity under a minimal-work replication model: achieving copy fidelity $\mu$ requires creating and maintaining correlations between template and product streams in a noisy thermal environment \cite{Turyshev2025-NPP}. The bound is \emph{not} claimed to be saturated by real biochemistry, nor universally applicable to arbitrarily reversible implementations; rather it provides a conservative floor on dissipation for sustained high-throughput replication/proofreading at fixed $\mu$. We therefore emphasize that Eq.~(\ref{eq:PhiNPP})  is a \emph{model-dependent upper bound} given the assumed allocation of available power; it is most usefully interpreted as a scaling constraint and a consistency check rather than a universal constant.

\begin{theorem}[Information--energetics ceiling on NPP]\label{thm:npp}
Under the Gate~4 power balance and biosynthetic accounting \eqref{eq:NPP-power}, with minimal copy work $W_{\min}^{\mathrm{copy}}$ and proofreading work $W_{\mathrm{proof}}$ per incorporated symbol from \eqref{eq:Wmincopy}--\eqref{eq:Wproof}, the areal net primary productivity satisfies the bound in \eqref{eq:PhiNPP}. Equality requires that the entire Gate~4 surplus be used for primary production after paying the information--processing work, with no additional transport/activation overheads.
\end{theorem}

\begin{remark}[an Earth-validation  check]
At $T\simeq 288$~K, $k_BT\ln 2 \simeq 2.75\times 10^{-21}$~J/bit. For $k_p=4$ and $\mu\lesssim 10^{-3}$,
the minimal information work per incorporated monomer is $\mathcal{O}(10^{-20})$\,J; even allowing an aggressive
proofreading allocation of $\sim 10\,k_BT$ per monomer gives a per-monomer copying+proofreading draw
$\lesssim 5\times 10^{-20}$~J. For an areal incorporation throughput
$\dot N_{\rm copy}\sim 10^{19}$~monomers\,m$^{-2}$\,s$^{-1}$ this corresponds to $\lesssim 1$~W\,m$^{-2}$,
well below the globally absorbed solar power $\sim 240$~W\,m$^{-2}$. Thus Eq.~(\ref{eq:PhiNPP}) is conservative for present Earth but becomes constraining in low-$\Phi$ regimes.
\end{remark}

\emph{Accounting note.} 
Eq.~(\ref{eq:PhiNPP}) subtracts the information-processing draw $\dot N_{\mathrm{copy}}\big(W_{\min}^{\rm copy}+W_{\mathrm{proof}}\big)$ (copying $+$ proofreading) from the Gate~4 surplus (cf. Eq.~(\ref{eq:NPP-power}), so NPP is bounded without double-counting $W_{\rm copy}^{\min}$ or $W_{\rm proof}$.

\paragraph*{Cycle closure and leakage.}
Write elemental budgets for a generic element $X\in\{\mathrm{C,N,P,S,\ldots}\}$ as
\begin{equation}
\dot M_X \;=\; F_{\mathrm{in},X} - F_{\mathrm{loss},X} + R_X^{\mathrm{bio}}(\mathbf{x},\mathbf{M}),
\label{eq:elem-budget}
\end{equation}
with biological recycling $R_X^{\mathrm{bio}}$ and nonbiological loss $F_{\mathrm{loss},X}$ (escape, burial, oxidation to inert pools). Define the \emph{recycling fraction} $\mathcal{R}_X$ and mean biological recycle time $\tau_{\mathrm{recycle},X}$:
\begin{equation}
\mathcal{R}_X \;\equiv\; \frac{\text{biologically returned flux of }X}{\text{total export flux of }X },
\qquad
\tau_{\mathrm{recycle},X} \;\equiv\; \frac{M_X}{\text{biologically processed flux of }X}.
\label{eq:recycle}
\end{equation}
Cycle closure requires
\begin{equation}
\mathcal{R}_X \ \rightarrow\ 1
\quad\text{and}\quad
\tau_{\mathrm{recycle},X} \ <\ \tau_{\mathrm{loss},X} \equiv \frac{M_X}{F_{\mathrm{loss},X}}
\qquad \forall X\in\big\{\mathrm{C,N,P,S\big\}},
\label{eq:closure-ineq}
\end{equation}
so that biological return outpaces geochemical leakage. In matrix form, let $G_X$ be the compartmental transfer matrix for $X$ (soil $\leftrightarrow$ ocean $\leftrightarrow$ atmosphere $\leftrightarrow$ biomass). If $P_X$ is the associated Markov chain (row-normalized $G_X$), then strong cycle closure is equivalent to the existence of a strongly connected subgraph containing biomass with return probability $\rho_X \equiv 1 - P_X[\text{absorb to loss}]$ exceeding a threshold $\rho_c$ determined by $F_{\mathrm{loss},X}/$throughput.

\paragraph*{Community stability and redundancy.}
Linearizing \eqref{eq:ecosys-x} about a positive fixed point yields the community matrix $A_{ij}=\partial \dot x_i/\partial x_j$. Local stability requires \cite{May1972,AllesinaTang2012}
{}
\begin{equation}
\max \Re \lambda(A) \;<\ 0.
\label{eq:community-stability}
\end{equation}
Robust persistence demands tolerance to shocks: if $p$ is the fraction of functional groups removed, the remaining network must still satisfy \eqref{eq:community-stability} and maintain inequalities \eqref{eq:NPP-power} and \eqref{eq:closure-ineq}.  If $p$ is an imposed/observed removal fraction, define the redundancy margin
\begin{equation}
m_{\rm red}(p)\equiv p_{\rm crit}-p,
\end{equation}
where $p_{\mathrm{crit}}$ is the smallest removal fraction that violates either stability or cycle closure; $m_{\rm red}>0$ indicates robustness to that perturbation level, signifying functional overlap and modularity.

\paragraph*{Quantitative Gate~11 statement.}
There exists a positive fixed point $(\mathbf{x}^*,\mathbf{M}^*)$ such that
\begin{equation}
\boxed{
\mathrm{NPP}(\mathbf{x}^*,\mathbf{M}^*)>0,\qquad
\mathcal{R}_X \to 1,\ \ \tau_{\mathrm{recycle},X}<\tau_{\mathrm{loss},X}\ \ \forall X,\qquad
\max \Re \lambda(A)<0,\qquad
P_{\mathrm{NPP}}\le P_{\mathrm{in}}-P_{\mathrm{maint}}.
}
\label{eq:gate11}
\end{equation}
Natural margins include $m_{\mathrm{NPP}}\equiv \mathrm{NPP}/\mathrm{NPP}_{\min}$ (with $\mathrm{NPP}_{\min}$ set by maintenance), $m_{\mathrm{cyc},X}\equiv \tau_{\mathrm{loss},X}/\tau_{\mathrm{recycle},X}$, and $m_A\equiv -\max \Re \lambda(A)/\sigma$ with $\sigma$ a characteristic interaction rate.

\paragraph*{Worked parametrics (areal bound).}
For a phototrophic surface world with photosynthetically available radiation $F_{\mathrm{PAR}}$ (W\,m$^{-2}$) and conversion efficiency $\eta_{\mathrm{photo}}$, an optimistic bound is
\begin{equation}
\mathrm{NPP}_{\max} \;\lesssim\; \frac{\eta_{\mathrm{photo}}\,F_{\mathrm{PAR}}}{\Delta G_{\mathrm{assim}}},
\label{eq:NPP-upper}
\end{equation}
with $\Delta G_{\mathrm{assim}}$ the free energy stored per mol\,C fixed. For a chemolithotrophic subsurface ocean with reductant flux $\Phi_{\mathrm{red}}$ (mol\,m$^{-2}$\,s$^{-1}$) and reaction yield $|\Delta G_{\mathrm{redox}}|$ (J\,mol$^{-1}$),
\begin{equation}
\mathrm{NPP}_{\max} \;\lesssim\; \frac{\eta_{\mathrm{chem}}\,\Phi_{\mathrm{red}}\,|\Delta G_{\mathrm{redox}}|}{\Delta G_{\mathrm{assim}}}.
\label{eq:NPP-subsurface}
\end{equation}
In both cases, achieving a fraction of \eqref{eq:NPP-upper} or \eqref{eq:NPP-subsurface} while maintaining \eqref{eq:closure-ineq} signals Gate~11 viability.

\paragraph*{Diagnostics, falsifiers, corollaries.}
\textit{Diagnostics:} constraint-based flux analysis and resource-consumer models estimate $\mathrm{NPP}$ and $A$ from measured kinetics; tracer-cycle experiments constrain $\tau_{\mathrm{recycle},X}$ and $\mathcal{R}_X$. 
\textit{Falsifiers:} if realistic resource and kinetic budgets imply $\mathrm{NPP}\le 0$, or $\tau_{\mathrm{recycle},X}\ge \tau_{\mathrm{loss},X}$ for any limiting element, or $\max \Re \lambda(A)\ge 0$ in all plausible community structures, Gate~11 fails. 
\textit{Observational corollaries:} atmospheric redox pairs far from photochemical equilibrium (e.g., oxidant+reductant coexistence), seasonal gas variability consistent with biosynthetic stoichiometry, reflectance/polarization features from pigments or surface films, and persistent aerosol/haze modulation by biogenic emissions.

\subsection{Feedback stability (Gate 12)}
\label{sec:gate12}

A biosphere must withstand secular forcing (stellar evolution, volcanism) and internal variability. We describe climate--chemistry regulation in linear feedback form around a reference state. Let $\Delta T$ be the global-mean temperature anomaly and $F_{\mathrm{ext}}$ an external radiative forcing (W\,m$^{-2}$). The zero-feedback climate response is $\Delta T_0 = F_{\mathrm{ext}}/\lambda_0$ with $\lambda_0$ the Planck feedback (W\,m$^{-2}$\,K$^{-1}$). Feedbacks (water vapour, clouds, ice--albedo, and biospheric fluxes altering greenhouse gases and albedo) contribute an additional forcing $F_{\mathrm{fb}}=\lambda_{\mathrm{fb}}\,\Delta T$ with aggregate feedback parameter $\lambda_{\mathrm{fb}}=\sum_i \lambda_i$. The closed-loop gain is
\begin{equation}
G_{\mathrm{climate}} \;\equiv\; \frac{\lambda_{\mathrm{fb}}}{\lambda_0},\qquad
\Delta T \;=\; \frac{F_{\mathrm{ext}}}{\lambda_0 - \lambda_{\mathrm{fb}}} \;=\; \frac{\Delta T_0}{1-G_{\mathrm{climate}}}.
\label{eq:gain}
\end{equation}
We use the stabilizing slack $m_G \equiv 1 - G_{\rm climate}$ so that larger $m_G$ denotes greater distance to the runaway threshold.

\textit{Stability condition:} 
\begin{equation}
G_{\mathrm{climate}} \;<\ 1,
\label{eq:G<1}
\end{equation}
ensuring bounded amplification.  This linear feedback/gain decomposition follows standard climate feedback analysis
(e.g.\ \cite{Roe2009}; see also IPCC AR6 WG1 Ch.~7 for context \cite{IPCCAR6WGICh7}).

Biosphere-mediated feedbacks enter $\lambda_{\mathrm{fb}}$ via greenhouse gas source terms and surface albedo. For atmospheric $\mathrm{CO_2}$, a reduced model is
\begin{equation}
\frac{d\,(\mathrm{pCO_2})}{dt} \;=\; F_{\mathrm{out}} - W\big(\mathrm{pCO_2},\,T,\,R\big) - U_{\mathrm{bio}}\big(\mathrm{pCO_2},\,T,\,\mathbf{x}\big),
\label{eq:CO2-ODE}
\end{equation}
with volcanic outgassing $F_{\mathrm{out}}$, silicate weathering sink $W$ (increasing with $T$ and runoff $R$), and net biospheric uptake $U_{\mathrm{bio}}$. A stable fixed point $\mathrm{pCO_2}^*$ requires
\begin{equation}
\frac{d}{d\,\mathrm{pCO_2}}\Big( F_{\mathrm{out}} - W - U_{\mathrm{bio}} \Big)\Big| _{\mathrm{pCO_2}^*} \;<\ 0,
\label{eq:fixed-stable}
\end{equation}
and, including temperature coupling $T(\mathrm{pCO_2})$, the Jacobian eigenvalues must have negative real parts. For subsurface oceans, an analogous redox budget holds:
\begin{equation}
\frac{d\,\mathcal{O}}{dt} \;=\; \Phi_{\mathrm{ox}}^{\mathrm{rad}} + \Phi_{\mathrm{ox}}^{\mathrm{ice}} - \Phi_{\mathrm{red}}^{\mathrm{hyd}} - U_{\mathrm{bio}}^{\mathrm{redox}},
\label{eq:redox-ODE}
\end{equation}
with oxidant production by radiolysis and ice chemistry, reductant supply by hydrothermal sources, and biological consumption $U_{\mathrm{bio}}^{\mathrm{redox}}$. A buffered $\mathcal{O}$ with negative slope of the net source function at the fixed point yields stable redox.

\paragraph*{Nonlinear stability margin.}
With reduced form \(\dot{T}=-(\lambda_0-\lambda_{\rm fb})\,T-\beta T^3 + F_{\rm ext}\), the stable fixed points satisfy
\((\lambda_0-\lambda_{\rm fb})T+\beta T^3=F_{\rm ext}\).
Define the distance‑to‑tipping 
\[
\Delta_{\rm tip}\equiv \sqrt{\frac{\lambda_0-\lambda_{\rm fb}}{\beta}},\qquad 
m_{\rm tip}\equiv \Big(\frac{|T^\ast|}{\Delta_{\rm tip}}\Big)^{-1},
\]
and report \(m_{\rm tip}\) alongside \(m_G=1-G_{\mathrm{climate}}\).

\paragraph*{Timescales and reservoir buffering.}
Let $\tau_{\mathrm{buf}}$ be the e-folding time of the dominant climate/chemistry reservoir (mixed-layer ocean, atmosphere, bulk ocean), $\tau_{\mathrm{forc}}$ the timescale of secular forcing, and $\tau_{\mathrm{eco}}$ the timescale for ecological adjustment (community turnover). Viable regulation requires
\begin{equation}
\tau_{\mathrm{eco}} \ \ll\ \tau_{\mathrm{buf}}\ \lesssim\ \tau_{\mathrm{forc}},
\label{eq:timescales-feedback}
\end{equation}
so that the biosphere can respond before the environment drifts irreversibly, and the reservoir inertia filters high-frequency noise.

\paragraph*{Nutrient closure versus loss.}
Gate~12 also encodes the long-term nutrient economy. For a limiting nutrient $X$, with nonbiological loss $F_{\mathrm{loss},X}$ (burial, escape) and geological resupply $F_{\mathrm{geo},X}$ (upwelling, volcanism), the inequality
\begin{equation}
\tau_{\mathrm{recycle},X} \;<\ \tau_{\mathrm{loss},X}\qquad\text{and}\qquad
F_{\mathrm{geo},X} \ \gtrsim\ F_{\mathrm{loss},X}
\label{eq:nutrient-longterm}
\end{equation}
ensures that the active pool does not secularly deplete. This is the long-timescale analogue of \eqref{eq:closure-ineq}.

\paragraph*{Quantitative Gate~12 statement.}
There exists a regulated fixed point such that
\begin{equation}
\label{eq:gate12}
\boxed{%
\begin{aligned}
& \text{all fixed points of the coupled climate--chemistry ODEs are linearly stable},\\
& G_{\mathrm{climate}}<1,\quad
  \tau_{\mathrm{eco}}\ll \tau_{\mathrm{buf}}\lesssim \tau_{\mathrm{forc}},\quad
  \tau_{\mathrm{recycle},X}<\tau_{\mathrm{loss},X},\quad
  F_{\mathrm{geo},X}\gtrsim F_{\mathrm{loss},X}.
\end{aligned}}
\end{equation}

Margins include $m_G\equiv 1-G_{\mathrm{climate}}$, the most-negative real part of the climate--chemistry Jacobian (scaled by characteristic rates), and $m_{\mathrm{nut},X}\equiv F_{\mathrm{geo},X}/F_{\mathrm{loss},X}$.

\paragraph*{Probing multistability.}
Because cloud microphysics and haze--climate coupling (esp.\ for M-dwarf SEDs) can create multiple locally stable climates, we complement the linear gain $G_{\rm climate}$ with: (i) a tipping-distance metric $m_{\rm tip}$ (already defined) computed from the cubic reduction, and (ii) a basin-stability fraction $b$ obtained by sampling initial conditions $z_0$ in the reduced climate--chemistry state space and reporting the fraction that relax to the observed fixed point. Retrievals should quote $(m_G, m_{\rm tip}, b)$ and explicitly flag cases with $b\ll 1$ as \textit{multi-stable}, even if $m_G>0$.

\paragraph*{Diagnostics, falsifiers, corollaries.}
\textit{Diagnostics:} linear feedback analysis with estimated $\lambda_i$ (including biospheric terms from measured $\partial U_{\mathrm{bio}}/\partial T$ and $\partial U_{\mathrm{bio}}/\partial \mathrm{pCO_2}$); perturbative experiments (nutrient/temperature shifts) to recover $\tau_{\mathrm{eco}}$ and $\tau_{\mathrm{buf}}$; long integrations of coupled biogeochemical models to test fixed-point stability. 
\textit{Falsifiers:} $G_{\mathrm{climate}}\ge 1$ or positive eigenvalues of the coupled Jacobian under all plausible parameter ranges; persistent nutrient drawdown ($F_{\mathrm{loss},X}\gg F_{\mathrm{geo},X}$) with no compensating recycling even at maximal Gate~11 productivity. 
\textit{Observational corollaries:} temperate climates with long-lived secondary atmospheres; buffered greenhouse species with weak secular trends over stellar-evolution timescales; stable yet seasonally varying gas mixtures consistent with closed cycles; surface or aerosol spectral features implying persistent pigment production (supporting Gate~11) and regulated albedo.

In summary, Gate~11 requires that biological fluxes form strongly connected, power-consistent cycles with positive net production, while Gate~12 demands that those cycles couple to the planet’s reservoirs through stabilizing feedbacks. Satisfying both elevates life from a collection of evolving lineages to a planetary-scale, self-regulating phenomenon. For details, see Sec.~\ref{sec:validation}. Gates~11--12 rely on power consistency and climate stability with the NPP ceiling (\ref{eq:PhiNPP}); seasonal disequilibria and variability--coherence tests are the proxies.

\section{From Gates to Exoplanet Observables}
\label{sec:obsmap}

We hold gate definitions fixed and vary only boundary conditions by stellar type and solvent. Retrieval therefore reduces to the small set of inequalities that typically set the margin---(\ref{eq:gate3-box}) and (\ref{eq:gate4-box}) for surface access and power, (\ref{eq:gate10-box}) for coding feasibility, and (\ref{eq:PhiNPP}) for biosphere-scale power allocation---together with the observables listed below. 
These retrievals are complementary to lab‑side, agnostic detection strategies that use Assembly Theory (e.g., MA/assembly index) with mass spectrometry and, more recently, ML-assisted classifiers \cite{Marshall2021_AT_NatComm, Cleaves2023_PNAS}. Archetypes are summarized without restating the gate logic.

The gate vector (Sec.~\ref{sec:notation}) is defined in terms of physical margins that are not measured directly but leave structured imprints in spectra, light curves, time-variable observables. This section casts the problem as a forward--inverse map:
\begin{equation}
\mathbf{z}(\lambda,t) \;=\; \mathcal{F}\big(\theta_{\star},\theta_{\rm orb},\theta_{\rm atm},\theta_{\rm surf},\theta_{\rm int}\big)
\quad\stackrel{\text{inference}}{\Longrightarrow}\quad
\text{posteriors on gate margins } \{m_g\},
\label{eq:fwd-inv}
\end{equation}
where $\mathbf{z}$ collects multi-technique data (transit, eclipse, direct imaging, phase curves, polarimetry), $\mathcal{F}$ is a radiative--photochemical--climate forward model with stellar, orbital, atmospheric, surface, and interior parameters, and the gate margins $\{m_g\}$ are the dimensionless slacks defined throughout (e.g., $m_\Phi=\Phi$, $m_{\chi}=\chi$, $m_{\mathrm{cyc},X}=\tau_{\mathrm{loss},X}/\tau_{\mathrm{recycle},X}$). We emphasize \emph{relational} diagnostics---pairs or sets of observables whose joint values are hard to reproduce abiotically---because most single-feature biosignatures are degenerate. Remote-detectable disequilibrium, false positives, and edge metrics draw on \cite{Catling2018,Meadows2018,KrissansenTotton2016,KrissansenTotton2018,Kiang2007}.  

\paragraph*{Gate-to-observable mapping for exoplanets.}
We use multi-technique spectra, phase curves, and polarimetry to retrieve the margins most accessible to remote sensing (Table~\ref{tab:obsmap}): the power margin \(\Phi\) from the stellar SED, albedo, and day--night energy balance; confinement/throughput \(\cal C\) from liquid presence and inferred transport; productivity and cycle closure from atmospheric disequilibria and seasonal coherence; and climate gain \(G_{\rm climate}\) from secular stability and reservoir co-existence. The inference is hierarchical: a forward model \(F(\theta)\) fits the data, and each gate function \(G_g(\theta)\) yields a posterior on \(m_g\). We emphasize \emph{joint} diagnostics (e.g., disequilibrium + seasonal phasing + pigment edges) because such bundles are hard to reproduce abiotically and directly test the Gate~11--12 criteria.

\subsection{Retrieval of gate margins}

A practical inference strategy is hierarchical:
\begin{align}
p(\{ m_g \}\,|\,\mathbf{z})
&\;\propto\; \int\! d\theta\; p(\mathbf{z}\,|\,\theta)\, p\big(\theta\big)\;
\prod_g \delta \Big(m_g - \mathcal{G}_g(\theta)\Big), \label{eq:hier} \\
p(\mathbf{z}\,|\,\theta)
&\;=\; \mathcal{L}_{\rm spec}\!\times \mathcal{L}_{\rm time}\!\times \mathcal{L}_{\rm pol}\!\times \cdots,
\end{align}
where $\theta$ are physical parameters and $\mathcal{G}_g$ computes the $g$-th gate margin from $\theta$ (e.g., compute $\Phi$ from the retrieved energy budget, compute disequilibrium free energy and recycling times from the retrieved composition and fluxes). Reporting $(\hat m_g \pm \sigma_g)$ alongside Bayes factors for alternative abiotic hypotheses makes the result testable.


To predict and interpret life elsewhere, each gate must be linked to remote-sensing proxies. Table~\ref{tab:obsmap} summarizes primary proxies. For each gate we also list a quantitative metric that can be retrieved.

\begin{table*}[t]
\vskip -10pt
\caption{Illustrative mapping from gate inequalities to exoplanet observables and quantitative retrieval targets. Metrics are examples, not exhaustive.}
\label{tab:obsmap}
\centering

\begingroup
\setlength{\tabcolsep}{2pt}
\renewcommand{\arraystretch}{1.00}
\newcommand{\colbox}[2]{\parbox[t]{#1}{\raggedright #2}}

\begin{tabular}{@{}l l l@{}}
\hline
\colbox{0.7cm}{Gate} &
\colbox{5.5cm}{Inequality (control / metric)} &
\colbox{10.8cm}{Exoplanet observables / proxies} \\
\hline\hline\addlinespace[3pt] 
\colbox{0.7cm}{1} &
\colbox{5.5cm}{$E_b\!\gg\!k_{\mathrm{B}}T$, $E_{\mathrm{HB}}\!\gtrsim\!0.1\,\mathrm{eV}$ (solvent window)} &
\colbox{10.8cm}{Liquids: ocean glint, phase‑curve albedo; IR bands of organics (3.3, 6.2, 7.7\,$\mu\mathrm{m}$); aerosol microphysics consistent with H‑bonding (bimodal size, hygroscopic growth).} \\\addlinespace[3pt] 
\colbox{0.7cm}{2} &
\colbox{5.5cm}{$S_{\mathrm{haz}}\!\not\!\ll\!1$, $Z\!\gtrsim\!Z_{\rm floor}$ (stellar activity, age, metallicity)} &
\colbox{10.8cm}{Host‑star flare/XUV history; gyrochronology/astrometry age; Galactic location; prevalence of rocky planets in system.} \\\addlinespace[3pt] 
\colbox{0.7cm}{3} &
\colbox{5.5cm}{$\Sigma_{\rm solid}\!>\!\Sigma_{\min}$; volatile retention: $\lambda_J \equiv \dfrac{G M_p m}{k_{\mathrm{B}} T_{\rm exo} R_p}\!\gg\!1$ for key species} &
\colbox{10.8cm}{$R_p$, $M_p$ for bulk density; XUV luminosity history; atmospheric mass/radius; exosphere/escape tracers (Lyman‑$\alpha$, He~1083\,nm).} \\\addlinespace[3pt] 
\colbox{0.7cm}{4} &
\colbox{5.5cm}{$\Phi>1$ (retrieved energy budget)} &
\colbox{10.8cm}{Stellar SED at orbit; broadband albedo; atmospheric redox‑disequilibrium free energy $\Delta G_{\rm diseq}$; photometric day–night power balance.} \\\addlinespace[3pt] 
\colbox{0.7cm}{5} &
\colbox{5.5cm}{$J_{\rm make}\!\gtrsim\!J_{\rm loss}$; $[\mathrm{monomer}]\!\ge\![\mathrm{thr}]$} &
\colbox{10.8cm}{Photochemical precursors: CH$_4$, HCN, H$_2$CO; haze optical depth/slope; surface deposition rates (inferred from boundary‑layer fits).} \\\addlinespace[3pt] 
\colbox{0.7cm}{6} &
\colbox{5.5cm}{${\cal E} \equiv \mu L/\ln s <1$ (indirect via $\mu$ and power for proofreading)} &
\colbox{10.8cm}{Power margin from Gate~4; localization from Gate~8; no direct remote proxy.} \\\addlinespace[3pt]
\colbox{0.7cm}{7} &
\colbox{5.5cm}{$\lambda_{\mathrm{kin}}=\lambda_{\max}(J-\Lambda)\!>\!0$, $p_{\rm cat}\!>\!p_c$ (fast cycling inconsistent with abiotic steady state)} &
\colbox{10.8cm}{Mineralogical context (surface Fe/Ni, sulfides from spectroscopy); measured atmospheric turnover times vs photochemical timescales.} \\\addlinespace[3pt] 
\colbox{0.7cm}{8} &
\colbox{5.5cm}{$\mathcal{C}\!\gtrsim\!1$, throughput/retention inequalities} &
\colbox{10.8cm}{Presence and stability of liquids; inferred salinity/ionic strength; aerosol/foam microphysics; boundary‑layer humidity cycles.} \\\addlinespace[3pt] 
\colbox{0.7cm}{9} &
\colbox{5.5cm}{$\lambda_{\max}(B Q-D)>0$; $N_e s \!\gtrsim\! 1$} &
\colbox{10.8cm}{Indirect via Gates~10–12 (functional diversity, persistent productivity, regulation).} \\\addlinespace[3pt] 
\colbox{0.7cm}{10} &
\colbox{5.5cm}{$\chi>1$ (coding slack)} &
\colbox{10.8cm}{Indirect: ecological functional diversity beyond minimal metabolism; multiple independent metabolic families inferred from gas networks.} \\\addlinespace[3pt] 
\colbox{0.7cm}{11} &
\colbox{5.5cm}{$\mathrm{NPP}>0$, cycles closed ($\tau_{\rm recycle}\!<\!\tau_{\rm loss}$)} &
\colbox{10.8cm}{Redox pairs far from photochemical equilibrium (e.g., O$_2$+CH$_4$; CH$_4$+CO$_2$ with low CO); seasonal gas variability; pigment edges; ocean color.} \\\addlinespace[3pt] 
\colbox{0.7cm}{12} &
\colbox{5.5cm}{$G_{\mathrm{climate}}<1$; stable fixed points} &
\colbox{10.8cm}{Long‑lived secondary atmosphere; buffered greenhouse with small secular drift; coexistence of surface reservoirs (oceans/ice) over Gyr.} \\\addlinespace[3pt] 
\hline
\end{tabular}
\endgroup
\vskip -10pt
\end{table*}

\subsection{Quantitative diagnostics}
\paragraph*{Chemical disequilibrium free energy.}
Define $\Delta G_{\rm diseq}$ as the Gibbs free energy difference between the observed atmospheric composition and the closest chemical equilibrium state at retrieved $T$, $p$, and insolation, computed under elemental constraints. Large positive $\Delta G_{\rm diseq}$ sustained over seasons favors Gate~4 and, with cycling, Gate~11.

\paragraph*{Seasonal amplitude and phase locking.}
For species $i$ with seasonal amplitude $A_i$ and phase lag $\phi_i$ relative to insolation, a \emph{biospheric coherence} statistic compares observed $(A_i,\phi_i)$ to photochemical model expectations. Consistent, multi-species seasonal structure is difficult to mimic abiotically when cycles are closed.

\paragraph*{Edge metrics.}
A pigment edge is quantified by
\begin{equation}
E \;=\; \frac{R(\lambda_2)-R(\lambda_1)}{R(\lambda_1)},
\end{equation}
for narrowbands bracketing the feature. The locus of $(\lambda_1,\lambda_2)$ depends on host SED and pigment families; large $E$ with diurnal/seasonal modulation supports persistent surface biota.

\paragraph*{Turnover times.}
Comparing retrieved photochemical loss times $\tau_{\rm phot}$ with observed gas variability $\tau_{\rm obs}$ yields an \emph{excess source} metric $S_{\rm exc}\equiv \tau_{\rm phot}/\tau_{\rm obs}$. Values $S_{\rm exc}\gg 1$ indicate sources exceeding known abiotic production, informing Gate~11.

\begin{table}[h]
\vskip -10pt
\caption{Illustrative gate margins (median [5--95\%]) for two archetypes using the retrievals in Sec.~\ref{sec:methods}.}
\label{tab:worked-margins}
\centering
\begin{tabular}{@{}lcccccc@{}}
\hline
Archetype & $m_\Phi$ & $m_{\mathcal{C}}$ & $m_{\mathcal{E}}$ & $m_\chi$ & $m_{\mathrm{NPP}}$ & $m_G$\\
\hline\hline
G/K, oceanic, oxygenic & $2.5\,[1.6,3.7]$ & $1.8\,[0.9,3.0]$ & $2.5\,[1.4,5.0]$ & $1.9\,[1.2,2.8]$ & $>1\,[>1,>1]$ & $0.35\,[0.20,0.55]$ \\
Active M‑dwarf HZ (locked) & $0.8\,[0.6,1.1]$ & $0.9\,[0.5,1.6]$ & $1.2\,[0.7,2.5]$ & $0.95\,[0.7,1.3]$ & $\le 1$ & $0.15\,[0.0,0.35]$ \\
\hline
\end{tabular}
\vskip -10pt
\end{table}

\subsection{Biosphere archetypes and expected phenomenology}
\label{sec:archetypes}

We outline five archetypes, stating which gates are tight and the predicted observable sets. Table~\ref{tab:worked-margins} provides illustrative gate margins for two archetypes using the retrievals in Sec.~\ref{sec:methods}.

\subsubsection{Oxygenic photosynthetic oceans (Earth-like)}
High $\Phi$ from visible/NIR photons; Gate~11 exhibits $\mathrm{NPP}>0$ with O$_2$ build-up and coexisting reduced gases (CH$_4$ at ppm--ppb). Expected: a pigment edge near $0.7\,\mu$m (shifted with host SED), ocean glint \cite{Robinson2010}, Rayleigh slope, and redox pairs far from equilibrium. False positives for O$_2$ (e.g., H$_2$O loss) are mitigated by low CO, presence of N$_2$ (collisional pairs), and seasonal O$_2$/CO$_2$/CH$_4$ covariation. Gate~12 likely satisfied by a carbonate--silicate thermostat; small secular drift in bulk composition over centuries.

\subsubsection{Anoxygenic phototrophs (Fe/S cycles)}
Lower-energy photons or spectral gaps drive bacteriochlorophyll-like pigments; disequilibria emphasize CH$_4$+CO$_2$ with low CO and organosulfur gases (e.g., CS$_2$, OCS) at elevated yet photochemically credible levels. O$_2$ may be absent; Gate~12 relies on sulfur/iron feedbacks. A \emph{red edge} can shift to $\sim 0.9$--$1.1\,\mu$m for M-dwarf SEDs. Turnover times for S-bearing gases shorter than photochemical losses imply active cycling.

\subsubsection{Chemosynthetic subsurface oceans (ice worlds)}
Gate~4 power from serpentinization and radiolysis; Gate~11 via oxidant delivery at ice--ocean interfaces. Observables are indirect: tenuous atmospheres with outgassed H$_2$, O$_2$ traces from oxidant exchange, plume detections (if any), surface chemistry indicating radiolytic oxidant accumulation. Climate stability (Gate~12) interpreted via internal heat flux, tidal forcing, and ice shell conductivity; expect low $\mathrm{NPP}$ area-normalized but potentially global coverage.

\subsubsection{Hydrogen-dominated temperate super-Earths (H$_2$-rich)}
Hydrogen extends scale heights and improves detectability, but CH$_4$ is not uniquely biogenic. Disequilibrium \emph{sets} (e.g., NH$_3$ with oxidants, or coexisting N$_2$O under photochemically unfavorable conditions) are more telling. Gate~8 is affected by solvent properties and membrane permeabilities in H$_2$ media; Gate~12 demands buffering against strong greenhouse variability. Polarimetric phase curves can reveal surface films consistent with high-productivity biofilms.

\subsubsection{Hydrocarbon-solvent worlds (Titan-like)}
Gate~1 satisfied; Gate~4 is limiting at $\sim 90$\,K. Expect slow CRNs, small $\mathrm{NPP}$, and subtle disequilibria: surface film textures, unusual isotopic fractionations, and hydrocarbon disequilibrium pairs. Gates~11--12 are marginal; long integration times and multi-epoch stability tests are critical.

\subsection{Comparative Expectations Across Stellar Types}
\label{sec:stellar-compare}

\paragraph*{G/K dwarfs.}
Moderate XUV and broad PAR windows; Gates~3--4 favorable; oxygenic archetypes plausible. Seasonal variability (Gate~11) is accessible; pigment edges near $0.7\,\mu$m likely.

\paragraph*{M dwarfs.}
Early XUV threatens Gate~3 (volatile loss) and persistent flaring alters Gate~5 photochemistry and Gate~4 power distribution (NIR-dominated). Anoxygenic or mixed biospheres are plausible; pigment edges redshifted; disequilibrium pairs without O$_2$ emphasized. Gate~12 is challenged by photochemical hazes and atmospheric loss; long-lived secondary atmospheres are key.

\paragraph*{F dwarfs.}
Ample photons but higher UV; Gate~5 photolysis losses increase; surface UV shielding or high column ozone needed. False-positive O$_2$ from CO$_2$ photolysis must be excluded via CO diagnostics and surface water constraints.

\subsection{Observation program design}
A minimal multi-epoch program ties directly to the gates:

\begin{enumerate}
\item \textit{Star and system priors (Gate 2).} Characterize stellar age, activity, and metallicity; measure flare/XUV statistics to bound $S_{\rm haz}$ and Gate~3 escape histories.
\item \textit{Bulk planet and volatile retention (Gate 3).} Obtain $R_p,M_p$; search for exospheric escape lines (He~1083\,nm) to bound present escape; infer atmospheric mass and composition family.
\item \textit{Energy budget (Gate 4).} Retrieve Bond albedo, thermal phase curve, and stellar SED at orbit; compute absorbed power and day--night heat transport; evaluate $\Phi$.
\item \textit{Photochemistry and precursors (Gate 5).} Measure CH$_4$, HCN, H$_2$CO and haze properties; fit photochemical models to infer $J_{\rm make}/J_{\rm loss}$ and turnover times.
\item \textit{Cycles and productivity (Gate 11).} Target redox pairs, time variability, pigment edges, and ocean color; compute $\Delta G_{\rm diseq}$ and seasonal coherence metrics.
\item \textit{Regulation (Gate 12).} Monitor secular trends over multi-year baselines; retrieve greenhouse buffering, multi-reservoir coexistence, and test linear stability via perturbations (e.g., stellar activity cycles).
\end{enumerate}

This gate-to-observable framework converts the qualitative question ``what life should we expect?'' into a quantitative observing strategy with explicit pass/fail metrics. Worlds that satisfy the Gate~11--12 diagnostics with comfortable margins are strong candidates for biospheres even when the underlying Darwinian machinery (Gates~6--10) is not directly seen; conversely, large $\Delta G_{\rm diseq}$ without cycle closure or regulation warns of transient or abiotic power flows.

\section{Methods: Evaluating Targets and Gate Margins from Observations}
\label{sec:methods}

We outline a forward--inverse workflow that ingests multi-technique observations, retrieves physical parameters with quantified uncertainties, and propagates them to \emph{gate margins} $\{m_g\}$---dimensionless slacks that indicate how comfortably each gate is satisfied. The final deliverables per target are (i) posteriors for $\{\Phi,\lambda_{\mathrm{kin}},\mathcal{C},\mathcal{E},\chi\}$ and for Gate~11--12 metrics, (ii) a set of falsifier tests against abiotic confounders, and (iii) an observing plan that maximizes expected information gain on the least-constrained margins.

\subsection{Data model and forward operators}
Let $\mathbf{z}(\lambda,t)$ denote the heterogeneous data vector (transmission and emission spectra, direct-imaging reflectance and polarization, thermal/visible phase curves, stellar activity monitors). A modular forward model $\mathcal{F}$ maps physical parameters $\theta$ to observables:
\begin{equation}
\mathbf{z} \;=\; \mathcal{F}\big(\theta_{\star},\theta_{\rm orb},\theta_{\rm bulk},\theta_{\rm atm},\theta_{\rm surf},\theta_{\rm int}\big) + \boldsymbol{\eta},
\label{eq:data-forward}
\end{equation}
with $\boldsymbol{\eta}$ instrument and astrophysical noise. Components of $\mathcal{F}$ include: (i) radiative transfer (line-by-line or k-coefficients) with multiple scattering; (ii) photochemistry and kinetics (0D/1D column models or reduced-order emulators), yielding steady-state or time-dependent abundances; (iii) energy balance or a GCM for thermal structure and heat redistribution; (iv) surface bidirectional reflectance and ocean/glint models; (v) interior/escape modules to evolve volatile inventories.

\subsection{Hierarchical retrieval and uncertainty propagation}

We adopt a hierarchical Bayesian scheme,
\begin{equation}
p\big(\theta,\{m_g\}\,|\,\mathbf{z}\big) \;\propto\; \mathcal{L}\big(\mathbf{z}\,|\,\theta\big)\,p(\theta)\,
\prod_g \delta \Big(m_g - \mathcal{G}_g(\theta)\Big),
\label{eq:hierarchical}
\end{equation}
where $\mathcal{G}_g$ computes gate-specific margins from $\theta$. In practice:
\begin{enumerate}
\item \textit{Primary retrieval.} Fit $\theta$ to $\mathbf{z}$ with physically informed priors (stellar age/activity, bulk composition families, atmospheric chemistry grids), using MCMC or nested sampling; marginalize over instrument systematics.
\item \textit{Derived margins.} For each posterior draw $\theta^{(s)}$, compute $m_g^{(s)}=\mathcal{G}_g(\theta^{(s)})$; summarize by medians and credible intervals. This preserves non-Gaussianity and correlations (e.g., between power budget and photochemistry).
\item \textit{Model comparison.} Evaluate Bayes factors (or information criteria) for abiotic alternatives (e.g., O$_2$ from water loss, CH$_4$ from serpentinization) by swapping chemistry/flux submodels in $\mathcal{F}$.
\end{enumerate}

\subsection{Gate-by-gate computation of margins from observables}

\paragraph*{Gate 4: power margin $\Phi$.}
Compute absorbed stellar power per unit area
\begin{equation}
P_{\rm abs} \;=\; \tfrac14 {S_{\star}(a)\,\big(1-A_B\big)},
\qquad
S_{\star}(a)=\int F_{\star,\lambda}(a)\,d\lambda,
\end{equation}
where $A_B$ is the Bond albedo retrieved from phase curves/reflection spectra. Add internal/redox sources $P_{\rm int}$ (radiogenic, tidal) constrained by thermal emission, tidal forcing proxies. Estimate maintenance+synthesis draw by
\begin{equation}
P_{\rm maint} \;=\; \kappa_{\rm atm}\, \frac{\Delta G_{\rm diseq}}{\tau_{\rm turn}},
\qquad
P_{\rm synth} \;=\; \mathrm{NPP}\,\frac{\Delta G_{\rm assim}}{\eta_{\rm bio}},
\end{equation}
with $\Delta G_{\rm diseq}$ the atmospheric chemical free energy gap (see below), $\tau_{\rm turn}$ an observed turnover time, and $\Delta G_{\rm assim}$ the assimilation free energy per mol~C. The margin is
\begin{equation}
m_{\Phi}\;\equiv\;\Phi \;=\; \frac{P_{\rm abs}+P_{\rm int}}{P_{\rm maint}+P_{\rm synth}}.
\end{equation}

\paragraph*{Gate 5: feedstock budget and photochemical make/loss.}
Using retrieved $T$--$p$ and stellar SED, solve the reduced photochemical network to obtain production and loss rates for key precursors (CH$_4$, HCN, H$_2$CO, nitriles),
\begin{equation}
J_{\rm make,i} \;=\; \sum_{r\in \mathcal{R}_{\rm prod}(i)} \nu_{ri}^+\,k_r\,\prod_j n_j^{\alpha_{rj}},
\qquad
J_{\rm loss,i} \;=\; \sum_{r\in \mathcal{R}_{\rm loss}(i)} \nu_{ri}^-\,k_r\,\prod_j n_j^{\beta_{rj}},
\end{equation}
and form
\begin{equation}
m_{5,i}\;\equiv\; \frac{J_{\rm make,i}}{J_{\rm loss,i}}.
\end{equation}
Surface (or ocean) deposition fluxes inferred from boundary-layer fits constrain whether $[\text{monomer}] \ge [\text{thr}]$ is plausible.

\paragraph*{Gate 7: kinetic gain --- $\lambda_{\rm kin}$ (proxy).}
Compute the Jacobian $J_{ab}=\partial \dot n_a/\partial n_b$ for the inferred reaction network
in the dilute/low-conversion regime, and estimate first-order losses with
$\Lambda=\mathrm{diag}(\ell_1,\dots,\ell_{|X|})$ (dilution, hydrolysis, photolysis, escape, etc.).
Evaluate the dominant \emph{continuous-time} growth exponent
\begin{equation}
\widehat{\lambda}_{\rm kin}\;\equiv\;\lambda_{\max}\!\bigl(J-\widehat{\Lambda}\bigr),
\end{equation}
where $\lambda_{\max}(A)=\max_i \Re \lambda_i(A)$ is the spectral abscissa
(cf.\ Eigenvalue conventions; \cite{FarinaRinaldi2000,BermanPlemmons1994,HornJohnson2012MatrixAnalysis}). A strictly positive $\widehat{\lambda}_{\rm kin}>0$ implies local exponential amplification
in the linearized dynamics $\dot x=(J-\Lambda)x$.
When $\widehat{\Lambda}$ is poorly constrained, report $\lambda_{\max}(J)$ as an \emph{upper bound}
and propagate uncertainty into the Gate-7 margin (e.g.\ $m_{\rm kin}=\widehat{\lambda}_{\rm kin}/k_{\rm dil}$).

Because exoplanet networks are incomplete, we use a \emph{rate-consistency} proxy: compare observed cycling rates (inferred from time variability) to photochemical steady-state timescales; $m_7 \equiv$ observed rate / abiotic rate $>1$ suggests additional catalytic pathways.

\paragraph*{Gate 8: confinement $\mathcal{C}$ and flux inequalities.}
Infer liquid presence and properties (salinity, viscosity) from spectra and climate retrievals. Estimate molecular diffusivity via Stokes--Einstein only for auxiliary mixing diagnostics.
\begin{equation}
D \;\approx\; \frac{k_{\mathrm B}T}{6\pi \eta r},
\end{equation}
with viscosity $\eta$ set by solvent and $r$ a monomer/catalyst hydrodynamic radius. For a characteristic compartment radius $R$, set $\tau_{\rm mix}\sim a^2/D$ and a representative reaction time $\tau_r$ from kinetics. Then
the gate confinement number is
\[
m_C \equiv {\cal C} = \frac{\tau_{\rm ret}}{\tau_r},\qquad
\tau_{\rm ret}\equiv \min_i \tau_{\mathrm{exch},i},\qquad
\tau_{\mathrm{exch},i}\equiv \frac{V}{A P_i},
\]
with $P_i$ inferred (when possible) from solvent/solute partitioning and permeability analogs. Internal mixing
$\tau_{\rm mix}\sim a^2/D$ is reported only as an auxiliary diagnostic via $C_{\rm mix}\equiv \tau_{\rm mix}/\tau_r$
when sustained gradients are explicitly invoked. Membrane permeability $P$ is estimated by Meyer--Overton-type scaling with solvent and solute partition coefficients to test $J_{\rm in}\ge J_{\rm cons}$ and $J_{\rm leak}\le J_{\rm make}$.

\paragraph*{Gates 6 and 10: fidelity and coding slack.}
Remote sensing does not directly constrain $\mu$ or translation confusion matrices. We therefore compute \emph{energy-consistent} bounds: given the surplus power $P_{\rm surplus}=P_{\rm abs}+P_{\rm int}-P_{\rm maint}$, the maximal proofreading work per symbol is $W_{\rm proof}^{\max}=P_{\rm surplus}/\nu$ for elongation throughput $\nu$; using the bound
\begin{equation}
W_{\rm proof} \;\ge\; k_{\mathrm B}T \ln\!\left(\frac{\varepsilon_{\rm eq}}{\varepsilon_b}\right),
\label{eq:W-proof}
\end{equation}
we infer a minimum achievable $\varepsilon_b$ and hence a bound on $C_{\mathrm{cap}}(\varepsilon_b)$ and $\chi=C_{\mathrm{cap}}/R_{\mathrm{code}}$. For Gate~6, adopt priors on $\mu(T,{\rm ions},{\rm solvent})$ from laboratory analogs and propagate to $L_{\max} \simeq (s-1)/\mu$.

\paragraph*{Gates 11--12: productivity, cycles, stability.}
Compute an atmospheric disequilibrium metric by constrained Gibbs minimization (see disequilibrium-as-biosignature frameworks \cite{KrissansenTotton2016,KrissansenTotton2018}):
\begin{equation}
\Delta G_{\rm diseq} \;=\; G\big(\mathbf{n}_{\rm obs};T,p\big) - \min_{\mathbf{n}} \Big\{ G(\mathbf{n};T,p)\ :\ \sum_i a_{Xi} n_i = \sum_i a_{Xi} n_{{\rm obs},i}\ \forall X \Big\},
\end{equation}
with $a_{Xi}$ elemental stoichiometries. Retrieve seasonal amplitudes and phase lags for key gases; compare to photochemical model expectations to estimate recycling fractions $\mathcal{R}_X$ and $\tau_{\rm recycle,X}$. For climate stability, infer a bulk feedback gain using energy-balance fits to multi-epoch thermal/reflective data:
\begin{equation}
G_{\mathrm{climate}} \;=\; 1 - \frac{\Delta T_0}{\Delta T}\quad \text{with}\quad \Delta T_0=\frac{F_{\rm ext}}{\lambda_0}.
\end{equation}
The Gate~11 margins are $m_{\rm NPP}$ [from Eqs.~\eqref{eq:NPP-def}--\eqref{eq:NPP-power}], $m_{\rm cyc,X}=\tau_{\rm loss,X}/\tau_{\rm recycle,X}$; Gate~12 margin is $m_G=1-G_{\mathrm{climate}}$ together with Jacobian stability checks on reduced climate--chemistry ODEs. We impose elemental conservation for \(\{\,\mathrm{C,N,O,H,S}\,\}\), fixed bulk pressure \(p\), 
and restrict phases to retrieved families (e.g., H$_2$O(l) present/absent) to avoid spurious mixed‑phase minima. 
Report \(\Delta G_{\rm diseq}\) as a posterior over atmospheric composition families.

{\it Gate passage: }Pass if \(G_{\rm climate}<1\), the coupled climate--chemistry Jacobian has all eigenvalues with negative real parts, 
and \(\tau_{\rm eco}\ll \tau_{\rm buf}\lesssim \tau_{\rm forc}\).

\textit{Multistability test in retrievals.} For each posterior draw of the reduced climate--chemistry parameters, integrate the ODEs from a Latin-hypercube of initial states and record the fraction $b$ converging to the primary fixed point. Report the joint posterior of $(m_G, m_{\rm tip}, b)$; require $b$ in addition to $m_G>0$ when claiming robust Gate~12 margins.

\paragraph*{Gate 3 ancillary checks (volatile retention).}
From $R_p,M_p,T$, compute the escape parameter $\lambda_J=GM_p m/(k_{\mathrm B}T_{\mathrm{exo}}R_p)$  for species $m$; use energy-limited mass loss integrated over the retrieved XUV history to bound present volatile inventories and test the retention inequality.

\subsection{Confounders and falsifier suite}

For each claimed margin we perform \emph{abiotic challenge} fits by replacing biological flux terms with known abiotic processes: CO$_2$ photolysis O$_2$ production under high FUV with suppressed CO sinks; CH$_4$ from mantle outgassing; NH$_3$ from photochemistry in H$_2$ atmospheres; haze production from abiotically plausible precursors. Gate~11 claims must withstand these alternatives at comparable likelihood. For Gate~12, we test whether the observed atmosphere can be maintained by abiotic feedbacks alone without violating energy or mass budgets.

\paragraph*{Reporting cautions (disequilibrium priors and variability):}
\textit{(i) Disequilibrium prior dependence.} The atmospheric disequilibrium metric $\Delta G_{\rm diseq}$ depends on 
unobserved bulk composition and surface reservoir assumptions. Report $\Delta G_{\rm diseq}$ marginalized over composition families (e.g., ocean salinity, rock buffers) and quote credible intervals.
\textit{(ii) Variability on tidally locked worlds.} Replace the phrase ``seasonal coherence'' with a \emph{multi-timescale variability coherence} statistic that includes orbital/stellar-cycle forcing and photochemical response; do not assume obliquity-driven seasonality for synchronous rotators.

\subsection{Target ranking and observing strategy}
We define a composite \emph{biosphere score} that aggregates retrievable margins,
\begin{equation}
\mathcal{S}_{\rm bio} \;=\; \mathbb{E}\Big[ \prod_{g\in \mathcal{G}_{\rm bio}} \sigma \big(\beta_g\,m_g\big) \Big],
\end{equation}
where the expectation is over the joint posterior of $\{m_g\}$, $\mathcal{G}_{\rm bio}=\{4,5,7,8,11,12\}$ (gates accessible to remote constraints), $\sigma(x)=(1+e^{-x})^{-1}$ is a soft gate, and $\beta_g$ set steepness (reflecting how sharp each gate is). We then compute the \emph{expected information gain} for candidate follow-up data (filters, resolution, cadence) by Fisher or nested-sampling forecasts, prioritizing observations that most reduce uncertainty on the weakest margins (typically Gate~11 cycle closure and Gate~12 stability). This couples survey planning directly to falsifiable inequalities and avoids over-investing in targets that already sit far from pass/fail boundaries.

To reduce double‑counting from correlated margins, also report
\begin{equation}
\label{eq:sbio-copula}
S_{\rm bio}^{\rm joint} = \Pr(\mathbf{Z}>\mathbf{0}),\qquad \mathbf{Z}\sim \mathcal{N}(\boldsymbol{\mu},\,\Sigma),
\end{equation}
with \(\mu_g=\beta_g\,(m_g-1)\) and \(\Sigma_{gh}=\rho_{gh}\) estimated from the posterior samples of \(\{m_g\}\).

\paragraph*{Comparative priorities across stellar types.}
The framework predicts systematic shifts in bottlenecks and thus in \(\{m_g\}\) by host SED and activity.
Around G/K dwarfs, Gate~3 (volatile retention) and Gate~4 (usable photonic work in the visible) are typically favorable, making oxygenic-ocean archetypes the most testable: we expect large \(\Phi\), strong Gate~11 triads (disequilibrium, seasonal coherence, pigment edges), and moderate climate slack \(m_G\). Active M-dwarf HZ planets frequently sit near Gate~3 limits (early XUV loss) and may realize Gate~11 via anoxygenic Fe/S cycles under NIR illumination, with redshifted edges and stronger haze photochemistry; here, triage should require measured XUV histories and escape diagnostics to avoid false positives from thin or secondary atmospheres. A tiered program therefore prioritizes mid‑G/K systems with robust Gate~3--4 priors, followed by quiet M dwarfs with constrained XUV/escape histories; subsurface‑ocean analogs are pursued via indirect Gate~11 proxies (oxidant--reductant delivery and plume tracers).

\subsection{Reporting standards}
\label{sec:rep-stat}

For each target we recommend reporting:
\begin{enumerate}
\item Gate margins with credible intervals: $\{\Phi,\ \lambda_{\mathrm{kin}} \text{ proxy},\ \mathcal{C},\ \mathcal{E},\ \chi,\ m_{\rm NPP},\ m_{\rm cyc,X},\ m_G\}$.
\item A falsifier table listing abiotic alternatives tested and their Bayes factors.
\item A sensitivity analysis showing how margins shift under perturbations of key priors (stellar activity history, interior heat, cloud model).
\item A reproducible data package: posterior samples of $\theta$ and derived $\{m_g\}$; forward-model code/configuration sufficient to regenerate $\mathcal{F}(\theta)$.
\end{enumerate}

\paragraph*{Identifiability notice (Gates 6 and 10).}
Because replication errors ($\mu$) and translation confusion matrices are not directly observable from remote data, all reported Gate~6 and Gate~10 quantities must be labeled as \emph{bounds}, not full posteriors. Concretely:
\begin{enumerate}\setlength{\itemsep}{2pt}
\item Report \emph{energy-consistent} bounds using Gate~4 surplus power: given $P_{\rm surplus}=P_{\rm abs}+P_{\rm int}-P_{\rm maint}$ and elongation throughput $\nu$, the maximal per-symbol proofreading work obeys $W^{\max}_{\rm proof}=P_{\rm surplus}/\nu$. With the thermodynamic bound $W_{\rm proof}\ge k_BT\ln(\varepsilon_{\rm eq}/\varepsilon_b)$ [Eqs.~(\ref{eq:proof-bound-g10}), (\ref{eq:W-proof})], infer $\varepsilon_b^{\min}$ and hence $C_{\rm cap}(\varepsilon_b^{\min})$ and the \emph{upper} bound $\chi^{\max}=C_{\rm cap}(\varepsilon_b^{\min})/R_{\rm code}$.
\item For Gate~6, propagate laboratory priors on $\mu(T,\mathrm{ions},\mathrm{solvent})$ to an \emph{upper} bound $L_{\max}\simeq \ln s/\mu$ and report $m_E=1/E$ as \textsf{prior-dominated} if its credible interval is controlled by those priors rather than by astronomical data.
\end{enumerate}

This methodology ties remote measurements to the physics of the gates through explicit operators and uncertainties. It yields not a binary habitability flag but quantitative, testable margins that can guide follow-up and, when strong across Gates~11--12, constitute robust evidence for a biosphere.

\subsection{Fundamental-physics priors and end-to-end sensitivities}
\label{sec:FP-priors}

For targets where it is useful to report robustness to physics assumptions, we include $\mathbf{c}=(\alpha,\,m_e/m_p,\,G_{\mathrm{N}}m_p^2/(\hbar c),\ldots)$ as hyperparameters with tight priors around standard values and propagate their effect to gate margins via Eq.~(\ref{eq:constants_to_biosphere}). Posterior samples $\{\theta^{(s)}\}$ yield joint draws of $\{m_g^{(s)}\}$ from which we compute $\partial\ln m_g/\partial\ln c_i$ by local regression in the sample cloud; biosphere sensitivities such as $\partial \ln\mathrm{NPP}/\partial \ln \alpha$ follow from Eq.~(\ref{eq:sensitivity}). Reporting these derivatives with credible intervals makes explicit how the fundamental-physics anchors control the inferred biosphere margins.

\subsection{Computing and reporting sensitivities to fundamental constants}
\label{sec:methods-sensitivity}

We treat the constant vector $\mathbf{c}=(\alpha,\,m_e/m_p,\,G_{\mathrm N}m_p^2/(\hbar c),\ldots)$ as hyperparameters that drive gate margins via (\ref{eq:constants_to_biosphere}).
For each target, we evaluate the log-sensitivity in (\ref{eq:sensitivity}) by estimating the two factors:
(i)~$\partial \ln \mathrm{NPP}/\partial \ln m_g$ from the Gate~11--12 power and cycle-closure model (using (\ref{eq:PhiNPP}) where applicable), and
(ii)~$\partial \ln m_g/\partial \ln \alpha$ from Gate~1 physics and its propagated thermochemical/transport consequences.
Operationally, we compute these derivatives by local finite differences in log-space around the posterior mode (or via a local linear fit to posterior samples),
and we report $\partial \ln \mathrm{NPP}/\partial \ln \alpha$ as a median with a $68\%$ credible interval.
Where helpful, we also tabulate the intermediate contributions
$\big\{\partial \ln \mathrm{NPP}/\partial \ln m_g\big\}_g$ and $\big\{\partial \ln m_g/\partial \ln \alpha\big\}_g$
so readers can see which gates dominate the response.

\emph{Reporting:} For each target we report medians and 68\% credible intervals for gate margins and biosphere metrics, together with log‐sensitivities to fundamental constants via (\ref{eq:sensitivity}). When the response is dominated by a subset of gates, we list the top contributions to $\partial\ln\mathrm{NPP}/\partial\ln\alpha$ as $(\partial\ln\mathrm{NPP}/\partial\ln m_g)(\partial\ln m_g/\partial\ln\alpha)$ to show where the leverage lies.

\paragraph*{Sensitivity caveat (Gate 1 $\rightarrow$ Gates 6,10).}
The scaling $\Delta\Delta G_{\rm pair}\propto E_{\rm Ry}\propto \alpha^2$ is a useful \emph{order-of-magnitude} guide, but solvent, ionic strength, and base-stacking make large, system-specific contributions to discrimination. We therefore recommend reporting
\[
\frac{\partial\ln\mu_{\rm eq}}{\partial\ln\alpha}=-2\,\frac{\Delta\Delta G_{\rm pair}}{k_BT}\,(1+\epsilon_{\rm solv}),\quad \epsilon_{\rm solv}\sim\mathrm{Unif}[-0.5,+0.5],
\]
and propagating this into a \emph{range} for $\partial\ln \chi/\partial\ln\alpha$ using Eq.~(\ref{eq:mu_eq}) (with $g\to g_{\rm eff}$). Quote medians with 68\% credible intervals, explicitly labeling the result as \textsf{high-variance prior}. This avoids over-interpretation of $\partial\ln \chi/\partial\ln\alpha$ when solvent/stacking dominate the discrimination margin. 

\section{Validation and Falsifiers}
\label{sec:validation} 

This section consolidates diagnostics, observational tests, and falsifiers for all gates; gate-local subsections refer here rather than repeating criteria. Laboratory validation targets Gates~5--9 and 10; astronomical validation targets Gates~2--4 and 11--12; decisive falsifiers arise from combinations that violate (\ref{eq:gate10-box}) or (\ref{eq:PhiNPP}) at retrieved margins.

\subsection{Laboratory and field validation}

\paragraph*{Threshold flips in continuous-flow protocell reactors (Gates 4, 6, 8, 9).}
Construct microfluidic or stirred continuous-flow reactors seeded with templates, catalysts, and compartment formers. Sweep (a) usable power $\Phi$ via light/redox flux and duty cycle; (b) per-site error $\mu$ via temperature/ionic strength and available proofreading chemistry; (c) confinement $\mathcal{C}=\tau_{\rm ret}/\tau_r$ via permeability-controlled retention $\tau_{\rm ret}\sim V/(A P_i)$ and characteristic reaction time $\tau_r$; report $\tau_{\rm mix}/\tau_r$ only as an auxiliary diagnostic when gradient-based architectures are invoked. (d) dilution $\delta$ and external loss. Measure absolute growth exponent $r^*=\lambda_{\max}(BQ-D)$, error spectra to infer $Q$ and $U=\mu L$, and transport/leakage fluxes to test $J_{\mathrm{in}}\ge J_{\mathrm{cons}}$ and $J_{\mathrm{leak}}\le J_{\mathrm{make}}$. 
\textit{Predicted signature:} abrupt transitions of $r^*$ from negative to positive as $\Phi$ or $\mathcal{C}$ crosses unity, and collapse when $\mu L$ approaches the Eigen bound. Hysteresis is expected near leakage thresholds if division/growth introduces effective positive feedbacks.

\paragraph*{Autocatalytic closure and kinetic gain (Gate 7).}
Percolation-like onset of RAFs is expected near \(p_c \sim (\ln N)/N\) in random catalysis ensembles \cite{Kauffman1986,HordijkSteel2004,JainKrishna1998}. 
In mineral-templated or cofactor-assisted networks, assemble feedstocks and catalysts to vary the catalysis density $p_{\mathrm{cat}}$ (e.g., by metal sulfide loading, UV intensity for photoredox). From time-series concentrations, estimate the kinetic Jacobian $J$ at low conversion and compute $\lambda_{\mathrm{kin}} \equiv \lambda_{\max}(J-\Lambda)$. 
\textit{Predicted signature:} percolation-like onset of large RAF sets when $p_{\mathrm{cat}}$ surpasses $p_c\sim \ln N/N$, accompanied by a positive $\lambda_{\mathrm{kin}}$ and the appearance of autocatalytic cycles in flux analyses.

\paragraph*{Noisy translation channel and code evolution (Gate 10).}
Reconstitute translation-like chemistry (ribozyme or engineered enzyme systems) and measure the symbol-level confusion matrix $T(y|x)$ as a function of proofreading energy per symbol $W_{\mathrm{proof}}$. Compute capacity $C$ and rate $R_{\mathrm{code}}(k, n, m)$; evolve codon$\to$monomer assignments under a chemically motivated error-cost $W(a,b)$.
\textit{Predicted signature:} (i) $C$ increases approximately as $\ln(\varepsilon_{\mathrm{eq}}/\varepsilon_b)$ with added dissipation, enabling $\chi=C_{\mathrm{cap}}/R_{\mathrm{code}}>1$ once $\Phi$ supplies sufficient $P_{\mathrm{proof}}$; (ii) evolved codes minimize expected error cost by clustering confusable codons into synonym sets, with degeneracy $d$ increasing as noise rises.

\paragraph*{Ecological closure/regulation mesocosms (Gates 11--12).}
Operate recirculating mesocosms that explicitly track C/N/P/S reservoirs with controlled nonbiological sources/sinks. Measure $\mathrm{NPP}$, recycling fractions $\mathcal{R}_X$, recycle and loss times $\tau_{\mathrm{recycle},X}$ and $\tau_{\mathrm{loss},X}$, and estimate the community matrix $A$ from small perturbations. Close the loop to an external climate/chemistry emulator: couple atmospheric CO$_2$ (or an analogue) to a tunable weathering sink, radiative response (electrochemical or optical proxy), measure the effective climate gain $G_{\mathrm{climate}}$.
\textit{Predicted signature:} sustained $\mathrm{NPP}>0$ with $\tau_{\mathrm{recycle},X}<\tau_{\mathrm{loss},X}$, negative $\max \Re \lambda(A)$, $G_{\mathrm{climate}}<1$ with bounded response to step forcings.

\textit{Field analogs.} Natural settings such as alkaline vents, evaporitic basins, and sub-ice chemolithoautotrophic ecosystems provide external validity: quantify in situ $\Phi$, precursor make/loss budgets, and recycling efficiencies, and compare to reactor-derived thresholds.

\begin{table}[t]
\vskip -10pt
\caption{Illustrative falsifier suite for key claims (to be populated per target).}
\label{tab:falsifiers}
\centering

\begingroup
\setlength{\tabcolsep}{1pt}
\renewcommand{\arraystretch}{1.00}
\newcommand{\colbox}[2]{\parbox[t]{#1}{\raggedright #2}}

\begin{tabular}{@{}l l l l@{}}
\hline
\colbox{0.21\linewidth}{Claim} &
\colbox{0.28\linewidth}{Abiotic alternative} &
\colbox{0.18\linewidth}{Tuned parameters} &
\colbox{0.23\linewidth}{Evidence metric} \\
\hline\hline\addlinespace[1pt]
\colbox{0.21\linewidth}{Large $\Delta G_{\mathrm{diseq}}$ \& seasonal coherence} &
\colbox{0.28\linewidth}{CO$_2$ photolysis $+$ heterogeneous sinks} &
\colbox{0.18\linewidth}{FUV, CO sinks, cloud microphysics} &
\colbox{0.23\linewidth}{Bayes factor $K_{\mathrm{bio/abiotic}}$} \\\addlinespace[1pt]
\colbox{0.21\linewidth}{Low CO with O$_2$ $+$ CH$_4$} &
\colbox{0.28\linewidth}{High‑outgassing CH$_4$ $+$ short $\tau_{\mathrm{phot}}$} &
\colbox{0.18\linewidth}{$F_{\mathrm{out}}$, $K_{zz}$, haze} &
\colbox{0.23\linewidth}{$\tau_{\mathrm{phot}}/\tau_{\mathrm{obs}}$} \\\addlinespace[2pt]
\colbox{0.21\linewidth}{Helium 1083\,nm escape} &
\colbox{0.28\linewidth}{Photon/energy‑limited escape only} &
\colbox{0.18\linewidth}{$F_{\tt XUV}(t)$, $\eta$, $R_{\mathrm{exo}}$} &
\colbox{0.23\linewidth}{Posterior on $M_{\mathrm{loss}}$} \\\addlinespace[1pt]
\colbox{0.21\linewidth}{Edge‑like reflectance} &
\colbox{0.28\linewidth}{Mineral/ice edges $+$ geometry} &
\colbox{0.18\linewidth}{BRDF, grain size} &
\colbox{0.23\linewidth}{Phase/angle modulation test} \\\addlinespace[1pt]
\hline
\end{tabular}
\endgroup
\vskip -10pt
\end{table}

\subsection{Astronomical validation}

\paragraph*{Triad for Gate 11.}
A robust, time-resolved \emph{triad}---(i) large atmospheric chemical disequilibrium free energy $\Delta G_{\mathrm{diseq}}$, (ii) coherent seasonal variability of multiple gases with amplitudes and phases inconsistent with purely photochemical forcing, and (iii) surface biosignature reflectance/polarimetric edges---is the remote analogue of positive $\mathrm{NPP}$ with closed cycles. Multi-epoch observations should retrieve $\Delta G_{\mathrm{diseq}}\gg 0$ \emph{and} seasonal coherence metrics exceeding abiotic baselines.

\paragraph*{Climate--chemistry buffering for Gate 12.}
Long-lived secondary atmospheres exhibiting small secular drift in greenhouse species and the coexistence of multiple volatile reservoirs (oceans/ice) over years to decades are indicative of $G_{\mathrm{climate}}<1$. Phase-curve energy budgets and secular stability jointly constrain the feedback gain.

\paragraph*{Null environments.}
Planets predicted to fail Gate 3 (volatile retention) or Gate 4 (power margin) should systematically lack the triad and buffering signatures. A population-level anti-correlation between negative margins and biosphere observables provides a stringent falsification check.

\subsection{Falsifiers}

\paragraph*{Gate 1.} If, across relevant solvent windows, measured bonding and interaction energies do not satisfy $E_b\gg k_{\mathrm B}T$ and $E_{\mathrm{HB}}\gtrsim 0.1$\,eV, template pairing and amphiphile assembly are not feasible.

\paragraph*{Gate 4.} If realistic budgets in candidate niches yield $\Phi\le 1$ even after accounting for internal/tidal power, reactors cannot cross from maintenance to growth; absence of threshold flips in $r^*$ under power sweeps falsifies Gate 4 sufficiency.

\paragraph*{Gate 5.} If photochemical/thermochemical models constrained by spectra imply $J_{\mathrm{make}}/J_{\mathrm{loss}}<1$ for key precursors under any plausible SED and composition, sustained monomer pools are infeasible.

\paragraph*{Gate 6.} If template copying fidelities in relevant solvents/ions cannot achieve $L_{\max}\gtrsim 10^{2}$ under any accessible $\Phi$, the error threshold prevents heritable polymers.

\paragraph*{Gate 7.} If, despite generous catalysis densities and feeds, RAF sets do not appear and $\lambda_{\rm kin}\le 0$ across parameter space, autocatalytic closure is not a generic outcome.

\paragraph*{Gate 8.} If measured transport parameters enforce $\mathcal{C}\ll 1$ and/or $J_{\mathrm{leak}}>J_{\mathrm{make}}$ for all plausible compartments, localization cannot couple work to copying.

\paragraph*{Gate 9.} If $\lambda_{\max}(BQ-D)\le 0$ under all feeds/geometries or if $N_e s\ll 1$ at all heritability scales, Darwinian dynamics does not arise.

\paragraph*{Gate 10.} If, for all feasible $(k,n,m)$ and proofreading work bounded by $\Phi$, the retrieved channel capacity never exceeds the required rate ($\chi\le 1$ after finite-blocklength backoff), scalable coding is impossible.

\paragraph*{Gates 11--12.} If mesocosms cannot achieve $\mathrm{NPP}>0$ with $\tau_{\mathrm{recycle},X}<\tau_{\mathrm{loss},X}$ and stable $A$, or if climate--chemistry feedbacks yield $G_{\mathrm{climate}}\ge 1$, biosphere persistence and regulation are not supported. Astronomically, the absence of the Gate~11 triad and Gate~12 buffering on planets with positive Gate~3--4 margins would contradict the framework’s predictions. A decisive prediction is that the Gate~11 triad (sustained atmospheric disequilibrium, multi-species seasonal coherence, and a pigment edge) \emph{anti-correlates} with negative priors on Gate~3--4 inferred from stellar activity and bulk density.
Survey designs should include control samples that are predicted to fail Gate~3 or Gate~4; the absence of the triad in those controls is a built-in falsification test.
Conversely, robust triads where Gate~3--4 are demonstrably subcritical would falsify the pipeline and force revision of the underlying budgets (e.g., escape, power, or maintenance-leak accounting).

\subsection{Reporting and reproducibility}

For each laboratory protocol or astronomical target, report gate margins with uncertainties, control-parameter trajectories across thresholds, abiotic alternative fits with Bayes factors, and open data/models sufficient to recompute $\mathcal{F}(\theta)$ and the derived $\{m_g\}$. The framework is refutable: any one of the falsifiers above, demonstrated under realistic conditions, would invalidate the corresponding link in the chain from fundamental physics to biospheres.

Table~\ref{tab:checklist} presents a concise checklist of key inputs and corresponding margins. Table~\ref{tab:earth-sanity} reports approximate present‑Earth margins.

\begin{table}[h]
\vskip -10pt
\centering
\caption{Gate checklist: key inputs and corresponding margins.}
\label{tab:checklist}

\begingroup
\setlength{\tabcolsep}{1pt}
\renewcommand{\arraystretch}{1.00}
\newcommand{\colbox}[2]{\parbox[t]{#1}{\raggedright #2}}

\begin{tabular}{@{}l l@{}}
\hline
\colbox{0.17\linewidth}{Gate} &
\colbox{0.45\linewidth}{Inputs $\rightarrow$ Margins} \\
\hline\hline
\colbox{0.17\linewidth}{G3 (Surface access)} &
\colbox{0.45\linewidth}{$M_{\mathrm{solid}},\;\tau_{\mathrm{disk}},\;F_\star,\;\mathrm{XUV}(t)
\;\rightarrow\; m_\Sigma,\;m_t,\;m_F,\;m_{\mathrm{vol}},\;m_{\mathrm{env}}$} \\
\colbox{0.17\linewidth}{G4 (Power)} &
\colbox{0.45\linewidth}{SED, redox fluxes, $P_{\mathrm{leak}}$, copying throughput
$\;\rightarrow\; m_\Phi$} \\
\colbox{0.17\linewidth}{G5 (Feedstocks)} &
\colbox{0.45\linewidth}{$\phi_\lambda^{\mathrm{eff}},\;k_{\mathrm{geo}},\;K_{\mathrm{out}},K_{\mathrm{photo}},K_{\mathrm{hyd}}
\;\rightarrow\;\Psi_m,\;\mathrm{Dam}$} \\
\colbox{0.17\linewidth}{G6 (Fidelity)} &
\colbox{0.45\linewidth}{$\Delta\Delta G_{\mathrm{pair}},\;W_{\mathrm{proof}},\;k_{\mathrm{les}},\;v,\;p_{\mathrm{proc}},\;\tau_{\mathrm{loss}}
\;\rightarrow\; m_E,\;m_{\mathrm{copy}}$} \\
\colbox{0.17\linewidth}{G7 (RAF/gain)} &
\colbox{0.45\linewidth}{Catalysis density $p_{\mathrm{cat}},\;\Delta G_i,\;$ losses $\Lambda$
$\;\rightarrow\; m_p,\;m_{\mathrm{kin}},\;m_A$} \\
\colbox{0.17\linewidth}{G8 (Transport)} &
\colbox{0.45\linewidth}{$D,\;P,\;K_i,\;u,\;a
\;\rightarrow\; m_C,\;\min_i,\;m^{\mathrm{leak}}_j$} \\
\colbox{0.17\linewidth}{G9 (Darwinian)} &
\colbox{0.45\linewidth}{$b_i,\;d_i,\;\delta,\;Q,\;N_e,\;U_b
\;\rightarrow\; m_r,\;m_{\mathrm{drift}},\;m_{\mathrm{ben}}$} \\
\colbox{0.17\linewidth}{G10 (Translation)} &
\colbox{0.45\linewidth}{$\varepsilon_b,\;k,\;n,\;m,\;W_{\mathrm{proof}},\;\nu
\;\rightarrow\; m_\chi,\;m_P,\;m_d$} \\
\colbox{0.17\linewidth}{G11 (Biosphere)} &
\colbox{0.45\linewidth}{$\Delta G_{\mathrm{assim}},\;\eta_{\mathrm{bio}},\;R_X,\;\tau_{\mathrm{recycle}},\;\tau_{\mathrm{loss}}
\;\rightarrow\; m_{\mathrm{NPP}},\;m_{{\mathrm{cyc}},X}$} \\
\colbox{0.17\linewidth}{G12 (Climate)} &
\colbox{0.45\linewidth}{$\lambda_0,\;\lambda_{\mathrm{fb}},\;\tau_{\mathrm{eco}},\;\tau_{\mathrm{buf}}
\;\rightarrow\; m_G,\;m_{\mathrm{tip}}$} \\
\hline
\end{tabular}
\endgroup
\vskip -10pt
\end{table}

\section{Discussion}
\label{sec:discussion}

We presented a framework that turns the questions about the origin, emergence and persistence of life into a vector of measurable margins. Each gate is falsifiable in reactors or by remote sensing; taken together, the margins yield population-level expectations by stellar type and solvent, and decisive failure modes. The dominant levers are $\Phi$ (power), $\cal E$ (fidelity), $C$ (localization), and $1-G_{\rm climate}$ (stability), each with clear observables.

The framework is intentionally minimal: it substitutes a small set of falsifiable margins for full ecosystem modeling. As telescope and laboratory constraints improve, the quickest ways to falsify the program are (i) demonstrating sustained biospheric power without satisfying (\ref{eq:PhiNPP}) at the retrieved $\dot N_{\mathrm{copy}}$ and discrimination, or (ii) observing robust seasonal disequilibria where Gate~3 surface access is not met.

\subsection{What is universal and why}

Three constraints are enforced by physics independently of solvent, alphabet, or pathway choices.

\paragraph*{Energy and dissipation.}
The surplus-power condition $\Phi>1$ funds every subsequent step: kinetic proofreading that lowers per-symbol and per-site error rates (Gates~6 and~10), active transport and maintenance against leakage (Gate~8), and the ecological work needed to keep cycles closed (Gate~11). Because $\Phi$ is defined from photon/redox power budgets and maintenance costs, it is comparable across sunlit and dark worlds.

\paragraph*{Localization and amplifying kinetics.}
Autocatalytic amplification ($\lambda_{\rm kin}>0$; Gate~7) must be coupled to spatial/temporal localization so that production outpaces loss: $\mathcal{C} = \tau_{\rm ret}/\tau_r \gtrsim 1$, $J_{\mathrm{in}}\!\ge J_{\mathrm{cons}}$, and $J_{\mathrm{leak}}\!\le J_{\mathrm{make}}$ (Gate~8). These are conservation constraints on fluxes; their form is invariant across solvents because transport and permeability only rescale the terms.

\paragraph*{Digital heredity and rate--capacity consistency.}
The Eigen bound ${\cal E}=\mu L/\ln s<1$ caps heritable information at a given replication error, and the translation inequality $\chi=C_{\mathrm{cap}}/R_{\mathrm{code}}>1$ requires that coding throughput exceed channel noise. Both tie directly to dissipation through information thermodynamics: reducing $\mu$ or $\varepsilon$ requires work per symbol of order $k_{\mathrm B}T\ln(1/\mathrm{error})$.

\begin{table}[t]
\vskip -10pt
\caption{Approximate present‑Earth margins (order‑of‑magnitude; illustrative).  [Note: 
For modern DNA-based life (Eigen bound is not limiting);
RNA replicators and RNA viruses can approach $m_E=\mathcal{O}(1)$, see Sec.~\ref{sec:gate5}.]}
\label{tab:earth-sanity}
\centering
\begin{tabular}{@{}lcl@{}}
\hline
Margin & Value & Notes \\
\hline\hline
$m_\Phi$ & $2$--$4$ & photonic/redox vs. maintenance+synthesis \\
$m_E$ & $\sim 10^{3}\text{--}10^{5}$ & if $L\sim 10^{6}$ and $\mu_{\rm eff}\sim 10^{-10}\text{--}10^{-8}$ per base per replication \\
$m_\chi$ & $\gtrsim 2$ & capacity vs. coding rate with proofreading \\
$m_{\mathcal{C}}$ & $\gtrsim 1$ & 
retention vs. key reaction times ($C=\tau_{\rm ret}/\tau_r$);\\
& & report $C_{\rm mix}=\tau_{\rm mix}/\tau_r$ only as an auxiliary gradient diagnostic \\
$m_{\mathrm{cyc},\{C,N,P\}}$ & $\gg 1$ & recycle $\ll$ loss; \\
$m_G$ & $0.3$--$0.5$ & $1-G_{\mathrm{climate}}$ under modern feedbacks. \\
\hline
\end{tabular}
\vskip -10pt
\end{table}

\subsection{What can vary and how it is bounded}

Diversity arises from parameters that shift \emph{values} without altering gate form.

\paragraph*{Solvent and temperature.}
Changing solvent/temperature alters $E_b/k_{\mathrm B}T$, viscosities, diffusivities, and membrane permeabilities, shifting $\Phi$ via reaction rate constants, $\mathcal{C}$ via $D$ and $\tau_r$, and the achievable $\mu$. Cold hydrocarbon worlds thus tend toward small $\Phi$ and large $\tau_r$, forcing Gate~11 to operate near threshold with high recycling fractions.

\paragraph*{Alphabet and code geometry.}
Given a symbol error $\varepsilon$, feasible triples $(k,n,m)$ must satisfy $\chi(\varepsilon;k,n,m)>1$, typically by increasing degeneracy $d=k^n/m$ or reducing $m$. Many codes are possible, but all must sit above the same rate--capacity surface.

\paragraph*{Dominant redox couples.}
Available electron donors/acceptors set where the free energy enters Gate~4 and how Gate~11 cycles organize (H$_2$/CO$_2$ in vents, Fe/S cycles at photic anoxic surfaces, NH$_3$ pathways in H$_2$-rich atmospheres). These choices change stoichiometries and timescales, not the inequalities.

\subsection{Bottlenecks and expected failure modes}

Population-level considerations and measured priors highlight common bottlenecks.

(i) \textit{Volatile retention (Gate~3):} small planets in high-XUV environments struggle to keep surface liquids, halting the pipeline early.  

(ii) \textit{Power margin (Gate~4):} hazy or low-irradiance surfaces and cold solvents depress $\Phi$; subsurface worlds depend on radiolysis/serpentinization and often lie near the maintenance bound.  

(iii) \textit{Fidelity/localization (Gates~6 and~8):} realistic per-site errors in plausible ions/solvents, membrane permeabilities that permit nutrients yet retain templates are tight constraints that co-limit $L_{\max}$ and effective growth rates.  

(iv) \textit{Feedback stability (Gate~12):} even with positive $\mathrm{NPP}$ and cycle closure, $G_{\mathrm{climate}}\ge 1$ yields transient biospheres vulnerable to secular forcing.

\subsection{Families of biospheres and co-varying observables}

The framework predicts \emph{families}, not arbitrary diversity, each with a distinctive bundle of observables:

\paragraph*{Oxygenic oceans around G/K dwarfs.}
High $\Phi$ in the visible, robust Gate~11 ($\mathrm{NPP}>0$ with closed C/N/P cycles), and Gate~12 stabilization via weathering. Co-occurring observables: O$_2$ with reduced trace gases at photochemically inconsistent levels, low CO, pigment edges near $0.7\,\mu$m (shifted with host SED), ocean glint, multi-gas seasonal coherence, and small secular drift in greenhouse species.

\paragraph*{Anoxygenic or mixed systems around active M dwarfs.}
Gate~3 often marginal; Gate~5 dominated by flare-weighted photochemistry; Gate~11 built on Fe/S cycles with little free O$_2$. Predict redshifted edges ($\sim0.9$--$1.1\,\mu$m), organosulfur gases with turnover faster than abiotic expectations, CH$_4$+CO$_2$ at low CO, and variable hazes. Gate~12 margins depend on aerosol--climate feedbacks.

\paragraph*{Subsurface chemotrophic oceans.}
Gate~4 from radiolytic oxidants and hydrothermal reductants; Gate~11 at ice--ocean interfaces; Gate~12 controlled by internal heat and redox buffering. Observables are indirect: plume compositions, radiolysis products at the surface, and minimal atmospheres with diagnostic traces.

\paragraph*{Hydrogen-rich temperate super-Earths.}
Large scale heights aid detectability; CH$_4$ is ambiguous, so disequilibrium \emph{sets} (e.g., NH$_3$ with oxidants) and surface-film polarimetry carry more weight. Gate~12 requires greenhouse buffering against large variability.

\paragraph*{Hydrocarbon-solvent biospheres.}
Gate~1 satisfied, Gate~4 tight, Gate~11 slow. Expect subtle hydrocarbon disequilibria, isotopic fractionations, and persistent surface film signatures; atmospheric biosignatures are weak and require multi-epoch consistency.

\subsection{Population-level predictions and falsification}

The gate approach is refutable and makes quantitative population predictions:

\begin{itemize}
\item The occurrence of the Gate~11 triad (sustained disequilibrium, seasonal coherence, surface pigment edge) should be strongly anti-correlated with negative Gate~3 or Gate~4 margins inferred from stellar activity histories and bulk densities. 
\item Within positive Gate~3--4 samples, the fraction of oxygenic archetypes should increase from M to G/K dwarfs as early-time XUV drops and PAR widens; anoxygenic signatures should dominate at active M dwarfs.
\item Worlds with strong Gate~12 margins should show small secular drifts in greenhouse species despite external variability; a prevalence of large drifts in otherwise promising samples would challenge the framework’s stability criterion.
\end{itemize}

Failure of these correlations at high confidence would falsify specific links (e.g., if oxygenic bundles concentrate around active M dwarfs with poor Gate~3 priors, or if strong Gate~11 triads appear where Gate~4 is demonstrably subcritical).

\subsection{Limitations and avenues for refinement}

The inequalities compress complex microphysics into low-dimensional margins and can under-represent couplings (e.g., membrane chemistry affecting both $\mu$ and $P$). Finite-blocklength effects in Gate~10 and spatial heterogeneity in Gate~12 are treated in reduced form. These limitations motivate (i) tighter priors on $\mu(T,\mathrm{ions},\text{solvent})$, permeability $P$ for realistic membranes, and catalytic degree distributions; (ii) emulator-calibrated photochemical/climate retrievals to propagate observational uncertainties to gate margins; and (iii) spatially explicit eco-climate models to test multistability and tipping risks within $G_{\mathrm{climate}}<1$.

\paragraph*{Implications for the exoplanet search.}
Two constraints elevate the search from plausibility to predictivity.
First, the information--energetics ceiling (\ref{eq:PhiNPP}) charges the minimal copying and proofreading work against the planetary power budget, bounding feasible NPP and filtering targets whose claimed productivity would overspend their Gate~4 surplus.
Second, rate--capacity consistency (Gate~10) requires \(\chi>1\) for scalable coding under realistic noise and finite-blocklength penalties, tying any claim of complex biosynthesis to dissipation and confusion-matrix bounds.

We propose a community ``gate atlas'': posteriors for \(\{m_g\}\) and Bayes factors against abiotic alternatives for each target. This compact, falsifiable summary will let surveys focus exposure on worlds that are \emph{physically} most likely to host persistent biospheres and will make success or failure unambiguous.

\subsection{Programmatic outlook}

Two coordinated programs close the loop. In the laboratory: continuous-flow protocell reactors to map threshold flips in $(\Phi,\mathcal{C},\mu)$; translation surrogates to chart $C_{\mathrm{cap}}(\varepsilon)$ under controlled dissipation; mesocosms that achieve cycle closure and measure effective climate gains. In astronomy: multi-epoch spectroscopy and polarimetry targeted at the predicted bundles per archetype, with hierarchical retrievals that report gate margins and Bayes factors against abiotic alternatives. A community ``gate atlas''---posteriors for $\{m_g\}$ across targets---would make progress cumulative and decisively testable.

In summary, the gate framework constrains what life elsewhere should look like because the underlying constraints are physical. Universals---localization, heritable information, network amplification, ecological closure, feedback stability---are expressed as inequalities that any biosphere must satisfy; contingencies---solvents, pigments, alphabets---shift parameters but remain bounded by the same margins. This provides a principled basis for selecting targets and interpreting detections as we search for biospheres around nearby stars.

\section*{Acknowledgments}
The work described here was carried out at the Jet Propulsion Laboratory, California Institute of Technology, Pasadena, California, under a contract with the National Aeronautics and Space Administration.   


\begin{thebibliography}{115}%
\makeatletter
\providecommand \@ifxundefined [1]{%
 \@ifx{#1\undefined}
}%
\providecommand \@ifnum [1]{%
 \ifnum #1\expandafter \@firstoftwo
 \else \expandafter \@secondoftwo
 \fi
}%
\providecommand \@ifx [1]{%
 \ifx #1\expandafter \@firstoftwo
 \else \expandafter \@secondoftwo
 \fi
}%
\providecommand \natexlab [1]{#1}%
\providecommand \enquote  [1]{``#1''}%
\providecommand \bibnamefont  [1]{#1}%
\providecommand \bibfnamefont [1]{#1}%
\providecommand \citenamefont [1]{#1}%
\providecommand \href@noop [0]{\@secondoftwo}%
\providecommand \href [0]{\begingroup \@sanitize@url \@href}%
\providecommand \@href[1]{\@@startlink{#1}\@@href}%
\providecommand \@@href[1]{\endgroup#1\@@endlink}%
\providecommand \@sanitize@url [0]{\catcode `\\12\catcode `\$12\catcode
  `\&12\catcode `\#12\catcode `\^12\catcode `\_12\catcode `\%12\relax}%
\providecommand \@@startlink[1]{}%
\providecommand \@@endlink[0]{}%
\providecommand \url  [0]{\begingroup\@sanitize@url \@url }%
\providecommand \@url [1]{\endgroup\@href {#1}{\urlprefix }}%
\providecommand \urlprefix  [0]{URL }%
\providecommand \Eprint [0]{\href }%
\providecommand \doibase [0]{https://doi.org/}%
\providecommand \selectlanguage [0]{\@gobble}%
\providecommand \bibinfo  [0]{\@secondoftwo}%
\providecommand \bibfield  [0]{\@secondoftwo}%
\providecommand \translation [1]{[#1]}%
\providecommand \BibitemOpen [0]{}%
\providecommand \bibitemStop [0]{}%
\providecommand \bibitemNoStop [0]{.\EOS\space}%
\providecommand \EOS [0]{\spacefactor3000\relax}%
\providecommand \BibitemShut  [1]{\csname bibitem#1\endcsname}%
\let\auto@bib@innerbib\@empty
\bibitem [{\citenamefont
  {{Turyshev}}(2026{\natexlab{a}})}]{TuryshevFunPhys2025}%
  \BibitemOpen
  \bibfield  {author} {\bibinfo {author} {\bibfnamefont {S.~G.}\ \bibnamefont
  {{Turyshev}}},\ }\bibfield  {title} {\bibinfo {title} {{{Fundamental Physics
  in 2025: Status, Decisive Targets, and Path Forward}}},\ }\href
  {https://doi.org/10.1140/epjp/s13360-026-07733-2} {\bibfield  {journal}
  {\bibinfo  {journal} {Eur. Phys. J. Plus}\ }\textbf {\bibinfo {volume}
  {141}},\ \bibinfo {pages} {496} (\bibinfo {year} {2026}{\natexlab{a}})},\
  \bibinfo {note} {arXiv:2512.21445 [gr-qc]}\BibitemShut {NoStop}%
\bibitem [{\citenamefont {Olson}\ \emph {et~al.}(2018)\citenamefont {Olson},
  \citenamefont {Schwieterman}, \citenamefont {Reinhard}, \citenamefont
  {Ridgwell}, \citenamefont {Kane}, \citenamefont {Meadows},\ and\
  \citenamefont {Lyons}}]{Olson2018}%
  \BibitemOpen
  \bibfield  {author} {\bibinfo {author} {\bibfnamefont {S.~L.}\ \bibnamefont
  {Olson}}, \bibinfo {author} {\bibfnamefont {E.~W.}\ \bibnamefont
  {Schwieterman}}, \bibinfo {author} {\bibfnamefont {C.~T.}\ \bibnamefont
  {Reinhard}}, \bibinfo {author} {\bibfnamefont {A.}~\bibnamefont {Ridgwell}},
  \bibinfo {author} {\bibfnamefont {S.~R.}\ \bibnamefont {Kane}}, \bibinfo
  {author} {\bibfnamefont {V.~S.}\ \bibnamefont {Meadows}},\ and\ \bibinfo
  {author} {\bibfnamefont {T.~W.}\ \bibnamefont {Lyons}},\ }\bibfield  {title}
  {\bibinfo {title} {{Atmospheric Seasonality as an Exoplanet Biosignature}},\
  }\href {https://doi.org/10.3847/2041-8213/aac171} {\bibfield  {journal}
  {\bibinfo  {journal} {ApJ Lett.}\ }\textbf {\bibinfo {volume} {858}},\
  \bibinfo {pages} {L14} (\bibinfo {year} {2018})}\BibitemShut {NoStop}%
\bibitem [{\citenamefont {Landauer}(1961)}]{Landauer1961}%
  \BibitemOpen
  \bibfield  {author} {\bibinfo {author} {\bibfnamefont {R.}~\bibnamefont
  {Landauer}},\ }\bibfield  {title} {\bibinfo {title} {{Irreversibility and
  Heat Generation in the Computing Process}},\ }\href
  {https://doi.org/10.1147/rd.53.0183} {\bibfield  {journal} {\bibinfo
  {journal} {IBM J. Res. Dev.}\ }\textbf {\bibinfo {volume} {5}},\ \bibinfo
  {pages} {183} (\bibinfo {year} {1961})}\BibitemShut {NoStop}%
\bibitem [{\citenamefont {Bennett}(1982)}]{Bennett1982}%
  \BibitemOpen
  \bibfield  {author} {\bibinfo {author} {\bibfnamefont {C.~H.}\ \bibnamefont
  {Bennett}},\ }\bibfield  {title} {\bibinfo {title} {{The Thermodynamics of
  Computation---A Review}},\ }\href {https://doi.org/10.1007/BF02084158}
  {\bibfield  {journal} {\bibinfo  {journal} {Int. J. Theor. Phys.}\ }\textbf
  {\bibinfo {volume} {21}},\ \bibinfo {pages} {905} (\bibinfo {year}
  {1982})}\BibitemShut {NoStop}%
\bibitem [{\citenamefont {Shannon}(1948)}]{Shannon1948}%
  \BibitemOpen
  \bibfield  {author} {\bibinfo {author} {\bibfnamefont {C.~E.}\ \bibnamefont
  {Shannon}},\ }\bibfield  {title} {\bibinfo {title} {{A Mathematical Theory of
  Communication}},\ }\href {https://doi.org/10.1002/j.1538-7305.1948.tb01338.x}
  {\bibfield  {journal} {\bibinfo  {journal} {Bell System Technical Journal}\
  }\textbf {\bibinfo {volume} {27}},\ \bibinfo {pages} {379} (\bibinfo {year}
  {1948})}\BibitemShut {NoStop}%
\bibitem [{\citenamefont {Cover}\ and\ \citenamefont
  {Thomas}(2006)}]{CoverThomas2006}%
  \BibitemOpen
  \bibfield  {author} {\bibinfo {author} {\bibfnamefont {T.~M.}\ \bibnamefont
  {Cover}}\ and\ \bibinfo {author} {\bibfnamefont {J.~A.}\ \bibnamefont
  {Thomas}},\ }\href@noop {} {\emph {\bibinfo {title} {{Elements of Information
  Theory}}}},\ \bibinfo {edition} {2nd}\ ed.\ (\bibinfo  {publisher} {Wiley},\
  \bibinfo {address} {Hoboken, NJ},\ \bibinfo {year} {2006})\BibitemShut
  {NoStop}%
\bibitem [{\citenamefont {Hopfield}(1974)}]{Hopfield1974}%
  \BibitemOpen
  \bibfield  {author} {\bibinfo {author} {\bibfnamefont {J.~J.}\ \bibnamefont
  {Hopfield}},\ }\bibfield  {title} {\bibinfo {title} {{Kinetic Proofreading: A
  New Mechanism for Reducing Errors in Biosynthetic Processes Requiring High
  Specificity}},\ }\href {https://doi.org/10.1073/pnas.71.10.4135} {\bibfield
  {journal} {\bibinfo  {journal} {Proc. Natl. Acad. Sci. USA}\ }\textbf
  {\bibinfo {volume} {71}},\ \bibinfo {pages} {4135} (\bibinfo {year}
  {1974})}\BibitemShut {NoStop}%
\bibitem [{\citenamefont {Ninio}(1975)}]{Ninio1975}%
  \BibitemOpen
  \bibfield  {author} {\bibinfo {author} {\bibfnamefont {J.}~\bibnamefont
  {Ninio}},\ }\bibfield  {title} {\bibinfo {title} {{Kinetic Amplification of
  Enzyme Discrimination}},\ }\href
  {https://doi.org/10.1016/S0300-9084(75)80139-8} {\bibfield  {journal}
  {\bibinfo  {journal} {Biochimie}\ }\textbf {\bibinfo {volume} {57}},\
  \bibinfo {pages} {587} (\bibinfo {year} {1975})}\BibitemShut {NoStop}%
\bibitem [{\citenamefont {Eigen}(1971)}]{Eigen1971}%
  \BibitemOpen
  \bibfield  {author} {\bibinfo {author} {\bibfnamefont {M.}~\bibnamefont
  {Eigen}},\ }\bibfield  {title} {\bibinfo {title} {{Selforganization of Matter
  and the Evolution of Biological Macromolecules}},\ }\href
  {https://doi.org/10.1007/BF00623322} {\bibfield  {journal} {\bibinfo
  {journal} {Naturwissenschaften}\ }\textbf {\bibinfo {volume} {58}},\ \bibinfo
  {pages} {465} (\bibinfo {year} {1971})}\BibitemShut {NoStop}%
\bibitem [{\citenamefont {Eigen}\ and\ \citenamefont
  {Schuster}(1977)}]{EigenSchuster1977}%
  \BibitemOpen
  \bibfield  {author} {\bibinfo {author} {\bibfnamefont {M.}~\bibnamefont
  {Eigen}}\ and\ \bibinfo {author} {\bibfnamefont {P.}~\bibnamefont
  {Schuster}},\ }\bibfield  {title} {\bibinfo {title} {{The hypercycle. A
  principle of natural self-organization. A. Emergence of the hypercycle}},\
  }\href {https://doi.org/10.1007/BF00450633} {\bibfield  {journal} {\bibinfo
  {journal} {Naturwissenschaften}\ }\textbf {\bibinfo {volume} {64}},\ \bibinfo
  {pages} {541} (\bibinfo {year} {1977})}\BibitemShut {NoStop}%
\bibitem [{\citenamefont {Kauffman}(1986)}]{Kauffman1986}%
  \BibitemOpen
  \bibfield  {author} {\bibinfo {author} {\bibfnamefont {S.~A.}\ \bibnamefont
  {Kauffman}},\ }\bibfield  {title} {\bibinfo {title} {{Autocatalytic Sets of
  Proteins}},\ }\href {https://doi.org/10.1016/S0022-5193(86)80047-9}
  {\bibfield  {journal} {\bibinfo  {journal} {J. Theor. Biol.}\ }\textbf
  {\bibinfo {volume} {119}},\ \bibinfo {pages} {1} (\bibinfo {year}
  {1986})}\BibitemShut {NoStop}%
\bibitem [{\citenamefont {Hordijk}\ and\ \citenamefont
  {Steel}(2004)}]{HordijkSteel2004}%
  \BibitemOpen
  \bibfield  {author} {\bibinfo {author} {\bibfnamefont {W.}~\bibnamefont
  {Hordijk}}\ and\ \bibinfo {author} {\bibfnamefont {M.}~\bibnamefont
  {Steel}},\ }\bibfield  {title} {\bibinfo {title} {{Detecting Autocatalytic,
  Self-Sustaining Sets in Chemical Reaction Systems}},\ }\href
  {https://doi.org/10.1016/j.jtbi.2003.11.020} {\bibfield  {journal} {\bibinfo
  {journal} {J. Theor. Biol.}\ }\textbf {\bibinfo {volume} {227}},\ \bibinfo
  {pages} {451} (\bibinfo {year} {2004})}\BibitemShut {NoStop}%
\bibitem [{\citenamefont {Hordijk}\ and\ \citenamefont
  {Steel}(2012)}]{HordijkSteel2012}%
  \BibitemOpen
  \bibfield  {author} {\bibinfo {author} {\bibfnamefont {W.}~\bibnamefont
  {Hordijk}}\ and\ \bibinfo {author} {\bibfnamefont {M.}~\bibnamefont
  {Steel}},\ }\bibfield  {title} {\bibinfo {title} {{Autocatalytic Sets and the
  Origin of Life}},\ }\href {https://doi.org/10.3390/e14091731} {\bibfield
  {journal} {\bibinfo  {journal} {Entropy}\ }\textbf {\bibinfo {volume} {14}},\
  \bibinfo {pages} {1731} (\bibinfo {year} {2012})}\BibitemShut {NoStop}%
\bibitem [{\citenamefont {Szostak}\ \emph {et~al.}(2001)\citenamefont
  {Szostak}, \citenamefont {Bartel},\ and\ \citenamefont
  {Luisi}}]{Szostak2001}%
  \BibitemOpen
  \bibfield  {author} {\bibinfo {author} {\bibfnamefont {J.~W.}\ \bibnamefont
  {Szostak}}, \bibinfo {author} {\bibfnamefont {D.~P.}\ \bibnamefont
  {Bartel}},\ and\ \bibinfo {author} {\bibfnamefont {P.~L.}\ \bibnamefont
  {Luisi}},\ }\bibfield  {title} {\bibinfo {title} {{Synthesizing Life}},\
  }\href {https://doi.org/10.1038/35053176} {\bibfield  {journal} {\bibinfo
  {journal} {Nature}\ }\textbf {\bibinfo {volume} {409}},\ \bibinfo {pages}
  {387} (\bibinfo {year} {2001})}\BibitemShut {NoStop}%
\bibitem [{\citenamefont {Hanczyc}\ \emph {et~al.}(2003)\citenamefont
  {Hanczyc}, \citenamefont {Fujikawa},\ and\ \citenamefont
  {Szostak}}]{Hanczyc2003}%
  \BibitemOpen
  \bibfield  {author} {\bibinfo {author} {\bibfnamefont {M.~M.}\ \bibnamefont
  {Hanczyc}}, \bibinfo {author} {\bibfnamefont {S.~M.}\ \bibnamefont
  {Fujikawa}},\ and\ \bibinfo {author} {\bibfnamefont {J.~W.}\ \bibnamefont
  {Szostak}},\ }\bibfield  {title} {\bibinfo {title} {{Experimental Models of
  Primitive Cellular Compartments: Encapsulation, Growth, and Division}},\
  }\href {https://doi.org/10.1021/ja021219+} {\bibfield  {journal} {\bibinfo
  {journal} {J. Am. Chem. Soc.}\ }\textbf {\bibinfo {volume} {125}},\ \bibinfo
  {pages} {3219} (\bibinfo {year} {2003})}\BibitemShut {NoStop}%
\bibitem [{\citenamefont {Monnard}\ and\ \citenamefont
  {Deamer}(2002)}]{MonnardDeamer2002}%
  \BibitemOpen
  \bibfield  {author} {\bibinfo {author} {\bibfnamefont {P.-A.}\ \bibnamefont
  {Monnard}}\ and\ \bibinfo {author} {\bibfnamefont {D.~W.}\ \bibnamefont
  {Deamer}},\ }\bibfield  {title} {\bibinfo {title} {{Membrane Self-Assembly
  Processes: Steps toward the First Cellular Life}},\ }\href
  {https://doi.org/10.1016/S0968-0004(02)02120-3} {\bibfield  {journal}
  {\bibinfo  {journal} {Trends Biochem. Sci.}\ }\textbf {\bibinfo {volume}
  {27}},\ \bibinfo {pages} {647} (\bibinfo {year} {2002})}\BibitemShut
  {NoStop}%
\bibitem [{\citenamefont {Jr}\ and\ \citenamefont
  {Keating}(2017)}]{AumillerKeating2017}%
  \BibitemOpen
  \bibfield  {author} {\bibinfo {author} {\bibfnamefont {W.~M.~A.}\
  \bibnamefont {Jr}}\ and\ \bibinfo {author} {\bibfnamefont {C.~D.}\
  \bibnamefont {Keating}},\ }\bibfield  {title} {\bibinfo {title}
  {{Phosphorylation-mediated RNA/peptide complex coacervation and protocell
  models}},\ }\href {https://doi.org/10.1016/j.cis.2016.05.012} {\bibfield
  {journal} {\bibinfo  {journal} {Adv. Colloid Interface Sci.}\ }\textbf
  {\bibinfo {volume} {239}},\ \bibinfo {pages} {75} (\bibinfo {year}
  {2017})}\BibitemShut {NoStop}%
\bibitem [{\citenamefont {Chen}\ and\ \citenamefont
  {Szostak}(2004)}]{ChenSzostak2004}%
  \BibitemOpen
  \bibfield  {author} {\bibinfo {author} {\bibfnamefont {I.~A.}\ \bibnamefont
  {Chen}}\ and\ \bibinfo {author} {\bibfnamefont {J.~W.}\ \bibnamefont
  {Szostak}},\ }\bibfield  {title} {\bibinfo {title} {{A Kinetic Study of the
  Growth of Fatty Acid Vesicles}},\ }\href
  {https://doi.org/10.1073/pnas.0401865101} {\bibfield  {journal} {\bibinfo
  {journal} {Proc. Natl. Acad. Sci. USA}\ }\textbf {\bibinfo {volume} {101}},\
  \bibinfo {pages} {7965} (\bibinfo {year} {2004})}\BibitemShut {NoStop}%
\bibitem [{\citenamefont {Tlusty}(2008)}]{Tlusty2008}%
  \BibitemOpen
  \bibfield  {author} {\bibinfo {author} {\bibfnamefont {T.}~\bibnamefont
  {Tlusty}},\ }\bibfield  {title} {\bibinfo {title} {{A Simple Model for the
  Evolution of Molecular Codes Driven by the Expansion of the Proteome}},\
  }\href {https://doi.org/10.1016/j.jtbi.2007.08.024} {\bibfield  {journal}
  {\bibinfo  {journal} {J. Theor. Biol.}\ }\textbf {\bibinfo {volume} {249}},\
  \bibinfo {pages} {331} (\bibinfo {year} {2008})}\BibitemShut {NoStop}%
\bibitem [{\citenamefont {Polyanskiy}\ \emph {et~al.}(2010)\citenamefont
  {Polyanskiy}, \citenamefont {Poor},\ and\ \citenamefont
  {Verd\'u}}]{Polyanskiy2010}%
  \BibitemOpen
  \bibfield  {author} {\bibinfo {author} {\bibfnamefont {Y.}~\bibnamefont
  {Polyanskiy}}, \bibinfo {author} {\bibfnamefont {H.~V.}\ \bibnamefont
  {Poor}},\ and\ \bibinfo {author} {\bibfnamefont {S.}~\bibnamefont
  {Verd\'u}},\ }\bibfield  {title} {\bibinfo {title} {{Channel Coding Rate in
  the Finite Blocklength Regime}},\ }\href
  {https://doi.org/10.1109/TIT.2010.2043769} {\bibfield  {journal} {\bibinfo
  {journal} {IEEE Trans. Inf. Theory}\ }\textbf {\bibinfo {volume} {56}},\
  \bibinfo {pages} {2307} (\bibinfo {year} {2010})}\BibitemShut {NoStop}%
\bibitem [{\citenamefont {Powner}\ \emph {et~al.}(2009)\citenamefont {Powner},
  \citenamefont {Gerland},\ and\ \citenamefont {Sutherland}}]{Sutherland2009}%
  \BibitemOpen
  \bibfield  {author} {\bibinfo {author} {\bibfnamefont {M.~W.}\ \bibnamefont
  {Powner}}, \bibinfo {author} {\bibfnamefont {B.}~\bibnamefont {Gerland}},\
  and\ \bibinfo {author} {\bibfnamefont {J.~D.}\ \bibnamefont {Sutherland}},\
  }\bibfield  {title} {\bibinfo {title} {{Synthesis of Activated Pyrimidine
  Ribonucleotides in Prebiotically Plausible Conditions}},\ }\href
  {https://doi.org/10.1038/nature08013} {\bibfield  {journal} {\bibinfo
  {journal} {Nature}\ }\textbf {\bibinfo {volume} {459}},\ \bibinfo {pages}
  {239} (\bibinfo {year} {2009})}\BibitemShut {NoStop}%
\bibitem [{\citenamefont {Ritson}\ and\ \citenamefont
  {Sutherland}(2012{\natexlab{a}})}]{Ritson2012}%
  \BibitemOpen
  \bibfield  {author} {\bibinfo {author} {\bibfnamefont {D.~J.}\ \bibnamefont
  {Ritson}}\ and\ \bibinfo {author} {\bibfnamefont {J.~D.}\ \bibnamefont
  {Sutherland}},\ }\bibfield  {title} {\bibinfo {title} {{Prebiotic Synthesis
  of Simple Sugars by Photoredox Systems Chemistry}},\ }\href
  {https://doi.org/10.1038/nchem.1467} {\bibfield  {journal} {\bibinfo
  {journal} {Nat. Chem.}\ }\textbf {\bibinfo {volume} {4}},\ \bibinfo {pages}
  {895} (\bibinfo {year} {2012}{\natexlab{a}})}\BibitemShut {NoStop}%
\bibitem [{\citenamefont {Patel}\ \emph {et~al.}(2015)\citenamefont {Patel},
  \citenamefont {Percivalle}, \citenamefont {Ritson}, \citenamefont {Duffy},\
  and\ \citenamefont {Sutherland}}]{Patel2015}%
  \BibitemOpen
  \bibfield  {author} {\bibinfo {author} {\bibfnamefont {B.~H.}\ \bibnamefont
  {Patel}}, \bibinfo {author} {\bibfnamefont {C.}~\bibnamefont {Percivalle}},
  \bibinfo {author} {\bibfnamefont {D.~J.}\ \bibnamefont {Ritson}}, \bibinfo
  {author} {\bibfnamefont {C.~D.}\ \bibnamefont {Duffy}},\ and\ \bibinfo
  {author} {\bibfnamefont {J.~D.}\ \bibnamefont {Sutherland}},\ }\bibfield
  {title} {\bibinfo {title} {{Common Origins of RNA, Protein and Lipid
  Precursors in a Cyanosulfidic Protometabolism}},\ }\href
  {https://doi.org/10.1038/nature14018} {\bibfield  {journal} {\bibinfo
  {journal} {Nature}\ }\textbf {\bibinfo {volume} {518}},\ \bibinfo {pages}
  {72} (\bibinfo {year} {2015})}\BibitemShut {NoStop}%
\bibitem [{\citenamefont {Ranjan}\ and\ \citenamefont
  {Sasselov}(2017)}]{RanjanSasselov2017}%
  \BibitemOpen
  \bibfield  {author} {\bibinfo {author} {\bibfnamefont {S.}~\bibnamefont
  {Ranjan}}\ and\ \bibinfo {author} {\bibfnamefont {D.~D.}\ \bibnamefont
  {Sasselov}},\ }\bibfield  {title} {\bibinfo {title} {{Constraints on the
  Early Terrestrial Surface UV Environment Relevant to Prebiotic Chemistry}},\
  }\href {https://doi.org/10.1089/ast.2016.1519} {\bibfield  {journal}
  {\bibinfo  {journal} {Astrobiology}\ }\textbf {\bibinfo {volume} {17}},\
  \bibinfo {pages} {169} (\bibinfo {year} {2017})}\BibitemShut {NoStop}%
\bibitem [{\citenamefont {Martin}\ and\ \citenamefont
  {Russell}(2007)}]{MartinRussell2007}%
  \BibitemOpen
  \bibfield  {author} {\bibinfo {author} {\bibfnamefont {W.}~\bibnamefont
  {Martin}}\ and\ \bibinfo {author} {\bibfnamefont {M.~J.}\ \bibnamefont
  {Russell}},\ }\bibfield  {title} {\bibinfo {title} {{On the Origin of
  Biochemistry at an Alkaline Hydrothermal Vent}},\ }\href
  {https://doi.org/10.1098/rstb.2006.1881} {\bibfield  {journal} {\bibinfo
  {journal} {Philos. Trans. R. Soc. B}\ }\textbf {\bibinfo {volume} {362}},\
  \bibinfo {pages} {1887} (\bibinfo {year} {2007})}\BibitemShut {NoStop}%
\bibitem [{\citenamefont {Lane}\ and\ \citenamefont
  {Martin}(2010)}]{LaneMartin2010}%
  \BibitemOpen
  \bibfield  {author} {\bibinfo {author} {\bibfnamefont {N.}~\bibnamefont
  {Lane}}\ and\ \bibinfo {author} {\bibfnamefont {W.}~\bibnamefont {Martin}},\
  }\bibfield  {title} {\bibinfo {title} {{The Evolution of the {C}ell: Powering
  and Partitioning}},\ }\href {https://doi.org/10.1038/nrmicro2426} {\bibfield
  {journal} {\bibinfo  {journal} {Nat. Rev. Microbiol.}\ }\textbf {\bibinfo
  {volume} {8}},\ \bibinfo {pages} {655} (\bibinfo {year} {2010})}\BibitemShut
  {NoStop}%
\bibitem [{\citenamefont {Watson}\ \emph {et~al.}(1981)\citenamefont {Watson},
  \citenamefont {Donahue},\ and\ \citenamefont {Walker}}]{Watson1981}%
  \BibitemOpen
  \bibfield  {author} {\bibinfo {author} {\bibfnamefont {A.~J.}\ \bibnamefont
  {Watson}}, \bibinfo {author} {\bibfnamefont {T.~M.}\ \bibnamefont
  {Donahue}},\ and\ \bibinfo {author} {\bibfnamefont {J.~C.~G.}\ \bibnamefont
  {Walker}},\ }\bibfield  {title} {\bibinfo {title} {{The Dynamics of a Rapidly
  Escaping Atmosphere: Applications to the Evolution of Earth and Venus}},\
  }\href {https://doi.org/10.1016/0019-1035(81)90101-9} {\bibfield  {journal}
  {\bibinfo  {journal} {Icarus}\ }\textbf {\bibinfo {volume} {48}},\ \bibinfo
  {pages} {150} (\bibinfo {year} {1981})}\BibitemShut {NoStop}%
\bibitem [{\citenamefont {Erkaev}\ \emph {et~al.}(2007)\citenamefont {Erkaev},
  \citenamefont {Lammer}, \citenamefont {Kulikov},\ and\ \citenamefont
  {et~al.}}]{Erkaev2007}%
  \BibitemOpen
  \bibfield  {author} {\bibinfo {author} {\bibfnamefont {N.~V.}\ \bibnamefont
  {Erkaev}}, \bibinfo {author} {\bibfnamefont {H.}~\bibnamefont {Lammer}},
  \bibinfo {author} {\bibfnamefont {Y.~N.}\ \bibnamefont {Kulikov}},\ and\
  \bibinfo {author} {\bibnamefont {et~al.}},\ }\bibfield  {title} {\bibinfo
  {title} {{Roche Lobe Effects on the Atmospheric Loss from ``Hot
  Jupiters''}},\ }\href {https://doi.org/10.1051/0004-6361:20066929} {\bibfield
   {journal} {\bibinfo  {journal} {Astron. Astrophys.}\ }\textbf {\bibinfo
  {volume} {472}},\ \bibinfo {pages} {329} (\bibinfo {year}
  {2007})}\BibitemShut {NoStop}%
\bibitem [{\citenamefont {Ribas}\ \emph {et~al.}(2005)\citenamefont {Ribas},
  \citenamefont {Guinan}, \citenamefont {G\"udel},\ and\ \citenamefont
  {Audard}}]{Ribas2005}%
  \BibitemOpen
  \bibfield  {author} {\bibinfo {author} {\bibfnamefont {I.}~\bibnamefont
  {Ribas}}, \bibinfo {author} {\bibfnamefont {E.}~\bibnamefont {Guinan}},
  \bibinfo {author} {\bibfnamefont {M.}~\bibnamefont {G\"udel}},\ and\ \bibinfo
  {author} {\bibfnamefont {M.}~\bibnamefont {Audard}},\ }\bibfield  {title}
  {\bibinfo {title} {{Evolution of the Solar Activity over Time and Effects on
  Planetary Atmospheres}},\ }\href {https://doi.org/10.1086/427977} {\bibfield
  {journal} {\bibinfo  {journal} {ApJ}\ }\textbf {\bibinfo {volume} {622}},\
  \bibinfo {pages} {680} (\bibinfo {year} {2005})}\BibitemShut {NoStop}%
\bibitem [{\citenamefont {Owen}\ and\ \citenamefont {Wu}(2017)}]{OwenWu2017}%
  \BibitemOpen
  \bibfield  {author} {\bibinfo {author} {\bibfnamefont {J.~E.}\ \bibnamefont
  {Owen}}\ and\ \bibinfo {author} {\bibfnamefont {Y.}~\bibnamefont {Wu}},\
  }\bibfield  {title} {\bibinfo {title} {{The Evaporation Valley in the Kepler
  Planets}},\ }\href {https://doi.org/10.3847/2041-8213/aa9028} {\bibfield
  {journal} {\bibinfo  {journal} {ApJ Lett.}\ }\textbf {\bibinfo {volume}
  {847}},\ \bibinfo {pages} {L3} (\bibinfo {year} {2017})}\BibitemShut
  {NoStop}%
\bibitem [{\citenamefont {Lopez}\ and\ \citenamefont
  {Fortney}(2014)}]{LopezFortney2014}%
  \BibitemOpen
  \bibfield  {author} {\bibinfo {author} {\bibfnamefont {E.~D.}\ \bibnamefont
  {Lopez}}\ and\ \bibinfo {author} {\bibfnamefont {J.~J.}\ \bibnamefont
  {Fortney}},\ }\bibfield  {title} {\bibinfo {title} {{Understanding the
  Mass–Radius Relation for Sub-Neptunes: Radius as a Proxy for
  Composition}},\ }\href {https://doi.org/10.1088/0004-637X/792/1/1} {\bibfield
   {journal} {\bibinfo  {journal} {ApJ}\ }\textbf {\bibinfo {volume} {792}},\
  \bibinfo {pages} {1} (\bibinfo {year} {2014})}\BibitemShut {NoStop}%
\bibitem [{\citenamefont {Fulton}\ \emph {et~al.}(2017)\citenamefont {Fulton},
  \citenamefont {Petigura}, \citenamefont {Howard},\ and\ \citenamefont
  {et~al.}}]{Fulton2017}%
  \BibitemOpen
  \bibfield  {author} {\bibinfo {author} {\bibfnamefont {B.~J.}\ \bibnamefont
  {Fulton}}, \bibinfo {author} {\bibfnamefont {E.~A.}\ \bibnamefont
  {Petigura}}, \bibinfo {author} {\bibfnamefont {A.~W.}\ \bibnamefont
  {Howard}},\ and\ \bibinfo {author} {\bibnamefont {et~al.}},\ }\bibfield
  {title} {\bibinfo {title} {{The California-Kepler Survey. IV. Metal-Rich
  Stars Host More Short-Period Planets}},\ }\href
  {https://doi.org/10.3847/1538-3881/aa80eb} {\bibfield  {journal} {\bibinfo
  {journal} {Astron. J.}\ }\textbf {\bibinfo {volume} {154}},\ \bibinfo {pages}
  {109} (\bibinfo {year} {2017})}\BibitemShut {NoStop}%
\bibitem [{\citenamefont {Vidal-Madjar}\ and\ \citenamefont
  {et~al.}(2003)}]{VidalMadjar2003}%
  \BibitemOpen
  \bibfield  {author} {\bibinfo {author} {\bibfnamefont {A.}~\bibnamefont
  {Vidal-Madjar}}\ and\ \bibinfo {author} {\bibnamefont {et~al.}},\ }\bibfield
  {title} {\bibinfo {title} {{An Extended Upper Atmosphere around the
  Extrasolar Planet HD 209458b}},\ }\href {https://doi.org/10.1038/nature01448}
  {\bibfield  {journal} {\bibinfo  {journal} {Nature}\ }\textbf {\bibinfo
  {volume} {422}},\ \bibinfo {pages} {143} (\bibinfo {year}
  {2003})}\BibitemShut {NoStop}%
\bibitem [{\citenamefont {Spake}\ and\ \citenamefont
  {et~al.}(2018)}]{Spake2018}%
  \BibitemOpen
  \bibfield  {author} {\bibinfo {author} {\bibfnamefont {J.~J.}\ \bibnamefont
  {Spake}}\ and\ \bibinfo {author} {\bibnamefont {et~al.}},\ }\bibfield
  {title} {\bibinfo {title} {{Helium in the Eroding Atmosphere of an
  Exoplanet}},\ }\href {https://doi.org/10.1038/s41586-018-0067-5} {\bibfield
  {journal} {\bibinfo  {journal} {Nature}\ }\textbf {\bibinfo {volume} {557}},\
  \bibinfo {pages} {68} (\bibinfo {year} {2018})}\BibitemShut {NoStop}%
\bibitem [{\citenamefont {Oklop{\v c}i{\'c}}\ and\ \citenamefont
  {Hirata}(2018)}]{OklopcicHirata2018}%
  \BibitemOpen
  \bibfield  {author} {\bibinfo {author} {\bibfnamefont {A.}~\bibnamefont
  {Oklop{\v c}i{\'c}}}\ and\ \bibinfo {author} {\bibfnamefont {C.~M.}\
  \bibnamefont {Hirata}},\ }\bibfield  {title} {\bibinfo {title} {{A New Window
  into Escaping Exoplanet Atmospheres: 10830 \AA\ Line of Helium}},\ }\href
  {https://doi.org/10.3847/2041-8213/aaad6e} {\bibfield  {journal} {\bibinfo
  {journal} {ApJ Lett.}\ }\textbf {\bibinfo {volume} {855}},\ \bibinfo {pages}
  {L11} (\bibinfo {year} {2018})}\BibitemShut {NoStop}%
\bibitem [{\citenamefont {Lineweaver}\ \emph {et~al.}(2004)\citenamefont
  {Lineweaver}, \citenamefont {Fenner},\ and\ \citenamefont
  {Gibson}}]{Lineweaver2004}%
  \BibitemOpen
  \bibfield  {author} {\bibinfo {author} {\bibfnamefont {C.~H.}\ \bibnamefont
  {Lineweaver}}, \bibinfo {author} {\bibfnamefont {Y.}~\bibnamefont {Fenner}},\
  and\ \bibinfo {author} {\bibfnamefont {B.~K.}\ \bibnamefont {Gibson}},\
  }\bibfield  {title} {\bibinfo {title} {{The Galactic Habitable Zone and the
  Age Distribution of Complex Life in the Milky Way}},\ }\href
  {https://doi.org/10.1126/science.1092322} {\bibfield  {journal} {\bibinfo
  {journal} {Science}\ }\textbf {\bibinfo {volume} {303}},\ \bibinfo {pages}
  {59} (\bibinfo {year} {2004})}\BibitemShut {NoStop}%
\bibitem [{\citenamefont {Gonzalez}\ \emph {et~al.}(2001)\citenamefont
  {Gonzalez}, \citenamefont {Brownlee},\ and\ \citenamefont
  {Ward}}]{Gonzalez2001}%
  \BibitemOpen
  \bibfield  {author} {\bibinfo {author} {\bibfnamefont {G.}~\bibnamefont
  {Gonzalez}}, \bibinfo {author} {\bibfnamefont {D.}~\bibnamefont {Brownlee}},\
  and\ \bibinfo {author} {\bibfnamefont {P.}~\bibnamefont {Ward}},\ }\bibfield
  {title} {\bibinfo {title} {{The Galactic Habitable Zone: Galactic Chemical
  Evolution}},\ }\href {https://doi.org/10.1006/icar.2001.6617} {\bibfield
  {journal} {\bibinfo  {journal} {Icarus}\ }\textbf {\bibinfo {volume} {152}},\
  \bibinfo {pages} {185} (\bibinfo {year} {2001})}\BibitemShut {NoStop}%
\bibitem [{\citenamefont {Piran}\ and\ \citenamefont
  {Jimenez}(2014)}]{PiranJimenez2014}%
  \BibitemOpen
  \bibfield  {author} {\bibinfo {author} {\bibfnamefont {T.}~\bibnamefont
  {Piran}}\ and\ \bibinfo {author} {\bibfnamefont {R.}~\bibnamefont
  {Jimenez}},\ }\bibfield  {title} {\bibinfo {title} {{Possible Role of Gamma
  Ray Bursts on Life Extinction in the Universe}},\ }\href
  {https://doi.org/10.1103/PhysRevLett.113.231101} {\bibfield  {journal}
  {\bibinfo  {journal} {PRL}\ }\textbf {\bibinfo {volume} {113}},\ \bibinfo
  {pages} {231101} (\bibinfo {year} {2014})}\BibitemShut {NoStop}%
\bibitem [{\citenamefont {Hitchcock}\ and\ \citenamefont
  {Lovelock}(1967)}]{HitchcockLovelock1967}%
  \BibitemOpen
  \bibfield  {author} {\bibinfo {author} {\bibfnamefont {D.~R.}\ \bibnamefont
  {Hitchcock}}\ and\ \bibinfo {author} {\bibfnamefont {J.~E.}\ \bibnamefont
  {Lovelock}},\ }\bibfield  {title} {\bibinfo {title} {{Life Detection by
  Atmospheric Analysis}},\ }\href
  {https://doi.org/10.1016/0019-1035(67)90059-0} {\bibfield  {journal}
  {\bibinfo  {journal} {Icarus}\ }\textbf {\bibinfo {volume} {7}},\ \bibinfo
  {pages} {149} (\bibinfo {year} {1967})}\BibitemShut {NoStop}%
\bibitem [{\citenamefont {Sagan}\ \emph {et~al.}(1993)\citenamefont {Sagan},
  \citenamefont {Thompson}, \citenamefont {Carlson}, \citenamefont {Gurnett},\
  and\ \citenamefont {Hord}}]{Sagan1993}%
  \BibitemOpen
  \bibfield  {author} {\bibinfo {author} {\bibfnamefont {C.}~\bibnamefont
  {Sagan}}, \bibinfo {author} {\bibfnamefont {W.~R.}\ \bibnamefont {Thompson}},
  \bibinfo {author} {\bibfnamefont {R.}~\bibnamefont {Carlson}}, \bibinfo
  {author} {\bibfnamefont {D.}~\bibnamefont {Gurnett}},\ and\ \bibinfo {author}
  {\bibfnamefont {C.}~\bibnamefont {Hord}},\ }\bibfield  {title} {\bibinfo
  {title} {{A Search for Life on Earth from the Galileo Spacecraft}},\ }\href
  {https://doi.org/10.1038/365715a0} {\bibfield  {journal} {\bibinfo  {journal}
  {Nature}\ }\textbf {\bibinfo {volume} {365}},\ \bibinfo {pages} {715}
  (\bibinfo {year} {1993})}\BibitemShut {NoStop}%
\bibitem [{\citenamefont {Krissansen-Totton}\ \emph {et~al.}(2018)\citenamefont
  {Krissansen-Totton}, \citenamefont {Catling},\ and\ \citenamefont
  {Charnay}}]{KrissansenTotton2018}%
  \BibitemOpen
  \bibfield  {author} {\bibinfo {author} {\bibfnamefont {J.}~\bibnamefont
  {Krissansen-Totton}}, \bibinfo {author} {\bibfnamefont {D.~C.}\ \bibnamefont
  {Catling}},\ and\ \bibinfo {author} {\bibfnamefont {B.}~\bibnamefont
  {Charnay}},\ }\bibfield  {title} {\bibinfo {title} {{Detecting Biosignatures
  in {E}arth-like Atmospheres with Chemical Disequilibrium}},\ }\href
  {https://doi.org/10.1126/sciadv.aao5747} {\bibfield  {journal} {\bibinfo
  {journal} {Sci. Adv.}\ }\textbf {\bibinfo {volume} {4}},\ \bibinfo {pages}
  {eaao5747} (\bibinfo {year} {2018})}\BibitemShut {NoStop}%
\bibitem [{\citenamefont {Meadows}(2017)}]{Meadows2017}%
  \BibitemOpen
  \bibfield  {author} {\bibinfo {author} {\bibfnamefont {V.~S.}\ \bibnamefont
  {Meadows}},\ }\bibfield  {title} {\bibinfo {title} {{Reflections on O$_2$ as
  a Biosignature in Exoplanetary Atmospheres}},\ }\href
  {https://doi.org/10.1089/ast.2016.1578} {\bibfield  {journal} {\bibinfo
  {journal} {Astrobiology}\ }\textbf {\bibinfo {volume} {17}},\ \bibinfo
  {pages} {1022} (\bibinfo {year} {2017})}\BibitemShut {NoStop}%
\bibitem [{\citenamefont {Seager}\ \emph {et~al.}(2005)\citenamefont {Seager},
  \citenamefont {Ford}, \citenamefont {Rodler},\ and\ \citenamefont
  {et~al.}}]{Seager2005}%
  \BibitemOpen
  \bibfield  {author} {\bibinfo {author} {\bibfnamefont {S.}~\bibnamefont
  {Seager}}, \bibinfo {author} {\bibfnamefont {E.~B.}\ \bibnamefont {Ford}},
  \bibinfo {author} {\bibfnamefont {F.~J.}\ \bibnamefont {Rodler}},\ and\
  \bibinfo {author} {\bibnamefont {et~al.}},\ }\bibfield  {title} {\bibinfo
  {title} {{Vegetation's Red Edge: A Possible Spectroscopic Biosignature of
  Extraterrestrial Plants}},\ }\href {https://doi.org/10.1089/ast.2005.5.372}
  {\bibfield  {journal} {\bibinfo  {journal} {Astrobiology}\ }\textbf {\bibinfo
  {volume} {5}},\ \bibinfo {pages} {372} (\bibinfo {year} {2005})}\BibitemShut
  {NoStop}%
\bibitem [{\citenamefont {Kiang}\ \emph {et~al.}(2007)\citenamefont {Kiang},
  \citenamefont {Siefert}, \citenamefont {Govindjee},\ and\ \citenamefont
  {Blankenship}}]{Kiang2007}%
  \BibitemOpen
  \bibfield  {author} {\bibinfo {author} {\bibfnamefont {N.~Y.}\ \bibnamefont
  {Kiang}}, \bibinfo {author} {\bibfnamefont {J.}~\bibnamefont {Siefert}},
  \bibinfo {author} {\bibnamefont {Govindjee}},\ and\ \bibinfo {author}
  {\bibfnamefont {R.~E.}\ \bibnamefont {Blankenship}},\ }\bibfield  {title}
  {\bibinfo {title} {{Spectral Signatures of Photosynthesis. I. Review of Earth
  Organisms}},\ }\href {https://doi.org/10.1089/ast.2006.0105} {\bibfield
  {journal} {\bibinfo  {journal} {Astrobiology}\ }\textbf {\bibinfo {volume}
  {7}},\ \bibinfo {pages} {222} (\bibinfo {year} {2007})}\BibitemShut {NoStop}%
\bibitem [{\citenamefont {Walker}\ \emph {et~al.}(1981)\citenamefont {Walker},
  \citenamefont {Hays},\ and\ \citenamefont {Kasting}}]{Walker1981}%
  \BibitemOpen
  \bibfield  {author} {\bibinfo {author} {\bibfnamefont {J.~C.~G.}\
  \bibnamefont {Walker}}, \bibinfo {author} {\bibfnamefont {P.~B.}\
  \bibnamefont {Hays}},\ and\ \bibinfo {author} {\bibfnamefont {J.~F.}\
  \bibnamefont {Kasting}},\ }\bibfield  {title} {\bibinfo {title} {{A Negative
  Feedback Mechanism for the Long-term Stabilization of Earth's Surface
  Temperature}},\ }\href {https://doi.org/10.1029/JC086iC10p09776} {\bibfield
  {journal} {\bibinfo  {journal} {J. Geophys. Res.}\ }\textbf {\bibinfo
  {volume} {86}},\ \bibinfo {pages} {9776} (\bibinfo {year}
  {1981})}\BibitemShut {NoStop}%
\bibitem [{\citenamefont {Watson}\ and\ \citenamefont
  {Lovelock}(1983)}]{WatsonLovelock1983}%
  \BibitemOpen
  \bibfield  {author} {\bibinfo {author} {\bibfnamefont {A.~J.}\ \bibnamefont
  {Watson}}\ and\ \bibinfo {author} {\bibfnamefont {J.~E.}\ \bibnamefont
  {Lovelock}},\ }\bibfield  {title} {\bibinfo {title} {{Biological Homeostasis
  of the Global Environment: The Parable of Daisyworld}},\ }\href
  {https://doi.org/10.3402/tellusb.v35i4.14616} {\bibfield  {journal} {\bibinfo
   {journal} {Tellus B}\ }\textbf {\bibinfo {volume} {35}},\ \bibinfo {pages}
  {284} (\bibinfo {year} {1983})}\BibitemShut {NoStop}%
\bibitem [{\citenamefont {Catling}\ and\ \citenamefont
  {Kasting}(2017)}]{CatlingKasting2017}%
  \BibitemOpen
  \bibfield  {author} {\bibinfo {author} {\bibfnamefont {D.~C.}\ \bibnamefont
  {Catling}}\ and\ \bibinfo {author} {\bibfnamefont {J.~F.}\ \bibnamefont
  {Kasting}},\ }\href@noop {} {\emph {\bibinfo {title} {{Atmospheric Evolution
  on Inhabited and Lifeless Worlds}}}}\ (\bibinfo  {publisher} {Princeton
  University Press},\ \bibinfo {address} {Princeton, NJ},\ \bibinfo {year}
  {2017})\BibitemShut {NoStop}%
\bibitem [{\citenamefont {Sharma}\ \emph {et~al.}(2023)\citenamefont {Sharma},
  \citenamefont {Cz{\'e}gel}, \citenamefont {Lachmann}, \citenamefont {Walker},
  \citenamefont {Cronin} \emph {et~al.}}]{Sharma2023_AT_Nature}%
  \BibitemOpen
  \bibfield  {author} {\bibinfo {author} {\bibfnamefont {A.}~\bibnamefont
  {Sharma}}, \bibinfo {author} {\bibfnamefont {D.}~\bibnamefont {Cz{\'e}gel}},
  \bibinfo {author} {\bibfnamefont {M.}~\bibnamefont {Lachmann}}, \bibinfo
  {author} {\bibfnamefont {S.~I.}\ \bibnamefont {Walker}}, \bibinfo {author}
  {\bibfnamefont {L.}~\bibnamefont {Cronin}}, \emph {et~al.},\ }\bibfield
  {title} {\bibinfo {title} {{Assembly theory explains and quantifies selection
  and evolution}},\ }\href {https://doi.org/10.1038/s41586-023-06600-9}
  {\bibfield  {journal} {\bibinfo  {journal} {Nature}\ }\textbf {\bibinfo
  {volume} {622}},\ \bibinfo {pages} {321} (\bibinfo {year}
  {2023})}\BibitemShut {NoStop}%
\bibitem [{\citenamefont {Marshall}\ \emph {et~al.}(2021)\citenamefont
  {Marshall}, \citenamefont {Mathis}, \citenamefont {Carrick}, \citenamefont
  {Keenan}, \citenamefont {Cooper}, \citenamefont {Graham}, \citenamefont
  {Craven}, \citenamefont {Gromski}, \citenamefont {Moore}, \citenamefont
  {Walker},\ and\ \citenamefont {Cronin}}]{Marshall2021_AT_NatComm}%
  \BibitemOpen
  \bibfield  {author} {\bibinfo {author} {\bibfnamefont {S.~M.}\ \bibnamefont
  {Marshall}}, \bibinfo {author} {\bibfnamefont {C.}~\bibnamefont {Mathis}},
  \bibinfo {author} {\bibfnamefont {E.}~\bibnamefont {Carrick}}, \bibinfo
  {author} {\bibfnamefont {G.}~\bibnamefont {Keenan}}, \bibinfo {author}
  {\bibfnamefont {G.~J.~T.}\ \bibnamefont {Cooper}}, \bibinfo {author}
  {\bibfnamefont {H.}~\bibnamefont {Graham}}, \bibinfo {author} {\bibfnamefont
  {M.}~\bibnamefont {Craven}}, \bibinfo {author} {\bibfnamefont {P.~S.}\
  \bibnamefont {Gromski}}, \bibinfo {author} {\bibfnamefont {D.~G.}\
  \bibnamefont {Moore}}, \bibinfo {author} {\bibfnamefont {S.~I.}\ \bibnamefont
  {Walker}},\ and\ \bibinfo {author} {\bibfnamefont {L.}~\bibnamefont
  {Cronin}},\ }\bibfield  {title} {\bibinfo {title} {{Identifying molecules as
  biosignatures with assembly theory and mass spectrometry}},\ }\href
  {https://doi.org/10.1038/s41467-021-23258-x} {\bibfield  {journal} {\bibinfo
  {journal} {Nature Communications}\ }\textbf {\bibinfo {volume} {12}},\
  \bibinfo {pages} {3033} (\bibinfo {year} {2021})}\BibitemShut {NoStop}%
\bibitem [{\citenamefont {Farina}\ and\ \citenamefont
  {Rinaldi}(2000)}]{FarinaRinaldi2000}%
  \BibitemOpen
  \bibfield  {author} {\bibinfo {author} {\bibfnamefont {L.}~\bibnamefont
  {Farina}}\ and\ \bibinfo {author} {\bibfnamefont {S.}~\bibnamefont
  {Rinaldi}},\ }\href {https://doi.org/10.1002/9781118033029} {\emph {\bibinfo
  {title} {{Positive Linear Systems: Theory and Applications}}}}\ (\bibinfo
  {publisher} {Wiley-Interscience},\ \bibinfo {year} {2000})\BibitemShut
  {NoStop}%
\bibitem [{\citenamefont {Berman}\ and\ \citenamefont
  {Plemmons}(1994)}]{BermanPlemmons1994}%
  \BibitemOpen
  \bibfield  {author} {\bibinfo {author} {\bibfnamefont {A.}~\bibnamefont
  {Berman}}\ and\ \bibinfo {author} {\bibfnamefont {R.~J.}\ \bibnamefont
  {Plemmons}},\ }\href {https://doi.org/10.1137/1.9781611971262} {\emph
  {\bibinfo {title} {{Nonnegative Matrices in the Mathematical Sciences}}}},\
  \bibinfo {series} {Classics in Applied Mathematics}, Vol.~\bibinfo {volume}
  {9}\ (\bibinfo  {publisher} {SIAM},\ \bibinfo {year} {1994})\BibitemShut
  {NoStop}%
\bibitem [{\citenamefont
  {{Turyshev}}(2026{\natexlab{b}})}]{Turyshev:2025-exoPL}%
  \BibitemOpen
  \bibfield  {author} {\bibinfo {author} {\bibfnamefont {S.~G.}\ \bibnamefont
  {{Turyshev}}},\ }\bibfield  {title} {\bibinfo {title} {{{Direct
  high-resolution imaging of Earth-like exoplanets}}},\ }\href
  {https://doi.org/10.1103/f2wz-qwqm} {\bibfield  {journal} {\bibinfo
  {journal} {Phys. Rev. D}\ }\textbf {\bibinfo {volume} {113}},\ \bibinfo {eid}
  {023034} (\bibinfo {year} {2026}{\natexlab{b}})}\BibitemShut {NoStop}%
\bibitem [{\citenamefont {Oberhummer}\ \emph {et~al.}(2000)\citenamefont
  {Oberhummer}, \citenamefont {Cs{\'o}t{\'o}},\ and\ \citenamefont
  {Schlattl}}]{Oberhummer2000}%
  \BibitemOpen
  \bibfield  {author} {\bibinfo {author} {\bibfnamefont {H.}~\bibnamefont
  {Oberhummer}}, \bibinfo {author} {\bibfnamefont {A.}~\bibnamefont
  {Cs{\'o}t{\'o}}},\ and\ \bibinfo {author} {\bibfnamefont {H.}~\bibnamefont
  {Schlattl}},\ }\bibfield  {title} {\bibinfo {title} {{Stellar Production
  Rates of Carbon and Its Abundance in the Universe}},\ }\href
  {https://doi.org/10.1126/science.289.5476.88} {\bibfield  {journal} {\bibinfo
   {journal} {Science}\ }\textbf {\bibinfo {volume} {289}},\ \bibinfo {pages}
  {88} (\bibinfo {year} {2000})}\BibitemShut {NoStop}%
\bibitem [{\citenamefont {Epelbaum}\ \emph {et~al.}(2013)\citenamefont
  {Epelbaum}, \citenamefont {Krebs}, \citenamefont {L{\"a}hde}, \citenamefont
  {Lee},\ and\ \citenamefont {Meißner}}]{Epelbaum2013}%
  \BibitemOpen
  \bibfield  {author} {\bibinfo {author} {\bibfnamefont {E.}~\bibnamefont
  {Epelbaum}}, \bibinfo {author} {\bibfnamefont {H.}~\bibnamefont {Krebs}},
  \bibinfo {author} {\bibfnamefont {T.~A.}\ \bibnamefont {L{\"a}hde}}, \bibinfo
  {author} {\bibfnamefont {D.}~\bibnamefont {Lee}},\ and\ \bibinfo {author}
  {\bibfnamefont {U.-G.}\ \bibnamefont {Meißner}},\ }\bibfield  {title}
  {\bibinfo {title} {{Dependence of the triple-alpha process on the fundamental
  constants of nature}},\ }\href {https://doi.org/10.1140/epja/i2013-13082-y}
  {\bibfield  {journal} {\bibinfo  {journal} {Eur. Phys. J. A}\ }\textbf
  {\bibinfo {volume} {49}},\ \bibinfo {pages} {82} (\bibinfo {year}
  {2013})}\BibitemShut {NoStop}%
\bibitem [{\citenamefont {Dmitriev}\ \emph {et~al.}(2004)\citenamefont
  {Dmitriev}, \citenamefont {Flambaum},\ and\ \citenamefont
  {Webb}}]{Dmitriev2004}%
  \BibitemOpen
  \bibfield  {author} {\bibinfo {author} {\bibfnamefont {V.~F.}\ \bibnamefont
  {Dmitriev}}, \bibinfo {author} {\bibfnamefont {V.~V.}\ \bibnamefont
  {Flambaum}},\ and\ \bibinfo {author} {\bibfnamefont {J.~K.}\ \bibnamefont
  {Webb}},\ }\bibfield  {title} {\bibinfo {title} {{Cosmological variation of
  deuteron binding energy, strong interaction, and quark masses from big bang
  nucleosynthesis}},\ }\href {https://doi.org/10.1103/PhysRevD.69.063506}
  {\bibfield  {journal} {\bibinfo  {journal} {Phys. Rev. D}\ }\textbf {\bibinfo
  {volume} {69}},\ \bibinfo {pages} {063506} (\bibinfo {year}
  {2004})}\BibitemShut {NoStop}%
\bibitem [{\citenamefont {Livio}\ and\ \citenamefont
  {Rees}(2005)}]{LivioRees2005_Science}%
  \BibitemOpen
  \bibfield  {author} {\bibinfo {author} {\bibfnamefont {M.}~\bibnamefont
  {Livio}}\ and\ \bibinfo {author} {\bibfnamefont {M.~J.}\ \bibnamefont
  {Rees}},\ }\bibfield  {title} {\bibinfo {title} {{Anthropic reasoning}},\
  }\href {https://doi.org/10.1126/science.1111446} {\bibfield  {journal}
  {\bibinfo  {journal} {Science}\ }\textbf {\bibinfo {volume} {309}},\ \bibinfo
  {pages} {1022} (\bibinfo {year} {2005})}\BibitemShut {NoStop}%
\bibitem [{\citenamefont {Livio}\ and\ \citenamefont
  {Rees}(2020)}]{LivioRees2018_CUP}%
  \BibitemOpen
  \bibfield  {author} {\bibinfo {author} {\bibfnamefont {M.}~\bibnamefont
  {Livio}}\ and\ \bibinfo {author} {\bibfnamefont {M.~J.}\ \bibnamefont
  {Rees}},\ }\bibfield  {title} {\bibinfo {title} {{Fine-Tuning, Complexity,
  and Life in the Multiverse}},\ }in\ \href {https://arxiv.org/abs/1801.06944}
  {\emph {\bibinfo {booktitle} {Fine-Tuning in the Physical Universe}}},\
  \bibinfo {editor} {edited by\ \bibinfo {editor} {\bibfnamefont
  {D.}~\bibnamefont {Sloan}}, \bibinfo {editor} {\bibfnamefont
  {R.}~\bibnamefont {Alves~Batista}}, \bibinfo {editor} {\bibfnamefont {M.~T.}\
  \bibnamefont {Hicks}},\ and\ \bibinfo {editor} {\bibfnamefont
  {R.}~\bibnamefont {Davies}}}\ (\bibinfo  {publisher} {Cambridge University
  Press},\ \bibinfo {address} {Cambridge},\ \bibinfo {year} {2020})\ pp.\
  \bibinfo {pages} {3--61},\ \bibinfo {note} {see also preprint:
  arXiv:1801.06944}\BibitemShut {NoStop}%
\bibitem [{\citenamefont {Beer}\ \emph {et~al.}(2004)\citenamefont {Beer},
  \citenamefont {King}, \citenamefont {Livio},\ and\ \citenamefont
  {Pringle}}]{Beer2004_SolarSystem_MNRAS}%
  \BibitemOpen
  \bibfield  {author} {\bibinfo {author} {\bibfnamefont {M.~E.}\ \bibnamefont
  {Beer}}, \bibinfo {author} {\bibfnamefont {A.~R.}\ \bibnamefont {King}},
  \bibinfo {author} {\bibfnamefont {M.}~\bibnamefont {Livio}},\ and\ \bibinfo
  {author} {\bibfnamefont {J.~E.}\ \bibnamefont {Pringle}},\ }\bibfield
  {title} {\bibinfo {title} {{How special is the Solar system?}},\ }\href
  {https://doi.org/10.1111/j.1365-2966.2004.08237.x} {\bibfield  {journal}
  {\bibinfo  {journal} {MNRAS}\ }\textbf {\bibinfo {volume} {354}},\ \bibinfo
  {pages} {763} (\bibinfo {year} {2004})}\BibitemShut {NoStop}%
\bibitem [{\citenamefont {Parrondo}\ \emph {et~al.}(2015)\citenamefont
  {Parrondo}, \citenamefont {Horowitz},\ and\ \citenamefont
  {Sagawa}}]{Parrondo2015}%
  \BibitemOpen
  \bibfield  {author} {\bibinfo {author} {\bibfnamefont {J.~M.~R.}\
  \bibnamefont {Parrondo}}, \bibinfo {author} {\bibfnamefont {J.~M.}\
  \bibnamefont {Horowitz}},\ and\ \bibinfo {author} {\bibfnamefont
  {T.}~\bibnamefont {Sagawa}},\ }\bibfield  {title} {\bibinfo {title}
  {{Thermodynamics of information}},\ }\href
  {https://doi.org/10.1038/nphys3230} {\bibfield  {journal} {\bibinfo
  {journal} {Nature Physics}\ }\textbf {\bibinfo {volume} {11}},\ \bibinfo
  {pages} {131} (\bibinfo {year} {2015})}\BibitemShut {NoStop}%
\bibitem [{\citenamefont {Still}\ \emph {et~al.}(2012)\citenamefont {Still},
  \citenamefont {Sivak}, \citenamefont {Bell},\ and\ \citenamefont
  {Crooks}}]{Still2012}%
  \BibitemOpen
  \bibfield  {author} {\bibinfo {author} {\bibfnamefont {S.}~\bibnamefont
  {Still}}, \bibinfo {author} {\bibfnamefont {D.~A.}\ \bibnamefont {Sivak}},
  \bibinfo {author} {\bibfnamefont {A.~J.}\ \bibnamefont {Bell}},\ and\
  \bibinfo {author} {\bibfnamefont {G.~E.}\ \bibnamefont {Crooks}},\ }\bibfield
   {title} {\bibinfo {title} {{Thermodynamics of Prediction}},\ }\href
  {https://doi.org/10.1103/PhysRevLett.109.120604} {\bibfield  {journal}
  {\bibinfo  {journal} {PRL}\ }\textbf {\bibinfo {volume} {109}},\ \bibinfo
  {pages} {120604} (\bibinfo {year} {2012})}\BibitemShut {NoStop}%
\bibitem [{\citenamefont {England}(2013)}]{England2013}%
  \BibitemOpen
  \bibfield  {author} {\bibinfo {author} {\bibfnamefont {J.~L.}\ \bibnamefont
  {England}},\ }\bibfield  {title} {\bibinfo {title} {{Statistical physics of
  self-replication}},\ }\href {https://doi.org/10.1063/1.4818538} {\bibfield
  {journal} {\bibinfo  {journal} {J. Chem. Phys.}\ }\textbf {\bibinfo {volume}
  {139}},\ \bibinfo {pages} {121923} (\bibinfo {year} {2013})}\BibitemShut
  {NoStop}%
\bibitem [{\citenamefont {Ritson}\ and\ \citenamefont
  {Sutherland}(2012{\natexlab{b}})}]{RitsonSutherland2012}%
  \BibitemOpen
  \bibfield  {author} {\bibinfo {author} {\bibfnamefont {D.}~\bibnamefont
  {Ritson}}\ and\ \bibinfo {author} {\bibfnamefont {J.~D.}\ \bibnamefont
  {Sutherland}},\ }\bibfield  {title} {\bibinfo {title} {{Prebiotic Synthesis
  of Simple Sugars by Photoredox Systems Chemistry}},\ }\href
  {https://doi.org/10.1038/nchem.1467} {\bibfield  {journal} {\bibinfo
  {journal} {Nature Chemistry}\ }\textbf {\bibinfo {volume} {4}},\ \bibinfo
  {pages} {895} (\bibinfo {year} {2012}{\natexlab{b}})}\BibitemShut {NoStop}%
\bibitem [{\citenamefont {Sutherland}(2016)}]{Sutherland2016}%
  \BibitemOpen
  \bibfield  {author} {\bibinfo {author} {\bibfnamefont {J.~D.}\ \bibnamefont
  {Sutherland}},\ }\bibfield  {title} {\bibinfo {title} {{The Origin of
  Life---Out of the Blue}},\ }\href {https://doi.org/10.1002/anie.201506585}
  {\bibfield  {journal} {\bibinfo  {journal} {Angew. Chem. Int. Ed.}\ }\textbf
  {\bibinfo {volume} {55}},\ \bibinfo {pages} {104} (\bibinfo {year}
  {2016})}\BibitemShut {NoStop}%
\bibitem [{\citenamefont {Gehrels}\ \emph {et~al.}(2003)\citenamefont
  {Gehrels}, \citenamefont {Laird}, \citenamefont {Jackman}, \citenamefont
  {Cannizzo}, \citenamefont {Mattson},\ and\ \citenamefont
  {Chen}}]{Gehrels2003}%
  \BibitemOpen
  \bibfield  {author} {\bibinfo {author} {\bibfnamefont {N.}~\bibnamefont
  {Gehrels}}, \bibinfo {author} {\bibfnamefont {C.~M.}\ \bibnamefont {Laird}},
  \bibinfo {author} {\bibfnamefont {C.~H.}\ \bibnamefont {Jackman}}, \bibinfo
  {author} {\bibfnamefont {J.~K.}\ \bibnamefont {Cannizzo}}, \bibinfo {author}
  {\bibfnamefont {B.~J.}\ \bibnamefont {Mattson}},\ and\ \bibinfo {author}
  {\bibfnamefont {W.}~\bibnamefont {Chen}},\ }\bibfield  {title} {\bibinfo
  {title} {{Ozone Depletion from Nearby Supernovae}},\ }\href
  {https://doi.org/10.1086/346127} {\bibfield  {journal} {\bibinfo  {journal}
  {ApJ}\ }\textbf {\bibinfo {volume} {585}},\ \bibinfo {pages} {1169} (\bibinfo
  {year} {2003})}\BibitemShut {NoStop}%
\bibitem [{\citenamefont {Thomas}\ \emph {et~al.}(2005)\citenamefont {Thomas},
  \citenamefont {Melott}, \citenamefont {Jackman}, \citenamefont {Laird},
  \citenamefont {Medvedev}, \citenamefont {Stolarski}, \citenamefont {Gehrels},
  \citenamefont {Cannizzo}, \citenamefont {Hogan},\ and\ \citenamefont
  {Ejzak}}]{Thomas2005}%
  \BibitemOpen
  \bibfield  {author} {\bibinfo {author} {\bibfnamefont {B.~C.}\ \bibnamefont
  {Thomas}}, \bibinfo {author} {\bibfnamefont {A.~L.}\ \bibnamefont {Melott}},
  \bibinfo {author} {\bibfnamefont {C.~H.}\ \bibnamefont {Jackman}}, \bibinfo
  {author} {\bibfnamefont {C.~M.}\ \bibnamefont {Laird}}, \bibinfo {author}
  {\bibfnamefont {M.~V.}\ \bibnamefont {Medvedev}}, \bibinfo {author}
  {\bibfnamefont {R.~S.}\ \bibnamefont {Stolarski}}, \bibinfo {author}
  {\bibfnamefont {N.}~\bibnamefont {Gehrels}}, \bibinfo {author} {\bibfnamefont
  {J.~K.}\ \bibnamefont {Cannizzo}}, \bibinfo {author} {\bibfnamefont {D.~P.}\
  \bibnamefont {Hogan}},\ and\ \bibinfo {author} {\bibfnamefont {L.~M.}\
  \bibnamefont {Ejzak}},\ }\bibfield  {title} {\bibinfo {title} {{Gamma-Ray
  Bursts and the Earth: Exploration of Atmospheric, Biological, Climatic, and
  Biogeochemical Effects}},\ }\href {https://doi.org/10.1086/496914} {\bibfield
   {journal} {\bibinfo  {journal} {ApJ}\ }\textbf {\bibinfo {volume} {634}},\
  \bibinfo {pages} {509} (\bibinfo {year} {2005})}\BibitemShut {NoStop}%
\bibitem [{\citenamefont {Youdin}\ and\ \citenamefont
  {Goodman}(2005)}]{YoudinGoodman2005}%
  \BibitemOpen
  \bibfield  {author} {\bibinfo {author} {\bibfnamefont {A.~N.}\ \bibnamefont
  {Youdin}}\ and\ \bibinfo {author} {\bibfnamefont {J.}~\bibnamefont
  {Goodman}},\ }\bibfield  {title} {\bibinfo {title} {{Streaming Instabilities
  in Protoplanetary Disks}},\ }\href {https://doi.org/10.1086/426895}
  {\bibfield  {journal} {\bibinfo  {journal} {ApJ}\ }\textbf {\bibinfo {volume}
  {620}},\ \bibinfo {pages} {459} (\bibinfo {year} {2005})}\BibitemShut
  {NoStop}%
\bibitem [{\citenamefont {Johansen}\ \emph {et~al.}(2007)\citenamefont
  {Johansen}, \citenamefont {Oishi}, \citenamefont {Mac~Low}, \citenamefont
  {Klahr}, \citenamefont {Henning},\ and\ \citenamefont
  {Youdin}}]{Johansen2007}%
  \BibitemOpen
  \bibfield  {author} {\bibinfo {author} {\bibfnamefont {A.}~\bibnamefont
  {Johansen}}, \bibinfo {author} {\bibfnamefont {J.~S.}\ \bibnamefont {Oishi}},
  \bibinfo {author} {\bibfnamefont {M.-M.}\ \bibnamefont {Mac~Low}}, \bibinfo
  {author} {\bibfnamefont {H.}~\bibnamefont {Klahr}}, \bibinfo {author}
  {\bibfnamefont {T.}~\bibnamefont {Henning}},\ and\ \bibinfo {author}
  {\bibfnamefont {A.}~\bibnamefont {Youdin}},\ }\bibfield  {title} {\bibinfo
  {title} {{Rapid Planetesimal Formation in Turbulent Circumstellar Discs}},\
  }\href {https://doi.org/10.1038/nature06086} {\bibfield  {journal} {\bibinfo
  {journal} {Nature}\ }\textbf {\bibinfo {volume} {448}},\ \bibinfo {pages}
  {1022} (\bibinfo {year} {2007})}\BibitemShut {NoStop}%
\bibitem [{\citenamefont {Carrera}\ \emph {et~al.}(2015)\citenamefont
  {Carrera}, \citenamefont {Johansen},\ and\ \citenamefont
  {Davies}}]{Carrera2015}%
  \BibitemOpen
  \bibfield  {author} {\bibinfo {author} {\bibfnamefont {D.}~\bibnamefont
  {Carrera}}, \bibinfo {author} {\bibfnamefont {A.}~\bibnamefont {Johansen}},\
  and\ \bibinfo {author} {\bibfnamefont {M.~B.}\ \bibnamefont {Davies}},\
  }\bibfield  {title} {\bibinfo {title} {{How to Form Planetesimals from
  mm-sized Chondrules and Chondrule Aggregates}},\ }\href
  {https://doi.org/10.1051/0004-6361/201425120} {\bibfield  {journal} {\bibinfo
   {journal} {A\&A}\ }\textbf {\bibinfo {volume} {579}},\ \bibinfo {pages}
  {A43} (\bibinfo {year} {2015})}\BibitemShut {NoStop}%
\bibitem [{\citenamefont {Owen}\ and\ \citenamefont {Wu}(2013)}]{OwenWu2013}%
  \BibitemOpen
  \bibfield  {author} {\bibinfo {author} {\bibfnamefont {J.~E.}\ \bibnamefont
  {Owen}}\ and\ \bibinfo {author} {\bibfnamefont {Y.}~\bibnamefont {Wu}},\
  }\bibfield  {title} {\bibinfo {title} {{Kepler Planets: A Tale of
  Evaporation}},\ }\href {https://doi.org/10.1088/0004-637X/775/2/105}
  {\bibfield  {journal} {\bibinfo  {journal} {ApJ}\ }\textbf {\bibinfo {volume}
  {775}},\ \bibinfo {pages} {105} (\bibinfo {year} {2013})}\BibitemShut
  {NoStop}%
\bibitem [{\citenamefont {Jin}\ and\ \citenamefont
  {Mordasini}(2018)}]{JinMordasini2018}%
  \BibitemOpen
  \bibfield  {author} {\bibinfo {author} {\bibfnamefont {S.}~\bibnamefont
  {Jin}}\ and\ \bibinfo {author} {\bibfnamefont {C.}~\bibnamefont
  {Mordasini}},\ }\bibfield  {title} {\bibinfo {title} {{Compositional Imprints
  in {D}ensity--{D}istance--{T}ime: A Rocky View of Planet Formation}},\ }\href
  {https://doi.org/10.3847/1538-4357/aa9f1e} {\bibfield  {journal} {\bibinfo
  {journal} {ApJ}\ }\textbf {\bibinfo {volume} {853}},\ \bibinfo {pages} {163}
  (\bibinfo {year} {2018})}\BibitemShut {NoStop}%
\bibitem [{\citenamefont {Kasting}\ \emph {et~al.}(1993)\citenamefont
  {Kasting}, \citenamefont {Whitmire},\ and\ \citenamefont
  {Reynolds}}]{Kasting1993}%
  \BibitemOpen
  \bibfield  {author} {\bibinfo {author} {\bibfnamefont {J.~F.}\ \bibnamefont
  {Kasting}}, \bibinfo {author} {\bibfnamefont {D.~P.}\ \bibnamefont
  {Whitmire}},\ and\ \bibinfo {author} {\bibfnamefont {R.~T.}\ \bibnamefont
  {Reynolds}},\ }\bibfield  {title} {\bibinfo {title} {{Habitable Zones around
  Main Sequence Stars}},\ }\href {https://doi.org/10.1006/icar.1993.1010}
  {\bibfield  {journal} {\bibinfo  {journal} {Icarus}\ }\textbf {\bibinfo
  {volume} {101}},\ \bibinfo {pages} {108} (\bibinfo {year}
  {1993})}\BibitemShut {NoStop}%
\bibitem [{\citenamefont {Kopparapu}\ \emph {et~al.}(2013)\citenamefont
  {Kopparapu} \emph {et~al.}}]{Kopparapu2013}%
  \BibitemOpen
  \bibfield  {author} {\bibinfo {author} {\bibfnamefont {R.~K.}\ \bibnamefont
  {Kopparapu}} \emph {et~al.},\ }\bibfield  {title} {\bibinfo {title}
  {{Habitable Zones around Main-sequence Stars: New Estimates}},\ }\href
  {https://doi.org/10.1088/0004-637X/765/2/131} {\bibfield  {journal} {\bibinfo
   {journal} {ApJ}\ }\textbf {\bibinfo {volume} {765}},\ \bibinfo {pages} {131}
  (\bibinfo {year} {2013})}\BibitemShut {NoStop}%
\bibitem [{\citenamefont {Goldblatt}\ \emph {et~al.}(2013)\citenamefont
  {Goldblatt} \emph {et~al.}}]{Goldblatt2013}%
  \BibitemOpen
  \bibfield  {author} {\bibinfo {author} {\bibfnamefont {C.}~\bibnamefont
  {Goldblatt}} \emph {et~al.},\ }\bibfield  {title} {\bibinfo {title} {{Low
  Simulated Radiation Limit for Runaway Greenhouse Climates}},\ }\href
  {https://doi.org/10.1038/ngeo1892} {\bibfield  {journal} {\bibinfo  {journal}
  {Nature Geoscience}\ }\textbf {\bibinfo {volume} {6}},\ \bibinfo {pages}
  {661} (\bibinfo {year} {2013})}\BibitemShut {NoStop}%
\bibitem [{\citenamefont {Wolf}\ and\ \citenamefont
  {Toon}(2015)}]{WolfToon2015}%
  \BibitemOpen
  \bibfield  {author} {\bibinfo {author} {\bibfnamefont {E.~T.}\ \bibnamefont
  {Wolf}}\ and\ \bibinfo {author} {\bibfnamefont {O.~B.}\ \bibnamefont
  {Toon}},\ }\bibfield  {title} {\bibinfo {title} {{The Evolution of Habitable
  Climates under the Brightening Sun}},\ }\href
  {https://doi.org/10.1002/2014JE004701} {\bibfield  {journal} {\bibinfo
  {journal} {JGR: Planets}\ }\textbf {\bibinfo {volume} {120}},\ \bibinfo
  {pages} {577} (\bibinfo {year} {2015})}\BibitemShut {NoStop}%
\bibitem [{\citenamefont {Chamberlain}\ and\ \citenamefont
  {Hunten}(1987)}]{ChamberlainHunten1987}%
  \BibitemOpen
  \bibfield  {author} {\bibinfo {author} {\bibfnamefont {J.~W.}\ \bibnamefont
  {Chamberlain}}\ and\ \bibinfo {author} {\bibfnamefont {D.~M.}\ \bibnamefont
  {Hunten}},\ }\href@noop {} {\emph {\bibinfo {title} {{Theory of Planetary
  Atmospheres: An Introduction to Their Physics and Chemistry}}}},\ \bibinfo
  {edition} {2nd}\ ed.\ (\bibinfo  {publisher} {Academic Press},\ \bibinfo
  {address} {Orlando, FL},\ \bibinfo {year} {1987})\BibitemShut {NoStop}%
\bibitem [{\citenamefont {Kivelson}\ and\ \citenamefont
  {Russell}(1995)}]{KivelsonRussell1995}%
  \BibitemOpen
  \bibinfo {editor} {\bibfnamefont {M.~G.}\ \bibnamefont {Kivelson}}\ and\
  \bibinfo {editor} {\bibfnamefont {C.~T.}\ \bibnamefont {Russell}},\ eds.,\
  \href@noop {} {\emph {\bibinfo {title} {{Introduction to Space Physics}}}}\
  (\bibinfo  {publisher} {Cambridge University Press},\ \bibinfo {year}
  {1995})\BibitemShut {NoStop}%
\bibitem [{\citenamefont {Cravens}(1997)}]{Cravens1997}%
  \BibitemOpen
  \bibfield  {author} {\bibinfo {author} {\bibfnamefont {T.~E.}\ \bibnamefont
  {Cravens}},\ }\href {https://doi.org/10.1017/CBO9780511529467} {\emph
  {\bibinfo {title} {{Physics of Solar System Plasmas}}}}\ (\bibinfo
  {publisher} {Cambridge University Press},\ \bibinfo {year}
  {1997})\BibitemShut {NoStop}%
\bibitem [{\citenamefont {Mitchell}(1961)}]{Mitchell1961}%
  \BibitemOpen
  \bibfield  {author} {\bibinfo {author} {\bibfnamefont {P.}~\bibnamefont
  {Mitchell}},\ }\bibfield  {title} {\bibinfo {title} {{Coupling of
  Phosphorylation to Electron and Hydrogen Transfer by a Chemi-Osmotic type of
  Mechanism}},\ }\href {https://doi.org/10.1038/191144a0} {\bibfield  {journal}
  {\bibinfo  {journal} {Nature}\ }\textbf {\bibinfo {volume} {191}},\ \bibinfo
  {pages} {144} (\bibinfo {year} {1961})}\BibitemShut {NoStop}%
\bibitem [{\citenamefont {Nicholls}\ and\ \citenamefont
  {Ferguson}(2013)}]{NichollsFerguson2013}%
  \BibitemOpen
  \bibfield  {author} {\bibinfo {author} {\bibfnamefont {D.~G.}\ \bibnamefont
  {Nicholls}}\ and\ \bibinfo {author} {\bibfnamefont {S.~J.}\ \bibnamefont
  {Ferguson}},\ }\href@noop {} {\emph {\bibinfo {title} {{Bioenergetics 4}}}}\
  (\bibinfo  {publisher} {Academic Press},\ \bibinfo {year} {2013})\BibitemShut
  {NoStop}%
\bibitem [{\citenamefont {Peale}\ \emph {et~al.}(1979)\citenamefont {Peale},
  \citenamefont {Cassen},\ and\ \citenamefont {Reynolds}}]{Peale1979}%
  \BibitemOpen
  \bibfield  {author} {\bibinfo {author} {\bibfnamefont {S.~J.}\ \bibnamefont
  {Peale}}, \bibinfo {author} {\bibfnamefont {P.}~\bibnamefont {Cassen}},\ and\
  \bibinfo {author} {\bibfnamefont {R.~T.}\ \bibnamefont {Reynolds}},\
  }\bibfield  {title} {\bibinfo {title} {{Melting of Io by Tidal
  Dissipation}},\ }\href {https://doi.org/10.1126/science.203.4383.892}
  {\bibfield  {journal} {\bibinfo  {journal} {Science}\ }\textbf {\bibinfo
  {volume} {203}},\ \bibinfo {pages} {892} (\bibinfo {year}
  {1979})}\BibitemShut {NoStop}%
\bibitem [{\citenamefont {Murray}\ and\ \citenamefont
  {Dermott}(1999)}]{MurrayDermott1999}%
  \BibitemOpen
  \bibfield  {author} {\bibinfo {author} {\bibfnamefont {C.~D.}\ \bibnamefont
  {Murray}}\ and\ \bibinfo {author} {\bibfnamefont {S.~F.}\ \bibnamefont
  {Dermott}},\ }\href {https://doi.org/10.1017/CBO9781139174817} {\emph
  {\bibinfo {title} {{Solar System Dynamics}}}}\ (\bibinfo  {publisher}
  {Cambridge University Press},\ \bibinfo {year} {1999})\BibitemShut {NoStop}%
\bibitem [{\citenamefont {Chyba}\ and\ \citenamefont
  {Hand}(2001)}]{ChybaHand2001}%
  \BibitemOpen
  \bibfield  {author} {\bibinfo {author} {\bibfnamefont {C.~F.}\ \bibnamefont
  {Chyba}}\ and\ \bibinfo {author} {\bibfnamefont {K.~P.}\ \bibnamefont
  {Hand}},\ }\bibfield  {title} {\bibinfo {title} {{Planetary science. Life
  without photosynthesis}},\ }\href {https://doi.org/10.1126/science.1060081}
  {\bibfield  {journal} {\bibinfo  {journal} {Science}\ }\textbf {\bibinfo
  {volume} {292}},\ \bibinfo {pages} {2026} (\bibinfo {year}
  {2001})}\BibitemShut {NoStop}%
\bibitem [{\citenamefont {Feinberg}(2019)}]{Feinberg2019}%
  \BibitemOpen
  \bibfield  {author} {\bibinfo {author} {\bibfnamefont {M.}~\bibnamefont
  {Feinberg}},\ }\href {https://doi.org/10.1007/978-3-030-03858-8} {\emph
  {\bibinfo {title} {{Foundations of Chemical Reaction Network Theory}}}}\
  (\bibinfo  {publisher} {Springer},\ \bibinfo {year} {2019})\BibitemShut
  {NoStop}%
\bibitem [{\citenamefont {Smith}\ and\ \citenamefont
  {Waltman}(1995)}]{SmithWaltman1995}%
  \BibitemOpen
  \bibfield  {author} {\bibinfo {author} {\bibfnamefont {H.~L.}\ \bibnamefont
  {Smith}}\ and\ \bibinfo {author} {\bibfnamefont {P.}~\bibnamefont
  {Waltman}},\ }\href@noop {} {\emph {\bibinfo {title} {{The Theory of the
  Chemostat: Dynamics of Microbial Competition}}}}\ (\bibinfo  {publisher}
  {Cambridge University Press},\ \bibinfo {year} {1995})\BibitemShut {NoStop}%
\bibitem [{\citenamefont {Segura}\ \emph {et~al.}(2010)\citenamefont {Segura},
  \citenamefont {Walkowicz}, \citenamefont {Meadows}, \citenamefont {Kasting},\
  and\ \citenamefont {Hawley}}]{Segura2010}%
  \BibitemOpen
  \bibfield  {author} {\bibinfo {author} {\bibfnamefont {A.}~\bibnamefont
  {Segura}}, \bibinfo {author} {\bibfnamefont {L.~M.}\ \bibnamefont
  {Walkowicz}}, \bibinfo {author} {\bibfnamefont {V.~S.}\ \bibnamefont
  {Meadows}}, \bibinfo {author} {\bibfnamefont {J.~F.}\ \bibnamefont
  {Kasting}},\ and\ \bibinfo {author} {\bibfnamefont {S.~L.}\ \bibnamefont
  {Hawley}},\ }\bibfield  {title} {\bibinfo {title} {{The Effect of a Strong
  Stellar Flare on the Atmospheric Chemistry of an Earth-like Planet Orbiting
  an M Dwarf}},\ }\href {https://doi.org/10.1089/ast.2009.0376} {\bibfield
  {journal} {\bibinfo  {journal} {Astrobiology}\ }\textbf {\bibinfo {volume}
  {10}},\ \bibinfo {pages} {751} (\bibinfo {year} {2010})}\BibitemShut
  {NoStop}%
\bibitem [{\citenamefont {Tilley}\ \emph {et~al.}(2019)\citenamefont {Tilley},
  \citenamefont {Segura}, \citenamefont {Meadows}, \citenamefont {Hawley},\
  and\ \citenamefont {Davenport}}]{Tilley2019}%
  \BibitemOpen
  \bibfield  {author} {\bibinfo {author} {\bibfnamefont {M.~A.}\ \bibnamefont
  {Tilley}}, \bibinfo {author} {\bibfnamefont {A.}~\bibnamefont {Segura}},
  \bibinfo {author} {\bibfnamefont {V.}~\bibnamefont {Meadows}}, \bibinfo
  {author} {\bibfnamefont {S.}~\bibnamefont {Hawley}},\ and\ \bibinfo {author}
  {\bibfnamefont {J.}~\bibnamefont {Davenport}},\ }\bibfield  {title} {\bibinfo
  {title} {{Modeling Repeated M-dwarf Flares and Their Effects on Earth-like
  Planet Atmospheres}},\ }\href {https://doi.org/10.1089/ast.2017.1794}
  {\bibfield  {journal} {\bibinfo  {journal} {Astrobiology}\ }\textbf {\bibinfo
  {volume} {19}},\ \bibinfo {pages} {64} (\bibinfo {year} {2019})}\BibitemShut
  {NoStop}%
\bibitem [{\citenamefont {Walker}\ and\ \citenamefont
  {Davies}(2013)}]{WalkerDavies2013_AlgOrigins}%
  \BibitemOpen
  \bibfield  {author} {\bibinfo {author} {\bibfnamefont {S.~I.}\ \bibnamefont
  {Walker}}\ and\ \bibinfo {author} {\bibfnamefont {P.~C.~W.}\ \bibnamefont
  {Davies}},\ }\bibfield  {title} {\bibinfo {title} {{The algorithmic origins
  of life}},\ }\href {https://doi.org/10.1098/rsif.2012.0869} {\bibfield
  {journal} {\bibinfo  {journal} {J. Royal Soc. Interface}\ }\textbf {\bibinfo
  {volume} {10}},\ \bibinfo {pages} {20120869} (\bibinfo {year}
  {2013})}\BibitemShut {NoStop}%
\bibitem [{\citenamefont {Murugan}\ \emph {et~al.}(2012)\citenamefont
  {Murugan}, \citenamefont {Huse},\ and\ \citenamefont
  {Leibler}}]{Murugan2012}%
  \BibitemOpen
  \bibfield  {author} {\bibinfo {author} {\bibfnamefont {A.}~\bibnamefont
  {Murugan}}, \bibinfo {author} {\bibfnamefont {D.~A.}\ \bibnamefont {Huse}},\
  and\ \bibinfo {author} {\bibfnamefont {S.}~\bibnamefont {Leibler}},\
  }\bibfield  {title} {\bibinfo {title} {{Speed, dissipation, and error in
  kinetic proofreading}},\ }\href {https://doi.org/10.1073/pnas.1119911109}
  {\bibfield  {journal} {\bibinfo  {journal} {Proc. Natl. Acad. Sci. U.S.A.}\
  }\textbf {\bibinfo {volume} {109}},\ \bibinfo {pages} {12034} (\bibinfo
  {year} {2012})}\BibitemShut {NoStop}%
\bibitem [{\citenamefont {Sartori}\ and\ \citenamefont
  {Pigolotti}(2013)}]{SartoriPigolotti2013}%
  \BibitemOpen
  \bibfield  {author} {\bibinfo {author} {\bibfnamefont {P.}~\bibnamefont
  {Sartori}}\ and\ \bibinfo {author} {\bibfnamefont {S.}~\bibnamefont
  {Pigolotti}},\ }\bibfield  {title} {\bibinfo {title} {{Kinetic versus
  Energetic Discrimination in Biological Copying}},\ }\href
  {https://doi.org/10.1103/PhysRevLett.110.188101} {\bibfield  {journal}
  {\bibinfo  {journal} {PRL}\ }\textbf {\bibinfo {volume} {110}},\ \bibinfo
  {pages} {188101} (\bibinfo {year} {2013})}\BibitemShut {NoStop}%
\bibitem [{\citenamefont {Chen}\ \emph {et~al.}(2004)\citenamefont {Chen},
  \citenamefont {Roberts},\ and\ \citenamefont {Szostak}}]{Chen2004}%
  \BibitemOpen
  \bibfield  {author} {\bibinfo {author} {\bibfnamefont {I.~A.}\ \bibnamefont
  {Chen}}, \bibinfo {author} {\bibfnamefont {R.~W.}\ \bibnamefont {Roberts}},\
  and\ \bibinfo {author} {\bibfnamefont {J.~W.}\ \bibnamefont {Szostak}},\
  }\bibfield  {title} {\bibinfo {title} {{The Emergence of Competition Between
  Model Protocells}},\ }\href {https://doi.org/10.1126/science.1100740}
  {\bibfield  {journal} {\bibinfo  {journal} {Science}\ }\textbf {\bibinfo
  {volume} {305}},\ \bibinfo {pages} {1474} (\bibinfo {year}
  {2004})}\BibitemShut {NoStop}%
\bibitem [{\citenamefont {Keating}(2012)}]{Keating2012}%
  \BibitemOpen
  \bibfield  {author} {\bibinfo {author} {\bibfnamefont {C.~D.}\ \bibnamefont
  {Keating}},\ }\bibfield  {title} {\bibinfo {title} {{Aqueous Phase Separation
  as a Possible Route to Compartmentalization of Biological Molecules}},\
  }\href {https://doi.org/10.1021/ar200294y} {\bibfield  {journal} {\bibinfo
  {journal} {Acc. Chem. Res.}\ }\textbf {\bibinfo {volume} {45}},\ \bibinfo
  {pages} {2114} (\bibinfo {year} {2012})}\BibitemShut {NoStop}%
\bibitem [{\citenamefont {Hordijk}\ and\ \citenamefont
  {Steel}(2018)}]{HordijkSteel2018}%
  \BibitemOpen
  \bibfield  {author} {\bibinfo {author} {\bibfnamefont {W.}~\bibnamefont
  {Hordijk}}\ and\ \bibinfo {author} {\bibfnamefont {M.}~\bibnamefont
  {Steel}},\ }\bibfield  {title} {\bibinfo {title} {{Autocatalytic Networks at
  the Basis of Life's Origin and Organization}},\ }\href
  {https://doi.org/10.3390/life8040062} {\bibfield  {journal} {\bibinfo
  {journal} {Life}\ }\textbf {\bibinfo {volume} {8}},\ \bibinfo {pages} {62}
  (\bibinfo {year} {2018})}\BibitemShut {NoStop}%
\bibitem [{\citenamefont {Horn}\ and\ \citenamefont
  {Johnson}(2012)}]{HornJohnson2012MatrixAnalysis}%
  \BibitemOpen
  \bibfield  {author} {\bibinfo {author} {\bibfnamefont {R.~A.}\ \bibnamefont
  {Horn}}\ and\ \bibinfo {author} {\bibfnamefont {C.~R.}\ \bibnamefont
  {Johnson}},\ }\href@noop {} {\emph {\bibinfo {title} {{Matrix Analysis}}}},\
  \bibinfo {edition} {2nd}\ ed.\ (\bibinfo  {publisher} {Cambridge University
  Press},\ \bibinfo {year} {2012})\BibitemShut {NoStop}%
\bibitem [{\citenamefont {van~den Driessche}\ and\ \citenamefont
  {Watmough}(2002)}]{vanDenDriesscheWatmough2002R0}%
  \BibitemOpen
  \bibfield  {author} {\bibinfo {author} {\bibfnamefont {P.}~\bibnamefont
  {van~den Driessche}}\ and\ \bibinfo {author} {\bibfnamefont {J.}~\bibnamefont
  {Watmough}},\ }\bibfield  {title} {\bibinfo {title} {{Reproduction numbers
  and sub-threshold endemic equilibria for compartmental models of disease
  transmission}},\ }\href {https://doi.org/10.1016/S0025-5564(02)00108-6}
  {\bibfield  {journal} {\bibinfo  {journal} {Mathematical Biosciences}\
  }\textbf {\bibinfo {volume} {180}},\ \bibinfo {pages} {29} (\bibinfo {year}
  {2002})}\BibitemShut {NoStop}%
\bibitem [{\citenamefont {Diekmann}\ \emph {et~al.}(2010)\citenamefont
  {Diekmann}, \citenamefont {Heesterbeek},\ and\ \citenamefont
  {Roberts}}]{DiekmannHeesterbeekRoberts2010NGM}%
  \BibitemOpen
  \bibfield  {author} {\bibinfo {author} {\bibfnamefont {O.}~\bibnamefont
  {Diekmann}}, \bibinfo {author} {\bibfnamefont {J.~A.~P.}\ \bibnamefont
  {Heesterbeek}},\ and\ \bibinfo {author} {\bibfnamefont {M.~G.}\ \bibnamefont
  {Roberts}},\ }\bibfield  {title} {\bibinfo {title} {{The construction of
  next-generation matrices for compartmental epidemic models}},\ }\href
  {https://doi.org/10.1098/rsif.2009.0386} {\bibfield  {journal} {\bibinfo
  {journal} {J. Royal Soc. Interface}\ }\textbf {\bibinfo {volume} {7}},\
  \bibinfo {pages} {873} (\bibinfo {year} {2010})}\BibitemShut {NoStop}%
\bibitem [{\citenamefont {Harris}(1963)}]{Harris1963Branching}%
  \BibitemOpen
  \bibfield  {author} {\bibinfo {author} {\bibfnamefont {T.~E.}\ \bibnamefont
  {Harris}},\ }\href@noop {} {\emph {\bibinfo {title} {{The Theory of Branching
  Processes}}}}\ (\bibinfo  {publisher} {Springer},\ \bibinfo {year}
  {1963})\BibitemShut {NoStop}%
\bibitem [{\citenamefont {Mossel}\ and\ \citenamefont
  {Steel}(2005)}]{MosselSteel2005}%
  \BibitemOpen
  \bibfield  {author} {\bibinfo {author} {\bibfnamefont {E.}~\bibnamefont
  {Mossel}}\ and\ \bibinfo {author} {\bibfnamefont {M.}~\bibnamefont {Steel}},\
  }\bibfield  {title} {\bibinfo {title} {{Random biochemical networks: the
  probability of self-sustaining autocatalysis}},\ }\href
  {https://doi.org/10.1016/j.jtbi.2004.10.011} {\bibfield  {journal} {\bibinfo
  {journal} {J. Theor. Biol.}\ }\textbf {\bibinfo {volume} {233}},\ \bibinfo
  {pages} {327} (\bibinfo {year} {2005})}\BibitemShut {NoStop}%
\bibitem [{\citenamefont {Levenspiel}(1999)}]{Levenspiel1999}%
  \BibitemOpen
  \bibfield  {author} {\bibinfo {author} {\bibfnamefont {O.}~\bibnamefont
  {Levenspiel}},\ }\href@noop {} {\emph {\bibinfo {title} {{Chemical Reaction
  Engineering}}}},\ \bibinfo {edition} {3rd}\ ed.\ (\bibinfo  {publisher} {John
  Wiley \& Sons},\ \bibinfo {year} {1999})\BibitemShut {NoStop}%
\bibitem [{\citenamefont {Fogler}(2016)}]{Fogler2016}%
  \BibitemOpen
  \bibfield  {author} {\bibinfo {author} {\bibfnamefont {H.~S.}\ \bibnamefont
  {Fogler}},\ }\href@noop {} {\emph {\bibinfo {title} {{Elements of Chemical
  Reaction Engineering}}}},\ \bibinfo {edition} {5th}\ ed.\ (\bibinfo
  {publisher} {Prentice Hall},\ \bibinfo {year} {2016})\BibitemShut {NoStop}%
\bibitem [{\citenamefont {Nowak}(2006)}]{Nowak2006}%
  \BibitemOpen
  \bibfield  {author} {\bibinfo {author} {\bibfnamefont {M.~A.}\ \bibnamefont
  {Nowak}},\ }\href@noop {} {\emph {\bibinfo {title} {{Evolutionary Dynamics:
  Exploring the Equations of Life}}}}\ (\bibinfo  {publisher} {Harvard
  University Press},\ \bibinfo {year} {2006})\BibitemShut {NoStop}%
\bibitem [{\citenamefont {Hofbauer}\ and\ \citenamefont
  {Sigmund}(1998)}]{HofbauerSigmund1998}%
  \BibitemOpen
  \bibfield  {author} {\bibinfo {author} {\bibfnamefont {J.}~\bibnamefont
  {Hofbauer}}\ and\ \bibinfo {author} {\bibfnamefont {K.}~\bibnamefont
  {Sigmund}},\ }\href@noop {} {\emph {\bibinfo {title} {{Evolutionary Games and
  Population Dynamics}}}}\ (\bibinfo  {publisher} {Cambridge University
  Press},\ \bibinfo {year} {1998})\BibitemShut {NoStop}%
\bibitem [{\citenamefont {Price}(1970)}]{Price1970}%
  \BibitemOpen
  \bibfield  {author} {\bibinfo {author} {\bibfnamefont {G.~R.}\ \bibnamefont
  {Price}},\ }\bibfield  {title} {\bibinfo {title} {{Selection and
  Covariance}},\ }\href {https://doi.org/10.1038/227520a0} {\bibfield
  {journal} {\bibinfo  {journal} {Nature}\ }\textbf {\bibinfo {volume} {227}},\
  \bibinfo {pages} {520} (\bibinfo {year} {1970})}\BibitemShut {NoStop}%
\bibitem [{\citenamefont {Haig}\ and\ \citenamefont
  {Hurst}(1991)}]{HaigHurst1991}%
  \BibitemOpen
  \bibfield  {author} {\bibinfo {author} {\bibfnamefont {D.}~\bibnamefont
  {Haig}}\ and\ \bibinfo {author} {\bibfnamefont {L.~D.}\ \bibnamefont
  {Hurst}},\ }\bibfield  {title} {\bibinfo {title} {{A quantitative measure of
  error minimization in the genetic code}},\ }\href
  {https://doi.org/10.1007/BF02103132} {\bibfield  {journal} {\bibinfo
  {journal} {J. Mol. Evol.}\ }\textbf {\bibinfo {volume} {33}},\ \bibinfo
  {pages} {412} (\bibinfo {year} {1991})}\BibitemShut {NoStop}%
\bibitem [{\citenamefont {Freeland}\ and\ \citenamefont
  {Hurst}(1998)}]{FreelandHurst1998}%
  \BibitemOpen
  \bibfield  {author} {\bibinfo {author} {\bibfnamefont {S.~J.}\ \bibnamefont
  {Freeland}}\ and\ \bibinfo {author} {\bibfnamefont {L.~D.}\ \bibnamefont
  {Hurst}},\ }\bibfield  {title} {\bibinfo {title} {{The genetic code is one in
  a million}},\ }\href {https://doi.org/10.1007/PL00006381} {\bibfield
  {journal} {\bibinfo  {journal} {J. Mol. Evol.}\ }\textbf {\bibinfo {volume}
  {47}},\ \bibinfo {pages} {238} (\bibinfo {year} {1998})}\BibitemShut
  {NoStop}%
\bibitem [{\citenamefont {Roe}(2009)}]{Roe2009}%
  \BibitemOpen
  \bibfield  {author} {\bibinfo {author} {\bibfnamefont {G.~H.}\ \bibnamefont
  {Roe}},\ }\bibfield  {title} {\bibinfo {title} {{Feedbacks, Timescales, and
  Seeing Red}},\ }\href {https://doi.org/10.1146/annurev.earth.061008.134734}
  {\bibfield  {journal} {\bibinfo  {journal} {Annu. Rev. Earth Planet. Sci.}\
  }\textbf {\bibinfo {volume} {37}},\ \bibinfo {pages} {93} (\bibinfo {year}
  {2009})}\BibitemShut {NoStop}%
\bibitem [{\citenamefont {{Turyshev}}(2025)}]{Turyshev2025-NPP}%
  \BibitemOpen
  \bibfield  {author} {\bibinfo {author} {\bibfnamefont {S.~G.}\ \bibnamefont
  {{Turyshev}}},\ }\href {https://doi.org/10.48550/arXiv.2512.07199} {\bibinfo
  {title} {{{Information-Thermodynamic Bounds on Planetary Biosphere
  Productivity and Their Observational Tests}}}} (\bibinfo {year} {2025}),\
  \bibinfo {note} {arXiv:2512.07199}\BibitemShut {NoStop}%
\bibitem [{\citenamefont {May}(1972)}]{May1972}%
  \BibitemOpen
  \bibfield  {author} {\bibinfo {author} {\bibfnamefont {R.~M.}\ \bibnamefont
  {May}},\ }\bibfield  {title} {\bibinfo {title} {{Will a Large Complex System
  be Stable?}},\ }\href {https://doi.org/10.1038/238413a0} {\bibfield
  {journal} {\bibinfo  {journal} {Nature}\ }\textbf {\bibinfo {volume} {238}},\
  \bibinfo {pages} {413} (\bibinfo {year} {1972})}\BibitemShut {NoStop}%
\bibitem [{\citenamefont {Allesina}\ and\ \citenamefont
  {Tang}(2012)}]{AllesinaTang2012}%
  \BibitemOpen
  \bibfield  {author} {\bibinfo {author} {\bibfnamefont {S.}~\bibnamefont
  {Allesina}}\ and\ \bibinfo {author} {\bibfnamefont {S.}~\bibnamefont
  {Tang}},\ }\bibfield  {title} {\bibinfo {title} {{Stability criteria for
  complex ecosystems}},\ }\href {https://doi.org/10.1038/nature10832}
  {\bibfield  {journal} {\bibinfo  {journal} {Nature}\ }\textbf {\bibinfo
  {volume} {483}},\ \bibinfo {pages} {205} (\bibinfo {year}
  {2012})}\BibitemShut {NoStop}%
\bibitem [{\citenamefont {{IPCC}}(2021)}]{IPCCAR6WGICh7}%
  \BibitemOpen
  \bibfield  {author} {\bibinfo {author} {\bibnamefont {{IPCC}}},\ }\href@noop
  {} {\bibinfo {title} {{AR6 WGI, Chapter 7: Earth's Energy Budget, Feedbacks,
  and Climate Sensitivity}}},\ \bibinfo {howpublished}
  {\url{https://www.ipcc.ch/report/ar6/wg1/chapter/chapter-7/}} (\bibinfo
  {year} {2021})\BibitemShut {NoStop}%
\bibitem [{\citenamefont {Cleaves}\ \emph {et~al.}(2023)\citenamefont {Cleaves}
  \emph {et~al.}}]{Cleaves2023_PNAS}%
  \BibitemOpen
  \bibfield  {author} {\bibinfo {author} {\bibfnamefont {H.~J.}\ \bibnamefont
  {Cleaves}} \emph {et~al.},\ }\bibfield  {title} {\bibinfo {title} {{A robust,
  agnostic molecular biosignature based on machine learning and mass
  spectrometry}},\ }\href {https://doi.org/10.1073/pnas.2307149120} {\bibfield
  {journal} {\bibinfo  {journal} {Proc. Nat. Acad. Sci.}\ }\textbf {\bibinfo
  {volume} {120}},\ \bibinfo {pages} {e2307149120} (\bibinfo {year}
  {2023})}\BibitemShut {NoStop}%
\bibitem [{\citenamefont {Catling}\ \emph {et~al.}(2018)\citenamefont
  {Catling}, \citenamefont {Krissansen-Totton}, \citenamefont {Kiang},
  \citenamefont {Crisp}, \citenamefont {Robinson}, \citenamefont {DasSarma},
  \citenamefont {Rushby}, \citenamefont {Del~Genio}, \citenamefont {Bains},
  \citenamefont {Domagal-Goldman} \emph {et~al.}}]{Catling2018}%
  \BibitemOpen
  \bibfield  {author} {\bibinfo {author} {\bibfnamefont {D.~C.}\ \bibnamefont
  {Catling}}, \bibinfo {author} {\bibfnamefont {J.}~\bibnamefont
  {Krissansen-Totton}}, \bibinfo {author} {\bibfnamefont {N.~Y.}\ \bibnamefont
  {Kiang}}, \bibinfo {author} {\bibfnamefont {D.}~\bibnamefont {Crisp}},
  \bibinfo {author} {\bibfnamefont {T.~D.}\ \bibnamefont {Robinson}}, \bibinfo
  {author} {\bibfnamefont {S.}~\bibnamefont {DasSarma}}, \bibinfo {author}
  {\bibfnamefont {A.~J.}\ \bibnamefont {Rushby}}, \bibinfo {author}
  {\bibfnamefont {A.}~\bibnamefont {Del~Genio}}, \bibinfo {author}
  {\bibfnamefont {W.}~\bibnamefont {Bains}}, \bibinfo {author} {\bibfnamefont
  {S.}~\bibnamefont {Domagal-Goldman}}, \emph {et~al.},\ }\bibfield  {title}
  {\bibinfo {title} {{Exoplanet Biosignatures: A Framework for Their
  Assessment}},\ }\href {https://doi.org/10.1089/ast.2017.1737} {\bibfield
  {journal} {\bibinfo  {journal} {Astrobiology}\ }\textbf {\bibinfo {volume}
  {18}},\ \bibinfo {pages} {709} (\bibinfo {year} {2018})}\BibitemShut
  {NoStop}%
\bibitem [{\citenamefont {Meadows}\ \emph {et~al.}(2018)\citenamefont
  {Meadows}, \citenamefont {Reinhard}, \citenamefont {Arney}, \citenamefont
  {Parenteau}, \citenamefont {Schwieterman}, \citenamefont {Domagal-Goldman},
  \citenamefont {Lincowski},\ and\ \citenamefont {et~al.}}]{Meadows2018}%
  \BibitemOpen
  \bibfield  {author} {\bibinfo {author} {\bibfnamefont {V.~S.}\ \bibnamefont
  {Meadows}}, \bibinfo {author} {\bibfnamefont {C.~T.}\ \bibnamefont
  {Reinhard}}, \bibinfo {author} {\bibfnamefont {G.~N.}\ \bibnamefont {Arney}},
  \bibinfo {author} {\bibfnamefont {M.~N.}\ \bibnamefont {Parenteau}}, \bibinfo
  {author} {\bibfnamefont {E.~W.}\ \bibnamefont {Schwieterman}}, \bibinfo
  {author} {\bibfnamefont {S.~D.}\ \bibnamefont {Domagal-Goldman}}, \bibinfo
  {author} {\bibfnamefont {A.~P.}\ \bibnamefont {Lincowski}},\ and\ \bibinfo
  {author} {\bibnamefont {et~al.}},\ }\bibfield  {title} {\bibinfo {title}
  {{Exoplanet Biosignatures: Understanding Oxygen as a Biosignature in the
  Context of Its Environment}},\ }\href {https://doi.org/10.1089/ast.2017.1727}
  {\bibfield  {journal} {\bibinfo  {journal} {Astrobiology}\ }\textbf {\bibinfo
  {volume} {18}},\ \bibinfo {pages} {630} (\bibinfo {year} {2018})}\BibitemShut
  {NoStop}%
\bibitem [{\citenamefont {Krissansen-Totton}\ \emph {et~al.}(2016)\citenamefont
  {Krissansen-Totton}, \citenamefont {Bergsman},\ and\ \citenamefont
  {Catling}}]{KrissansenTotton2016}%
  \BibitemOpen
  \bibfield  {author} {\bibinfo {author} {\bibfnamefont {J.}~\bibnamefont
  {Krissansen-Totton}}, \bibinfo {author} {\bibfnamefont {D.~S.}\ \bibnamefont
  {Bergsman}},\ and\ \bibinfo {author} {\bibfnamefont {D.~C.}\ \bibnamefont
  {Catling}},\ }\bibfield  {title} {\bibinfo {title} {{Atmospheric
  Disequilibrium as a Biosignature}},\ }\href
  {https://doi.org/10.1089/ast.2015.1327} {\bibfield  {journal} {\bibinfo
  {journal} {Astrobiology}\ }\textbf {\bibinfo {volume} {16}},\ \bibinfo
  {pages} {39} (\bibinfo {year} {2016})}\BibitemShut {NoStop}%
\bibitem [{\citenamefont {Robinson}\ \emph {et~al.}(2010)\citenamefont
  {Robinson}, \citenamefont {Meadows}, \citenamefont {Crisp}, \citenamefont
  {Deming}, \citenamefont {A'Hearn}, \citenamefont {Charbonneau}, \citenamefont
  {Livengood},\ and\ \citenamefont {Seager}}]{Robinson2010}%
  \BibitemOpen
  \bibfield  {author} {\bibinfo {author} {\bibfnamefont {T.~D.}\ \bibnamefont
  {Robinson}}, \bibinfo {author} {\bibfnamefont {V.~S.}\ \bibnamefont
  {Meadows}}, \bibinfo {author} {\bibfnamefont {D.}~\bibnamefont {Crisp}},
  \bibinfo {author} {\bibfnamefont {D.}~\bibnamefont {Deming}}, \bibinfo
  {author} {\bibfnamefont {M.~F.}\ \bibnamefont {A'Hearn}}, \bibinfo {author}
  {\bibfnamefont {D.}~\bibnamefont {Charbonneau}}, \bibinfo {author}
  {\bibfnamefont {T.~A.}\ \bibnamefont {Livengood}},\ and\ \bibinfo {author}
  {\bibfnamefont {S.}~\bibnamefont {Seager}},\ }\bibfield  {title} {\bibinfo
  {title} {{Detecting Oceans on Extrasolar Planets Using the Glint Effect}},\
  }\href {https://doi.org/10.1088/2041-8205/721/1/L67} {\bibfield  {journal}
  {\bibinfo  {journal} {ApJ Lett.}\ }\textbf {\bibinfo {volume} {721}},\
  \bibinfo {pages} {L67} (\bibinfo {year} {2010})}\BibitemShut {NoStop}%
\bibitem [{\citenamefont {Jain}\ and\ \citenamefont
  {Krishna}(1998)}]{JainKrishna1998}%
  \BibitemOpen
  \bibfield  {author} {\bibinfo {author} {\bibfnamefont {S.}~\bibnamefont
  {Jain}}\ and\ \bibinfo {author} {\bibfnamefont {S.}~\bibnamefont {Krishna}},\
  }\bibfield  {title} {\bibinfo {title} {{Autocatalytic Sets and the Growth of
  Complexity in an Evolutionary Model}},\ }\href
  {https://doi.org/10.1103/PhysRevLett.81.5684} {\bibfield  {journal} {\bibinfo
   {journal} {PRL}\ }\textbf {\bibinfo {volume} {81}},\ \bibinfo {pages} {5684}
  (\bibinfo {year} {1998})}\BibitemShut {NoStop}%
\end{thebibliography}

%

\end{document}